\documentclass[aps,prb,onecolumn,nofootinbib,citeautoscript,10pt]{revtex4-2}

\usepackage{amsmath,amssymb}
\usepackage{comment}
\usepackage[tight]{subfigure}
\usepackage[dvipsnames]{xcolor}
\usepackage[papersize={8.5in,11in}]{geometry}
\usepackage[colorlinks=true]{hyperref}
\hypersetup{
    bookmarks=true,
    unicode=false,
    pdftoolbar=true,
    pdfmenubar=true,
    pdffitwindow=false,
    pdfstartview={FitH},
    pdfkeywords={plasmons} {triple-point semimetal} {random phase approximation},
    pdfnewwindow=true,
    colorlinks=true,
    linkcolor=magenta,
    citecolor=blue,
    filecolor=magenta,
    urlcolor=blue
}
\usepackage{dcolumn}
\usepackage{color}
\usepackage{amssymb,amsmath}
\usepackage{tabularx,graphicx,makecell}
\usepackage{epstopdf}
\usepackage{latexsym}
\usepackage{colortbl}
\usepackage{psfrag}
\usepackage{bbm,bm,array,physics}
\usepackage{dsfont}
\usepackage{float,mathrsfs}
\newcommand{\bs}[1]{\boldsymbol{#1}}
\def\nn{\nonumber \\}
\def\*#1{\boldsymbol{#1}}
\newcommand{\at}[2][]{#1|_{#2}}

\begin{document}

\title{Plasmon modes in three-dimensional multifold semimetals. I. Triple-point semimetal, its quadratic cousin, and Hopf semimetal}

\author{Ipsita Mandal}
\email{ipsita.mandal@snu.edu.in}
\affiliation{Department of Physics, Shiv Nadar Institution of Eminence (SNIoE), Gautam Buddha Nagar, Uttar Pradesh 201314, India}

\begin{abstract}
We study collective charge excitations in pseudospin-1 triple-point semimetals within the random-phase approximation, considering both the linearised model with a flat-band and its quadratic-corrected counterpart. We analyse their density response, Drude weights, screening, plasmon dispersion and damping, particle-hole continua, and Friedel oscillations from the long-wavelength to the arbitrary-momentum regime. The quadratic correction renders the flat-band dispersive and endows it with a Fermi surface, thereby affecting the overall response. Comparisons with Weyl and Hopf semimetals reveal how bandstructure and quantum geometry jointly govern collective charge dynamics. While the Drude weights are determined by the Fermi surface and band-velocities, interband screening, plasmon dispersion, and Friedel oscillations reflect the quantum geometry of the eigenstates.
\end{abstract}

\maketitle
\tableofcontents

\section{Introduction}
\label{secint}

Three-dimensional (3d) semimetals have been studied intensively on both the theoretical and the experimental fronts. They host symmetry-protected band-crossing points in the Brillouin zone (BZ) near the Fermi level~\cite{burkov11_Weyl, armitage_review, yan17_topological, bernevig, ips-biref, grushin-multifold, ips-hermann-review, ips-kush-review, ips-internode}. The density-of-states (DOS) vanishes exactly at a nodal point, which distinguishes semimetals from insulators, whose bands are always gapped, and from conventional metals, in which the bands overlap over finite regions of the BZ and the DOS remains finite. The simplest nodal points are the twofold band crossings of the Weyl semimetals (WSMs). Each Weyl node is a singular point of the Berry-curvature (BC) field and carries a BC-monopole~\cite{polash-review, xiao_review}. The monopole charge is the Chern number, and its sign is the chirality $\chi$ of the node, which distinguishes right-handed ($\chi = 1$) from left-handed ($\chi = -1$) quasiparticles. The chiral charges of all nodes in the BZ add up to zero, as required by the Nielsen-Ninomiya theorem~\cite{nielsen81_no}. The nodes therefore occur in groups of vanishing net chirality, the simplest being a pair of opposite chirality.

Nodes of higher degeneracy are described by a low-energy Hamiltonian $\bs{d} (\bs{k}) \cdot \bs{\mathcal{S}}$, where $\bs{d} (\bs{k}) = \{d_x (\bs{k}), d_y (\bs{k}), d_z (\bs{k})\}$ is a function of the momentum $\bs{k} = \{k_x, k_y, k_z\}$ and $\bs{\mathcal{S}}$ collects the generators of SU(2) in the spin-$\varsigma$ representation. The resulting $(2\,\varsigma + 1)$-fold node is said to carry pseudospin-$\varsigma$~\cite{bernevig, grushin-multifold, tang2017_multiple}, a terminology that distinguishes this quantum number, which arises from crystal symmetries, from the relativistic spin, which arises from Lorentz invariance. Multifold crossing points have been identified in the 65 chiral space groups, which contain only orientation-preserving symmetries~\cite{grushin-multifold}. For dominant terms linear in momentum, $\bs{d} (\bs{k}) \propto \bs{k}$, and the WSM corresponds to $\varsigma = 1/2$. Quasiparticles with an integer $\varsigma$ have an odd number of bands crossing at the node and no analogues in high-energy physics, since the spin-statistics theorem prohibits relativistic fermions of integer spin. The simplest integer case, $\varsigma = 1$, defines the pseudospin-1 triple-point semimetals (TSMs), which host threefold-degenerate nodes with Chern number $\pm 2$~\cite{bernevig, ips3by2, ady-spin1, krish-spin1, ips-cd1, prb108035428, ips-abs-spin1, ips-spin1-ph, ips-internode} and are sometimes called Maxwell fermions~\cite{cold-atom}. The minimal low-energy model of a TSM is linear in momentum, and its spectrum consists of a Dirac cone, formed by two dispersive bands of opposite energy sign, and a flat-band at the nodal energy. The flat-band is an artefact of the linear approximation. A quadratic-in-momentum correction, which the symmetries of the little group at the node allow, renders it dispersive and is necessary for a realistic description~\cite{ni2021_giant, ips-exact-spin1}.

A different threefold node is realised in the Hopf semimetal (HSM), the gapless descendant of the 3d Hopf insulator~\cite{graf-hopf, hopf2, hopf3, hopf-ins-surface-states}. Its topology is encoded in the Hopf invariant, which is the linking number of the preimages of the Bloch-state map~\cite{graf-hopf, ips-fa, ips-fa-lett}. The linearised HSM has the same spectrum as the TSM, namely a Dirac cone and a flat-band. The Chern number, however, vanishes on every surface enclosing the node, and the BC around the node is an ideal BC-dipole in place of a BC-monopole~\cite{graf-hopf, ips-dipole-vnr, ips-firdous-vnr}. The TSM and the HSM therefore share the dispersion and differ in their eigenstates and BC.

Plasmons, the collective oscillations of the charge density, probe these band structures. In the doped regime, the long-wavelength plasmon frequency is controlled by the Drude weight, which is fixed by the Fermi surfaces (FSs) and band-velocities, whereas the damping is governed by the interband continuum, as in graphene~\cite{hwang-dassarma-graphene}. Collective modes have been studied for multi-Weyl~\cite{ahn_mWeyl} and Luttinger semimetals~\cite{ips-plasmon-luttinger-epjb, ips-plasmon-luttinger-aop}, and for nodal-ring semimetals (NRSs), whose band crossings form closed curves in the BZ~\cite{balents-nodal, fu_nlsm, yang_review_nlsm}, in two~\cite{rahmipoor-2d-nlsm} and three dimensions~\cite{rhim-kim-friedel, ips-nr-plasmon, ips-vnr-plasmon}. For the WSM, Ref.~\cite{uchoa2026} obtains the arbitrary-wavevector polarisability within the random-phase approximation (RPA), the particle-hole continuum, and the plasmon dispersion of the untilted node, and it examines the topology of the plasmon envelope function and the optical selection rules that follow from it. It does not address the static response, the Kohn anomaly, the Friedel oscillations, or the Landau damping. Ref.~\cite{ips-nr-plasmon} derives the long-wavelength plasmons of the untilted and type-I tilted nodes, which are isotropic without tilt and whose interband screening depends logarithmically on the momentum cut-off of the linearised description. For the HSM, the plasmons are anisotropic although the band structure is isotropic, and the anisotropy was attributed to the BC-dipole~\cite{hopf-plasmon}. The full dielectric function of the HSM, including the interband screening, was subsequently derived in Ref.~\cite{ips-vnr-plasmon} for low doping and small wavevectors. The flat-band and the Chern number $\pm 2$ distinguish the TSM from both the WSM and the HSM and motivate an analysis of its plasmon response.

\begin{figure}[t!]
\centering
\includegraphics[width= 0.75 \linewidth]{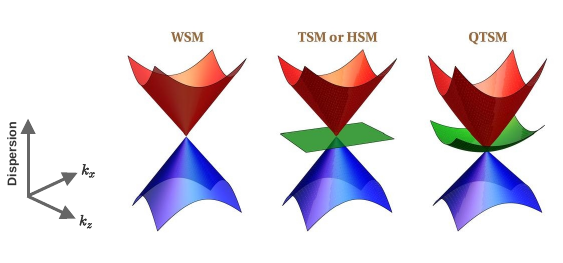}
\caption{\label{fig:combo-dis} Schematics of the band dispersions in the $k_x$-$k_z$ plane for the WSM, TSM/HSM, and QTSM. The blue and red curves represent the two dispersive bands, which are the same in all three systems. The green curve represents the flat-band in the TSM/HSM, which becomes a quadratically dispersing band in the QTSM.}
\end{figure}

In this paper, we study the collective charge excitations of a TSM within the RPA. We compute the density response, the Drude weights, the dielectric function, and the plasmon dispersion, first for the linear model, in which the flat-band contributes only to the interband response, and then for the quadratic-corrected model (QTSM), in which the flat-band also contributes to the intraband response once the chemical potential cuts it. The quadratic correction leaves the eigenstates unchanged and modifies only the FS and the band-velocity. Fig.~\ref{fig:combo-dis} compares the band structures of the three systems in the $k_x k_z$-plane. For both models, we go beyond the long-wavelength limit, which gives access to the particle-hole continua, the plasmon damping and its critical wavevector, the static screening, and the Friedel oscillations. We also treat the untilted WSM and the HSM at finite wavevectors. For the WSM, we recompute the polarisability of Ref.~\cite{uchoa2026} to derive the static response, the screening, the Kohn anomaly, the Friedel oscillations, and the damping rate. For the HSM, we extend the long-wavelength results of Ref.~\cite{ips-vnr-plasmon}. We then compare the TSM and the QTSM with both systems. Since the TSM and the HSM share the spectrum and differ in their eigenstates, this comparison separates the roles of the spectrum and the eigenstate structure in the plasmon response.

The notions of low doping and long wavelength differ between nodal rings and point nodes. In Refs.~\cite{ips-nr-plasmon} and~\cite{ips-vnr-plasmon}, the nodal manifold is a ring of radius $k_0$, which is a lattice scale. The linearisation about the ring requires low doping, for which the FS is a thin torus of minor radius much smaller than $k_0$. The mapping of the response onto decoupled two-dimensional Dirac cones, one for each point of the ring, requires in addition a momentum transfer much smaller than $k_0$. These restrictions exclude the Friedel oscillations, which require a momentum transfer comparable to $k_0$. The untilted WSM and the HSM were analysed in the same limits to allow a comparison with the nodal rings on an equal footing. A point node has no analogue of $k_0$. Its linearised Hamiltonian is limited only by the cut-off $\Lambda$ beyond which the linear dispersion fails, and low doping means $k_F \ll \Lambda$. The cut-off then enters the response only through logarithms of the ratio of $\Lambda$ to the chemical potential.

We distinguish two regimes of the momentum transfer $\bs{q}$, whose magnitude $q$ is measured relative to $k_F$. For $q \ll k_F$, the plasmon frequency is set by the Drude weight of the bands that cross the Fermi level, and the interband transitions reduce to a background dielectric constant. The overlap factors can be expanded in $q/k_F$, which brings the quantum metric of the relevant bands to the foreground. For arbitrary $q$ up to the cut-off, the overlap factors must be retained exactly, and the response is assembled channel by channel. In the linear models, the doped part of the polarisability then depends on $q$ and $\omega$ only through $q/k_F$ and $\omega/\mu$, up to an overall factor, and no additional condition on the doping is required. This regime gives access to the Kohn anomaly at $q = 2\,k_F$, the particle-hole continua, and the Landau damping of the plasmon, none of which the small-$q$ expansion can capture.

The paper is organised as follows. Sec.~\ref{secmodel} defines the overlap factors, the quantum geometry, and the density-response formalism for generic Bloch bands, and it presents the Hamiltonians of the TSM, the QTSM, the HSM, and the untilted WSM together with their BC, orbital magnetic moment, eigenstates, and overlap factors. The plasmon modes of the linear model are derived in Sec.~\ref{seclinear} and extended to finite wavevectors in Sec.~\ref{secfinite}. The quadratic correction is addressed at long wavelengths in Sec.~\ref{secquad} and at generic wavevectors in Sec.~\ref{sec-qtsm-full}. The HSM and the untilted WSM are treated at generic wavevectors in Secs.~\ref{sechfull} and~\ref{secwsmhigh}, respectively. Sec.~\ref{seccompare} compares the TSM and the QTSM with the HSM and the WSM and collects the results in tables, and Sec.~\ref{secsum} contains the summary and outlook. The appendices elucidate the intermediate steps. 

We use natural units and set $\hbar = k_B = e = 1$ and the speed of light to unity, but retain the symbol $e$ in the analytical expressions for book-keeping. The low-energy models are valid for momenta below the lattice cutoff $\Lambda \sim 1/a$ (where $a$ is the lattice constant) and for energies below $v_0\,\Lambda$ (where $v_0$ is Fermi velocity). For typical WSMs and TSMs, $\Lambda \sim 150$ to $200\,\text{eV}$~\cite{lv_weyl,spin13d1}. We set $v_0 = 0.0004$ and $\Lambda = 160\,\text{eV}$ for generating all the figures. These values are not repeated in the figure captions.

\section{Model}
\label{secmodel}

In this section, we first define the overlap factors, the quantum geometry, and the density-response formalism for a generic set of Bloch bands in Sec.~\ref{sec:geometry}, since these definitions are common to all the nodes. We then introduce the Hamiltonians of the TSM, the HSM, and the WSM, together with their BC, OMM, explicit eigenstates, and overlap factors, in Secs.~\ref{sec:ham},~\ref{sec:hsm}, and~\ref{sec:wsm}, respectively.

\subsection{Overlap factors, quantum geometry, and density response of Bloch bands}
\label{sec:geometry}

Consider a Bloch Hamiltonian $\mathcal{H} (\bs{k})$ with the eigenvalues $\varepsilon_s (\bs{k})$ and the normalised eigenstates $|s, \bs{k} \rangle$, where $s$ labels the bands. The Hamiltonians of this paper are low-energy continuum models, and $|s, \bs{k} \rangle$ denotes the pseudospin part of the Bloch state, with the plane-wave factor understood. The density response depends on the eigenstates only through the overlap factor
\begin{align}
\label{eq:overlap-def}
\mathcal{F}_{ss'} (\bs{k}, \bs{k}') = |\langle s', \bs{k}' | s, \bs{k} \rangle|^2,
\end{align}
which is the squared modulus of the inner product of two states and measures how much they resemble each other. The phase of each state drops out of Eq.~\eqref{eq:overlap-def}, and the overlap factor is therefore independent of the gauge. By completeness, $\sum_{s'} \mathcal{F}_{ss'} (\bs{k}, \bs{k}') = 1$ for each $s$. For $\bs{k}' = \bs{k}$, the overlap factor reduces to $\delta_{ss'}$, because the states of different bands are orthogonal at the same momentum.

For $\bs{k}' = \bs{k} + \bs{q}$ with $q \ll k$, the overlap factors can be expanded in $\bs{q}$. For $s' \neq s$, the orthogonality of the states at the same momentum gives $\langle s', \bs{k} + \bs{q} | s, \bs{k} \rangle \simeq -\,q_i\,\langle s', \bs{k} | \partial_{k_i} s, \bs{k} \rangle$, and hence
\begin{align}
\label{eq:ff-expansion}
\mathcal{F}_{ss'} (\bs{k}, \bs{k} + \bs{q}) \simeq q_i\,q_j\,Q^{(ss')}_{ij} (\bs{k})\,, \quad
Q^{(ss')}_{ij} (\bs{k}) = \mathrm{Re} \left[ \langle \partial_{k_i} s, \bs{k} | s', \bs{k} \rangle \langle s', \bs{k} | \partial_{k_j} s, \bs{k} \rangle \right], \quad s' \neq s\,,
\end{align}
with a sum over repeated indices. The intraband factor follows from the sum rule as $\mathcal{F}_{ss} \simeq 1 - q_i\,q_j\,Q^{(s)}_{ij}$, where
\begin{align}
\label{eq:qmetric}
Q^{(s)}_{ij} (\bs{k}) = \sum_{s' \neq s} Q^{(ss')}_{ij} (\bs{k}) = \mathrm{Re}\left[ \langle \partial_{k_i} s, \bs{k} | \left( 1 - |s, \bs{k}\rangle \langle s, \bs{k}| \right) | \partial_{k_j} s, \bs{k} \rangle \right].
\end{align}
The second equality follows from completeness. The tensor $Q^{(s)}_{ij}$ is the quantum metric~\cite{provost-vallee, berry} of the band $s$. It is real, symmetric, and positive semidefinite, and it measures the distance between the states at neighbouring momenta. The pair-resolved metric $Q^{(ss')}_{ij}$ is the part of $Q^{(s)}_{ij}$ that connects the band $s$ to the band $s'$, and it fixes the weight of the interband transitions $s \to s'$ at order $q^2$.

The metric is the real part of the quantum geometric tensor,
\begin{align}
\mathcal{T}^{(s)}_{ij} (\bs{k}) = \langle \partial_{k_i} s, \bs{k} | \left( 1 - |s, \bs{k}\rangle \langle s, \bs{k}| \right) | \partial_{k_j} s, \bs{k} \rangle = Q^{(s)}_{ij} (\bs{k}) - \frac{i}{2}\,\Omega^{(s)}_{ij} (\bs{k})\,,
\end{align}
whose imaginary part is the Berry curvature (BC) $\Omega^{(s)}_{ij} = \partial_{k_i} A^{(s)}_j - \partial_{k_j} A^{(s)}_i$, with the Berry connection $\bs{A}^{(s)} = i \langle s, \bs{k} | \bs{\nabla}_{\bs{k}} | s, \bs{k} \rangle$. In three dimensions, the BC is conveniently written as the vector $\bs{\Omega}_s = \bs{\nabla}_{\bs{k}} \times \bs{A}^{(s)}$, for which $\Omega^{(s)}_{ij} = \varepsilon_{ijl}\,\Omega_{s,l}$. The metric measures the distance between neighbouring states and enters the overlap factors, and hence the density response. The BC measures the phase that a state acquires around a closed loop in momentum space, and it does not enter the overlap factors, which are insensitive to the phases of the states. Consequently, neither the BC nor the orbital magnetic moment (OMM) appears in the calculations of this paper. We nevertheless quote them for each node below, because they distinguish the topology of the nodes. The TSM and the WSM carry a BC-monopole and the HSM a BC-dipole, and we use this contrast in Sec.~\ref{seccompare}.

We now collect the density-response formalism, which is common to all the nodes. At temperature $T = 0$, the noninteracting density-density response function at chemical potential $\mu$ is given by the retarded bare bubble diagram,
\begin{align}
\label{eq:pol}
\Pi(\bs{q},\omega) = g \int \frac{d^3\bs{k}}{(2\,\pi)^3} \sum_{s,s'} \mathcal{F}_{ss'}(\bs{k},\bs{k}')
\frac{f_0(\varepsilon_s(\bs{k})) - f_0(\varepsilon_{s'}(\bs{k}'))}{\omega + \varepsilon_s(\bs{k}) - \varepsilon_{s'}(\bs{k}') + i\,0^+}\,,
\quad \bs{k}' = \bs{k} + \bs{q}\,,
\end{align}
where $g$ is the degeneracy factor, $f_0$ is the Fermi-Dirac distribution, and $\mathcal{F}_{ss'}$ is the overlap factor. The polarisability splits as $\Pi = \Pi_{\rm intra} + \Pi_{\rm inter}$, where the first term collects the contributions with $s = s'$.
The imaginary part of the retarded polarisability,
\begin{align}
\Im \Pi (\bs{q}, \omega) \propto \sum_{s,s'} \int d^3 \bs{k}\;\mathcal{F}_{ss'}\left[f_0 (\varepsilon_s) - f_0 (\varepsilon_{s'})\right]\delta\!\left(\omega + \varepsilon_s (\bs{k}) - \varepsilon_{s'} (\bs{k}')\right),
\end{align}
is nonzero precisely when a particle--hole pair with momentum transfer $\bs{q}$ and energy $\omega$ can be created. The set of $(\bs{q}, \omega)$ for which such a pair exists is the particle--hole continuum. Outside this region, $\Im \Pi$ vanishes and a collective mode can propagate without damping. The edges of the continuum therefore mark the points at which the plasmon ceases to be undamped.
The real part of the polarisability determines the frequency of the collective mode. Within the RPA, the dielectric function is
\begin{align}
\label{eq:epsrpa}
\epsilon^{\rm RPA}(\bs{q},\omega) = 1 - \mathcal{V} (q)\,\Pi(\bs{q},\omega)\,, \quad \mathcal{V} (q) = \frac{4\,\pi\,e^2}{q^2}\,,
\end{align}
where $\mathcal{V} (q)$ is the Fourier transform of the three-dimensional Coulomb interaction. For the nodes considered below, we use the shorthand $\alpha \equiv g\,e^2/v_0$, where $v_0$ is the Fermi velocity. With $g = e = 1$ in natural units, $\alpha$ is fixed by $v_0$, which is held fixed in all figures, and it is not an independent parameter. An undamped collective mode is a zero of $\Re \epsilon^{\rm RPA}$ outside the particle--hole continuum, and the imaginary part of $\Pi$ vanishes there. Inside the continuum, $\Im \Pi$ is nonzero and the mode acquires a finite lifetime.

The Drude-weight tensor is
\begin{align}
\label{eq:drude}
D_{ij} = \pi\,g\,e^2 \sum_s \int \frac{d^3\bs{k}}{(2\,\pi)^3}\, \delta(\varepsilon_s - \mu)\, v_i^{(s)}\, v_j^{(s)}\,,
\end{align}
which fixes the Drude peak of the optical conductivity, $\Re\tilde\sigma_{ij}(\omega) = D_{ij}\,\delta(\omega)$. Taking the $q \to 0$ limit and then $\omega \to 0$, the intraband part gives
\begin{align}
\label{eq:intra}
\Pi_{\rm intra}(\bs{q},\omega) \simeq \frac{q_i\,q_j\,D_{ij}}{\pi\,e^2\,\omega^2}\,.
\end{align}
At long wavelengths, $\Re\Pi$ is dominated by the intraband part. Inserting Eq.~\eqref{eq:intra} into $\Re\epsilon^{\rm RPA} = 0$ gives
\begin{align}
\label{eq:plasmon}
\omega_{p,i}^2 = 4\,D_{ii}\,,
\end{align}
with no sum over $i$. The interband part is real and negative below the interband threshold. It acts as a background dielectric constant $\epsilon_{b,i}$ and lowers the bare frequencies to $\tilde\omega_{p,i}^2 = \omega_{p,i}^2 / \epsilon_{b,i}$.

\subsection{Triple-point Semimetal}
\label{sec:ham}

The effective low-energy Hamiltonian in the vicinity of a TSM node is~\cite{ips-exact-spin1}
\begin{align}
\label{eq:ham}
\mathcal{H}_\chi(\bs{k}) = \bs{d}(\bs{k}) \cdot \bs{S} + c\,v_0\,k^2\,\mathbb{I}_{3\times3}\,,
\quad
\bs{d}(\bs{k}) = v_0 \left\lbrace k_x, k_y, \chi\,k_z \right\rbrace,
\quad
k = \sqrt{k_x^2 + k_y^2 + k_z^2}\,,
\end{align}
where $\bs{S} = \{S_x, S_y, S_z\}$ are the spin-1 matrices, $\chi = \pm1$ is the chirality of the node, $v_0$ is the isotropic Fermi velocity, and $c$ parametrises the quadratic correction. 
The eigenvalue (see also Fig.~\ref{fig:combo-dis}) and group-velocity of the band $s$ are
\begin{align}
\label{eq:ev}
\varepsilon_s(\bs{k}) = s\,v_0\,k + c\,v_0\,k^2\,,
\bs{v}_s(\bs{k}) = \nabla_{\bs{k}} \varepsilon_s(\bs{k}) = 
\frac{v_0 \left( s + 2\,c\,k \right)}{k}\,\bs{k}\,,
\quad s \in \lbrace -1, 0, 1 \rbrace \,.
\end{align}
For $c=0$, the model reduces to the linear one with a flat-band. When we set $c$ to a nonzero value, we dub the system as a quadratic-corrected TSM, abbreviated as QTSM. Because the quadratic term is proportional to the identity in band space, it commutes with the linear Hamiltonian. The eigenvectors are therefore the same as those of the linear model.

For comparison with the HSM, we also quote the BC and the OMM of the band $s$~\cite{ips-exact-spin1}, which are
\begin{align}
\label{eq:bc}
\bs{\Omega}_s(\bs{k}) = -\,\frac{s}{k^3}\,\bs{k}\,,
\quad
\bs{m}_s (\bs{k}) = -\,\frac{e\,v_0\,\check{\mathcal{G}}_s}{2\,k^2}\,\bs{k}\,,
\quad
\check{\mathcal{G}}_s = \begin{cases} 1 & \text{ for } s = \pm1 \\ 2 & \text{ for } s = 0 \end{cases}.
\end{align}
The $s=0$ band has zero BC but a nonzero OMM. The total Chern number of the node is $2$, carried by the $s=\pm1$ bands.

The eigenstates of $\mathcal{H}_\chi$ are the eigenstates of the spin-$1$ operator $\bs{\tilde k} \cdot \bs{S}$ with $\bs{\tilde k} = \{k_x, k_y, \chi\,k_z\}$, which projects the spin along $\bs{\tilde k}$. Let $\vartheta$ and $\varphi$ be the polar and azimuthal angles of $\bs{\tilde k}$. In the basis of the eigenstates of $S_z$ with the eigenvalues $1$, $0$, and $-1$, the eigenstates with $\bs{\tilde k} \cdot \bs{S}\,|s, \bs{k} \rangle = s\,k\,|s, \bs{k} \rangle$ are the spin-$1$ coherent states
\begin{align}
\label{eq:t-states}
|1, \bs{k} \rangle = \begin{pmatrix} \cos^2 (\vartheta/2)\,e^{-\,i\,\varphi} \\ \sin\vartheta/\sqrt{2} \\ \sin^2 (\vartheta/2)\,e^{i\,\varphi} \end{pmatrix}, \quad
|0, \bs{k} \rangle = \begin{pmatrix} -\,\sin\vartheta\,e^{-\,i\,\varphi}/\sqrt{2} \\ \cos\vartheta \\ \sin\vartheta\,e^{i\,\varphi}/\sqrt{2} \end{pmatrix}, \quad
|-1, \bs{k} \rangle = \begin{pmatrix} \sin^2 (\vartheta/2)\,e^{-\,i\,\varphi} \\ -\,\sin\vartheta/\sqrt{2} \\ \cos^2 (\vartheta/2)\,e^{i\,\varphi} \end{pmatrix}.
\end{align}
The overlap factors of Eq.~\eqref{eq:overlap-def} are the squared Wigner matrix elements $|d^{\,1}_{s s'} (\theta)|^2$, where $\cos\theta = \hat{\bs{k}} \cdot \hat{\bs{k}}'$. Explicitly, they read
\begin{align}
\label{eq:lin-ff}
& \mathcal{F}_{1 1} = \mathcal{F}_{-1,-1} = \left( \frac{1 + \cos\theta}{2} \right)^2, \quad
\mathcal{F}_{0 0} = \cos^2\theta\,, \quad
\mathcal{F}_{1 0} = \mathcal{F}_{0 1} = \mathcal{F}_{0,-1} 
= \mathcal{F}_{-1,0} = \frac{\sin^2\theta}{2}\,, \quad
\mathcal{F}_{1,-1} = \mathcal{F}_{-1,1} = \left( \frac{1 - \cos\theta}{2} \right)^2.
\end{align}
For $\bs{k}' = \bs{k} + \bs{q}$ with $q \ll k$, we have $\sin\theta \simeq |\bs{q} \times \hat{\bs{k}}|/k$, and the overlap factors become
\begin{align}
\label{eq:lin-qm}
\mathcal{F}_{1 0} \simeq q_i\,q_j\,Q_{ij} (\bs{k})\,, \quad
Q_{ij} (\bs{k}) = \frac{\delta_{ij} - \hat k_i\,\hat k_j}{2\,k^2}\,, \quad
\mathcal{F}_{1,-1} = \mathcal{O} \left( q^4 \right), \quad
\mathcal{F}_{ss} = 1 - \mathcal{O} \left( q^2 \right).
\end{align}
In the notation of Eq.~\eqref{eq:ff-expansion}, we have $Q_{ij} = Q^{(10)}_{ij} = Q^{(-1,0)}_{ij}$, and Eq.~\eqref{eq:qmetric} then gives the quantum metric $Q_{ij}$ for each of the dispersive bands $s = \pm 1$. The flat-band carries the metric $Q^{(0)}_{ij} = 2\,Q_{ij}$, because it connects to both dispersive bands with the same weight. The transitions between the two dispersive bands carry no weight at order $q^2$. The derivation of Eqs.~\eqref{eq:lin-ff}--\eqref{eq:lin-qm} is provided in Appendix~\ref{app:tsm}.

The overlap factors, the quantum metric, and the angular average of Appendix~\ref{app:tsm} are therefore independent of $c$ and hold for the QTSM as well. The parameter $c$ enters the response only through the energies and the occupations.

\subsection{Three-band Hopf semimetal}
\label{sec:hsm}

For comparison, we also introduce the effective low-energy Hamiltonian of a threefold Hopf node, which is the gapless descendant of a 3d Hopf insulator~\cite{graf-hopf, hopf2, hopf3, ips-vnr-plasmon}. It is given by
\begin{align}
\label{eq:hhopf}
\mathcal{H}_{\rm H} (\bs{k}) = v_0 \left( k_x\,\mathcal{M}_x + k_y\,\mathcal{M}_y + k_z\,\mathcal{M}_z \right),
\end{align}
where $\mathcal{M}_x$, $\mathcal{M}_y$, and $\mathcal{M}_z$ are the Gell-Mann matrices given by
\begin{align}
\mathcal{M}_x = \begin{pmatrix} 0 & 1 & 0 \\ 1 & 0 & 0 \\ 0 & 0 & 0 \end{pmatrix}, \quad
\mathcal{M}_y = \begin{pmatrix} 0 & -\,i & 0 \\ i & 0 & 0 \\ 0 & 0 & 0 \end{pmatrix}, \quad
\mathcal{M}_z = \begin{pmatrix} 0 & 0 & -\,i \\ 0 & 0 & 0 \\ i & 0 & 0 \end{pmatrix}.
\end{align}
The eigenvalues are
\begin{align}
\label{eq:evhopf}
\varepsilon_s (\bs{k}) =
\begin{cases}
(-1)^s\,v_0\,k & \text{ for } s \in \lbrace 1, 2 \rbrace \\
0 & \text{ for } s = 0
\end{cases},
\end{align}
as illustrated in Fig.~\ref{fig:combo-dis}. The bands $s = 1$ and $s = 2$ form isotropic Dirac cones, and $s = 0$ is a flat-band in the linear model. Its existence follows from the chiral symmetry $\Gamma = \operatorname{diag} (1, -1, -1)$, which anticommutes with each of the three matrices. 

In contrast with the TSM, the HSM Hamiltonian is not rotationally-symmetric in 3d, which is reflected in the orientation of dipole-axis of the HSM's BC. The BC and the OMM of the three bands, which are quoted for comparison with the TSM, are~\cite{graf-hopf, ips-dipole-vnr, ips-vnr-plasmon}
\begin{align}
\label{eq:bcommhopf}
\bs{\Omega}_s (\bs{k}) =
\begin{cases}
\dfrac{k_z\,\bs{k}}{k^4} & \text{ for } s \in \lbrace 1, 2 \rbrace \\[6pt]
-\,\dfrac{2\,k_z\,\bs{k}}{k^4} & \text{ for } s = 0
\end{cases}
\text{ and } \quad
\bs{m}_s (\bs{k}) =
\begin{cases}
\dfrac{(-1)^{s+1}\,e\,v_0\,k_z\,\bs{k}}{2\,k^3} & \text{ for } s \in \lbrace 1, 2 \rbrace \\[6pt]
0 & \text{ for } s = 0
\end{cases}.
\end{align}
The two dispersive bands carry the same BC. The flat-band carries the opposite BC with twice the magnitude, and the total BC therefore vanishes at each $\bs{k}$. The flux through any sphere around the node is zero, and the Chern number is zero on every enclosing surface. The BC is an ideal BC-dipole rather than a BC-monopole.

We denote the upper and lower dispersive bands of Eq.~\eqref{eq:evhopf} by $+$ ($s = 2$) and $-$ ($s = 1$), respectively, and the flat-band by $0$. The eigenstates are
\begin{align}
\label{eq:h-states}
|\pm, \bs{k} \rangle = \frac{1}{\sqrt{2}} \begin{pmatrix} 1 \\ \pm\,k_+/k \\ \pm\,i\,k_z/k \end{pmatrix}, \quad
|0, \bs{k} \rangle = \frac{1}{k} \begin{pmatrix} 0 \\ i\,k_z \\ k_- \end{pmatrix},
\end{align}
where $k_\pm = k_x \pm i\,k_y$. For $\bs{k}' = \bs{k} + \bs{q}$, we define the complex number
\begin{align}
\label{eq:h-z}
\mathcal{Z} (\bs{k}, \bs{k}') = \frac{k'_-\,k_+ + k_z\,k'_z}{k\,k'}\,, \quad
\Re \mathcal{Z} = \cos\theta = \frac{\bs{k} \cdot \bs{k}'}{k\,k'}\,, \quad
\Im \mathcal{Z} = -\,\Upsilon\,, \quad
\Upsilon = \frac{(\bs{k} \times \bs{q})_z}{k\,k'}\,.
\end{align}
The overlap factors of Eq.~\eqref{eq:overlap-def} are then
\begin{align}
\label{eq:h-ff}
& \mathcal{F}_{\pm\pm} = \frac{|1 + \mathcal{Z}|^2}{4} = \frac{(1 + \cos\theta)^2 + \Upsilon^2}{4}\,, \quad
\mathcal{F}_{+-} = \mathcal{F}_{-+} = \frac{|1 - \mathcal{Z}|^2}{4} = \frac{(1 - \cos\theta)^2 + \Upsilon^2}{4}\,, \nn
& \mathcal{F}_{\pm 0} = \mathcal{F}_{0 \pm} = \frac{1 - |\mathcal{Z}|^2}{2} = \frac{\sin^2\theta - \Upsilon^2}{2}\,, \quad
\mathcal{F}_{00} = |\mathcal{Z}|^2 = \cos^2\theta + \Upsilon^2\,.
\end{align}
They differ from the spin-$1$ factors of Eq.~\eqref{eq:lin-ff} by the terms in $\Upsilon^2$, which vanish when $\bs{q} \parallel \bs{\hat z}$.

The kernel of the polarisability depends on $\bs{k}$ and $\bs{k}'$ only through $k$ and $k'$, apart from the overlap factors. We may therefore average $\Upsilon^2$ over the azimuthal angle of $\bs{k}$ about $\bs{q}$ at fixed $k$ and $k'$. The average is
\begin{align}
\label{eq:h-avg}
\langle \Upsilon^2 \rangle = \lambda_{\bs{q}}\,\sin^2\theta\,, \quad
\lambda_{\bs{q}} = \frac{\sin^2\vartheta_{\bs{q}}}{2}\,,
\end{align}
where $\vartheta_{\bs{q}}$ is the polar angle of $\bs{q}$ and $0 \leq \lambda_{\bs{q}} \leq 1/2$. With $ \cos\theta = (k^2 + k'^2 - q^2)/(2\,k\,k')$, the averaged factors become
\begin{align}
\label{eq:h-fl}
& \mathcal{F}_{\pm\pm} = \frac{(1 + \cos \theta)^2 + \lambda_{\bs{q}}\,\sin^2 \theta}{4}\,, \quad
\mathcal{F}_{+-} = \frac{(1 - \cos\theta)^2 + \lambda_{\bs{q}}\,\sin^2 \theta}{4}\,, \quad
 \mathcal{F}_{\pm 0} = \frac{(1 - \lambda_{\bs{q}}) \sin^2 \theta}{2}\,, \quad
\mathcal{F}_{00} = \cos^2 \theta + \lambda_{\bs{q}} \sin^2 \theta\,.
\end{align}
The direction of $\bs{q}$ thus enters only through $\lambda_{\bs{q}}$. For $\bs{q} \parallel \bs{\hat z}$, we have $\lambda_{\bs{q}} = 0$, and the factors and the spectrum coincide with those of the TSM. For $\bs{q}$ in the $xy$-plane, we have $\lambda_{\bs{q}} = 1/2$. For $q \ll k$, we have $\mathcal{F}_{+-} \simeq \Upsilon^2/4$ with $\Upsilon \simeq \bs{q} \cdot (\bs{\hat z} \times \hat{\bs{k}})/k$, and Eq.~\eqref{eq:ff-expansion} gives the pair-resolved metric $Q^{(+-)}_{ij} = (\bs{\hat z} \times \hat{\bs{k}})_i\,(\bs{\hat z} \times \hat{\bs{k}})_j/(4\,k^2)$. The averaged factors of Eq.~\eqref{eq:h-fl} reproduce the long-wavelength coefficients of Ref.~\cite{ips-vnr-plasmon}. In particular, the weight of the transitions $0 \to +$ at order $q^2$ is $(1 - \lambda_{\bs{q}})$ times its value for the TSM, and the transitions $- \to +$ acquire a weight proportional to $\lambda_{\bs{q}}$. In contrast to the TSM, where the transitions between the two dispersive bands carry no weight at order $q^2$, the pair $- \to +$ of the HSM has a nonvanishing metric and therefore contributes to the interband response.

\subsection{Weyl semimetal}
\label{sec:wsm}

For comparison, we also consider the low-energy Hamiltonian of a single untilted Weyl node~\cite{ips-nr-plasmon}. It is $\mathcal{H}  (\bs{k}) = v_0\,\bs{k} \cdot \bs{\sigma}$, with the eigenvalues $\varepsilon_s (\bs{k}) = s\,v_0\,k$ for $s = \pm 1$ (cf. Fig.~\ref{fig:combo-dis}). For comparison with the other two nodes, we also quote the BC and the OMM of the band $s$, which are
\begin{align}
\label{eq:w-bc}
\bs{\Omega}_s (\bs{k}) = -\,\frac{s}{2\,k^3}\,\bs{k}\,, \quad
\bs{m}_s (\bs{k}) = -\,\frac{e\,v_0}{2\,k^2}\,\bs{k}\,.
\end{align}
The BC is a monopole field with the Chern number $\pm 1$, and the OMM is the same for both bands. Let $\vartheta$ and $\varphi$ be the polar and azimuthal angles of $\bs{k}$. The eigenstates are
\begin{align}
\label{eq:w-states}
|+, \bs{k} \rangle = \begin{pmatrix} \cos (\vartheta/2) \\ \sin (\vartheta/2)\,e^{i\,\varphi} \end{pmatrix}, \quad
|-, \bs{k} \rangle = \begin{pmatrix} -\,\sin (\vartheta/2)\,e^{-\,i\,\varphi} \\ \cos (\vartheta/2) \end{pmatrix},
\end{align}
and the overlap factors of Eq.~\eqref{eq:overlap-def} are given by
\begin{align}
\label{eq:w-ff}
\mathcal{F}_{ss'} (\bs{k}, \bs{k}') = \frac{1 + s\,s'\,\cos\theta}{2}\,, \quad \cos\theta = \hat{\bs{k}} \cdot \hat{\bs{k}}'\,.
\end{align}
For $\bs{k}' = \bs{k} + \bs{q}$ with $q \ll k$, we have $\mathcal{F}_{+-} = (1 - \cos\theta)/2 \simeq \sin^2\theta/4$. Equation~\eqref{eq:qmetric} then gives the quantum metric $Q^{(\pm)}_{ij} = Q^{(+-)}_{ij} = (\delta_{ij} - \hat k_i\,\hat k_j)/(4\,k^2)$ for both bands, which is half the metric of the dispersive bands of the TSM.

\section{TSM at small momentum-transfers}
\label{seclinear}

In this section, we set $c = 0$ in Eq.~\eqref{eq:ham}. The Hamiltonian then becomes $\mathcal{H}_\chi (\bs{k}) = v_0\,\bs{\tilde k} \cdot \bs{S}$ with $\bs{\tilde k} = \{k_x, k_y, \chi\,k_z\}$, and its eigenvalues are $\varepsilon_s (\bs{k}) = s\,v_0\,k$ for $s \in \lbrace -1, 0, 1 \rbrace$, where the band $s = 0$ is flat. All quantities depend on the momenta only through scalar products of $\bs{\tilde k}$ and $\bs{\tilde k}'$ and are therefore independent of $\chi$, which we set to $\chi = 1$. The spectrum is symmetric under $\varepsilon \to -\,\varepsilon$ together with $s \to -\,s$. It is therefore sufficient to treat $\mu \geq 0$, and the results for $\mu < 0$ follow by replacing $\mu$ with $|\mu|$. The Hamiltonian is rotationally symmetric, and the response is isotropic. The derivations are given in Appendix~\ref{app:tsm}.

\subsection{Intrinsic limit}
\label{seclinintr}

In the limit $\mu \to 0^+$, the FS shrinks to the node, the bands $s = -1$ and $s = 0$ are filled, and the band $s = 1$ is empty. The transitions $-1 \to 0$ are therefore blocked, and only the transitions $0 \to 1$ with the energy $v_0\,k$ contribute at order $q^2$ because of Eq.~\eqref{eq:lin-qm}. The two time orderings of these transitions combine into $1/(\omega - v_0\,k) - 1/(\omega + v_0\,k) = 2\,v_0\,k/(\omega^2 - v_0^2\,k^2)$. For $v_0\,q \ll \omega \ll v_0\,\Lambda$, where $\Lambda$ is the cut-off of the linearised description, Eq.~\eqref{eq:pol} then gives
\begin{align}
\label{eq:lin-pi0}
\Pi_{\rm inter} (\bs{q}, \omega)
= \frac{g\,q_i\,q_j}{(2\,\pi)^3} \int d\tilde\Omega\; k^2\,Q_{ij} \int_0^{\Lambda} dk\, \frac{2\,v_0\,k}{(\omega + i\,0^+)^2 - v_0^2\,k^2} 
= \frac{g\,q^2}{6\,\pi^2} 
\int_0^{\Lambda} dk\, \frac{2\,v_0\,k}{(\omega + i\,0^+)^2 - v_0^2\,k^2}\,.
\end{align}
Evaluating the radial integral, we obtain
\begin{align}
\label{eq:lin-pi0b}
\Re \Pi (\bs{q}, \omega) = -\,\frac{g\,q^2}{6\,\pi^2\,v_0}
\ln \!\left( \frac{v_0^2\,\Lambda^2}{\omega^2} \right), \quad
\Im \Pi (\bs{q}, \omega) = -\,\frac{g\,q^2}{6\,\pi\,v_0}\,,
\end{align}
which do not depend on the direction of $\bs{q}$. The dielectric function then follows from Eq.~\eqref{eq:epsrpa} as
\begin{align}
\label{eq:lin-eps0}
\Re \epsilon^{\rm RPA} (\bs{q}, \omega) = 1 + \frac{2\,g\,e^2}{3\,\pi\,v_0}\,\ln \!\left( \frac{v_0^2\,\Lambda^2}{\omega^2} \right), \quad
\Im \epsilon^{\rm RPA} (\bs{q}, \omega) = \frac{2\,g\,e^2}{3\,v_0}\,.
\end{align}
The real part exceeds unity for $\omega < v_0\,\Lambda$, and the imaginary part is nonzero for all $\omega > 0$. The dielectric function therefore has no zero, and the intrinsic system supports no undamped RPA plasmon. This is because the interband onset vanishes, with the transitions $0 \to 1$ being unblocked at every $k > 0$.

\subsection{Doped regime: DOS and Drude weights}
\label{seclindrude}

For $\mu > 0$, only the band $s = 1$ crosses the Fermi level, and the FS is the sphere $k = k_F = \mu/v_0$. The flat-band is filled and does not contribute to the intraband response, because its velocity vanishes. The DOS, the density, and the Drude weights are
\begin{align}
\label{eq:lin-dos}
\rho (\mu) = \frac{g\,\mu^2}{2\,\pi^2\,v_0^3}\,, \quad
\tilde n = \frac{g\,\mu^3}{6\,\pi^2\,v_0^3}\,, \quad
D_{ij} = \pi\,e^2\,\rho\,\langle V_i\,V_j \rangle = \frac{g\,e^2\,\mu^2}{6\,\pi\,v_0}\,\delta_{ij}\,,
\end{align}
where $\bs{V}^{(1)} = v_0\,\hat{\bs{k}}$ and $\langle \hat k_i\,\hat k_j \rangle = \delta_{ij}/3$ have been used. According to Eq.~\eqref{eq:plasmon}, the bare plasmon frequencies are isotropic and given by
\begin{align}
\label{eq:lin-wp}
\omega_p^2 = \frac{2\,g\,e^2\,\mu^2}{3\,\pi\,v_0}\,.
\end{align}
These frequencies coincide with those of the untilted WSM and of the HSM, which share the same FS and band-velocity, and they scale as $\omega_p \propto \mu \propto \tilde n^{1/3}$.

\subsection{Interband screening and plasmons}
\label{seclinscreen}

For $\mu > 0$, the transitions $0 \to 1$ are allowed only for $k > k_F$, while the transitions $-1 \to 0$ remain blocked. The lower limit of the radial integral in Eq.~\eqref{eq:lin-pi0} therefore becomes $k_F$, which replaces $\omega^2$ in the logarithm by $|\mu^2 - \omega^2|$. After adding the Drude term, we obtain
\begin{align}
\label{eq:lin-eps}
\Re \epsilon^{\rm RPA} (\bs{q}, \omega) = 1 - \frac{\omega_p^2}{\omega^2} + \frac{2\,g\,e^2}{3\,\pi\,v_0}\,\ln \!\left( \frac{v_0^2\,\Lambda^2}{|\mu^2 - \omega^2|} \right), \quad
\Im \epsilon^{\rm RPA} (\bs{q}, \omega) = \frac{2\,g\,e^2}{3\,v_0}\,\Theta (\omega - \mu)\,.
\end{align}
The interband onset is $\omega_{\rm th} = \mu$, which is set by the flat-band. The background dielectric constant, defined by the limit $\omega \to 0$ taken after $q \to 0$, is
\begin{align}
\label{eq:lin-eb}
\epsilon_b = 1 + \frac{4\,g\,e^2}{3\,\pi\,v_0}\,\ln \!\left( \frac{v_0\,\Lambda}{\mu} \right).
\end{align}
This constant is isotropic, and only its dependence on $\mu$ is independent of the regularisation. The ratio $\omega_p^2/\omega_{\rm th}^2 = 2\,g\,e^2/(3\,\pi\,v_0)$ is fixed by $\alpha$ alone.

For $\omega < \mu$, both the background term and the Drude term $-\,\omega_p^2/\omega^2$ increase with $\omega$, and $\Re\epsilon^{\rm RPA}$ is therefore monotonic. It tends to $-\,\infty$ for $\omega \to 0$ and to $+\,\infty$ for $\omega \to \mu^-$, and it hence has a unique zero $\omega_*$ below the onset, since $\alpha > 0$. Since the background term is at least $\epsilon_b$, this zero obeys $\omega_* \leq \tilde\omega_p$ with
$\tilde\omega_p^2 = \frac{\omega_p^2}{\epsilon_b}$.
The imaginary part vanishes at $\omega_*$, and the plasmon is therefore undamped. It is isotropic, in contrast to the HSM, whose interband background is anisotropic.

\section{TSM at arbitrary momentum-transfers}
\label{secfinite}

In this section, we keep $c = 0$ and lift the restriction $q \ll k_F$ of Sec.~\ref{seclinear}, with $k_F = \mu/v_0$ and $\mu > 0$. We denote the frequency in momentum units by $\varpi = \omega/v_0$. The polarisability is then $g/v_0$ times a function of $q/k_F$ and $\varpi/k_F$, and $\alpha = g\,e^2/v_0$ enters only through Eq.~\eqref{eq:epsrpa}. The cut-off of the linearised description affects only the filled bands, and we implement it as an upper limit $\mathcal{R} = v_0\,\Lambda$ on the transition energy, which reproduces Eq.~\eqref{eq:lin-pi0b} at long wavelengths. A different regularisation adds a term proportional to $q^2$ to $\Pi$ and shifts $\epsilon_b$ by a constant, while every nonanalytic feature remains unchanged. The derivations are given in Appendix~\ref{app:fin}.

\subsection{Imaginary part of polarisability: Particle-hole continua and their edges}
\label{secfincont}

With $k' = |\bs{k} + \bs{q}|$, the overlap factors of Eq.~\eqref{eq:lin-ff} depend on $\cos\theta = (k^2 + k'^2 - q^2)/(2\,k\,k')$. The transitions $-1 \to 0$ are blocked because both bands are filled, and three other groups of transitions have a nonzero imaginary part. These are the intraband transitions within the band $s = 1$ across the FS, the transitions $0 \to 1$ into $k' > k_F$, and the transitions $-1 \to 1$ into $k' > k_F$. For $\varpi > 0$, the delta function of Eq.~\eqref{eq:pol} fixes one of the two momenta, and the remaining integrals are elementary. We obtain $\Im \Pi = \Im \Pi^{\rm L}_{11} + \Im \Pi^{\rm L}_{01} + \Im \Pi^{\rm L}_{-1,1}$, with
\begin{align}
\label{eq:fin-Ima}
& \Im \Pi^{\rm L}_{11} = -\,\frac{g}{4\,\pi\,v_0\,q} \left[ \mathcal{G}^{\rm L}_{11} (k_F) - \mathcal{G}^{\rm L}_{11} (\check{k}_-) \right], \quad
\check{k}_- = \max \left\lbrace \frac{q - \varpi}{2}, k_F - \varpi \right\rbrace, \\
\label{eq:fin-Imb}
& \Im \Pi^{\rm L}_{01} = -\,\frac{g}{32\,\pi\,v_0} \left[ 2 \left( \varpi^2 + q^2 \right) - \frac{\left( \varpi^2 - q^2 \right)^2}{\varpi\,q}\,\ln \!\left( \frac{\varpi + q}{|\varpi - q|} \right) \right], \\
\label{eq:fin-Imc}
& \Im \Pi^{\rm L}_{-1,1} = -\,\frac{g}{4\,\pi\,v_0\,q} \left[ \mathcal{G}^{\rm L}_{-1,1} \!\left( \frac{\varpi + q}{2} \right) - \mathcal{G}^{\rm L}_{-1,1} (\check{k}_+) \right], \quad
\check{k}_+ = \max \left\lbrace k_F, \frac{\varpi - q}{2} \right\rbrace.
\end{align}
The ranges are $\max \lbrace 0, q - 2\,k_F \rbrace < \varpi < q$ for $\Pi^{\rm L}_{11}$, $\varpi > k_F$ for $\Pi^{\rm L}_{01}$, and $\varpi > \max \lbrace q, 2\,k_F - q \rbrace$ for $\Pi^{\rm L}_{-1,1}$. Each part vanishes outside its range. The functions are
\begin{align}
\mathcal{G}^{\rm L}_{11} (p) &= \frac{p \left( 3\,\varpi^2 + 3\,\varpi\,p + 2\,p^2 - 3\,q^2 \right)}{6} + \frac{\left( \varpi^2 - q^2 \right)^2}{16\,\varpi}\,\ln \!\left( \frac{p}{p + \varpi} \right)\,, \notag \\
\label{eq:fin-G}
\mathcal{G}^{\rm L}_{-1,1} (p) &= \frac{p \left( 3\,q^2 - 3\,\varpi^2 + 3\,\varpi\,p - 2\,p^2 \right)}{6} + \frac{\left( \varpi^2 - q^2 \right)^2}{16\,\varpi}\,\ln \!\left( \frac{p}{\varpi - p} \right)\,.
\end{align}
The imaginary part is odd in $\omega$.

\begin{figure}[t!]
\centering
\subfigure[]{\includegraphics[width=0.75 \linewidth]{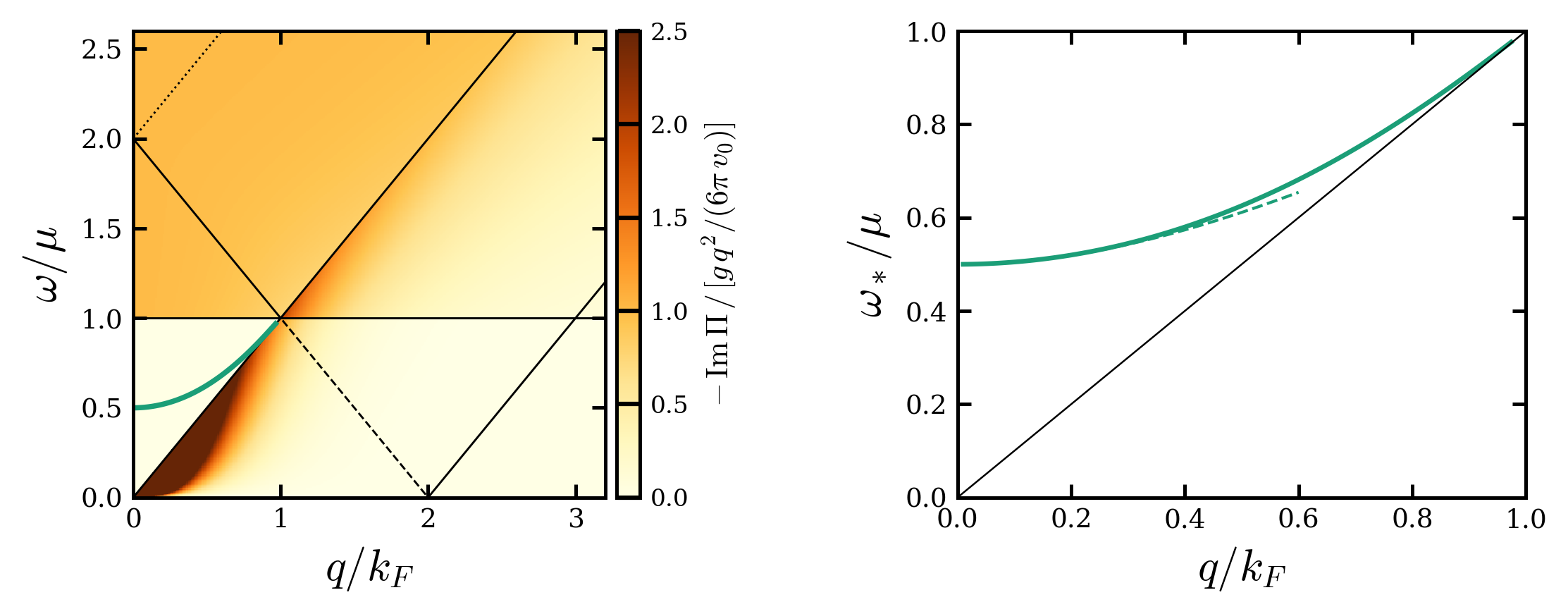}}
\subfigure[]{\includegraphics[width=0.75 \linewidth]{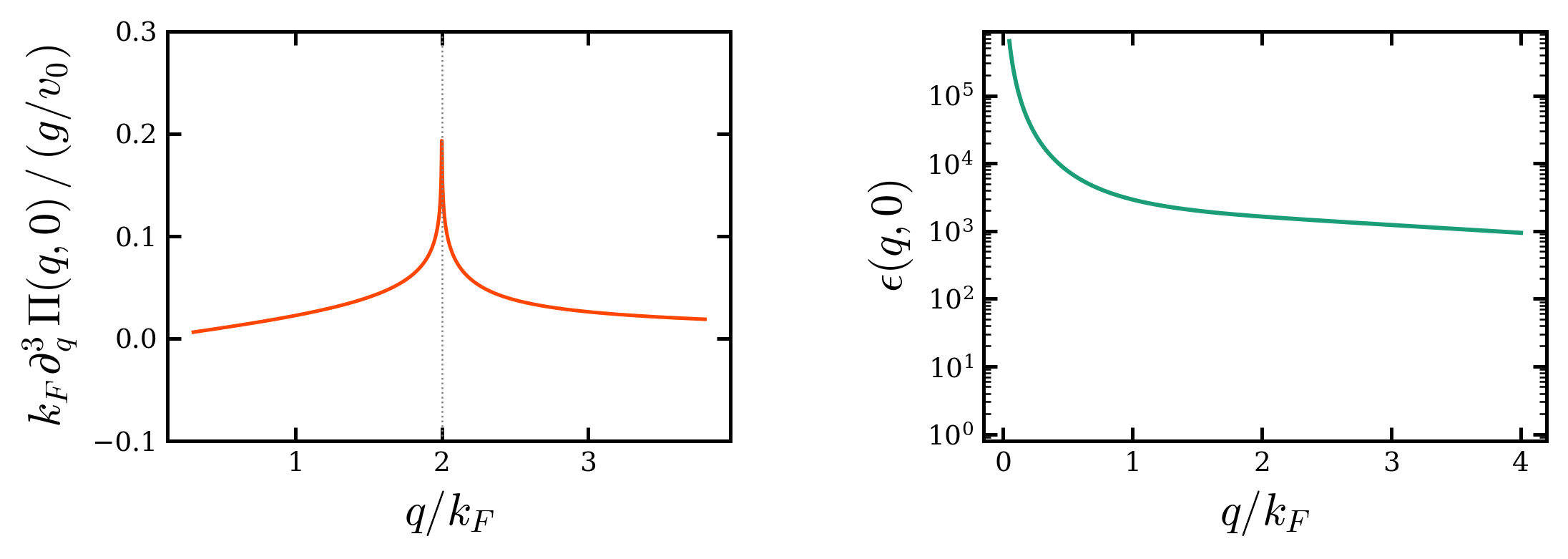}}
\caption{\label{fig:plasmon-static} TSM characteristics. (a) $-\,\Im\Pi$ normalised by its intrinsic value, with the undamped plasmon branch (left), and the plasmon branch (right). (b) Third derivative of $\Pi (q, 0)$ (left) and static dielectric function (right). Parameters are $\mu = 0.01\,\text{eV}$ and $c = 0$.}
\end{figure}

In the intrinsic limit $k_F \to 0$, only $\Pi^{\rm L}_{01}$ and $\Pi^{\rm L}_{-1,1}$ survive. For $\varpi > q$ they add up to $\Im \Pi (q, \omega) = -\,\frac{g\,q^2}{6\,\pi\,v_0}$.
This is Eq.~\eqref{eq:lin-pi0b} without any restriction on $q/\varpi$. For $\varpi < q$, only $\Pi^{\rm L}_{01}$ of Eq.~\eqref{eq:fin-Imb} remains. The transitions $-1 \to 1$ set in at $\varpi = q$ with a finite step of $-\,g\,q^2/(24\,\pi\,v_0)$, because the overlap factor equals unity when the two states have opposite momenta. In the doped system, the three ranges stated above define the edges of the continuum in the $(q, \varpi)$ plane. For $\varpi > q + 2\,k_F$, the imaginary part has reached its intrinsic value of Eq.~\eqref{eq:lin-pi0b}. Only two regions are free of single-particle excitations, namely $q < \varpi < k_F$ for $q < k_F$, and $\varpi < \min\lbrace k_F, q - 2\,k_F \rbrace$ for $q > 2\,k_F$. The first region hosts the plasmon, as shown in Fig.~\ref{fig:plasmon-static}(a).

Fig.~\ref{fig:plasmon-static}(a) shows $-\,\Im\Pi$ from Eqs.~\eqref{eq:fin-Ima}--\eqref{eq:fin-Imc} for $\varpi > q$, normalised by its intrinsic value, together with the undamped plasmon branch, which is drawn up to $q_{\rm max}$. The solid black lines are the edges of the continuum, namely $\omega = v_0\,q$, $\omega = v_0 \left( q - 2\,k_F \right)$ for $q > 2\,k_F$, $\omega = \mu$, and $\omega = 2\,\mu - v_0\,q$ for $q < k_F$. The dashed line $\omega = 2\,\mu - v_0\,q$ for $k_F < q < 2\,k_F$ marks a kink of the intraband weight, and the dotted line $\omega = v_0\,q + 2\,\mu$ marks the point beyond which $\Im\Pi$ has reached its intrinsic value. The right panel shows the branch from Eq.~\eqref{eq:fin-kk} (solid) and from Eq.~\eqref{eq:fin-wdisp} (dashed), which coincide at small $q$ and separate at larger $q$. The black lines are $\omega = v_0\,q$ and $\omega = \mu$. Fig.~\ref{fig:plasmon-static}(b) shows the third derivative of $\Pi (q, 0)$ from Eq.~\eqref{eq:fin-static}, scaled by $k_F\,v_0/g$, which diverges logarithmically at $q = 2\,k_F$ as in Eq.~\eqref{eq:fin-kohn}. The right panel shows the static dielectric function from Eq.~\eqref{eq:fin-static} (solid) and the Thomas-Fermi form of Eq.~\eqref{eq:fin-tf} (dashed), which lies on top of the solid curve over the range of $q$ in which it is drawn. The parameters are $\mu = 0.01\,\text{eV}$ and $c = 0$, which give $k_F = \mu/v_0 = 25\,\text{eV}$ and $v_0\,\Lambda/\mu = 6.4$, and hence $L = \ln \!\left( v_0\,\Lambda/\mu \right) = \ln 6.4$.

\subsection{Real part of polarisability and RPA dielectric function}
\label{secfinre}

The real part follows from the dispersion relation. The imaginary part approaches its intrinsic value for $\varpi > q + 2\,k_F$, where the difference from the intrinsic reference vanishes. The dispersion relation therefore reduces to an integral over a finite window,
\begin{align}
\label{eq:fin-kk}
\Re \Pi (q, \omega) = -\,\frac{g\,q^2}{6\,\pi^2\,v_0}\,\ln \!\left( \frac{v_0^2\,\Lambda^2}{|\omega^2 - v_0^2\,q^2|} \right)
+ \frac{2}{\pi}\,\mathrm{P} \!\! \int_0^{\omega_{\rm w}} d\omega'\, \frac{\omega' \left[ \Im \Pi (q, \omega') + \dfrac{g\,q^2}{6\,\pi\,v_0}\,\Theta (\omega' - v_0\,q) \right]}{\omega'^2 - \omega^2}\,,
\end{align}
with $\omega_{\rm w} = v_0\,q + 2\,\mu$. For $\omega > v_0\,q$, the intrinsic part of the window integral is elementary except for one dilogarithm. With $W = \omega/(v_0\,q)$, the intrinsic polarisability is
\begin{align}
\label{eq:fin-Re0}
 \Re \Pi^{(0)} (q, \omega) & = -\,\frac{g\,q^2}{6\,\pi^2\,v_0}\,\ln \!\left( \frac{v_0^2\,\Lambda^2}{\omega^2 - v_0^2\,q^2} \right) - \frac{g\,q^2}{16\,\pi^2\,v_0}\,\mathcal{K} (W)\,,
\nn  \mathcal{K} (W) &= 1 + \left( 1 + W^2 \right) 
\ln \!\left( \frac{W^2 - 1}{W^2} \right) - \frac{1 + 2\ln 2}{3} 
- 2 \left( W^2 - 2 \right) \ln 2
\nn & \qquad - \frac{\left( W^2 - 1 \right)^2}{2\,W} \left[ 
\mathrm{Li}_2 \!\left( -\,\frac{W + 1}{W - 1} \right) 
- \mathrm{Li}_2 \!\left( -\,\frac{W - 1}{W + 1} \right) \right].
\end{align}
The function $\mathcal{K}$ vanishes for $W \to \infty$, and Eq.~\eqref{eq:fin-Re0} then reduces to Eq.~\eqref{eq:lin-pi0b}. The doping-induced remainder is a finite integral over the window of Eq.~\eqref{eq:fin-kk}, and it is evaluated by a single quadrature.

\subsection{Static polarisability and screening}
\label{secfinstatic}

At $\omega = 0$, the dispersion relation can be integrated in closed form. For all $q$, we obtain
\begin{align}
\label{eq:fin-static}
\Pi (q, 0) = -\,\frac{g}{4\,\pi^2\,v_0} \left\lbrace \frac{4\,k_F^2}{3} + \frac{q^2}{6} + \frac{2\,q^2}{3}\,\ln \!\left( \frac{4\,\Lambda^2}{|4\,k_F^2 - q^2|} \right)
+ \frac{1}{q} \left( \frac{2\,k_F^3}{3} - k_F\,q^2 - \frac{q^4}{8\,k_F} \right) \ln \!\left( \frac{2\,k_F + q}{|2\,k_F - q|} \right) \right\rbrace.
\end{align}
For $q \to 0$, this gives $\Pi (q, 0) \to -\,\rho (\mu)$ with the DOS $\rho$ of Eq.~\eqref{eq:lin-dos}, and the static dielectric function takes the Thomas-Fermi form
\begin{align}
\label{eq:fin-tf}
\epsilon^{\rm RPA} (q, 0) \simeq \epsilon_b + \frac{q_{\rm TF}^2}{q^2}\,, \quad q_{\rm TF}^2 = 4\,\pi\,e^2\,\rho = \frac{2\,\alpha}{\pi}\,k_F^2\,.
\end{align}
The background is $\epsilon_b = 1 + (4\,\alpha/(3\,\pi)) \ln \!\left( v_0\,\Lambda/\mu \right) - 7\,\alpha/(9\,\pi)$. It agrees with Eq.~\eqref{eq:lin-eb} up to the constant, which depends on the regularisation. For $q \gg k_F$, the doped system approaches the intrinsic value:
\begin{align}
\label{eq:fin-large}
\Pi (q, 0) \to -\,\frac{g\,q^2}{3\,\pi^2\,v_0} \left[ \ln \!\left( \frac{2\,\Lambda}{q} \right) - \frac{1}{4} \right] + \frac{g\,k_F^2}{6\,\pi^2\,v_0}\,.
\end{align}
The second term is positive, since Pauli-blocking removes interband transitions. The static dielectric function is shown in Fig.~\ref{fig:plasmon-static}(b), together with its Thomas-Fermi form of Eq.~\eqref{eq:fin-tf}.

\subsection{Friedel oscillations}
\label{secfinfried}

The coefficient of $\ln \!\left( |2\,k_F - q| \right)$ in Eq.~\eqref{eq:fin-static} vanishes at $q = 2\,k_F$ together with its first two derivatives. The Kohn singularity is therefore
\begin{align}
\label{eq:fin-kohn}
\Pi (q, 0) \simeq \Pi_{\rm reg} (q) - \frac{g}{24\,\pi^2\,v_0\,k_F} \left( q - 2\,k_F \right)^3 \ln \!\left( |q - 2\,k_F| \right)\,,
\end{align}
where $\Pi_{\rm reg}$ is analytic. In an ordinary electron gas, the singularity is $(q - 2\,k_F) \ln \!\left( |q - 2\,k_F| \right)$. Here the intraband overlap factor $\mathcal{F}_{11} = \left( (1 + \cos\theta)/2 \right)^2$ vanishes quartically for backscattering, and this raises the power by two. The third derivative of $\Pi$ diverges logarithmically, as shown in Fig.~\ref{fig:plasmon-static}(b). The singularity determines the large-distance behaviour of the density induced by a static potential $\mathcal{U} (q)$. The singular part of $\delta n (q) = \Pi (q, 0)\,\mathcal{U} (q)/\epsilon^{\rm RPA} (q, 0)$ follows from Eq.~\eqref{eq:fin-kohn} and equals $\Pi_{\rm sing}\,\mathcal{U}/\left( \epsilon^{\rm RPA} \right)^2$, since $\epsilon^{\rm RPA}$ also depends on $\Pi$. We find
\begin{align}
\label{eq:fin-friedel}
\delta n (r) \simeq -\,\frac{g}{4\,\pi^3\,v_0}\,\frac{\mathcal{U} (2\,k_F)}{\left[ \epsilon^{\rm RPA} (2\,k_F, 0) \right]^2}\,\frac{\cos \!\left( 2\,k_F\,r \right)}{r^5}\,.
\end{align}
For a Coulomb impurity of charge $Z\,e$, with $\mathcal{U} (q) = -\,4\,\pi\,Z\,e^2/q^2$, the amplitude is $g\,Z\,e^2/(4\,\pi^2\,v_0\,k_F^2\,\left[ \epsilon^{\rm RPA} (2\,k_F, 0) \right]^2)$. The Friedel oscillations decay as $r^{-5}$, compared with $r^{-3}$ for an ordinary electron gas. The oscillations of all systems are compared in Fig.~\ref{fig:friedel-all}.

\subsection{Plasmon dispersion and Landau damping}
\label{secfinplasmon}

For $\omega < \mu$ and $q < \omega/v_0$, the expansion of $\Pi$ in powers of $q$ contains only even powers, $\Pi = q^2\,\Pi_2 + q^4\,\Pi_4 + \dots$, with
\begin{align}
\label{eq:fin-pi24}
\Pi_2 = \frac{g}{6\,\pi^2\,v_0} \left[ \frac{\mu^2}{\omega^2} - \ln \!\left( \frac{v_0^2\,\Lambda^2}{\mu^2 - \omega^2} \right) \right], \quad
\Pi_4 = \frac{g\,v_0}{\pi^2\,\omega^2} \left[ \frac{\mu^2}{10\,\omega^2} - \frac{1}{20} + \frac{1}{30}\,\ln \!\left( \frac{4\,\mu^2 - \omega^2}{4 \left( \mu^2 - \omega^2 \right)} \right) \right].
\end{align}
The first term reproduces Eq.~\eqref{eq:lin-eps}. Let $\epsilon_b (\omega) = 1 + (2\,\alpha/(3\,\pi)) \ln \!\left( v_0^2\,\Lambda^2/(\mu^2 - \omega^2) \right)$ and $\Phi (\omega) = \epsilon_b (\omega) + (2\,\alpha/(3\,\pi))\,\omega^2/(\mu^2 - \omega^2)$. To order $q^2$, the zero of $\Re\epsilon^{\rm RPA}$ is
\begin{align}
\label{eq:fin-wdisp}
\omega_*^2 (q) = \omega_*^2 (0) + \frac{4\,\alpha\,v_0^2\,q^2}{\pi}\,\frac{1}{\Phi (\omega_*)} \left[ \frac{\mu^2}{10\,\omega_*^2} - \frac{1}{20} + \frac{1}{30}\,\ln \!\left( \frac{4\,\mu^2 - \omega_*^2}{4 \left( \mu^2 - \omega_*^2 \right)} \right) \right]_{\omega_* = \omega_* (0)}.
\end{align}

The dispersion increases with $q$, since $\alpha > 0$.

At larger wavevectors, we solve $\Re \epsilon^{\rm RPA} = 0$ numerically using Eq.~\eqref{eq:fin-kk}, as shown in Fig.~\ref{fig:plasmon-static}(a). The plasmon is undamped as long as $v_0\,q < \omega_* < \mu$. The imaginary part jumps at $\omega = v_0\,q$, and $\Re \epsilon^{\rm RPA}$ therefore tends to $-\,\infty$ as $\omega \to v_0\,q$ from above, whereas it tends to $+\,\infty$ at $\omega \to \mu$ from below, since the interband logarithm diverges at the point $(q, \varpi) = (k_F, k_F)$ where the two boundaries meet. A zero exists in the window for every $q < k_F$, but its distance from the edge $\omega = v_0\,q$ decreases exponentially with increasing $q$. We define $q_{\rm max}$ as the wavevector beyond which this distance is below $2 \times 10^{-3}\,v_0\,q$. The branch is then indistinguishable from the edge on the scale of Fig.~\ref{fig:plasmon-static}(a). 

The doping enters only through $\ln \!\left( v_0\,\Lambda/\mu \right)$, and a large $\mu$ at fixed $\Lambda$ corresponds to a small logarithm. The background $\epsilon_b$ is then close to unity, and the plasmon is nearly unscreened. For $\omega_p > \mu$, which requires $\alpha > 3\,\pi/2$, the zero is pinned just below the interband onset. For $q > q_{\rm max}$, the zero of $\Re\epsilon^{\rm RPA}$ lies inside the intraband continuum, where $\Im\epsilon^{\rm RPA} = -\,\mathcal{V} (q)\,\Im\Pi^{\rm L}_{11} \neq 0$. The mode is Landau damped, with the rate $\gamma = \Im\epsilon^{\rm RPA}/\partial_\omega \Re\epsilon^{\rm RPA}$ at $\omega = \Re\,\omega_*$. For $k_F < q < 2\,k_F$, there is no region free of excitations. For $q > 2\,k_F$, the free region $\varpi < \min\lbrace k_F, q - 2\,k_F \rbrace$ contains no zero of $\Re\epsilon^{\rm RPA}$ for $q \ll \Lambda$, where $\Re\epsilon^{\rm RPA} > 1$.

\section{QTSM at small momentum-transfers}
\label{secquad}

We now restore the quadratic term of Eq.~\eqref{eq:ham}. Since the eigenstates and the overlap factors do not depend on $c$ (Sec.~\ref{sec:ham}), the parameter enters only through the energies and the occupations $n_s (\bs{k}) = f_0 \left( \varepsilon_s (\bs{k}) \right)$. The energy difference of two bands at the same momentum, $\varepsilon_{s'} (\bs{k}) - \varepsilon_s (\bs{k}) = (s' - s)\,v_0\,k$, is also independent of $c$. We introduce $\xi = c\,\mu/v_0$, which gives $\beta = \sqrt{1 + 4\,\xi}$. Sec.~\ref{secquadintr} treats the intrinsic limit, and Sec.~\ref{secquadtop} derives the DOS and the Drude weight of the top band. The regime in which the top band is the only band with an FS, which requires $c < 0$, follows in Sec.~\ref{secquadonly}. The case $c > 0$, in which the flat-band acquires a dispersion and an FS, is addressed in Sec.~\ref{secquadflat}.

For a pair of bands $s < s'$, the substitution $\bs{k} \to -\,\bs{k} - \bs{q}$ exchanges $k$ and $k'$ and leaves $\mathcal{F}_{ss'}$ invariant. The two time orderings in Eq.~\eqref{eq:pol} then combine exactly into
\begin{align}
\label{eq:q-pair}
\Pi_{ss'} (\bs{q}, \omega) = g \int \frac{d^3\bs{k}}{(2\,\pi)^3}\,\mathcal{F}_{ss'} (\bs{k}, \bs{k}') \left[ n_s (\bs{k}) - n_{s'} (\bs{k}') \right] \frac{2\,X_{ss'}}{\left( \omega + i\,0^+ \right)^2 - X_{ss'}^2}\,, \quad
X_{ss'} = \varepsilon_{s'} (\bs{k}') - \varepsilon_s (\bs{k})\,.
\end{align}
Among the interband pairs, only $(0, 1)$ and $(-1, 0)$ have an overlap factor of order $q^2$, namely $\mathcal{F}_{10} = \mathcal{F}_{0,-1} \simeq q_i\,q_j\,Q_{ij}$, whereas the pair $(-1, 1)$ is of order $q^4$. For the former, the coefficient of $q^2$ follows by setting $\bs{k}' = \bs{k}$ in the occupations and in $X_{ss'}$, which gives
\begin{align}
\label{eq:q-master}
\Pi_{ss'} (\bs{q}, \omega) = g\,q_i\,q_j \int \frac{d^3\bs{k}}{(2\,\pi)^3}\,Q_{ij} (\bs{k}) \left[ n_s (\bs{k}) - n_{s'} (\bs{k}) \right] \frac{2\,(s' - s)\,v_0\,k}{\left( \omega + i\,0^+ \right)^2 - (s' - s)^2\,v_0^2\,k^2} + o \left( q^2 \right).
\end{align}
The quadratic correction thus enters only through the occupation difference $n_s - n_{s'}$, which fixes the region of $k$ in which the transitions are unblocked. The derivation, together with an independent check of the imaginary part, is given in Appendix~\ref{app:quad-top}.

\subsection{Intrinsic limit}
\label{secquadintr}

For $\mu \to 0^+$, the FSs shrink to the node, and the Drude weights vanish as positive powers of $\mu$, as shown below. Only the interband response therefore remains. We assume $|c|\,\Lambda < 1$, which guarantees that the occupation of each band is the same at all $k < \Lambda$. The bands $s = -1$ and $s = 0$ are then filled and the band $s = 1$ is empty for $c < 0$, whereas for $c > 0$ the flat-band is lifted above the node and only the band $s = -1$ is filled. The unblocked transitions at order $q^2$ are $0 \to 1$ for $c < 0$ and $-1 \to 0$ for $c > 0$. Both carry the overlap factor $\mathcal{F}_{10}$ and the energy $v_0\,k$, and they reproduce the interband polarisability of Eq.~\eqref{eq:lin-pi0}. Hence Eqs.~\eqref{eq:lin-pi0b} and~\eqref{eq:lin-eps0} hold for $v_0\,q \ll \omega \ll v_0\,\Lambda$ and for both signs of $c$, and the intrinsic QTSM supports no undamped plasmon. The details are given in Appendix~\ref{app:quad-intr}.

\subsection{Doped regime: DOS and Drude weights}
\label{secquadtop}

For $\mu > 0$, the Fermi momentum of the band $s = 1$ is the root of $\varepsilon_1 (k_F) = \mu$ that vanishes for $\mu \to 0$, and it exists for $1 + 4\,\xi > 0$. Together with the band-velocity at the FS, it is given by
\begin{align}
\label{eq:q-kf}
k_F = \frac{\beta - 1}{2\,c} = \frac{2\,\mu}{v_0 \left( 1 + \beta \right)}\,, \quad
V_F = v_0 \left( 1 + 2\,c\,k_F \right) = v_0\,\beta\,.
\end{align}
We also introduce the energy $\mu_1 = v_0\,k_F = 2\,\mu/(1 + \beta)$. It differs from $\mu$ for $c \neq 0$ and coincides with it for $c = 0$. The cut-off enters the QTSM through the ratio $\Lambda/k_F = v_0\,\Lambda/\mu_1$, as in Eq.~\eqref{eq:q-eps}, and we therefore define $L_1 = \ln \!\left( v_0\,\Lambda/\mu_1 \right)$, which replaces the logarithm $L = \ln \!\left( v_0\,\Lambda/\mu \right)$ of the TSM. The DOS, the density, and the Drude weights are
\begin{align}
\label{eq:q-dos}
\rho_1 (\mu) = \frac{g\,\mu_1^2}{2\,\pi^2\,v_0^3\,\beta}\,, \quad
\tilde n_1 = \frac{g\,\mu_1^3}{6\,\pi^2\,v_0^3}\,, \quad
D^{(1)}_{ij} = \frac{g\,e^2\,\beta\,\mu_1^2}{6\,\pi\,v_0}\,\delta_{ij} = \frac{g\,e^2\,\mu^2}{6\,\pi\,v_0}\,\frac{4\,\beta}{\left( 1 + \beta \right)^2}\,\delta_{ij}\,,
\end{align}
which reduce to Eq.~\eqref{eq:lin-dos} for $c \to 0$. The Drude weight differs from its linear-model value by the factor $4\,\beta/(1 + \beta)^2$. This factor does not exceed unity and equals $1 - \xi^2 + \mathcal{O} (\xi^3)$ for small $\xi$, which shows that the term linear in $c$ cancels. The intraband polarisability of the band $s = 1$ is $q_i\,q_j\,D^{(1)}_{ij}/(\pi\,e^2\,\omega^2)$, since Eq.~\eqref{eq:intra} depends only on the FS and the band-velocity. The details are given in Appendix~\ref{app:quad-fs}.

\subsection{Regime with only the top band at the Fermi level}
\label{secquadonly}

For $c < 0$ and $\mu > 0$, the bands $s = 0$ and $s = -1$ lie below the Fermi level at every $\bs{k}$ and are filled. The band $s = 1$ is the only band with an FS, and the flat-band does not contribute to the intraband response. The condition $1 + 4\,\xi > 0$ reads $\mu < v_0/(4\,|c|)$, which is the maximum of $\varepsilon_1$. We assume $\Lambda < (1 + \beta)/(2\,|c|)$, which keeps the band $s = 1$ empty for $k_F < k < \Lambda$.

The occupations are those of the linear model with $k_F$ of Eq.~\eqref{eq:q-kf}. The only unblocked transitions at order $q^2$ are therefore $0 \to 1$ for $k > k_F$, with the energy $v_0\,k$ independent of $c$. The interband polarisability is given by Eq.~\eqref{eq:lin-pi0} with the lower limit $k_F$, and adding the Drude term of Eq.~\eqref{eq:q-dos} yields, for $v_0\,q \ll \omega \ll v_0\,\Lambda$ and $q \ll k_F$,
\begin{align}
\label{eq:q-eps}
\Re \epsilon^{\rm RPA} (\bs{q}, \omega) = 1 - \frac{\omega_p^2}{\omega^2} + \frac{2\,g\,e^2}{3\,\pi\,v_0}\,\ln \!\left( \frac{v_0^2\,\Lambda^2}{|\mu_1^2 - \omega^2|} \right), \quad
\Im \epsilon^{\rm RPA} (\bs{q}, \omega) = \frac{2\,g\,e^2}{3\,v_0}\,\Theta (\omega - \mu_1)\,,
\end{align}
with the bare plasmon frequency
\begin{align}
\label{eq:q-wp}
\omega_p^2 = 4\,D^{(1)}_{ii} = \frac{2\,g\,e^2\,\beta\,\mu_1^2}{3\,\pi\,v_0}\,.
\end{align}
The response is isotropic, and the interband onset is $\omega_{\rm th} = \mu_1$. The background dielectric constant, defined by the limit $\omega \to 0$ taken after $q \to 0$, is
\begin{align}
\label{eq:q-eb}
\epsilon_b = 1 + \frac{4\,g\,e^2}{3\,\pi\,v_0}\,\ln \!\left( \frac{v_0\,\Lambda}{\mu_1} \right).
\end{align}
The ratio $\omega_p^2/\omega_{\rm th}^2 = 2\,g\,e^2\,\beta/(3\,\pi\,v_0)$ is fixed by $\alpha$ and by $\beta$. The monotonicity argument of Sec.~\ref{seclinscreen} carries over with $\mu$ replaced by $\mu_1$. It gives a unique zero $\omega_*$ of $\Re \epsilon^{\rm RPA}$ below the onset, which obeys $\omega_*^2 \leq \tilde\omega_p^2 = \omega_p^2/\epsilon_b$, and the plasmon is undamped. The quadratic correction thus enters the plasmon only through $\mu_1$ and $\beta$. The details are given in Appendix~\ref{app:quad-inter}.

\subsection{Regime with the flat-band at the Fermi level}
\label{secquadflat}

For $c > 0$ and $\mu > 0$, the band $s = 0$ disperses as $\varepsilon_0 = c\,v_0\,k^2$ and acquires an FS, while the band $s = 1$ remains monotonic. The band $s = -1$, with $\varepsilon_{-1} = -\,v_0\,k + c\,v_0\,k^2$, reaches the energy $\mu$ only at $k = (1 + \beta)/(2\,c)$. We assume $k_F^{(0)} < \Lambda < (1 + \beta)/(2\,c)$, with $k_F^{(0)}$ defined below, which keeps this band filled throughout the range of the linearised description. The window for $\Lambda$ is nonempty, as shown in Appendix~\ref{app:quad-flat}. The Fermi momentum of the band $s = 1$ is the $k_F$ of Eq.~\eqref{eq:q-kf}, and that of the band $s = 0$ is
\begin{align}
\label{eq:q-kf0}
k_F^{(0)} = \sqrt{\frac{\mu}{c\,v_0}} = \frac{\sqrt{\xi}}{c}\,, \quad
V_0 = 2\,c\,v_0\,k_F^{(0)} = 2\,v_0\,\sqrt{\xi}\,,
\end{align}
where $V_0$ is the band-velocity at the FS. We also define the energy $\mu_0 = v_0\,k_F^{(0)} = \mu/\sqrt{\xi}$. Since $\varepsilon_1 - \varepsilon_0 = v_0\,k > 0$, the two Fermi spheres are concentric with $k_F < k_F^{(0)}$, and $\mu_1 < \mu_0$.

The DOS, the density, and the Drude weight of the flat-band are
\begin{align}
\label{eq:q-dos0}
\rho_0 (\mu) = \frac{g\,\mu_0^2}{4\,\pi^2\,v_0^3\,\sqrt{\xi}}\,, \quad
\tilde n_0 = \frac{g\,\mu_0^3}{6\,\pi^2\,v_0^3}\,, \quad
D^{(0)}_{ij} = \frac{g\,e^2\,\sqrt{\xi}\,\mu_0^2}{3\,\pi\,v_0}\,\delta_{ij} = \frac{g\,e^2\,\mu^2}{6\,\pi\,v_0}\,\frac{2}{\sqrt{\xi}}\,\delta_{ij}\,.
\end{align}
These expressions have the form of Eq.~\eqref{eq:q-dos}, with $\beta$ replaced by $V_0/v_0 = 2\,\sqrt{\xi}$ and $\mu_1$ by $\mu_0$. The ratio $D^{(0)}_{ii}/D^{(1)}_{ii} = (1 + \beta)^2/(2\,\beta\,\sqrt{\xi})$ exceeds unity for every $\xi > 0$ and tends to unity for $\xi \to \infty$. Each band with an FS contributes separately to Eq.~\eqref{eq:intra}, and the total Drude weight is therefore $D_{ij} = D^{(1)}_{ij} + D^{(0)}_{ij}$, which gives the bare plasmon frequency
\begin{align}
\label{eq:q-wp0}
\omega_p^2 = 4\,D_{ii} = \frac{2\,g\,e^2}{3\,\pi\,v_0} \left( \beta\,\mu_1^2 + 2\,\sqrt{\xi}\,\mu_0^2 \right).
\end{align}
For small $\xi$, the flat-band term dominates because $D^{(0)}_{ii}/D^{(1)}_{ii} \to \infty$ for $\xi \to 0$, and $\omega_p^2 \propto c^{-1/2}$ at fixed $\mu$. These expressions hold only while $k_F^{(0)} < \Lambda$.

Here the band $s = 0$ is filled for $k < k_F^{(0)}$ and the band $s = 1$ for $k < k_F$, with $k_F < k_F^{(0)}$. Two groups of transitions survive at order $q^2$, both with the overlap factor $\mathcal{F}_{10}$ and the energy $v_0\,k$. The transitions $0 \to 1$ are allowed in the shell $k_F < k < k_F^{(0)}$, and the transitions $-1 \to 0$ are allowed for $k > k_F^{(0)}$, where the band $s = 0$ is empty. For $v_0\,q \ll \omega \ll v_0\,\Lambda$, the two groups give
\begin{align}
\label{eq:q-pi-flat}
 \Pi_{0 \to 1} (\bs{q}, \omega) = \frac{ - g\,q^2}{6\,\pi^2\,v_0} 
 \left[ \ln \!\left( \frac{|\mu_0^2 - \omega^2|}{|\mu_1^2 - \omega^2|} \right) 
 + i\,\pi\,\Theta (\omega - \mu_1)\,\Theta (\mu_0 - \omega) \right], 
\quad \Pi_{-1 \to 0} (\bs{q}, \omega) = \frac{ - g\,q^2}{6\,\pi^2\,v_0} 
\left[ \ln \!\left( \frac{v_0^2\,\Lambda^2}{|\mu_0^2 - \omega^2|} \right) 
+ i\,\pi\,\Theta (\omega - \mu_0) \right].
\end{align}
Since the logarithms telescope and $\Theta (\omega - \mu_1)\,\Theta (\mu_0 - \omega) + \Theta (\omega - \mu_0) = \Theta (\omega - \mu_1)$, the sum of the two groups equals the interband polarisability of the case $c < 0$ with the Fermi momentum $k_F$. Eqs.~\eqref{eq:q-eps} and~\eqref{eq:q-eb} therefore hold unchanged, with $\omega_p^2$ given by Eq.~\eqref{eq:q-wp0}. For $\mu_1 < \omega < \mu_0$, the absorption comes entirely from the transitions $0 \to 1$, whereas for $\omega > \mu_0$ it comes entirely from the transitions $-1 \to 0$. The static background splits accordingly into $(4\,g\,e^2/3\,\pi\,v_0)\,\ln \!\left( \mu_0/\mu_1 \right)$ and $(4\,g\,e^2/3\,\pi\,v_0)\,\ln \!\left( v_0\,\Lambda/\mu_0 \right)$, which add up to $\epsilon_b - 1$ of Eq.~\eqref{eq:q-eb}. The interband onset remains $\omega_{\rm th} = \mu_1$, because the lowest transition energy belongs to the pair $(0, 1)$ at the FS of the top band. The ratio of the bare plasmon frequency to the onset is
\begin{align}
\frac{\omega_p^2}{\omega_{\rm th}^2} = \frac{2\,g\,e^2}{3\,\pi\,v_0} \left[ \beta + \frac{\left( 1 + \beta \right)^2}{2\,\sqrt{\xi}} \right].
\end{align}
The monotonicity argument of Sec.~\ref{secquadonly} carries over and gives a unique zero $\omega_*$ of $\Re\epsilon^{\rm RPA}$ below the onset, which obeys $\omega_*^2 \leq \tilde\omega_p^2 = \omega_p^2/\epsilon_b$, and the plasmon is undamped and isotropic. The flat-band thus enters only through the Drude weight, which raises the bare plasmon frequency while leaving the screening background unchanged. The details are given in Appendix~\ref{app:quad-flat}.

\subsection{Density dependence and static screening}
\label{secquaddens}

The bare plasmon frequency can be expressed through the carrier densities of the bands with an FS. Since $\tilde n_1 = g\,k_F^3/(6\,\pi^2)$, the Fermi momentum is $k_F = \left( 6\,\pi^2\,\tilde n_1/g \right)^{1/3}$, and $\beta = 1 + 2\,c\,k_F$. Eqs.~\eqref{eq:q-wp} and~\eqref{eq:q-wp0} are then equivalent to
\begin{align}
\label{eq:q-wp-dens}
\omega_p^2 = \frac{2\,g\,e^2\,v_0\,\beta}{3\,\pi} \left( \frac{6\,\pi^2\,\tilde n_1}{g} \right)^{2/3} + \Theta (c)\,8\,\pi\,e^2\,c\,v_0\,\tilde n_0\,.
\end{align}
The first term is the contribution of the top band, which scales as $\tilde n_1^{2/3}$ for $c \to 0$. The second term is the contribution of the flat-band, which is linear in $\tilde n_0$ and equals the conventional plasma frequency $4\,\pi\,e^2\,\tilde n_0/m_*$ of a parabolic band with the mass $m_* = 1/(2\,c\,v_0)$.

We now consider the static limit, in which $q \to 0$ is taken at $\omega = 0$. The intraband part of each band with an FS tends to the negative of its DOS, and the interband part is of order $q^2$. We therefore obtain
\begin{align}
\label{eq:q-static}
\lim_{q \to 0} \Pi (q, 0) = -\,\rho_{\rm tot}\,, \quad
\rho_{\rm tot} = \rho_1 + \Theta (c)\,\rho_0\,,
\end{align}
with $\rho_1$ and $\rho_0$ of Eqs.~\eqref{eq:q-dos} and~\eqref{eq:q-dos0}. The static dielectric function of Eq.~\eqref{eq:epsrpa} then obeys $\lim_{q \to 0} q^2\,\epsilon^{\rm RPA} (q, 0) = q_{\rm TF}^2$, with the Thomas-Fermi wavevector
\begin{align}
\label{eq:q-tf}
q_{\rm TF}^2 = 4\,\pi\,e^2\,\rho_{\rm tot} = \frac{2\,g\,e^2}{\pi\,v_0} \left[ \frac{k_F^2}{\beta} + \Theta (c)\,\frac{\left( k_F^{(0)} \right)^2}{2\,\sqrt{\xi}} \right].
\end{align}
The bare plasmon frequency is related to the band velocities at the FSs by
\begin{align}
\label{eq:q-wp-rho}
\omega_p^2 = \frac{4\,\pi\,e^2}{3} \left [ \rho_1\,V_F^2 + \Theta (c)\,\rho_0\,V_0^2 \right ].
\end{align}
For $c \leq 0$, only the top band contributes, and $\omega_p^2 = q_{\rm TF}^2\,V_F^2/3$ with $V_F = v_0\,\beta$. For $c \to 0$, this gives $q_{\rm TF}^2 = 3\,\omega_p^2/v_0^2$, which is the relation found for the HSM in Sec.~\ref{sechstatic}. The intermediate steps are given in Appendix~\ref{app:quad-dens}.

Fig.~\ref{fig:qtsm-plasmon-static}(a) shows $-\,\Im\Pi$ from Eq.~\eqref{eq:qf-imtot} for $\varpi > q$, normalised by its intrinsic value, together with the undamped plasmon branch, which is drawn up to $q_{\rm max}$. The solid black lines are the edges and onsets of the continuum, namely $\omega = v_0\,\varpi_\pm (q)$ of Eq.~\eqref{eq:qf-edges}, the onset $\omega = v_0\,\varpi_{\rm th}^{(0,1)} (q)$ of Eq.~\eqref{eq:qf-th-neg}, and the onset $\omega = v_0\,\varpi_{\rm th}^{(-1,1)} (q)$ of Eq.~\eqref{eq:qf-th-m1}, which for $q > k_F$ coincides with $\omega = v_0\,\varpi_+ (q)$. The dashed line $\omega = \mu - v_0\,(q - k_F) \left[ 1 + c\,(q - k_F) \right]$ for $k_F < q < 2\,k_F$ marks a non-analyticity of the intraband weight, and the dotted line $\omega = v_0\,(q + 2\,k_F)(1 - c\,q)$ marks the point beyond which $\Im\Pi_{-1,1}$ has reached its undoped value of Eq.~\eqref{eq:qf-imvac}. The right panel shows the branch from Eq.~\eqref{eq:qf-re} (solid) and from Eq.~\eqref{eq:qf-wdisp} (dashed), which coincide at small $q$ and separate at larger $q$. The black lines bound the free window, namely $\omega = v_0\,\varpi_+ (q)$ and the lowest interband onset. Fig.~\ref{fig:qtsm-plasmon-static}(b) shows the third derivative of $\Pi (q, 0)$ from Eq.~\eqref{eq:qf-static}, scaled by $k_F\,v_0/g$, which diverges logarithmically at $q = 2\,k_F$ as in Eq.~\eqref{eq:qf-kohn1}. The right panel shows the static dielectric function from Eq.~\eqref{eq:qf-static} (solid) and the Thomas-Fermi form of Eq.~\eqref{eq:qf-tf} (dashed), which lies on top of the solid curve over the range of $q$ in which it is drawn. The parameters are $\mu = 0.01\,\text{eV}$ and $c = -\,0.001\,\text{eV}^{-1}$. They give $\xi = c\,\mu/v_0 = -\,0.025$, $\beta = \sqrt{1 + 4\,\xi} \simeq 0.949$, $\mu_1 = 2\,\mu/(1 + \beta) \simeq 0.01026\,\text{eV}$, $k_F = \mu_1/v_0 \simeq 25.66\,\text{eV}$, and $L_1 \simeq 1.830$.

\begin{figure}[t!]
\centering
\subfigure[]{\includegraphics[width=0.75 \linewidth]{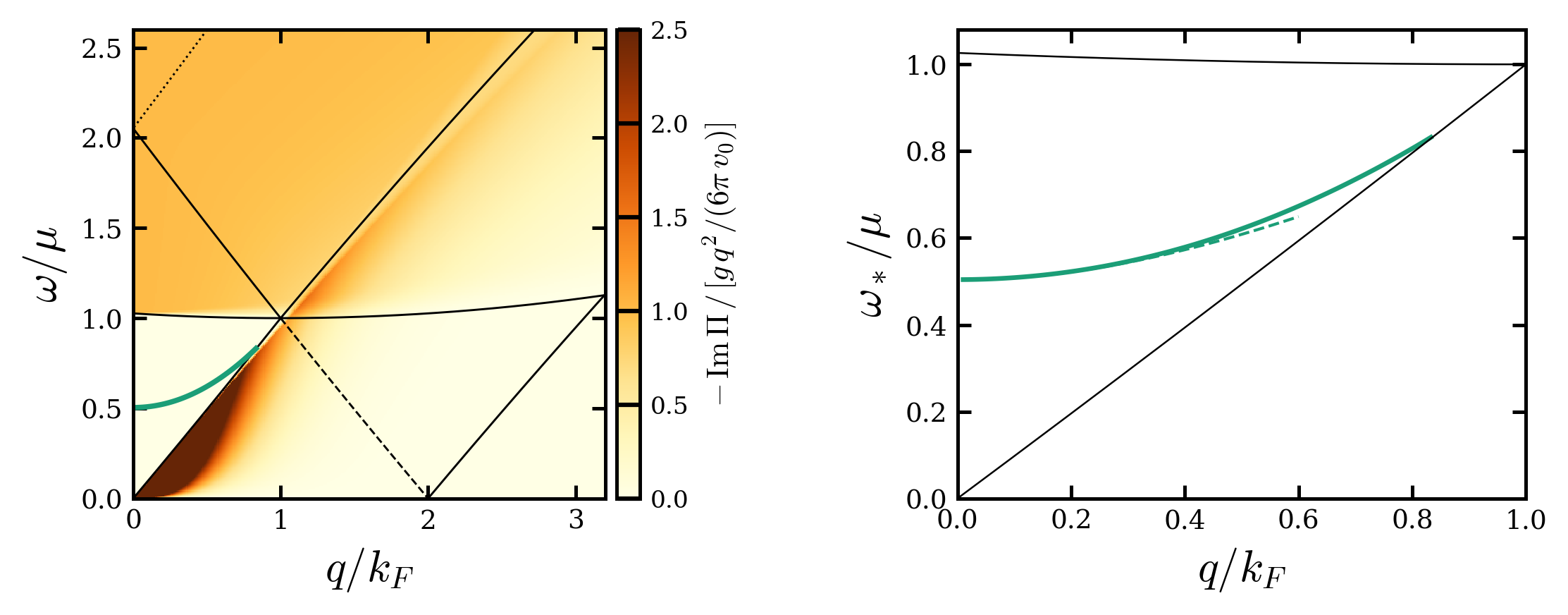}}
\subfigure[]{\includegraphics[width=0.75 \linewidth]{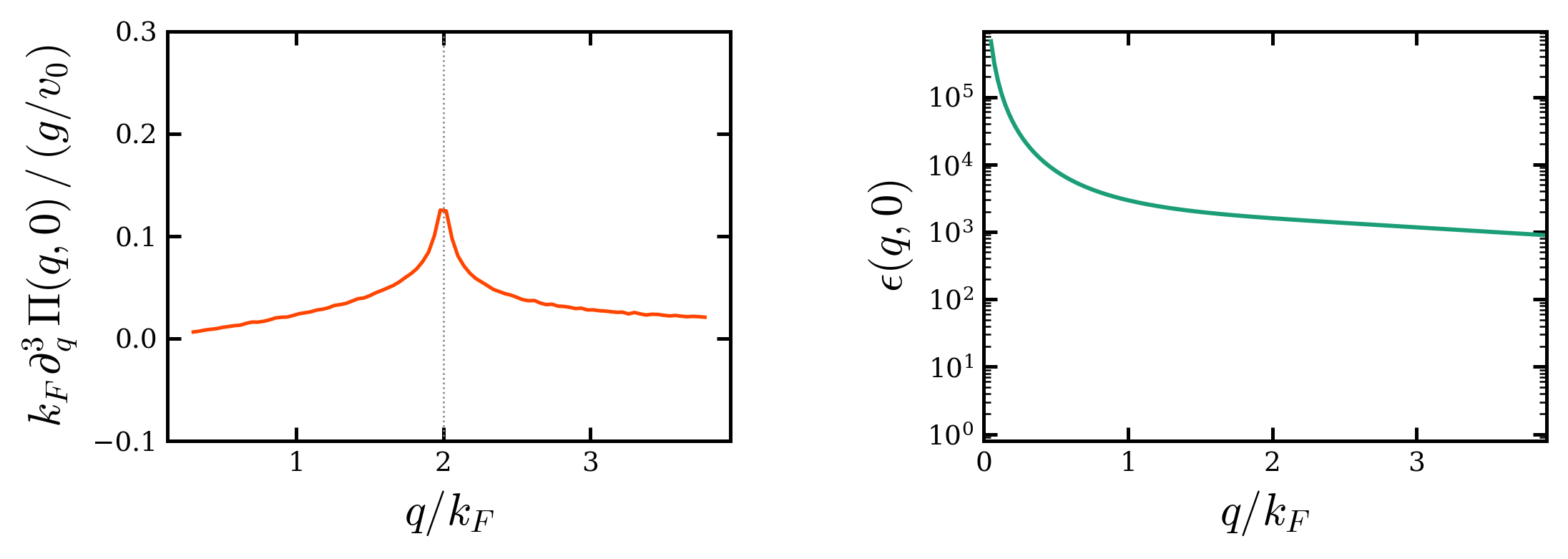}}
\caption{\label{fig:qtsm-plasmon-static} QTSM characteristics for $c < 0$. (a) $-\,\Im\Pi$ normalised by its intrinsic value, with the undamped plasmon branch (left), and the plasmon branch (right). (b) Third derivative of $\Pi (q, 0)$ (left) and static dielectric function (right). Parameters are $\mu = 0.01\,\text{eV}$ and $c = -\,0.001\,\text{eV}^{-1}$.}
\end{figure}


\section{QTSM at arbitrary momentum-transfers}
\label{sec-qtsm-full}

In this section, we lift the restriction $q \ll k_F$ of Secs.~\ref{secquadonly} and~\ref{secquadflat} for the QTSM, with $\mu > 0$ and $\varpi = \omega/v_0$. Since the eigenstates do not depend on $c$, the overlap factors of Eq.~\eqref{eq:lin-ff} remain functions of $\cos\theta = (k^2 + k'^2 - q^2)/(2\,k\,k')$ alone, and the quadratic term enters only through the energies. For $k' \neq k$, the transition energy between two bands is
\begin{align}
\label{eq:qf-trans}
\varepsilon_{s'} (k') - \varepsilon_s (k) = v_0 \left[ s'\,k' - s\,k + c \left( k'^2 - k^2 \right) \right],
\end{align}
which depends on $c$, in contrast to the limit $q \to 0$. We denote the Fermi momentum of the band $s = 1$ by $k_F$, as in Eq.~\eqref{eq:q-kf}, and recall from Sec.~\ref{secquad} that $\xi = c\,k_F \left( 1 + c\,k_F \right)$ and $\beta = 1 + 2\,c\,k_F$, with $c\,k_F > -1/2$. The polarisability is $g/v_0$ times a function of $q/k_F$, $\varpi/k_F$, and $c\,k_F$. The derivations are given in Appendix~\ref{app:qf}.

\subsection{Imaginary part of polarisability: Particle-hole continua and their edges}
\label{secqfim}

\subsubsection{Continuum edges}
\label{secqfedges}


The band $s = 1$ is filled for $k < k_F$ for both signs of $c$. It is monotonic, since $\partial_k \varepsilon_1 = v_0 \left( 1 + 2\,c\,k \right) > 0$ within the range of the linearised description. The intraband transitions $k \to k'$ with $k < k_F < k'$ have the energy
\begin{align}
\varepsilon_1 (k') - \varepsilon_1 (k) = v_0 \left( k' - k \right) \left[ 1 + c \left( k' + k \right) \right].
\end{align}
For $c = 0$, the continuum is the wedge $\max \lbrace 0, q - 2\,k_F \rbrace < \varpi < q$ of Sec.~\ref{secfincont}. For $c \neq 0$, its edges are
\begin{align}
\label{eq:qf-edges}
\varpi_- (q) = \left( q - 2\,k_F \right) \left( 1 + c\,q \right) \Theta \left( q - 2\,k_F \right), \quad
\varpi_+ (q) = q \left[ 1 + c \left( 2\,k_* + q \right) \right], \quad
k_* = \begin{cases} k_F & \text{ for } c > 0 \\ \max \lbrace 0, k_F - q \rbrace & \text{ for } c < 0 \end{cases}.
\end{align}
Both edges reduce to those of the linear model for $c \to 0$. The position of the upper edge relative to the line $\varpi = q$ depends on the sign of $c$.
\begin{itemize}
\item For $c > 0$, the upper edge lies above this line.
\item For $c < 0$, the upper edge lies below this line.
\end{itemize}
For $q \to 0$, both signs of $c$ give $\varpi_+ \to \beta\,q$, and the edge is thus tangent to the line $\omega = V_F\,q$, as expected for the intraband continuum at long wavelengths.


For $c > 0$, the band $s = 0$ has the energy $\varepsilon_0 = c\,v_0\,k^2$, which increases monotonically with $k$, and it is filled for $k < k_F^{(0)}$. The intraband transitions $k \to k'$ with $k < k_F^{(0)} < k'$ carry the overlap factor $\mathcal{F}_{00} = \cos^2\theta$ and the energy $c\,v_0 \left( k' - k \right) \left( k' + k \right)$. The edges of their continuum are
\begin{align}
\label{eq:qf-edges0}
\varpi_-^{(0)} (q) = c\,q \left( q - 2\,k_F^{(0)} \right) \Theta \left( q - 2\,k_F^{(0)} \right), \quad
\varpi_+^{(0)} (q) = c\,q \left( q + 2\,k_F^{(0)} \right).
\end{align}
These edges have the structure of Eq.~\eqref{eq:qf-edges}, with the Fermi momentum $k_F^{(0)}$ and the quadratic spectrum of the flat-band, and no case distinction arises because $\varepsilon_0$ is purely quadratic. For $q \to 0$, the upper edge is tangent to the line $\omega = V_0\,q$, where $V_0 = 2\,v_0\,\sqrt{\xi}$ is the band-velocity at the FS. Using $c\,k_F^{(0)} = \sqrt{\xi}$, the distance to the upper edge of the top-band continuum is
\begin{align}
\varpi_+ (q) - \varpi_+^{(0)} (q) = q \left( \beta - 2\,\sqrt{\xi} \right) > 0\,,
\end{align}
since $\beta^2 - 4\,\xi = 1$. The upper edge of all intraband excitations is therefore the edge $\varpi_+$ of Eq.~\eqref{eq:qf-edges}, for every $q$.


The transitions $0 \to 1$ connect a filled state of the band $s = 0$ at momentum $k$ to an empty state of the band $s = 1$ at $k' = |\bs{k} + \bs{q}| > k_F$. The filling of the band $s = 0$ depends on the sign of $c$.
\begin{itemize}
\item For $c < 0$, the band $s = 0$ is filled at every $k$.
\item For $c > 0$, the band $s = 0$ is filled for $k < k_F^{(0)} = \mathfrak{l}\,k_F$, with $\mathfrak{l} = \sqrt{\frac{1 + c\,k_F}{c\,k_F}} > 1$, which follows from Eqs.~\eqref{eq:q-kf0} and~\eqref{eq:q-kf}.
\end{itemize}
The transition energy is $\varepsilon_1 (k') - \varepsilon_0 (k) = v_0 \left[ k' + c \left( k'^2 - k^2 \right) \right]$. The onset $\varpi_{\rm th}^{(0, 1)} (q)$ of the continuum is its minimum over all allowed pairs $(k, k')$. For $c < 0$, we obtain
\begin{align}
\label{eq:qf-th-neg}
\varpi_{\rm th}^{(0, 1)} (q) = k_F \left( 1 + c\,k_F \right) - c \left( q - k_F \right)^2.
\end{align}
For $c > 0$, we define $q_1 = (\mathfrak{l} - 1)\,k_F$ and $q_2 = (\mathfrak{l} + 1)\,k_F$, and obtain
\begin{align}
\label{eq:qf-th-pos}
\varpi_{\rm th}^{(0, 1)} (q) = \begin{cases}
k_F \left( 1 + c\,k_F \right) - c \left( q + k_F \right)^2 & \text{ for } q \leq q_1 \\
0 & \text{ for } q_1 \leq q \leq q_2 \\
\left( q - k_F^{(0)} \right) \left[ 1 + c \left( q - k_F^{(0)} \right) \right] - c \left( k_F^{(0)} \right)^2 & \text{ for } q \geq q_2
\end{cases}.
\end{align}
For $q \to 0$, both expressions give $\varpi_{\rm th}^{(0, 1)} \to k_F$, which is the onset $\mu_1/v_0$ of Sec.~\ref{secquad}. For $c \to 0$, they give $\varpi_{\rm th}^{(0, 1)} = k_F$ at every $q$, as in Sec.~\ref{secfincont}. The continuum fills every $\varpi > \varpi_{\rm th}^{(0, 1)} (q)$ up to the cut-off.

For $c > 0$, the two FSs are concentric spheres with radii $k_F < k_F^{(0)}$ and the same energy $\mu$. A transition between them costs no energy, and it exists whenever $\bs{q}$ connects the two spheres, that is for $q_1 \leq q \leq q_2$. For these wavevectors, the interband continuum extends to $\varpi = 0$ and overlaps the intraband continuum of Eq.~\eqref{eq:qf-edges}. A region free of excitations can therefore exist only for $q < q_1$.


The band $s = -1$ is filled at every $k$. We assume $2\,|c|\,\max \lbrace q, \bar{k}_F \rbrace < 1$, with $\bar{k}_F = k_F$ for $c < 0$ and $\bar{k}_F = k_F^{(0)}$ for $c > 0$. This condition holds within the range of the linearised description. The transitions $-1 \to 1$ to the states $k' > k_F$ have the energy $v_0 \left( k + k' \right) \left[ 1 + c \left( k' - k \right) \right]$. Their onset is
\begin{align}
\label{eq:qf-th-m1}
\varpi_{\rm th}^{(-1, 1)} (q) = \begin{cases}
\left( 2\,k_F - q \right) \left( 1 + c\,q \right) & \text{ for } q \leq k_F \\
q \left[ 1 + c \left( 2\,k_F - q \right) \right] & \text{ for } q \geq k_F \text{ and } c > 0 \\
q \left( 1 + c\,q \right) & \text{ for } q \geq k_F \text{ and } c < 0
\end{cases}.
\end{align}
For $c \to 0$, this gives $\max \lbrace q, 2\,k_F - q \rbrace$ of Sec.~\ref{secfincont}. For $q \to 0$, it gives $2\,k_F$. For $q \geq k_F$, the minimum is attained by a transition that depends on the sign of $c$.
\begin{itemize}
\item For $c < 0$, it is the transition from $k = 0$ to $k' = q$.
\item For $c > 0$, it is the transition from $k = q - k_F$ to $k' = k_F$.
\end{itemize}

The transitions $-1 \to 0$ are blocked for $c < 0$, since both bands are filled.

For $c > 0$, the band $s = 0$ is empty for $k' > k_F^{(0)}$, and the transitions $-1 \to 0$ to these states have the energy $v_0 \left[ k + c \left( k'^2 - k^2 \right) \right]$. Their onset is
\begin{align}
\label{eq:qf-th-m0}
\varpi_{\rm th}^{(-1, 0)} (q) = \begin{cases}
\left( k_F^{(0)} - q \right) + c\,q \left( 2\,k_F^{(0)} - q \right) & \text{ for } q \leq k_F^{(0)} \\
c\,q^2 & \text{ for } q \geq k_F^{(0)}
\end{cases}.
\end{align}
For $q \to 0$, the onset is $k_F^{(0)} = \mu_0/v_0$, in agreement with Sec.~\ref{secquadflat}. It decreases with $q$ up to $q = k_F^{(0)}$, where it equals $\mu/v_0$, and it increases for larger $q$.

For $c > 0$, the transitions $1 \to 0$ from the filled states $k < k_F$ of the top band to the empty states $k' > k_F^{(0)}$ of the flat-band also contribute. They have the energy $v_0 \left[ c\,k'^2 - k - c\,k^2 \right]$ and exist only for $q \geq q_1$, since $k' \leq k + q$. Their onset is
\begin{align}
\label{eq:qf-th-10}
\varpi_{\rm th}^{(1, 0)} (q) = \begin{cases}
0 & \text{ for } q_1 \leq q \leq q_2 \\
c \left( q - k_F \right)^2 - k_F \left( 1 + c\,k_F \right) & \text{ for } q \geq q_2
\end{cases}.
\end{align}
For $q_1 \leq q \leq q_2$, these are the zero-energy transitions between two FSs, now with the initial state on the top band.

\subsubsection{Spectral weight of individual channels}


For $\omega > 0$, the intraband part of $\Im\Pi$ for the band $s = 1$ comes from the transitions $k < k_F < k'$, with $k' = |\bs{k} + \bs{q}|$. We denote this part by $\Im\Pi_{11}$. The delta function of Eq.~\eqref{eq:pol} fixes $k'$ as a function of $k$. We change the integration variable to $\Sigma = k + k'$ and denote $\varkappa (u) = \left( \sqrt{1 + 4\,c\,u} - 1 \right)/(2\,c)$, which solves $\varkappa + c\,\varkappa^2 = u$ and reduces to $\varkappa (u) = u$ for $c \to 0$. For $\varpi_- (q) < \varpi < \varpi_+ (q)$, we obtain
\begin{align}
\label{eq:qf-im11}
\Im \Pi_{11} (q, \omega) = -\,\frac{g}{32\,\pi\,v_0\,q} \int_{\Sigma_-}^{\Sigma_+} d\Sigma\; \frac{\left( 1 + c\,\Sigma \right) \left( \Sigma^2 - q^2 \right)^2}{\left( \Sigma + c\,\Sigma^2 \right)^2 - \varpi^2}\,,
\end{align}
and $\Im\Pi_{11} = 0$ outside this range. The upper limit is $\Sigma_+ = k_F + \varkappa \left( \mu/v_0 + \varpi \right)$ and the lower limit is $\Sigma_- = \max \lbrace q, k_F + \varkappa \left( \mu/v_0 - \varpi \right) \rbrace$, with the following modifications.
\begin{itemize}
\item For $c < 0$, the upper limit is replaced by $\min \lbrace \Sigma_+, (\varpi - q)/(c\,q) \rbrace$.
\item For $c > 0$, the lower limit is also bounded below by $(\varpi - q)/(c\,q)$.
\end{itemize}
The second entry of the maximum is absent for $\varpi > \mu/v_0$, where $k' > k_F$ is automatic. The integrand is regular on the range of integration. For $c \to 0$, we have $\Sigma = 2\,k + \varpi$, and Eq.~\eqref{eq:qf-im11} reduces to Eq.~\eqref{eq:fin-Ima}. The integral is elementary, and
\begin{align}
\label{eq:qf-im11-cf}
& \Im \Pi_{11} (q, \omega) = -\,\frac{g}{32\,\pi\,v_0\,q} \left[ \mathcal{G}_1 (\Sigma_+) - \mathcal{G}_1 (\Sigma_-) \right], \quad
\mathcal{G}_1 (\Sigma) = \frac{\Sigma^2}{2\,c} - \frac{\Sigma}{c^2} + \sum_{\tau, \check\sigma = \pm 1} \bar\varrho_{\tau\check\sigma}\,\ln \left| \Sigma - \Sigma_{\tau\check\sigma} \right|,
\nn & \Sigma_{\tau\check\sigma} = \frac{-\,1 + \check\sigma\,\check{r}_\tau}{2\,c}\,, \quad
\check{r}_\tau = \sqrt{1 + 4\,c\,\tau\,\varpi}\,, \quad
\bar\varrho_{\tau\check\sigma} = \frac{\tau\,\check\sigma 
\left( 1 + \check\sigma\,\check{r}_\tau \right) \left( \Sigma_{\tau\check\sigma}^2 - q^2 \right)^2}
{4\,\varpi\,\check{r}_\tau}\,.
\end{align}
For $c < 0$, the square roots are real for $\varpi < 1/(4\,|c|)$, which is the largest energy difference within the monotonic range of the band. The terms in $1/c$ and $1/c^2$ cancel against the logarithms for $c \to 0$, and $\mathcal{G}_1$ then reduces to $8\,\mathcal{G}^{\rm L}_{11}$ of Eq.~\eqref{eq:fin-G}, up to a constant that drops out of the difference in Eq.~\eqref{eq:qf-im11-cf}.


For $\omega > 0$, the intraband part of $\Im\Pi$ for the band $s = 0$ comes from the transitions $k < k_F^{(0)} < k'$. We denote this part by $\Im\Pi_{00}$. The overlap factor is $\mathcal{F}_{00} = \cos^2\theta$. With $\Sigma = k + k'$ and $\check{D} = k' - k$, the energy condition $c\,v_0 \left( k'^2 - k^2 \right) = v_0\,\varpi$ becomes $c\,\Sigma\,\check{D} = \varpi$, and gives $\check{D} = \varpi/(c\,\Sigma)$. We obtain
\begin{align}
\label{eq:qf-im00}
\Im \Pi_{00} (q, \omega) = \begin{cases}
-\,\frac{g}{32\,\pi\,c^3\,v_0\,q} \int_{\Sigma_-}^{\Sigma_+} d\Sigma\; 
\frac{\left( c^2\,\Sigma^4 - 2\,c^2\,q^2\,\Sigma^2 + \varpi^2 \right)^2}
{\Sigma^3 \left( c^2\,\Sigma^4 - \varpi^2 \right)} & \text{ for } 
\varpi_-^{(0)} (q) < \varpi < \varpi_+^{(0)} (q) \\
0 & \text{ otherwise }
\end{cases} .
\end{align}
The upper limit is $\Sigma_+ = k_F^{(0)} + \sqrt{\left( k_F^{(0)} \right)^2 + \varpi/c}$. The lower limit is $\Sigma_- = \max \lbrace q, \varpi/(c\,q), k_F^{(0)} + \sqrt{\left( k_F^{(0)} \right)^2 - \varpi/c} \rbrace$. The third entry of the maximum is absent for $\varpi > \mu/v_0 = c \left( k_F^{(0)} \right)^2$, where $k' > k_F^{(0)}$ is automatic. The first entry follows from $\Sigma \geq q$, and the second from $\check{D} \leq q$. The integrand is regular on the range of integration. The integral is elementary, and
\begin{align}
\label{eq:qf-im00-cf}
& \Im \Pi_{00} (q, \omega) = -\,\frac{g}{64\,\pi\,c\,v_0\,q} \left[ \mathcal{G}_{00} (\Sigma_+) - \mathcal{G}_{00} (\Sigma_-) \right], \nn &
\mathcal{G}_{00} (\Sigma) = \Sigma^2 + \frac{\varpi^2}{c^2\,\Sigma^2} 
+ 8\,q^2 \ln \Sigma + \frac{2}{c\,\varpi} \left[ \left( c\,q^2 - \varpi \right)^2 \ln \left| c\,\Sigma^2 - \varpi \right| - \left( c\,q^2 + \varpi \right)^2 
\ln \!\left( c\,\Sigma^2 + \varpi \right) \right].
\end{align}
All poles of the integrand in the variable $\Sigma^2$ are real, and no arctangent arises. The logarithm of $c\,\Sigma^2 - \varpi$ is regular, because $c\,\Sigma^2 > \varpi$ on the range of integration.

\begin{figure}[t!]
\centering
\subfigure[]{\includegraphics[width=0.75 \linewidth]{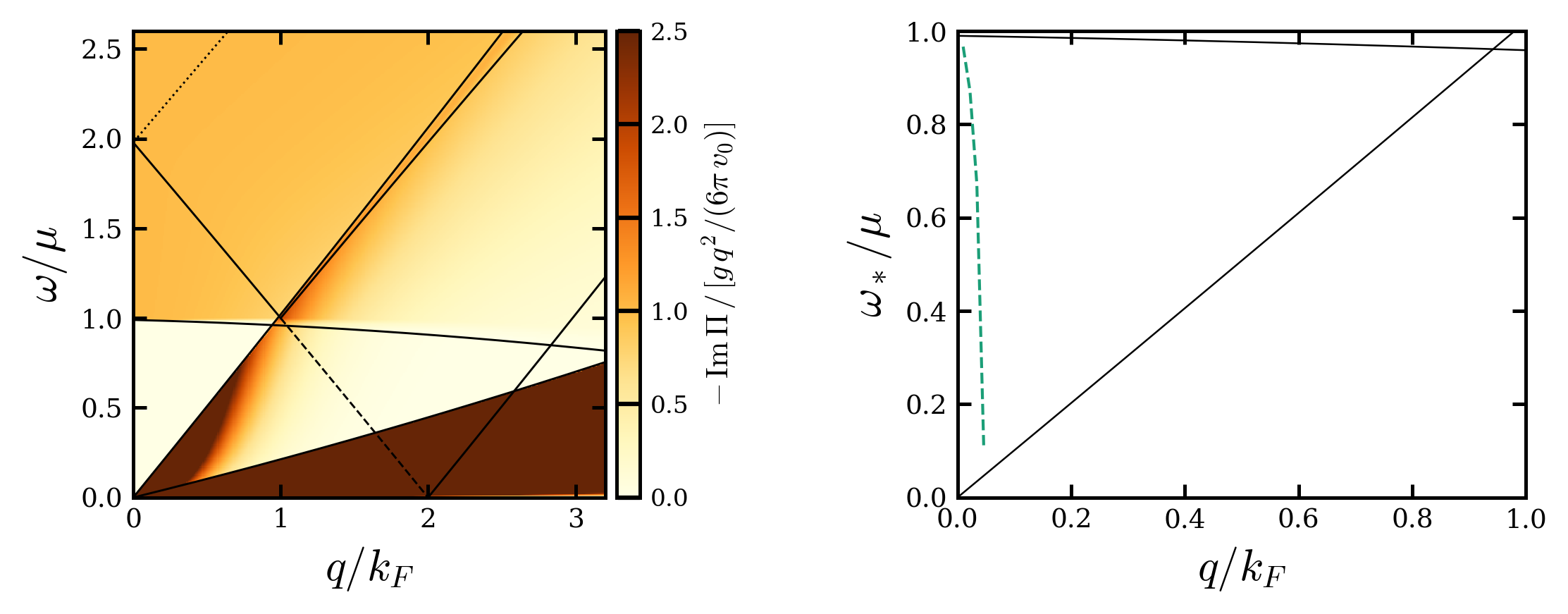}}
\subfigure[]{\includegraphics[width=0.75 \linewidth]{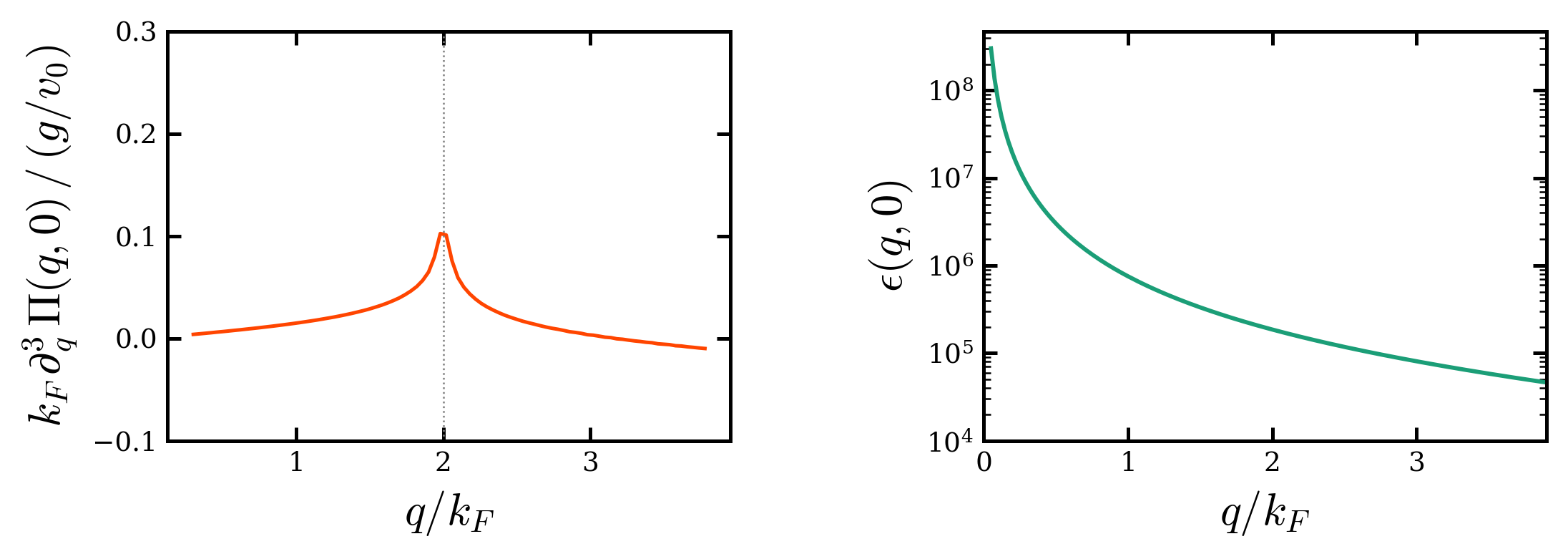}}
\caption{\label{fig:qtsm-cpos-plasmon-static} QTSM characteristics for $c > 0$. (a) $-\,\Im\Pi$ normalised by its intrinsic value (left), and the small-$q$ plasmon solution (right). (b) Third derivative of $\Pi (q, 0)$ (left) and static dielectric function (right). Parameters are $\mu = 0.003\,\text{eV}$ and $c = 0.0014\,\text{eV}^{-1}$.}
\end{figure}

Fig.~\ref{fig:qtsm-cpos-plasmon-static}(a) shows $-\,\Im\Pi$ from Eq.~\eqref{eq:qf-imtot} for $\varpi > q$, normalised by its intrinsic value. No undamped plasmon branch appears, because $\Re\epsilon^{\rm RPA}$ stays negative throughout the free window. The solid black lines are the edges and onsets of the continuum, namely $\omega = v_0\,\varpi_\pm (q)$ of Eq.~\eqref{eq:qf-edges}, $\omega = v_0\,\varpi_\pm^{(0)} (q)$ of Eq.~\eqref{eq:qf-edges0}, and the onsets $\omega = v_0\,\varpi_{\rm th}^{(0,1)} (q)$, $v_0\,\varpi_{\rm th}^{(-1,1)} (q)$, $v_0\,\varpi_{\rm th}^{(-1,0)} (q)$, and $v_0\,\varpi_{\rm th}^{(1,0)} (q)$ of Eqs.~\eqref{eq:qf-th-pos}, \eqref{eq:qf-th-m1}, \eqref{eq:qf-th-m0}, and~\eqref{eq:qf-th-10}. The dotted vertical lines mark $q_1$ and $q_2$. The dashed line $\omega = \mu - v_0\,(q - k_F) \left[ 1 + c\,(q - k_F) \right]$ for $k_F < q < 2\,k_F$ marks a non-analyticity of the intraband weight, and the dotted line $\omega = v_0\,(q + 2\,k_F)(1 - c\,q)$ marks the point beyond which $\Im\Pi_{-1,1}$ has reached its undoped value of Eq.~\eqref{eq:qf-imvac}. The right panel shows the small-$q$ solution of Eq.~\eqref{eq:qf-wdisp} (dashed), whereas Eq.~\eqref{eq:qf-re} has no solution inside the free window. The black lines bound the free window, namely $\omega = v_0\,\varpi_+ (q)$ and the lowest interband onset. Fig.~\ref{fig:qtsm-cpos-plasmon-static}(b) shows the third derivative of $\Pi (q, 0)$ from Eq.~\eqref{eq:qf-static}, scaled by $k_F\,v_0/g$, which diverges logarithmically at $q = 2\,k_F$ as in Eq.~\eqref{eq:qf-kohn1}. The right panel shows the static dielectric function from Eq.~\eqref{eq:qf-static} (solid) and the Thomas-Fermi form of Eq.~\eqref{eq:qf-tf} (dashed), which lies on top of the solid curve over the range of $q$ in which it is drawn. The parameters are $\mu = 0.003\,\text{eV}$ and $c = 0.0014\,\text{eV}^{-1}$. The value of $\mu$ is smaller than for the other systems. Hence $2\,k_F^{(0)} < \Lambda$ and $4\,c\,\Lambda < 1$ hold together. The parameters give $\xi = c\,\mu/v_0 = 0.0105$, $\beta = \sqrt{1 + 4\,\xi} \simeq 1.021$, $\mu_1 = 2\,\mu/(1 + \beta) \simeq 0.00297\,\text{eV}$, $k_F = \mu_1/v_0 \simeq 7.42\,\text{eV}$, $L_1 \simeq 3.07$, $\mu_0 = \mu/\sqrt{\xi} \simeq 0.0293\,\text{eV}$, and $k_F^{(0)} = \mu_0/v_0 \simeq 73.2\,\text{eV}$.

The overlap factor of the transitions $0 \to 1$ is $\mathcal{F}_{01} = \sin^2\theta/2$. The initial state is in the band $s = 0$, which is filled at every $k$ for $c < 0$ and for $k < k_F^{(0)}$ for $c > 0$. The final state is in the band $s = 1$ with $k' > k_F$. With $\Sigma = k + k'$ and $\check{D} = k' - k$, the energy condition $\varepsilon_1 (k') - \varepsilon_0 (k) = v_0\,\varpi$ is linear in $\check{D}$, and gives
$\check{D} = \frac{2\,\varpi - \Sigma}{1 + 2\,c\,\Sigma}$.
We obtain
\begin{align}
\label{eq:qf-im01}
\Im \Pi_{01} (q, \omega) = -\,\frac{g}{32\,\pi\,v_0\,q} \int_{\Sigma_-}^{\Sigma_+} d\Sigma\; 
\frac{\left( \Sigma^2 - q^2 \right) \left[ q - 2\,\varpi + \left( 1 + 2\,c\,q \right) \Sigma \right] 
\left[ q + 2\,\varpi - \left( 1 - 2\,c\,q \right) \Sigma \right]}{\left( 1 + 2\,c\,\Sigma \right) \left( c\,\Sigma^2 + \Sigma - \varpi \right) \left( c\,\Sigma^2 + \varpi \right)}\,.
\end{align}
The limits follow from the constraints on the pair $(k, k')$. The triangle inequality gives $\Sigma \geq q$ and $|\check{D}| \leq q$, that is $(2\,\varpi - q)/(1 + 2\,c\,q) \leq \Sigma \leq (2\,\varpi + q)/(1 - 2\,c\,q)$. The remaining bounds depend on the sign of $c$.
\begin{itemize}
\item For $c < 0$, the condition $k' > k_F$ gives $\varpi > \mu/v_0$ and the upper bound $\Sigma \leq k_F + \sqrt{(\varpi - \mu/v_0)/|c|}$.
\item For $c > 0$, the condition $k' > k_F$ gives, for $\varpi < \mu/v_0$, the lower bound $\Sigma \geq k_F + \sqrt{(\mu/v_0 - \varpi)/c}$. The occupation of the band $s = 0$ gives the upper bound $\Sigma \leq k_F^{(0)} + \varkappa \left( \mu/v_0 + \varpi \right)$.
\end{itemize}
The largest lower bound and the smallest upper bound define $\Sigma_-$ and $\Sigma_+$, and the contribution vanishes when $\Sigma_+ \leq \Sigma_-$. We assume $2\,|c|\,q < 1$, and $\varpi < 1/(4\,|c|)$ for $c < 0$. For $c \to 0$ and $\varpi > k_F$, Eq.~\eqref{eq:qf-im01} reduces to Eq.~\eqref{eq:fin-Imb}.

The integrand of Eq.~\eqref{eq:qf-im01} is a proper rational function of $\Sigma$ with simple poles, and the integral is elementary. We obtain
\begin{align}
\label{eq:qf-im01-cf}
& \Im \Pi_{01} (q, \omega) = -\,\frac{g}{32\,\pi\,v_0\,q} \left[ \mathcal{G}_{01} (\Sigma_+) - \mathcal{G}_{01} (\Sigma_-) \right], \quad
\mathcal{G}_{01} (\Sigma) = \varrho_{\rm W}\,\ln \left| 1 + 2\,c\,\Sigma \right| + \sum_{\check\sigma = \pm 1} \varrho_{\check\sigma}\,\ln \left| \Sigma - \Sigma_{+\check\sigma} \right| + \varrho_{\rm q}\,\ln \left| c\,\Sigma^2 + \varpi \right|,
\nn & \varrho_{\rm W} = \frac{1 - 4\,c^2\,q^2}{2\,c^3}\,, \quad
\varrho_{\check\sigma} = -\,\frac{\left( \Sigma_{+\check\sigma}^2 - q^2 \right)^2}{\check\sigma\,\check{r}_+\,\Sigma_{+\check\sigma}}\,, \quad
\varrho_{\rm q} = \frac{\left( c\,q^2 + \varpi \right)^2}{2\,c^2\,\varpi}\,,
\end{align}
where $\Sigma_{+\check\sigma}$ and $\check{r}_+$ are defined below Eq.~\eqref{eq:qf-im11-cf}. The roots of $c\,\Sigma^2 + \varpi$ are imaginary for $c > 0$ and real for $c < 0$. They have equal residues, and the two corresponding terms therefore combine into the logarithm of $c\,\Sigma^2 + \varpi$. No arctangent arises for $c > 0$. The logarithms are regular on the range of integration, because the product of the three denominator factors of Eq.~\eqref{eq:qf-im01} is positive there. The terms of $\mathcal{G}_{01}$ that are singular for $c \to 0$ cancel among the logarithms.


For $c > 0$, the band $s = 0$ is empty for $k' > k_F^{(0)}$, and the band $s = -1$ is filled at every $k$. The overlap factor is $\mathcal{F}_{-1,0} = \sin^2\theta/2$. With $\Sigma = k + k'$ and $\check{D} = k' - k$, the energy condition $\varepsilon_0 (k') - \varepsilon_{-1} (k) = v_0\,\varpi$ is linear in $\check{D}$, and gives $ \check{D} = \frac{\Sigma - 2\,\varpi}{1 - 2\,c\,\Sigma}$.
We assume $2\,c\,q < 1$ and $\varpi < 1/(4\,c)$. The constraint $|\check{D}| \leq q$ then requires $\Sigma < 1/(2\,c)$, and therefore $1 - 2\,c\,\Sigma > 0$. We obtain
\begin{align}
\label{eq:qf-imm10}
\Im \Pi_{-1,0} (q, \omega) = -\,\frac{g}{32\,\pi\,v_0\,q} \int_{\Sigma_-}^{\Sigma_+} d\Sigma\; \frac{\left( \Sigma^2 - q^2 \right) \left[ q - 2\,\varpi + \left( 1 - 2\,c\,q \right) \Sigma \right] \left[ q + 2\,\varpi - \left( 1 + 2\,c\,q \right) \Sigma \right]}{\left( 1 - 2\,c\,\Sigma \right) \left( \Sigma - c\,\Sigma^2 - \varpi \right) \left( \varpi - c\,\Sigma^2 \right)}\,.
\end{align}
The limits follow from the constraints on the pair $(k, k')$. The condition $|\check{D}| \leq q$ gives $(2\,\varpi - q)/(1 - 2\,c\,q) \leq \Sigma \leq (2\,\varpi + q)/(1 + 2\,c\,q)$, and the triangle inequality gives $\Sigma \geq q$. The condition $k' > k_F^{(0)}$ gives $\Sigma \geq k_F^{(0)} - \varkappa \left( \mu/v_0 - \varpi \right)$ for $\varpi > \mu/v_0$, and no transition exists for $\varpi \leq \mu/v_0$. The limits $\Sigma_-$ and $\Sigma_+$ are defined by these bounds as before. For $c \to 0$, we have $k_F^{(0)} \to \infty$ and the contribution vanishes, in agreement with Sec.~\ref{secfincont}.

The integrand of Eq.~\eqref{eq:qf-imm10} is that of Eq.~\eqref{eq:qf-im01} with $c$ replaced by $-\,c$. The integral is therefore elementary, and
\begin{align}
\label{eq:qf-imm10-cf}
& \Im \Pi_{-1,0} (q, \omega) 
= -\,\frac{g}{32\,\pi\,v_0\,q} \left[ \mathcal{G}_{-1,0} (\Sigma_+) - \mathcal{G}_{-1,0} (\Sigma_-) \right], \nn &
\mathcal{G}_{-1,0} (\Sigma) 
= -\,\varrho_{\rm W}\,\ln \left| 1 - 2\,c\,\Sigma \right| + \sum_{\check\sigma = \pm 1} \tilde\varrho_{\check\sigma}\,\ln \left| \Sigma + \Sigma_{-\check\sigma} \right| 
+ \tilde\varrho_{\rm q}\,\ln \left| \varpi - c\,\Sigma^2 \right|,
\nn & \tilde\varrho_{\check\sigma} = \frac{\left( \Sigma_{-\check\sigma}^2 - q^2 \right)^2}{\check\sigma\,\check{r}_-\,\Sigma_{-\check\sigma}}\,, \quad
\tilde\varrho_{\rm q} = \frac{\left( \varpi - c\,q^2 \right)^2}{2\,c^2\,\varpi}\,.
\end{align}
Here $\varrho_{\rm W}$ is given below Eq.~\eqref{eq:qf-im01-cf}, and $\Sigma_{-\check\sigma}$ and $\check{r}_-$ are given in Eq.~\eqref{eq:qf-im11-cf}. The assumption on $\varpi$ makes $\check{r}_-$ real. The numbers $-\Sigma_{-\check\sigma}$ are the roots of $c\,\Sigma^2 - \Sigma + \varpi$. The three denominator factors of Eq.~\eqref{eq:qf-imm10} are positive on the range of integration and the logarithms are regular.


The band $s = -1$ is filled at every $k$, and the final state is in the band $s = 1$ with $k' > k_F$. The overlap factor is $\mathcal{F}_{-1,1} = (1 - \cos\theta)^2/4$. With $\Sigma = k + k'$ and $\check{D} = k' - k$, the energy condition $\varepsilon_1 (k') - \varepsilon_{-1} (k) = v_0\,\varpi$ becomes $\Sigma \left( 1 + c\,\check{D} \right) = \varpi$. It fixes $\Sigma = \varpi/(1 + c\,\check{D})$, and we use $\check{D}$ as the integration variable. We obtain
\begin{align}
\label{eq:qf-imm11}
\Im \Pi_{-1,1} (q, \omega) = -\,\frac{g}{32\,\pi\,v_0\,q} \int_{\check{D}_-}^{\check{D}_+} d\check{D}\; \frac{\left( 1 + c\,\check{D} \right) \left( \check{D}^2 - q^2 \right)^2}{\varpi^2 - \left( \check{D} + c\,\check{D}^2 \right)^2}\,.
\end{align}
The limits follow from the constraints on the pair $(k, k')$. The triangle inequality gives $|\check{D}| \leq q$ and $\Sigma \geq q$. The latter is equivalent to $c\,q\,\check{D} \leq \varpi - q$, which is a bound on $\check{D}$ whose type depends on the sign of $c$.
\begin{itemize}
\item For $c < 0$, it is a lower bound.
\item For $c > 0$, it is an upper bound.
\end{itemize}
The condition $k' > k_F$ gives $\check{D} \geq k_F + \varkappa \left( \mu/v_0 - \varpi \right)$. This bound lies below $-q$ and is therefore inactive for $\varpi \geq \left( q + 2\,k_F \right) \left( 1 - c\,q \right)$, for both signs of $c$. Above this line, Pauli-blocking has no effect on the channel, and $\Im \Pi_{-1,1}$ equals its undoped value of Eq.~\eqref{eq:qf-imvac}. For $c = 0$, the line is $\varpi = q + 2\,k_F$. The largest lower bound and the smallest upper bound define $\check{D}_-$ and $\check{D}_+$ as before. We assume $2\,|c|\,q < 1$, and $\varpi < 1/(4\,|c|)$ for both signs of $c$. For $c \to 0$, Eq.~\eqref{eq:qf-imm11} reduces to Eq.~\eqref{eq:fin-Imc}.

The integrand of Eq.~\eqref{eq:qf-imm11} is that of Eq.~\eqref{eq:qf-im11} with $\Sigma$ replaced by $\check{D}$ and with the sign of the denominator reversed. The integral is therefore elementary, and
\begin{align}
\label{eq:qf-imm11-cf}
\Im \Pi_{-1,1} (q, \omega) = \frac{g}{32\,\pi\,v_0\,q} \left[ \mathcal{G}_1 (\check{D}_+) - \mathcal{G}_1 (\check{D}_-) \right],
\end{align}
where $\mathcal{G}_1$, $\Sigma_{\tau\check\sigma}$, $\check{r}_\tau$, and $\bar\varrho_{\tau\check\sigma}$ are those of Eq.~\eqref{eq:qf-im11-cf}. The assumption on $\varpi$ makes all $\check{r}_\tau$ real for both signs of $c$. The logarithms are regular on the range of integration, because the denominator of Eq.~\eqref{eq:qf-imm11} equals $\left( 1 + c\,\check{D} \right)^2 \left( \Sigma^2 - \check{D}^2 \right)$, which is positive there.

For $c > 0$ and $\varpi > 1/(4\,c)$, the quantity $\check{r}_- = i\sqrt{4\,c\,\varpi - 1}$ is imaginary, and the roots $\Sigma_{-,\pm}$ are complex conjugates of each other. We then assume $\varpi < \left( 1 - c\,q \right)^2/c$, which keeps $k$ below $1/(2\,c)$, the position of the extremum of $\varepsilon_{-1}$. Eq.~\eqref{eq:qf-imm11-cf} remains valid, with $\mathcal{G}_1$ replaced by
\begin{align}
\label{eq:qf-imm11-cf2}
\mathcal{G}_1 (\check{D}) = \frac{\check{D}^2}{2\,c} - \frac{\check{D}}{c^2} + \sum_{\check\sigma = \pm 1} \bar\varrho_{+\check\sigma}\,\ln \left| \check{D} - \Sigma_{+\check\sigma} \right| + \Re\bar\varrho_{-+}\,\ln \!\left( \frac{c\,\check{D}^2 + \check{D} + \varpi}{c} \right) - 2\,\Im\bar\varrho_{-+}\,\arctan \!\left( \frac{1 + 2\,c\,\check{D}}{\sqrt{4\,c\,\varpi - 1}} \right).
\end{align}
Here $\bar\varrho_{-+}$ is given by Eq.~\eqref{eq:qf-im11-cf} with the imaginary $\check{r}_-$, and $\bar\varrho_{--} = \bar\varrho_{-+}^*$. For $\varpi > \mu/v_0 + 1/(4\,c)$, the condition $k' > k_F$ is fulfilled for every pair, and the lower limit $k_F + \varkappa \left( \mu/v_0 - \varpi \right)$ is absent.


For $c > 0$, the initial state is in the band $s = 1$ with $k < k_F$, and the final state is in the band $s = 0$ with $k' > k_F^{(0)}$. The overlap factor is $\mathcal{F}_{10} = \sin^2\theta/2$. With $\Sigma = k + k'$ and $\check{D} = k' - k$, the energy condition $\varepsilon_0 (k') - \varepsilon_1 (k) = v_0\,\varpi$ is linear in $\check{D}$, and gives $\check{D} = \frac{\Sigma + 2\,\varpi}{1 + 2\,c\,\Sigma}$.
If we assume $2\,c\,q < 1$, the energy obeys $\varpi \leq c\,q^2 < 1/(4\,c)$, yielding
\begin{align}
\label{eq:qf-im10}
& \Im \Pi_{10} (q, \omega) = \frac{-\,g}{32\,\pi\,v_0\,q} 
\int_{\Sigma_-}^{\Sigma_+} d\Sigma\; \frac{\left( \Sigma^2 - q^2 \right) 
\left[ q - 2\,\varpi - \left( 1 - 2\,c\,q \right) \Sigma \right] 
\left[ q + 2\,\varpi + \left( 1 + 2\,c\,q \right) \Sigma \right]}
{\left( 1 + 2\,c\,\Sigma \right) \left( c\,\Sigma^2 + \Sigma + \varpi \right) 
\left( c\,\Sigma^2 - \varpi \right)}\,, \nn
& \Sigma_+ = \min \lbrace (q - 2\,\varpi)/(1 - 2\,c\,q), k_F 
+ \sqrt{\left( \mu/v_0 + \varpi \right)/c} \rbrace \,, 
\quad \Sigma_- = \max \lbrace q, k_F^{(0)} 
+ \varkappa \left( \mu/v_0 - \varpi \right) \rbrace\,. 
\end{align}
The second entry of the maximum is absent for $\varpi > \mu/v_0$, where $k' > k_F^{(0)}$ is automatic. The contribution vanishes when $\Sigma_+ \leq \Sigma_-$, which reproduces the onset of Eq.~\eqref{eq:qf-th-10}.
The integrand of Eq.~\eqref{eq:qf-im10} is that of Eq.~\eqref{eq:qf-im01} with $\varpi$ replaced by $-\,\varpi$. The integral is therefore elementary, and
\begin{align}
\label{eq:qf-im10-cf}
& \Im \Pi_{10} (q, \omega) = -\,\frac{g}{32\,\pi\,v_0\,q} \left[ \mathcal{G}_{10} (\Sigma_+) - \mathcal{G}_{10} (\Sigma_-) \right], \quad
\mathcal{G}_{10} (\Sigma) = \varrho_{\rm W}\,\ln \left| 1 + 2\,c\,\Sigma \right| + \sum_{\check\sigma = \pm 1} \hat\varrho_{\check\sigma}\,\ln \left| \Sigma - \Sigma_{-\check\sigma} \right| + \hat\varrho_{\rm q}\,\ln \left| c\,\Sigma^2 - \varpi \right|,
\nn & \hat\varrho_{\check\sigma} = 
-\,\frac{\left( \Sigma_{-\check\sigma}^2 - q^2 \right)^2}{\check\sigma\,\check{r}_-\,\Sigma_{-\check\sigma}}\,, \quad
\hat\varrho_{\rm q} = -\,\frac{\left( c\,q^2 - \varpi \right)^2}{2\,c^2\,\varpi}\,.
\end{align}
Here $\varrho_{\rm W}$ is given below Eq.~\eqref{eq:qf-im01-cf}, and $\Sigma_{-\check\sigma}$ and $\check{r}_-$ are given in Eq.~\eqref{eq:qf-im11-cf}. The quantity $\check{r}_-$ is real since $\varpi < 1/(4\,c)$. The three denominator factors of Eq.~\eqref{eq:qf-im10} are positive on the range of integration, and the logarithms are regular.

\subsubsection{Total spectral function and absorption}
\label{secqfimtot}

For $\omega > 0$, the imaginary part is the sum of the channels with a nonzero phase space. These are the intraband transitions of the top band, the transitions $0 \to 1$ and $-1 \to 1$, and, for $c > 0$, the transitions $1 \to 0$, $-1 \to 0$, and the intraband transitions of the flat-band. We obtain
\begin{align}
\label{eq:qf-imtot}
\Im \Pi (q, \omega) = \Im \Pi_{11} + \Im \Pi_{01} + \Im \Pi_{-1,1} + \Theta (c) \left[ \Im \Pi_{10} + \Im \Pi_{-1,0} + \Im \Pi_{00} \right].
\end{align}
The terms are given in Eqs.~\eqref{eq:qf-im11-cf}, \eqref{eq:qf-im01-cf}, \eqref{eq:qf-imm11-cf}, \eqref{eq:qf-im10-cf}, \eqref{eq:qf-imm10-cf}, and~\eqref{eq:qf-im00-cf}. Each term vanishes outside its range. The ranges are those of Eq.~\eqref{eq:qf-edges} for $\Im\Pi_{11}$, the onsets of Eqs.~\eqref{eq:qf-th-neg} and~\eqref{eq:qf-th-pos} for $\Im\Pi_{01}$, the onset of Eq.~\eqref{eq:qf-th-m1} for $\Im\Pi_{-1,1}$, the onsets of Eqs.~\eqref{eq:qf-th-10} and~\eqref{eq:qf-th-m0} for $\Im\Pi_{10}$ and $\Im\Pi_{-1,0}$, and the edges of Eq.~\eqref{eq:qf-edges0} for $\Im\Pi_{00}$. For $c < 0$, the band $s = 0$ is filled at every $k$, and the three channels with final states in this band are blocked. For $c \to 0$, we have $k_F^{(0)} \to \infty$, and the terms in the bracket vanish. The first three terms reduce to $\Im\Pi^{\rm L}_{11}$, $\Im\Pi^{\rm L}_{01}$, and $\Im\Pi^{\rm L}_{-1,1}$ of Eqs.~\eqref{eq:fin-Ima}--\eqref{eq:fin-Imc}, respectively. The result is shown in Figs.~\ref{fig:qtsm-plasmon-static}(a) and~\ref{fig:qtsm-cpos-plasmon-static}(a) for $c < 0$ and $c > 0$, respectively. The position of the Kohn singularity is fixed by the FS and does not depend on $\Lambda$, whereas the plasmon branches, the static background, and the large-$q$ behaviour depend on $\Lambda$.

\subsection{Real part of polarisability and RPA dielectric function}
\label{secqfre}

The real part of the polarisability follows from the imaginary part of Sec.~\ref{secqfimtot} by causality, as in Sec.~\ref{secfinre} for $c = 0$. We keep the regularisation of Sec.~\ref{secfinite}, in which only the transitions with an energy below $\mathcal{R} = v_0\,\Lambda$ are retained. The retarded function is then analytic in the upper half of the complex $\omega$-plane and decays at large $|\omega|$, and its imaginary part is odd in $\omega$. Hence
\begin{align}
\label{eq:qf-kk}
\Re \Pi (q, \omega) = \frac{2}{\pi}\,\mathrm{P} \!\! \int_0^{\mathcal{R}} d\omega'\, \frac{\omega'\,\Im \Pi (q, \omega')}{\omega'^2 - \omega^2}\,,
\end{align}
with $\Im\Pi$ of Eq.~\eqref{eq:qf-imtot}. This relation is exact within the regularisation, but it does not display the dependence on the doping, since $\Im\Pi$ extends up to $\mathcal{R}$. For $c = 0$, this difficulty was removed in Eq.~\eqref{eq:fin-kk} by subtracting the intrinsic reference. For $c \neq 0$, the reference must be the undoped QTSM with the same value of $c$, because the energies and the occupations depend on $c$.

We denote by $\Pi_{\rm vac}^{(c)}$ the polarisability of the undoped QTSM, which is obtained for $\mu \to 0^+$ at fixed $c$ and fixed $\mathcal{R}$. Its occupations are those of Sec.~\ref{secquadintr}, and only the channels with a nonzero phase space survive. These depend on the sign of $c$.
\begin{itemize}
\item For $c < 0$, the bands $s = -1$ and $s = 0$ are filled and the band $s = 1$ is empty. The channels $0 \to 1$ and $-1 \to 1$ survive, whereas the channel $-1 \to 0$ is blocked.
\item For $c > 0$, the band $s = -1$ is filled and the bands $s = 0$ and $s = 1$ are empty. The channels $-1 \to 0$ and $-1 \to 1$ survive.
\end{itemize}
Both cases are summarised by
\begin{align}
\label{eq:qf-imvac}
\Im \Pi_{\rm vac}^{(c)} (q, \omega) = \left[ \Im \Pi_{-1,1} + \Theta (-\,c)\,\Im \Pi_{01} + \Theta (c)\,\Im \Pi_{-1,0} \right]_{\mu \to 0^+}.
\end{align}
The subscript means that the integration limits of Eqs.~\eqref{eq:qf-imm11-cf}, \eqref{eq:qf-im01-cf}, and~\eqref{eq:qf-imm10-cf} are evaluated with $k_F \to 0$ and, for $c > 0$, with $k_F^{(0)} \to 0$. The doping-induced part of the imaginary part is
\begin{align}
\label{eq:qf-dim}
\Delta \Im \Pi (q, \omega) = \Im \Pi (q, \omega) - \Im \Pi_{\rm vac}^{(c)} (q, \omega)\,,
\end{align}
which has a finite support. Its upper bound is set by
\begin{align}
\label{eq:qf-omc}
\omega_{\rm w} = 2\,v_0 \left( k_m + q \right) \left( 1 + |c|\,q \right), \quad
k_m = \begin{cases} k_F & \text{ for } c \leq 0 \\ k_F^{(0)} & \text{ for } c > 0 \end{cases}.
\end{align}
The momentum $k_m$ is the largest Fermi momentum of the system. For $c \leq 0$, only the band $s = 1$ has an FS. For $c > 0$, the flat band has an FS as well, and $k_F < k_F^{(0)}$. We then have $\Delta \Im \Pi (q, \omega) = 0$ for $\omega > \omega_{\rm w}$. We assume $\omega_{\rm w} < \mathcal{R}$, which holds when $k_m + q$ lies within the range of the linearised description. The real part is therefore
\begin{align}
\label{eq:qf-re}
\Re \Pi (q, \omega) = \Re \Pi_{\rm vac}^{(c)} (q, \omega) + \frac{2}{\pi}\,\mathrm{P} \!\! \int_0^{\omega_{\rm w}} d\omega'\, \frac{\omega'\,\Delta \Im \Pi (q, \omega')}{\omega'^2 - \omega^2}\,,
\end{align}
where the integral is a finite window, as in Eq.~\eqref{eq:fin-kk}. The undoped part is related to the intrinsic polarisability $\Pi^{(0)}$ of the linear model, whose real part is given in Eq.~\eqref{eq:fin-Re0} for $\omega > v_0\,q$ and whose imaginary part is the sum of Eqs.~\eqref{eq:fin-Imb} and~\eqref{eq:fin-Imc} at $k_F = 0$. We obtain
\begin{align}
\label{eq:qf-revac}
\Re \Pi_{\rm vac}^{(c)} (q, \omega) = \Re \Pi^{(0)} (q, \omega) + \frac{2}{\pi}\,\mathrm{P} \!\! \int_0^{\mathcal{R}} d\omega'\, \frac{\omega' \left[ \Im \Pi_{\rm vac}^{(c)} (q, \omega') - \Im \Pi^{(0)} (q, \omega') \right]}{\omega'^2 - \omega^2}\,.
\end{align}
Both integrals are principal-value integrals of the closed forms of Sec.~\ref{sec-qtsm-full}, and each is evaluated by a single quadrature. The first term of Eq.~\eqref{eq:qf-re} is independent of the doping and carries the dependence on the cut-off, whereas the doping enters only through the finite window. The splitting is merely a reorganisation of Eq.~\eqref{eq:qf-kk} and does not change the result. For $c \to 0$, Eq.~\eqref{eq:qf-imtot} reduces to the imaginary part of Sec.~\ref{secfincont}, and Eq.~\eqref{eq:qf-kk} then reduces to the relation from which Eq.~\eqref{eq:fin-kk} was obtained.



\subsection{Static polarisability and screening}
\label{secqfstatic}

At $\omega = 0$, Eq.~\eqref{eq:qf-kk} gives $\Pi (q, 0) = (2/\pi) \int_0^{\mathcal{R}} d\omega'\,\Im\Pi (q, \omega')/\omega'$, with $\Im\Pi$ of Eq.~\eqref{eq:qf-imtot}. We give its behaviour for $q \ll k_F$ and for large $q$, and we express the dielectric function at the singular wavevectors of the following subsubsections through this integral. For $q \ll k_F$, we obtain
\begin{align}
\label{eq:qf-static}
\Pi (q, 0) = -\,\rho_{\rm tot} - \frac{g\,q^2}{4\,\pi^2\,v_0} \left[ \frac{4}{3} \ln \!\left( \frac{\Lambda}{k_F} \right) - \mathcal{S} \right] + \dots, \quad
\mathcal{S} = \frac{30\,c^2\,k_F^2 + 30\,c\,k_F + 7}{9\,\beta^3} + \Theta (c)\,\frac{3}{4\,\sqrt{\xi}}\,,
\end{align}
with $\rho_{\rm tot}$ of Eq.~\eqref{eq:q-static}, where the dots stand for terms of higher order in $q$. The logarithm comes from the interband transitions, and it is the same for both signs of $c$. The term $\mathcal{S}$ collects the intraband terms of order $q^2$.
\begin{itemize}
\item For $c < 0$, only the top band has an FS. It gives the first term of $\mathcal{S}$, which depends on $c\,k_F$.
\item For $c > 0$, the flat-band has an FS as well. It adds the term proportional to $1/\sqrt{\xi}$.
\end{itemize}
The static dielectric function of Eq.~\eqref{eq:epsrpa} takes the Thomas-Fermi form
\begin{align}
\label{eq:qf-tf}
\epsilon^{\rm RPA} (q, 0) \simeq \epsilon_b^{\rm st} + \frac{q_{\rm TF}^2}{q^2}\,, \quad
\epsilon_b^{\rm st} = \epsilon_b - \frac{\alpha}{\pi}\,\mathcal{S}\,,
\end{align}
with $q_{\rm TF}$ of Eq.~\eqref{eq:q-tf} and $\epsilon_b$ of Eq.~\eqref{eq:q-eb}. The limit $c \to 0$ depends on the sign of $c$.
\begin{itemize}
\item For $c \to 0^-$, we have $\mathcal{S} \to 7/9$, and Eq.~\eqref{eq:qf-tf} reduces to Eq.~\eqref{eq:fin-tf}.
\item For $c \to 0^+$, the term $\mathcal{S}$ diverges as $1/\sqrt{\xi}$, in the same way as $\rho_0$, because the FS of the flat-band moves to $k_F^{(0)} \to \infty$.
\end{itemize}

For $q \gg k_F$ and $q \gg k_F^{(0)}$ for $c > 0$, the occupied states have $k \ll q$, and the doped system approaches the intrinsic one. Let $\Pi_{\rm vac}^{(c)} (q, 0)$ be the static polarisability of the system with $\mu \to 0^+$ at fixed $c$ of Sec.~\ref{secqfre}. Under the conditions $|c|\,q^2 \gg k_F$ and $|c|\,q < 1$, we obtain
\begin{align}
\label{eq:qf-large}
\Pi (q, 0) = \Pi_{\rm vac}^{(c)} (q, 0) + \frac{g \left[ k_F^3 + \Theta (c) \left( k_F^{(0)} \right)^3 \right]}{9\,\pi^2\,v_0}\,\frac{3\,c^2\,q^2 - 1}{c\,q^2 \left( 1 - c^2\,q^2 \right)} \left[ 1 + \mathcal{O} (\delta) \right].
\end{align}
The relative correction $\delta$ is of the order of $k_F^{(0)}/q$ for $c > 0$, $k_F/q$ for $c < 0$, and $k_F/\left( |c|\,q^2 \right)$. The leading term vanishes at $c\,q = 1/\sqrt{3}$, where the corrections dominate. For $c\,q \ll 1$, the correction is $-\,g \left[ k_F^3 + \Theta (c) \left( k_F^{(0)} \right)^3 \right]/(9\,\pi^2\,v_0\,c\,q^2)$, and it decays as $q^{-2}$. In the linear model, the transitions $1 \to 0$ have the energy $v_0\,k$ instead of $c\,v_0\,q^2$, and they give the constant $g\,k_F^2/(6\,\pi^2\,v_0)$ of Eq.~\eqref{eq:fin-large}. The conditions of Eq.~\eqref{eq:qf-large} exclude this regime, and the limits $c \to 0$ and $q \to \infty$ do not commute.

For $c > 0$, the intraband part of the flat-band has a closed form for every $q$, up to one radial integral. The band has the spectrum $\varepsilon_0 = c\,v_0\,k^2$ and the overlap factor $\mathcal{F}_{00} = \cos^2\theta$, and it is filled for $k < k_F^{(0)}$. We obtain
\begin{align}
\label{eq:qf-pi00}
\Pi_{00} (q, 0) = -\,\frac{g}{16\,\pi^2\,c\,v_0\,q} \int_0^{k_F^{(0)}} \frac{dk}{k}\,\mathcal{B} (k, q)\,, \quad
\mathcal{B} (k, q) = 4\,k\,q - \frac{2 \left( k^2 - q^2 \right)^2}{k^2} \ln \!\left| \frac{k + q}{k - q} \right| + \frac{\left( 2\,k^2 - q^2 \right)^2}{k^2} \ln \!\left| \frac{2\,k + q}{2\,k - q} \right|.
\end{align}
The integrand has integrable logarithmic singularities at $k = q/2$ and $k = q$, and it behaves as $k^2$ for $k \to 0$. The remaining channels of $\Pi (q, 0)$ follow from the integral of Eq.~\eqref{eq:qf-kk} with $\Im\Pi_{00}$ omitted. For $q \ll k_F^{(0)}$, Eq.~\eqref{eq:qf-pi00} reduces to the flat-band terms of Eq.~\eqref{eq:qf-static}, and the term linear in $q$ vanishes. The intermediate steps, including the proof of this vanishing, are given in Appendix~\ref{app:qf-static}.

The static dielectric function is shown in Figs.~\ref{fig:qtsm-plasmon-static}(b) and~\ref{fig:qtsm-cpos-plasmon-static}(b), together with its Thomas-Fermi form of Eq.~\eqref{eq:qf-tf}. The dielectric function at the wavevectors $q = 2\,k_F$, $q = 2\,k_F^{(0)}$, and $q = q_2$, which enter the Friedel amplitudes below, follows from $\epsilon^{\rm RPA} (q, 0) = 1 - 4\,\pi\,e^2\,\Pi (q, 0)/q^2$ with the integral above and Eq.~\eqref{eq:qf-pi00}.

\subsection{Friedel oscillations}
\label{secqffried}

  \subsubsection{Kohn singularity and Friedel oscillations of the top band}
\label{secqfkohn1}

The intraband transitions of the top band are non-analytic at $q = 2\,k_F$ for both signs of $c$. The overlap factor $\mathcal{F}_{11}$ vanishes quartically for backscattering, as in the linear model. The quadratic correction enters the static response only through the factor $1 + c\,(k + k')$ of the energy denominator, which equals $\beta$ at backscattering. We find
\begin{align}
\label{eq:qf-kohn1}
\Pi (q, 0) \simeq \Pi_{\rm reg} (q) - \frac{g}{24\,\pi^2\,\beta\,v_0\,k_F} \left( q - 2\,k_F \right)^3 \ln \!\left( \left| q - 2\,k_F \right| \right),
\end{align}
where $\Pi_{\rm reg}$ is analytic. This is Eq.~\eqref{eq:fin-kohn} with $v_0$ replaced by the Fermi velocity $V_F = \beta\,v_0$. The third derivative of $\Pi (q, 0)$ diverges logarithmically at $q = 2\,k_F$, as shown in Figs.~\ref{fig:qtsm-plasmon-static}(b) and~\ref{fig:qtsm-cpos-plasmon-static}(b).

The overlap factors of the interband channels are discontinuous at $k' = 0$, and each channel $1 \to s'$ has a term of the order $\left( q - k_F \right)^3 \ln \left| q - k_F \right|$ at $q = k_F$. These terms cancel in the sum over $s'$, since the three bands are degenerate at $k' = 0$ and $\sum_{s'} \mathcal{F}_{1 s'} = 1$. The total $\Pi (q, 0)$ is therefore analytic at $q = k_F$. For $c > 0$, the same holds for the channels $0 \to s'$ at $q = k_F^{(0)}$.

For a static potential $\mathcal{U} (q)$ and $\delta n (q) = \Pi (q, 0)\,\mathcal{U} (q)/\epsilon^{\rm RPA} (q, 0)$, the induced density at distances $k_F\,r \gg 1$ is
\begin{align}
\label{eq:qf-friedel1}
\delta n (r) \simeq -\,\frac{g}{4\,\pi^3\,\beta\,v_0}\,\frac{\mathcal{U} (2\,k_F)}{\left[ \epsilon^{\rm RPA} (2\,k_F, 0) \right]^2}\,\frac{\cos \!\left( 2\,k_F\,r \right)}{r^5}\,.
\end{align}
For a Coulomb impurity of charge $Z\,e$, with $\mathcal{U} (q) = -\,4\,\pi\,Z\,e^2/q^2$, the amplitude is $g\,Z\,e^2/(4\,\pi^2\,\beta\,v_0\,k_F^2\,\left[ \epsilon^{\rm RPA} (2\,k_F, 0) \right]^2)$. The oscillations have the period $\pi/k_F$ and decay as $r^{-5}$.
\begin{itemize}
\item For $c < 0$, the top band has the only FS, and this is the leading Friedel term.
\item For $c > 0$, the flat-band term of Eq.~\eqref{eq:qf-friedel0} decays more slowly and dominates at large distances, as shown in Fig.~\ref{fig:friedel-all}.
\end{itemize}

\subsubsection{Kohn singularity and Friedel oscillations of the flat-band for $c > 0$}
\label{secqfkohn0}

The static polarisability is non-analytic where transitions connect states at the Fermi energy. For $c > 0$, the transitions within the flat-band connect states on the FS of radius $k_F^{(0)}$, and $\Pi (q, 0)$ is non-analytic at $q = 2\,k_F^{(0)}$. The overlap factor $\mathcal{F}_{00} = \cos^2\theta$ equals unity for backscattering, in contrast to $\mathcal{F}_{11}$, which vanishes quartically. The flat-band therefore has the Kohn singularity of an ordinary parabolic band with the mass $m = 1/(2\,c\,v_0)$,
\begin{align}
\label{eq:qf-kohn0}
\Pi (q, 0) \simeq \Pi_{\rm reg} (q) - \frac{g}{16\,\pi^2\,c\,v_0} \left( q - 2\,k_F^{(0)} \right) \ln \!\left( \left| q - 2\,k_F^{(0)} \right| \right),
\end{align}
where $\Pi_{\rm reg}$ is analytic, up to corrections of the order $\left( q - 2\,k_F^{(0)} \right)^2 \ln \left| q - 2\,k_F^{(0)} \right|$. No other channel is non-analytic at this momentum, since no other pair of states at the Fermi energy is connected by $q = 2\,k_F^{(0)}$. The transitions of the top band are singular at $q = 2\,k_F$, as treated in Sec.~\ref{secqfkohn1}. The transitions between the two FSs give a weaker singularity at $q_2$, which is treated in Sec.~\ref{secqfkohn01}.

The singularity of Eq.~\eqref{eq:qf-kohn0} gives the induced density, at distances $k_F^{(0)}\,r \gg 1$,
\begin{align}
\label{eq:qf-friedel0}
\delta n (r) \simeq \frac{g\,k_F^{(0)}}{16\,\pi^3\,c\,v_0}\,\frac{\mathcal{U} \left( 2\,k_F^{(0)} \right)}{\left[ \epsilon^{\rm RPA} \left( 2\,k_F^{(0)}, 0 \right) \right]^2}\,\frac{\cos \!\left( 2\,k_F^{(0)}\,r \right)}{r^3}\,.
\end{align}
For a Coulomb impurity, we have $c\,k_F^{(0)} = \sqrt{\xi}$, and the amplitude is $-\,g\,Z\,e^2/(16\,\pi^2\,v_0\,\sqrt{\xi}\,\left[ \epsilon^{\rm RPA} (2\,k_F^{(0)}, 0) \right]^2)$. The oscillations have the period $\pi/k_F^{(0)}$ and decay as $r^{-3}$, as in an ordinary electron gas. This is slower than the $r^{-5}$ decay of the linear model, and the flat-band dominates the Friedel oscillations at large distances.

\subsubsection{Kohn singularities from transitions between two Fermi surfaces for $c > 0$}
\label{secqfkohn01}

For $c > 0$, the transitions $0 \to 1$ and $1 \to 0$ connect the two FSs at zero energy for $q_1 \leq q \leq q_2$, with $q_1 = k_F^{(0)} - k_F$ and $q_2 = k_F^{(0)} + k_F$. At the end points of this interval, $\bs{q}$ connects the two spheres only at a single pair of collinear points, and the static polarisability can be non-analytic. At these points, the overlap factor $\mathcal{F}_{01} = \sin^2\theta/2$ vanishes. It is proportional to the distance of the pair $(k, k')$ from the edge of the allowed region, in contrast to the unit overlap of the flat-band at $q = 2\,k_F^{(0)}$. The overlap factor therefore raises the power of the singularity by one. For $q$ close to $q_2$, we obtain
\begin{align}
\label{eq:qf-kohn01}
\Pi (q, 0) \simeq \Pi_{\rm reg} (q) + \frac{g}{2\,\pi^2\,v_0 \left( 1 + 2\,c\,q_2 \right)} \left( q - q_2 \right)^2 \ln \!\left( \left| q - q_2 \right| \right),
\end{align}
where $\Pi_{\rm reg}$ is analytic. The factor $1 + 2\,c\,q_2$ is the sum of the velocities of the two bands at their FSs in units of $v_0$, which are $2\,c\,k_F^{(0)}$ and $\beta$. It equals $\left( \sqrt{c\,k_F} + \sqrt{1 + c\,k_F} \right)^2$. At $q_1$, the terms of the same order cancel between the two sides of the FSs, and the singularity at $q_2$ is the only one of this order. The singularity of Eq.~\eqref{eq:qf-kohn01} gives the induced density, at distances $k_F\,r \gg 1$,
\begin{align}
\label{eq:qf-friedel01}
\delta n (r) \simeq \frac{g\,q_2}{2\,\pi^3\,v_0 \left( 1 + 2\,c\,q_2 \right)}\,\frac{\mathcal{U} (q_2)}{\left[ \epsilon^{\rm RPA} (q_2, 0) \right]^2}\,\frac{\sin \!\left( q_2\,r \right)}{r^4}\,.
\end{align}
For a Coulomb impurity, the amplitude is $-\,2\,g\,Z\,e^2/\left[ \pi^2\,v_0\,q_2 \left( 1 + 2\,c\,q_2 \right) \left[ \epsilon^{\rm RPA} (q_2, 0) \right]^2 \right]$. The oscillations have the period $2\,\pi/\left( k_F + k_F^{(0)} \right)$ and decay as $r^{-4}$. This decay is faster than the $r^{-3}$ decay of Eq.~\eqref{eq:qf-friedel0} and slower than the $r^{-5}$ decay of Eq.~\eqref{eq:qf-friedel1}.


\subsection{Plasmon dispersion and Landau damping}
\label{secqfplasmon}

We consider $q \ll k_F$ and $\omega < \mu_1$, with $\omega$ above the upper edge $\varpi_+ (q)$ of Eq.~\eqref{eq:qf-edges}. For $c > 0$, we also assume $k_F^{(0)} < \Lambda$, as in Sec.~\ref{secquadflat}. The edge $\varpi_+$ tends to $\beta\,q$ for $q \to 0$, and the lowest interband onset tends to $k_F$. For small $q$, this window is therefore free of excitations, and the expansion of $\Pi$ contains only even powers of $q$, $\Pi = q^2\,\Pi_2 + q^4\,\Pi_4 + \dots$. We obtain
\begin{align}
\label{eq:qf-pi24}
\Pi_2 = \frac{g}{6\,\pi^2\,v_0} \left[ \frac{\beta\,k_F^2 + 2\,\sqrt{\xi}\,
\Theta (c) \left( k_F^{(0)} \right)^2}{\varpi^2} - \ln \!\left( \frac{\Lambda^2}{k_F^2 - \varpi^2} \right) \right], \quad
\Pi_4 = \frac{g}{\pi^2\,v_0\,\varpi^2} \left[ \mathcal{B}_1 + \Theta (c)\,\mathcal{B}_0 \right],
\end{align}
with
\begin{align}
\label{eq:qf-b1-b0}
\mathcal{B}_1 &= \frac{\beta^3\,k_F^2}{10\,\varpi^2} - \frac{3 + 4\,c\,k_F}{60} + \frac{1}{30} 
\ln \!\left( \frac{4\,k_F^2 - \varpi^2}{4 \left( k_F^2 - \varpi^2 \right)} \right) 
- \frac{2\,c^2\,\varpi^2}{15} \ln \!\left( \frac{\Lambda^2}{k_F^2 - \varpi^2} \right) 
 - \frac{2\,c^2\,k_F^2\,\varpi^2 \left( k_F^2 + \varpi^2 \right)}{15 \left( k_F^2 - \varpi^2 \right)^2} 
 - \frac{c\,k_F \left( 1 + 4\,c\,k_F \right) \varpi^2}{15 \left( k_F^2 - \varpi^2 \right)}\,, \nn 
\mathcal{B}_0 &= -\,\frac{2\,\sqrt{\xi}}{15} + \frac{4\,\xi^{3/2} \left( k_F^{(0)} \right)^2}
{5\,\varpi^2} + \frac{4\,\sqrt{\xi}\,\varpi^2 \left[ \left( k_F^{(0)} \right)^2 + \varpi^2 \right]}{15 \left[ \left( k_F^{(0)} \right)^2 - \varpi^2 \right]^2} + \frac{2\,\sqrt{\xi}\,\varpi^2}{15 \left[ \left( k_F^{(0)} \right)^2 - \varpi^2 \right]}\,.
\end{align}
The cut-off is implemented as in Sec.~\ref{secfinite}. The logarithm in Eq.~\eqref{eq:qf-pi24} proportional to $c^2\,k_F^2$ depends on the cut-off, and the constant that accompanies it is the one of this prescription. The limit $c \to 0$ depends on the sign of $c$.
\begin{itemize}
\item For $c \to 0^-$, we have $c\,k_F \to 0$ and $\beta \to 1$, and Eq.~\eqref{eq:qf-pi24} reduces to Eq.~\eqref{eq:fin-pi24}.
\item For $c \to 0^+$, we have $\mathcal{B}_0 \to 0$, because $\sqrt{\xi} \to 0$ and $\xi^{3/2} \left( k_F^{(0)} \right)^2 \to 0$ in this limit.
\end{itemize}
The first term of $\Pi_2$ is the Drude term of Eq.~\eqref{eq:q-wp0}, and the logarithm is the interband screening of Eq.~\eqref{eq:q-eb}.

Let $\epsilon_b (\omega) = 1 + (2\,\alpha/(3\,\pi)) \ln \!\left( v_0^2\,\Lambda^2/(\mu_1^2 - \omega^2) \right)$ and $\Phi (\omega) = \epsilon_b (\omega) + (2\,\alpha/(3\,\pi))\,\omega^2/(\mu_1^2 - \omega^2)$. The zero of $\Re\epsilon^{\rm RPA} = 1 - 4\,\pi\,e^2 \left( \Pi_2 + q^2\,\Pi_4 \right)$ obeys $\omega^2\,\epsilon_b (\omega) - \omega_p^2 = \left( 4\,\alpha\,v_0^2\,q^2/\pi \right) \left[ \mathcal{B}_1 + \Theta (c)\,\mathcal{B}_0 \right]$, with $\omega_p$ of Eq.~\eqref{eq:q-wp0}. To order $q^2$, it is
\begin{align}
\label{eq:qf-wdisp}
\omega_*^2 (q) = \omega_*^2 (0) + \frac{4\,\alpha\,v_0^2\,q^2}{\pi}\,\frac{\mathcal{B}_1 + \Theta (c)\,\mathcal{B}_0}{\Phi (\omega_*)} \Bigg|_{\omega_* = \omega_* (0)}\,,
\end{align}
where $\omega_* (0)$ is the unique zero below the onset of Secs.~\ref{secquadonly} and~\ref{secquadflat}, and $\mathcal{B}_{1,0}$ are evaluated at $\varpi = \omega_* (0)/v_0$. Since $\Lambda > k_F$ gives $\epsilon_b > 1$, we have $\Phi > 0$, and the zero is an analytic function of $q^2$ near $q = 0$. For small $q$, the zero therefore lies in the window above, and the plasmon is undamped.

For $\omega_p \ll \mu_1$, the terms of $\mathcal{B}_{1,0}$ proportional to $1/\varpi^2$ dominate, and Eq.~\eqref{eq:qf-wdisp} reduces to
\begin{align}
\label{eq:qf-bg}
\omega_*^2 (q) = \omega_p^2 + \frac{3}{5}\,V_{\rm BG}^2\,q^2\,, \quad
V_{\rm BG}^2 = v_0^2\,\frac{\beta^3\,k_F^2 + 8\,\xi^{3/2}\,\Theta (c) \left( k_F^{(0)} \right)^2}{\beta\,k_F^2 + 2\,\sqrt{\xi}\,\Theta (c) \left( k_F^{(0)} \right)^2}\,.
\end{align}
This is the Bohm-Gross dispersion \cite{bohm-gross}, which describes how the frequency of electrostatic plasma waves depends on their wavenumber. The velocity $V_{\rm BG}$ is the average of $V^2$ over the FSs, weighted with the Drude weights. Its value depends on the sign of $c$.
\begin{itemize}
\item For $c < 0$, it equals the band-velocity $V_F = \beta\,v_0$ at the FS of the top band.
\item For $c > 0$, it lies between $V_0 = 2\,\sqrt{\xi}\,v_0$ and $V_F$.
\end{itemize}
For $c \to 0$, it reduces to $v_0$ and Eq.~\eqref{eq:qf-bg} to the Bohm-Gross form of Sec.~\ref{secfinplasmon}.

For larger $q$, the plasmon leaves the free region when $\omega_* (q)$ reaches one of its boundaries, namely the edge $\omega = v_0\,\varpi_+ (q)$ of Eq.~\eqref{eq:qf-edges} or the lowest interband onset of Eqs.~\eqref{eq:qf-th-neg}--\eqref{eq:qf-th-m0}. We define $q_{\rm max}$ as the wavevector beyond which this relative distance is below $2 \times 10^{-3}$, or at which the zero leaves the free region through the lowest interband onset, whichever occurs first. The zero is found from Eq.~\eqref{eq:qf-re}. The solution and its expansion in Eq.~\eqref{eq:qf-wdisp} are shown in Fig.~\ref{fig:qtsm-plasmon-static}(a) for $c < 0$. For $c > 0$, Fig.~\ref{fig:qtsm-cpos-plasmon-static}(a) shows no undamped branch. For $c > 0$, no free region exists for $q_1 \leq q \leq q_2$, and hence $q_{\rm max} \leq q_1 = k_F^{(0)} - k_F$. In this range, the two FSs are connected by transitions of zero energy.

For $q > q_{\rm max}$, the zero of $\Re\epsilon^{\rm RPA}$ lies inside the continuum, where $\Im\epsilon^{\rm RPA} = -\,(4\,\pi\,e^2/q^2)\,\Im\Pi \neq 0$ with $\Im \Pi$ of Eq.~\eqref{eq:qf-imtot}. The mode is Landau damped with the rate $\gamma = \Im\epsilon^{\rm RPA}/\partial_\omega \Re\epsilon^{\rm RPA}$ at $\omega = \Re\,\omega_*$, which holds when $\gamma$ is small compared to the scale on which $\Im\epsilon^{\rm RPA}$ and $\Re\epsilon^{\rm RPA}$ vary. The damping is due to the following transitions.
\begin{itemize}
\item The intraband transitions of the top band for $\omega < v_0\,\varpi_+ (q)$, and for $c > 0$ also of the flat-band.
\item The interband transitions above the lowest onset.
\item For $c > 0$ and $q_1 \leq q \leq q_2$, the transitions between the two FSs, which add a damping channel at all frequencies.
\end{itemize}

\section{HSM at arbitrary momentum-transfers}
\label{sechfull}

The analysis of Secs.~\ref{seclinear} and~\ref{secquad} is restricted to long wavelengths, with $q \ll k_F$, where the overlap factors can be expanded in $q$. The same regime was treated for the HSM in Ref.~\cite{ips-vnr-plasmon}, where the plasmon was derived at small $q$. This expansion excludes the region $q \simeq 2\,k_F$, in which the static response is non-analytic, and wavevectors comparable to $k_F$, where the plasmon meets the particle-hole continua. In this section, we lift the restriction for the HSM. We keep the linear model of Eq.~\eqref{eq:hhopf} with $\mu > 0$ and $k_F = \mu/v_0$, and we impose no relation between $q$ and $k_F$ other than requiring both to be much smaller than the cut-off $\Lambda$. As in Sec.~\ref{secfinite}, the cut-off is implemented as an upper limit $\mathcal{R} = v_0\,\Lambda$ on the transition energy of the interband transitions. The overlap factors of the HSM, which are given exactly in Sec.~\ref{sec:hsm}, depend on the direction of $\bs{q}$ only through a single parameter, and the TSM of Sec.~\ref{secfinite} is recovered when this parameter vanishes, as commented on at the end of this section. We first obtain the imaginary part of the polarisability together with the particle-hole continua (Sec.~\ref{sechdamp}) and then its real part (Sec.~\ref{sechre}). The static polarisability with the screening and the Friedel oscillations follow in Secs.~\ref{sechstatic} and~\ref{sechfried}, and the plasmon is discussed in Sec.~\ref{sechplasmon}. The steps behind the results are collected in Appendix~\ref{app:hfull}.

For $\mu > 0$, the bands $-$ and $0$ are filled and the band $+$ is partially filled. Using Eq.~\eqref{eq:pol} and relabelling the integration variables, we obtain in Appendix~\ref{app:hfull} the decomposition
\begin{align}
\label{eq:h-split}
\Pi (\bs{q}, \omega) = 2\,g \int_{k < k_F} \frac{d^3 k}{(2\,\pi)^3}\; \frac{\mathcal{F}_{++}\,X_+}{X_+^2 - (\omega + i\,0^+)^2}
- 2\,g \int_{k > k_F} \frac{d^3 k}{(2\,\pi)^3} \sum_{s' \in \lbrace 0, - \rbrace} \frac{\mathcal{F}_{+ s'}\,X_{s'}\,\Theta (v_0\,\Lambda - X_{s'})}{X_{s'}^2 - (\omega + i\,0^+)^2}\,,
\end{align}
with $X_+ = v_0\,(k - k')$, $X_0 = v_0\,k$, and $X_- = v_0\,(k + k')$. The first term is the intraband response of the Fermi sea. The second term is the interband response, with the Pauli-blocked region $k < k_F$ removed. The step function in the second term implements the cut-off on the transition energy, as in Sec.~\ref{secfinite}.

\begin{figure}[t!]
\centering
\subfigure[]{\includegraphics[width=0.75\linewidth]{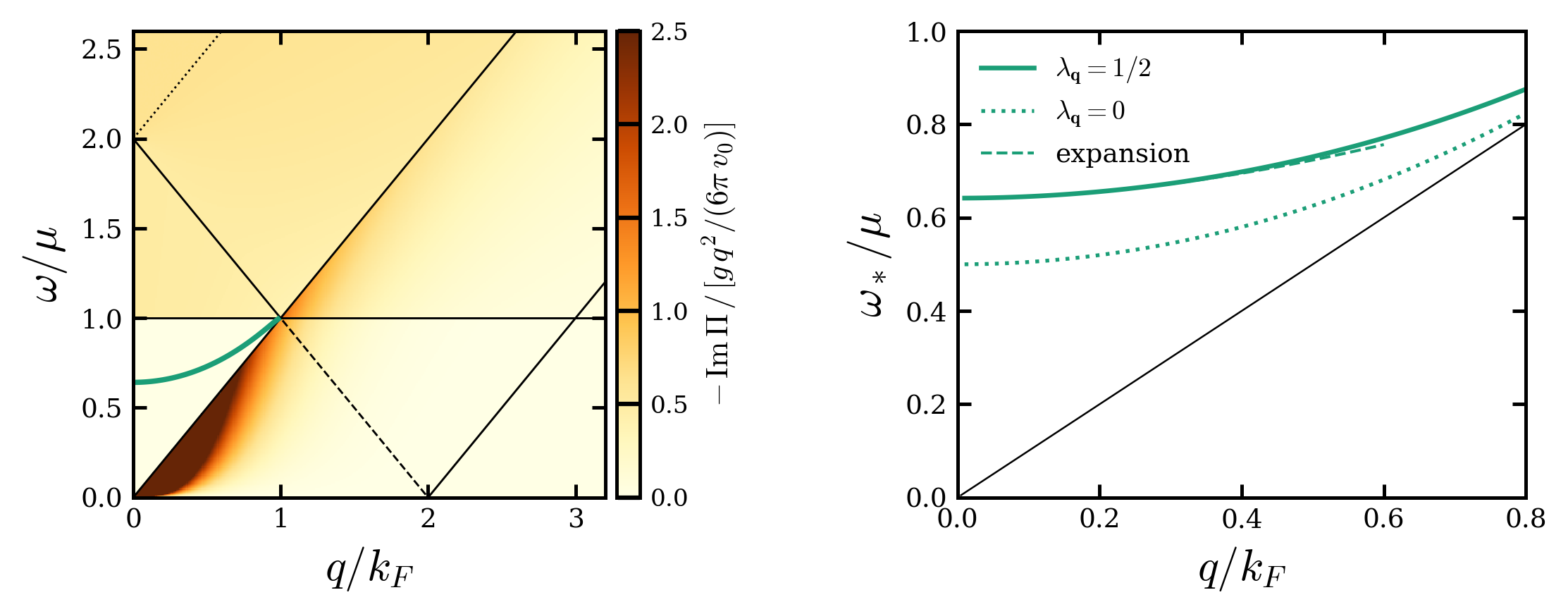}}
\subfigure[]{\includegraphics[width=0.75\linewidth]{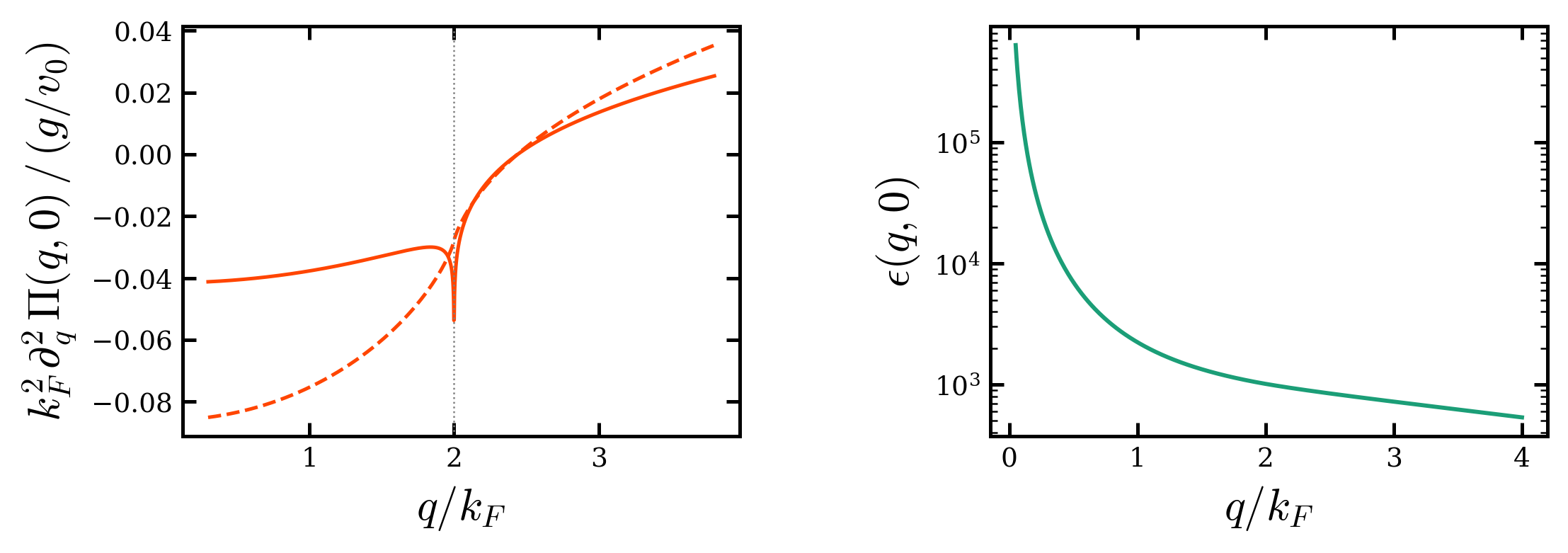}}
\caption{\label{fig:hsm-plasmon-static} HSM characteristics for $\lambda_{\bs{q}} = 1/2$ (in-plane). (a) $-\,\Im\Pi$ normalised by $g\,q^2/(6\,\pi\,v_0)$, with the undamped plasmon branch (left), and the plasmon branches (right). (b) Second derivative of $\Pi (q, 0)$ (left) and static dielectric function (right). Parameters are $\mu = 0.01\,\text{eV}$ and $\lambda_{\bs{q}} = 1/2$.}
\end{figure}

Fig.~\ref{fig:hsm-plasmon-static}(a) shows $-\,\Im\Pi$ from Eq.~\eqref{eq:h-impi} for $\lambda_{\bs{q}} = 1/2$, normalised by $g\,q^2/(6\,\pi\,v_0)$, together with the undamped plasmon branch. The branch is drawn up to $q_{\rm max}$, beyond which the zero of $\Re\epsilon^{\rm RPA}$ lies within $2 \times 10^{-3}\,v_0\,q$ of the edge $\omega = v_0\,q$. The solid black lines are the edges of the continuum, namely $\omega = v_0\,q$, $\omega = v_0 \left( q - 2\,k_F \right)$ for $q > 2\,k_F$, $\omega = \mu$, and $\omega = 2\,\mu - v_0\,q$ for $q < k_F$. The dashed line $\omega = 2\,\mu - v_0\,q$ for $k_F < q < 2\,k_F$ marks a kink of the intraband weight, and the dotted line $\omega = v_0\,q + 2\,\mu$ marks the point beyond which $\Im\Pi$ has reached its intrinsic value. The right panel shows the branches from Eq.~\eqref{eq:h-kk} for $\lambda_{\bs{q}} = 1/2$ (solid) and $\lambda_{\bs{q}} = 0$ (dotted), the latter being the TSM and the direction $\bs{\hat z}$, together with the expansion of Eq.~\eqref{eq:h-wdisp} for $\lambda_{\bs{q}} = 1/2$ (dashed), which coincides with the solid curve at small $q$ and separates at larger $q$. The black lines are $\omega = v_0\,q$ and $\omega = \mu$. Fig.~\ref{fig:hsm-plasmon-static}(b) shows the second derivative of $\Pi (q, 0)$ from Eq.~\eqref{eq:h-pistat}, scaled by $k_F^2\,v_0/g$, for $\lambda_{\bs{q}} = 1/2$ (solid) and $\lambda_{\bs{q}} = 0$ (dashed). The solid curve diverges logarithmically towards negative values at $q = 2\,k_F$ as in Eq.~\eqref{eq:h-sing}, whereas the dashed curve is finite. The right panel shows the static dielectric function from Eq.~\eqref{eq:h-pistat} (solid) and the Thomas-Fermi form of Eq.~\eqref{eq:h-tf} (dashed), which lies on top of the solid curve over the range of $q$ in which it is drawn. The parameters are $\mu = 0.01\,\text{eV}$ and $\lambda_{\bs{q}} = 1/2$, which give $k_F = \mu/v_0 = 25\,\text{eV}$ and $v_0\,\Lambda/\mu = 6.4$, and hence $L = \ln 6.4$.

\subsection{Imaginary part of polarisability: Particle-hole continua and their edges}
\label{sechdamp}

The imaginary part of $\Pi$ follows from Eq.~\eqref{eq:h-split} and Eq.~\eqref{eq:h-fl}. We define $\varpi = \omega/v_0 > 0$ and $\lambda = \lambda_{\bs{q}}$. The three contributions are
\begin{align}
\label{eq:h-impi}
& \Im \Pi_{++} = -\,\frac{g}{4\,\pi\,v_0\,q} \left[ \mathcal{P}_{++} (k_F) - \mathcal{P}_{++} (\check{k}_-) \right] \Theta (q - \varpi)\,\Theta (\varpi - q + 2\,k_F)\,, \nn
& \Im \Pi_{0+} = -\,\frac{g\,(1 - \lambda)}{32\,\pi\,v_0\,q\,\varpi} \left[ 2\,q\,\varpi\,(\varpi^2 + q^2) - (\varpi^2 - q^2)^2 \ln \!\left| \frac{\varpi + q}{\varpi - q} \right| \right] \Theta (\varpi - k_F)\,, \nn
& \Im \Pi_{-+} = -\,\frac{g}{4\,\pi\,v_0\,q} \left[ \mathcal{P}_{-+} (k_2) - \mathcal{P}_{-+} (k_1) \right] \Theta (\varpi - q)\,\Theta (\varpi - 2\,k_F + q)\,,
\end{align}
where $\check{k}_- = \max \lbrace k_F - \varpi, (q - \varpi)/2 \rbrace$, $k_1 = (\varpi - q)/2$, $k_2 = \min \lbrace (\varpi + q)/2, \varpi - k_F \rbrace$, and
\begin{align}
\label{eq:h-pab}
& \mathcal{P}_{++} (k) = \frac{k^3}{3} + \frac{\varpi\,k^2}{2} - \frac{(2 - \lambda)\,(q^2 - \varpi^2)\,k}{4} + \frac{(1 - \lambda)\,(q^2 - \varpi^2)^2}{16\,\varpi}\,\ln \!\left( \frac{k}{k + \varpi} \right), \nn
& \mathcal{P}_{-+} (k) = \frac{\varpi\,k^2}{2} - \frac{k^3}{3} - \frac{(2 - \lambda)\,(\varpi^2 - q^2)\,k}{4} + \frac{(1 - \lambda)\,(\varpi^2 - q^2)^2}{16\,\varpi}\,\ln \!\left( \frac{k}{\varpi - k} \right).
\end{align}
The imaginary part of the dielectric function is $\Im \epsilon^{\rm RPA} = -\,(4\,\pi\,e^2/q^2)\,\Im \Pi$, which is non-negative. For $q \to 0$, Eq.~\eqref{eq:h-impi} gives $\Im \Pi_{0+} \to -\,g\,(1 - \lambda)\,q^2/(6\,\pi\,v_0)$ and $\Im \Pi_{-+} \to -\,g\,\lambda\,q^2/(24\,\pi\,v_0)$. Hence
\begin{align}
\label{eq:h-imeps0}
\Im \epsilon^{\rm RPA} (\bs{q} \to 0, \omega) = \frac{2\,g\,e^2\,(1 - \lambda)}{3\,v_0}\,\Theta (\omega - \mu) + \frac{g\,e^2\,\lambda}{6\,v_0}\,\Theta (\omega - 2\,\mu)\,,
\end{align}
which reproduces the long-wavelength results quoted in Sec.~\ref{seccomphsm} for $\lambda = 0$ and $\lambda = 1/2$.

The supports of the three terms in the $(q, \omega)$ plane follow from the step functions in Eq.~\eqref{eq:h-impi}. The intraband continuum $\Im \Pi_{++}$ occupies $\max \lbrace 0, v_0\,(q - 2\,k_F) \rbrace < \omega < v_0\,q$. Its upper edge is the line $\omega = v_0\,q$ for all $q$, and its lower edge vanishes for $q < 2\,k_F$. The continuum of the transitions $0 \to +$ starts at $\omega = \mu$ for every $q$, since the flat-band is filled at all momenta. The continuum of the transitions $- \to +$ starts at $\omega = \max \lbrace v_0\,q, 2\,\mu - v_0\,q \rbrace$. At $q = k_F$, the edge $\omega = v_0\,q$ of the intraband continuum and the onsets of both interband continua coincide at $\omega = \mu$.

In the intrinsic limit $k_F \to 0$, only $\Pi_{0+}$ and $\Pi_{-+}$ survive, and the step function in $\Im \Pi_{0+}$ becomes unity. For $\varpi > q$, the logarithms of the two terms cancel, and we find
\begin{align}
\label{eq:h-Im0}
\Im \Pi^{(0)} (\bs{q}, \omega) = -\,\frac{g\,(4 - 3\,\lambda)\,q^2}{24\,\pi\,v_0}\,,
\end{align}
which is independent of $\varpi/q$. The prefactor is the sum of the long-wavelength weights $(1 - \lambda)/6$ and $\lambda/24$ of the two interband terms in Eq.~\eqref{eq:h-imeps0}. For $\varpi < q$, only the transitions $0 \to +$ remain, and
\begin{align}
\label{eq:h-Im0b}
\Im \Pi^{(0)} (\bs{q}, \omega) = -\,\frac{g\,(1 - \lambda)}{32\,\pi\,v_0\,q\,\varpi} \left[ 2\,q\,\varpi\,(\varpi^2 + q^2) - (\varpi^2 - q^2)^2 \ln \!\left( \frac{q + \varpi}{q - \varpi} \right) \right].
\end{align}
This is $(1 - \lambda)$ times the TSM expression of Eq.~\eqref{eq:fin-Imb}. The transitions $- \to +$ set in at $\varpi = q$ with a finite step of $-\,g\,q^2/(24\,\pi\,v_0)$. The step is independent of $\lambda$, because the overlap factor $\mathcal{F}_{+-}$ of Eq.~\eqref{eq:h-fl} equals unity at $\cos\theta = -1$, which corresponds to two states with opposite momenta. At this point, the transitions $0 \to +$ contribute $-\,g\,(1 - \lambda)\,q^2/(8\,\pi\,v_0)$, and the two contributions add up to Eq.~\eqref{eq:h-Im0}.

In the doped system, the imaginary part reaches its intrinsic value of Eq.~\eqref{eq:h-Im0} for $\varpi > q + 2\,k_F$. In this range, $\Im \Pi_{++}$ vanishes, the onset of $\Im \Pi_{0+}$ lies below $\varpi$, and the upper limit $k_2$ of $\Im \Pi_{-+}$ equals $(\varpi + q)/2$. The point $\varpi = q + 2\,k_F$ is thus the upper end of the region in which Pauli-blocking modifies the imaginary part, and it is the same for every direction of $\bs{q}$. Only the intrinsic value itself depends on $\lambda$.

Only two regions in the $(q, \omega)$ plane are free of single-particle excitations, namely $q < \varpi < k_F$ for $q < k_F$, and $\varpi < \min \lbrace k_F, q - 2\,k_F \rbrace$ for $q > 2\,k_F$. These regions follow from the supports of Eq.~\eqref{eq:h-impi}, and they do not depend on $\lambda$. The first region hosts the plasmon of Sec.~\ref{sechplasmon}. For $k_F \leq q \leq 2\,k_F$, the intraband continuum reaches $\varpi \to 0$ and the continuum of the transitions $0 \to +$ covers every $\varpi > q$. Hence there is no region free of excitations in this range of $q$. The particle-hole continua and the plasmon branches are shown in Fig.~\ref{fig:hsm-plasmon-static}(a).

\subsection{Real part of polarisability and RPA dielectric function}
\label{sechre}

The real part of $\Pi$ follows from the dispersion relation, with the cut-off $\mathcal{R} = v_0\,\Lambda$ on the transition energy. It requires the intrinsic imaginary part of Eq.~\eqref{eq:h-Im0} and the saturation of the doped imaginary part above $\varpi = q + 2\,k_F$, both of which are given in Sec.~\ref{sechdamp}. The difference between $\Im \Pi$ and its intrinsic reference vanishes above $\omega_{\rm w} = v_0\,q + 2\,\mu$, and the dispersion relation reduces to a finite window,
\begin{align}
\label{eq:h-kk}
\Re \Pi (\bs{q}, \omega) = -\,\frac{g\,(4 - 3\,\lambda)\,q^2}{24\,\pi^2\,v_0}\,\ln \!\left( \frac{v_0^2\,\Lambda^2}{|\omega^2 - v_0^2\,q^2|} \right)
+ \frac{2}{\pi}\,\mathrm{P} \!\! \int_0^{\omega_{\rm w}} d\omega'\, \frac{\omega' \left[ \Im \Pi (\bs{q}, \omega') + \dfrac{g\,(4 - 3\,\lambda)\,q^2}{24\,\pi\,v_0}\,\Theta (\omega' - v_0\,q) \right]}{\omega'^2 - \omega^2}\,.
\end{align}
For $\lambda = 0$, this is Eq.~\eqref{eq:fin-kk}. For $\omega > v_0\,q$ and $k_F \to 0$, the window integral is nonzero only for $\omega' < v_0\,q$, where the integrand is $\Im \Pi_{0+}^{(0)}$. It is thus $(1 - \lambda)$ times the corresponding TSM integral. With $W = \omega/(v_0\,q)$, the intrinsic polarisability is
\begin{align}
\label{eq:h-Re0}
\Re \Pi^{(0)} (\bs{q}, \omega) = -\,\frac{g\,(4 - 3\,\lambda)\,q^2}{24\,\pi^2\,v_0}\,\ln \!\left( \frac{v_0^2\,\Lambda^2}{\omega^2 - v_0^2\,q^2} \right) - \frac{g\,(1 - \lambda)\,q^2}{16\,\pi^2\,v_0}\,\mathcal{K} (W)\,,
\end{align}
where $\mathcal{K} (W)$ is the function of Eq.~\eqref{eq:fin-Re0}. Since $\mathcal{K}$ vanishes for $W \to \infty$, the first term gives the $\ln \!\left( v_0\,\Lambda/\omega \right)$ dependence of the intrinsic response at long wavelengths, with the coefficient $(4 - 3\,\lambda)$ of Eq.~\eqref{eq:h-tf}. The direction of $\bs{q}$ enters through the prefactors $(4 - 3\,\lambda)$ and $(1 - \lambda)$ only.

The doping-induced remainder is $\Re \Pi - \Re \Pi^{(0)}$. We write $\Im \Pi_{0+}^{(0)}$ and $\Im \Pi_{-+}^{(0)}$ for the expressions of Eq.~\eqref{eq:h-impi} at $k_F \to 0$. The remainder is then
\begin{align}
\label{eq:h-dop}
\Re \Pi^{\rm dop} (\bs{q}, \omega) = \frac{2}{\pi}\,\mathrm{P} \!\! \int_0^{\omega_{\rm w}} d\omega'\, \frac{\omega' \left\lbrace \Im \Pi_{++} - \Theta (k_F - \varpi')\,\Im \Pi_{0+}^{(0)} + \Theta (\varpi' - q) \left[ \Im \Pi_{-+} - \Im \Pi_{-+}^{(0)} \right] \right\rbrace}{\omega'^2 - \omega^2}\,.
\end{align}
The three terms in the braces are supported on $\max \lbrace 0, q - 2\,k_F \rbrace < \varpi' < q$, on $\varpi' < k_F$, and on $q < \varpi' < q + 2\,k_F$, respectively. All of them lie inside the window. The remainder is therefore a single quadrature of the elementary functions of Eq.~\eqref{eq:h-impi}.

At $\omega = 0$, the intrinsic part of Eq.~\eqref{eq:h-kk} gives the static polarisability of the undoped system,
\begin{align}
\label{eq:h-pivac}
\Pi_{\rm vac} (\bs{q}, 0) = -\,\frac{g\,q^2}{2\,\pi^2\,v_0} \left[ \frac{4 - 3\,\lambda}{6}\,\ln \!\left( \frac{\Lambda}{q} \right) + b_\lambda \right],
\end{align}
with
\begin{align}
\label{eq:h-b}
b_\lambda = \frac{(1 - \lambda)\,(4\,\ln 2 - 1)}{6}\,.
\end{align}
The coefficient of the logarithm does not depend on the way in which the cut-off is imposed. The constant $b_\lambda$ does, and Eq.~\eqref{eq:h-b} holds for the cut-off on the transition energy. For $\lambda = 0$, Eq.~\eqref{eq:h-b} reproduces the large-$q$ limit of Eq.~\eqref{eq:fin-large}.

\subsection{Static polarisability and screening}
\label{sechstatic}

The static limit $\omega = 0$ can be done in closed form. We introduce the function
\begin{align}
\label{eq:h-phi}
\Psi_\lambda (q, k_F) = -\,\frac{2\,k_F^2}{3\,q^2} - \frac{1 - \lambda}{4} + \frac{4 - 3\,\lambda}{12}\,\ln \!\left| 1 - \frac{q^2}{4\,k_F^2} \right| - \frac{q}{48\,k_F} \left[ \frac{16\,k_F^4}{q^4} + \frac{4\,(3\,\lambda - 6)\,k_F^2}{q^2} + 3\,\lambda - 3 \right] \ln \!\left| \frac{2\,k_F + q}{2\,k_F - q} \right|.
\end{align}
We then find, for a direction of $\bs{q}$ with $\lambda = \lambda_{\bs{q}}$,
\begin{align}
\label{eq:h-pistat}
\Pi (\bs{q}, 0) = \frac{g\,q^2}{2\,\pi^2\,v_0} \left[ \Psi_\lambda (q, k_F) - \frac{4 - 3\,\lambda}{6}\,\ln \!\left( \frac{\Lambda}{2\,k_F} \right) - b_\lambda \right].
\end{align}
Here $b_\lambda$ is the constant of Eq.~\eqref{eq:h-b}. Another regularisation changes it by a constant and leaves every nonanalytic feature unchanged. The terms containing $\Lambda$ and $b_\lambda$ are proportional to $q^2$ and cancel in all differences used below, and the function $\Psi_\lambda (q, k_F)$ is continuous at $q = 2\,k_F$.

The limits of Eq.~\eqref{eq:h-pistat} are as follows. For $q \ll k_F$, we have $\Psi_\lambda (q, k_F) = -\,k_F^2/q^2 + 2/9 + \mathcal{O} (q^2/k_F^2)$, and
\begin{align}
\label{eq:h-pilw}
\Pi (\bs{q}, 0) = -\,\frac{g\,k_F^2}{2\,\pi^2\,v_0} - \frac{g\,q^2}{2\,\pi^2\,v_0} \left[ \frac{4 - 3\,\lambda}{6}\,\ln \!\left( \frac{\Lambda}{2\,k_F} \right) + b_\lambda - \frac{2}{9} \right] + \mathcal{O} \left( \frac{q^4}{k_F^2} \right).
\end{align}
The first term is $-\,\rho (\mu)$, with $\rho (\mu) = g\,\mu^2/(2\,\pi^2\,v_0^3)$ of Eq.~\eqref{eq:lin-dos}, and the second term gives the background of Sec.~\ref{seclinscreen}. The static dielectric function $\epsilon^{\rm RPA} (\bs{q}, 0) = 1 - 4\,\pi\,e^2\,\Pi/q^2$ then takes the Thomas-Fermi form $\epsilon_b (\lambda) + q_{\rm TF}^2/q^2$ with
\begin{align}
\label{eq:h-tf}
& q_{\rm TF}^2 = \frac{2\,g\,e^2\,k_F^2}{\pi\,v_0} = \frac{3\,\omega_p^2}{v_0^2}\,, \quad
\nn & \epsilon_b (\lambda) = 
1 + \frac{2\,g\,e^2}{\pi\,v_0} 
\left[ \frac{4 - 3\,\lambda}{6} \ln \!\left( \frac{\Lambda}{2\,k_F} \right) + b_\lambda - \frac{2}{9} \right]
= 1 + \frac{(4 - 3\,\lambda)\,g\,e^2}{3\,\pi\,v_0}
\ln \!\left( \frac{v_0\,\Lambda}{\mu} \right) - \frac{g\,e^2}{\pi\,v_0} 
\left[ \frac{7}{9} - \frac{\lambda\,(1 - \ln 2)}{3} \right],
\end{align}
where the second form uses Eq.~\eqref{eq:h-b}. For $\lambda = 0$, it coincides with the background of Eq.~\eqref{eq:fin-tf}. For $\lambda = 0$ and $\lambda = 1/2$, the coefficient of $\ln \!\left( v_0\,\Lambda/\mu \right)$ in $\epsilon_b$ is $4\,g\,e^2/(3\,\pi\,v_0)$ and $5\,g\,e^2/(6\,\pi\,v_0)$, respectively, which are the values of the HSM at long wavelengths~\cite{ips-vnr-plasmon}. For $q \gg k_F$, Eq.~\eqref{eq:h-phi} gives $\Psi_\lambda (q, k_F) + (4 - 3\,\lambda)\,\ln \!\left( 2\,k_F/q \right)/6 = (1 - \lambda)\,k_F^2/(3\,q^2) + \mathcal{O} (k_F^4/q^4)$. With the intrinsic response of Eq.~\eqref{eq:h-pivac}, we have $\Pi (\bs{q}, 0) - \Pi_{\rm vac} (\bs{q}, 0) \to g\,(1 - \lambda)\,k_F^2/(6\,\pi^2\,v_0)$. The doping then only adds a constant to the intrinsic response. The static polarisability and the static dielectric function, together with the Thomas-Fermi form, are shown in Fig.~\ref{fig:hsm-plasmon-static}(b).

\subsection{Friedel oscillations}
\label{sechfried}

The non-analytic behaviour of $\Pi (\bs{q}, 0)$ determines the oscillations of the induced density at large distances. The function $\Psi_\lambda (q, k_F)$ contains the term $(q - 2\,k_F)^2 \left[ 4\,k_F^2 + 4\,k_F\,q + (3\,\lambda - 3)\,q^2 \right] \ln \!\left| (q - 2\,k_F)/q \right|/(48\,k_F\,q^3)$, which is the only source of non-analyticity (see Appendix~\ref{app:hfull-friedel}). Near $q = 2\,k_F$, it gives
\begin{align}
\label{eq:h-sing}
\Pi_{\rm sing} (\bs{q}, 0) = \frac{g\,(q - 2\,k_F)^2}{2\,\pi^2\,v_0} \left[ \frac{\lambda}{8} - \frac{4 - 3\,\lambda}{24}\,\frac{q - 2\,k_F}{q} + \mathcal{O} \left( (q - 2\,k_F)^2 \right) \right] \ln \!\left| \frac{q - 2\,k_F}{q} \right|\,.
\end{align}
For a generic direction of $\bs{q}$, the leading singularity is therefore $(q - 2\,k_F)^2\,\ln \!\left| q - 2\,k_F \right|$, with the amplitude $g\,\sin^2\vartheta_{\bs{q}}/(32\,\pi^2\,v_0)$. For $\bs{q} \parallel \bs{\hat z}$, the amplitude of this term vanishes, and the leading singularity is $(q - 2\,k_F)^3\,\ln \!\left| q - 2\,k_F \right|$. Since $\lambda_{\bs{q}} = 0$ along the $z$-axis, Eq.~\eqref{eq:h-sing} gives
\begin{align}
\label{eq:h-kohnz}
\Pi_{\rm sing} (q\,\bs{\hat z}, 0) = -\,\frac{g}{24\,\pi^2\,v_0\,k_F}\,(q - 2\,k_F)^3\,\ln \!\left| q - 2\,k_F \right|\,,
\end{align}
which is the Kohn singularity of the TSM in Eq.~\eqref{eq:fin-kohn}. For a generic direction, the second derivative of $\Pi (\bs{q}, 0)$ with respect to $q$ diverges logarithmically at $q = 2\,k_F$, with the divergent part $g\,\sin^2\vartheta_{\bs{q}}\,\ln \!\left| q - 2\,k_F \right|/(16\,\pi^2\,v_0)$, as shown in Fig.~\ref{fig:hsm-plasmon-static}(b). Along the $z$-axis, the second derivative is finite, and the third derivative diverges logarithmically, as in the TSM. By comparison, the Lindhard function of a parabolic band has the singularity $(q - 2\,k_F)\,\ln \!\left| q - 2\,k_F \right|$. The weaker singularity of the HSM reflects the suppression of backscattering in the band $+$, since $\mathcal{F}_{++} \to \Upsilon^2/4$ for $\bs{k}' \to -\,\bs{k}$, and this vanishes quadratically in the deviation from backscattering.

We consider a static point charge $Z\,e$ at the origin, with the bare potential energy $\mathcal{U} (q) = -\,4\,\pi\,Z\,e^2/q^2$ of an electron. Within the RPA, the induced density is $\delta n (\bs{q}) = \Pi (\bs{q}, 0)\,\mathcal{U} (q)/\epsilon^{\rm RPA} (\bs{q}, 0)$. Its non-analytic part at $q = 2\,k_F$ is $\Pi_{\rm sing}\,\mathcal{U}/\epsilon^2$, with $\epsilon$ evaluated at $q = 2\,k_F$ from the regular part of $\Pi$. Using Eq.~\eqref{eq:h-phi}, the static dielectric function at $q = 2\,k_F$ is
\begin{align}
\label{eq:h-eps2kf}
\epsilon_{2 \, k_F} (\lambda) = \epsilon_b (\lambda) + \frac{2\,g\,e^2}{\pi\,v_0} \left[ \frac{23}{36} - \frac{\lambda}{4} - \frac{4 - 3\,\lambda}{6}\,\ln 2 \right].
\end{align}
The Fourier transform of $\delta n (\bs{q})$ at large $r = |\bs{r}|$ is evaluated in Appendix~\ref{app:hfull-friedel}. For $k_F\,r \gg 1$ and for directions of $\bs{r}$ away from the $z$-axis, we obtain
\begin{align}
\label{eq:h-friedel}
\delta n (\bs{r}) \simeq -\,\frac{g\,Z\,e^2\,\sin^2\vartheta_{\bs{r}}}{16\,\pi^2\,v_0\,k_F\,\epsilon_{2 \, k_F}^2 (\lambda_{\bs{r}})}\;\frac{\sin \!\left( 2\,k_F\,r \right)}{r^4}\,,
\end{align}
where $\vartheta_{\bs{r}}$ is the polar angle of $\bs{r}$ and $\lambda_{\bs{r}} = \sin^2\vartheta_{\bs{r}}/2$. The oscillations have the period $\pi/k_F$ and decay as $r^{-4}$, faster than the $r^{-3}$ of a parabolic band. The amplitude is anisotropic and is proportional to $\sin^2\vartheta_{\bs{r}}$ up to the angular dependence of $\epsilon_{2 \, k_F}$. The anisotropy is inherited from the dipolar BC and from the lack of full rotational symmetry. The coefficient of the $r^{-4}$ term vanishes along the $z$-axis, and the terms of order $r^{-5}$ must therefore be kept near the axis. For $\vartheta_{\bs{r}} \ll 1$, we find
\begin{align}
\label{eq:h-friedel-axis}
\delta n (\bs{r}) \simeq -\,\frac{g\,Z\,e^2}{16\,\pi^2\,v_0\,k_F^2\,\epsilon_{2 \, k_F}^2 (0)} \left[ k_F\,\sin^2\vartheta_{\bs{r}}\,\frac{\sin \!\left( 2\,k_F\,r \right)}{r^4} - \left( 4 + P_2 (\cos\vartheta_{\bs{r}}) \right) \frac{\cos \!\left( 2\,k_F\,r \right)}{r^5} \right],
\end{align}
where $P_2 (\tilde x) = (3\,\tilde x^2 - 1)/2$ is the Legendre polynomial. The first term is Eq.~\eqref{eq:h-friedel} at small $\vartheta_{\bs{r}}$. In the second bracket, the term $4$ comes from the cubic singularity of Eq.~\eqref{eq:h-kohnz} and is the TSM term, whereas the term $P_2$ comes from the angular structure of the quadratic singularity. The two parts of Eq.~\eqref{eq:h-friedel-axis} are comparable for $\vartheta_{\bs{r}} \sim (k_F\,r)^{-1/2}$, and the $r^{-4}$ term dominates for larger angles. Along the $z$-axis, the oscillations decay as $r^{-5}$, as in the TSM, and the amplitude of $\cos \!\left( 2\,k_F\,r \right)/r^5$ is $5\,g\,Z\,e^2/(16\,\pi^2\,v_0\,k_F^2\,\epsilon_{2 \, k_F}^2 (0))$. The in-plane oscillations are shown in Fig.~\ref{fig:friedel-all}.

\subsection{Plasmon dispersion and Landau damping}
\label{sechplasmon}

For $v_0\,q \ll \omega$ and $q \ll k_F$, the intraband part of Eq.~\eqref{eq:h-split} has the expansion
\begin{align}
\label{eq:h-drude}
\Re \Pi_{\rm intra} (\bs{q}, \omega) = \frac{g\,v_0\,k_F^2\,q^2}{6\,\pi^2\,\omega^2} \left[ 1 + \frac{3}{5}\,\frac{v_0^2\,q^2}{\omega^2} + \mathcal{O} \left( \frac{v_0^4\,q^4}{\omega^4}, \frac{q^2}{k_F^2} \right) \right].
\end{align}
The leading term gives $-\,\omega_p^2/\omega^2$ in $\epsilon^{\rm RPA}$, as in Eq.~\eqref{eq:lin-wp}. The correction is isotropic and has the coefficient of a three-dimensional electron gas. 
For $\omega_p \ll \mu$, the zero of $\Re \epsilon^{\rm RPA}$ gives the dispersion
\begin{align}
\label{eq:h-disp}
\omega_{\rm pl}^2 (\bs{q}) = \tilde\omega_p^2 (\lambda_{\bs{q}}) + \frac{3}{5}\,v_0^2\,q^2 + \mathcal{O} \left( \alpha\,v_0^2\,q^2, \frac{v_0^2\,q^4}{k_F^2} \right), \quad
\tilde\omega_p^2 (\lambda) = \frac{\omega_p^2}{\epsilon_b (\lambda)}\,.
\end{align}
The anisotropy of the plasmon at small $q$ is thus entirely contained in $\epsilon_b (\lambda_{\bs{q}})$, and the dispersive correction is isotropic at this order.

We use the expansion of $\Pi$ in $q$ at fixed $\omega$. For $\omega < \mu$ and $q < \omega/v_0$, it contains only even powers, $\Pi = q^2\,\Pi_2 + q^4\,\Pi_4 + \dots$, with
\begin{align}
\label{eq:h-pi24}
\Pi_2 &= \frac{g}{6\,\pi^2\,v_0} \left[ \frac{\mu^2}{\omega^2} - (1 - \lambda)\,\ln \!\left( \frac{v_0^2\,\Lambda^2}{\mu^2 - \omega^2} \right) - \frac{\lambda}{4}\,\ln \!\left( \frac{v_0^2\,\Lambda^2}{4\,\mu^2 - \omega^2} \right) \right], \nonumber \\
\Pi_4 &= \frac{g\,v_0}{\pi^2\,\omega^2} \left[ \frac{\mu^2}{10\,\omega^2} - \frac{3 - \lambda}{60} + \frac{1 - \lambda}{30}\,\ln \!\left( \frac{4\,\mu^2 - \omega^2}{4 \left( \mu^2 - \omega^2 \right)} \right) - \frac{\lambda\,\omega^2 \left( 4\,\mu^2 + \omega^2 \right)}{120 \left( 4\,\mu^2 - \omega^2 \right)^2} \right],
\end{align}
where $\lambda = \lambda_{\bs{q}}$. For $\lambda = 0$, they reduce to Eq.~\eqref{eq:fin-pi24}. The first term of $\Pi_2$ is the Drude term of Eq.~\eqref{eq:h-drude}. The logarithms of $\Pi_2$ define
\begin{align}
\label{eq:h-einter}
\epsilon_{\rm inter} (\omega) = 1 + \frac{2\,\alpha}{\pi} \left[ \frac{1 - \lambda}{3}\,\ln \!\left( \frac{v_0^2\,\Lambda^2}{\mu^2 - \omega^2} \right) + \frac{\lambda}{12}\,\ln \!\left( \frac{v_0^2\,\Lambda^2}{4\,\mu^2 - \omega^2} \right) \right],
\end{align}
and we also define
\begin{align}
\label{eq:h-xi}
\Xi (\omega) = \epsilon_{\rm inter} (\omega) + \frac{2\,\alpha\,\omega^2}{\pi} \left[ \frac{1 - \lambda}{3 \left( \mu^2 - \omega^2 \right)} + \frac{\lambda}{12 \left( 4\,\mu^2 - \omega^2 \right)} \right].
\end{align}
To order $q^2$, the zero $\omega_*$ of $\Re \epsilon^{\rm RPA}$ is
\begin{align}
\label{eq:h-wdisp}
\omega_*^2 (q) = \omega_*^2 (0) + \frac{4\,\alpha\,v_0^2\,q^2}{\pi}\,\frac{1}{\Xi (\omega_*)} \left[ \frac{\mu^2}{10\,\omega_*^2} - \frac{3 - \lambda}{60} + \frac{1 - \lambda}{30}\,\ln \!\left( \frac{4\,\mu^2 - \omega_*^2}{4 \left( \mu^2 - \omega_*^2 \right)} \right) - \frac{\lambda\,\omega_*^2 \left( 4\,\mu^2 + \omega_*^2 \right)}{120 \left( 4\,\mu^2 - \omega_*^2 \right)^2} \right]_{\omega_* = \omega_* (0)}.
\end{align}

The shift depends on the direction of $\bs{q}$ through $\lambda_{\bs{q}}$ in $\Xi$ and in the bracket. The bracket and $\Xi$ are positive, and hence the dispersion increases with $q$ for every direction, since $\alpha > 0$.

The expansion requires $v_0\,q \ll \omega_*$ and $q \ll k_F$, since Eq.~\eqref{eq:h-drude} has corrections of relative order $v_0^2\,q^2/\omega^2$ and $q^2/k_F^2$. It also loses accuracy when $\omega_*$ approaches $\mu$, where $\Pi_4$ contains the logarithm of $\mu^2 - \omega^2$. At larger wavevectors, we solve $\Re \epsilon^{\rm RPA} = 0$ numerically using Eq.~\eqref{eq:h-kk}, as shown in Fig.~\ref{fig:hsm-plasmon-static}(a). The undamped window $v_0\,q < \omega < \mu$, the behaviour of $\Re \epsilon^{\rm RPA}$ at its edges, and the definition of $q_{\rm max}$ are the same as in Sec.~\ref{secfinplasmon}, and the branch beyond $q_{\rm max}$ is not drawn. We find $q_{\rm max} < k_F$. The critical wavevector depends on $\lambda_{\bs{q}}$ through the weights of the logarithms in Eqs.~\eqref{eq:h-pi24} and~\eqref{eq:h-einter}. The anisotropy of the plasmon is therefore visible beyond small $q$, where the branches for different directions of $\bs{q}$ end at different wavevectors.

The doping enters only through $\ln \!\left( v_0\,\Lambda/\mu \right)$ and through the scaling of $\omega$ and $q$ with $\mu$ and $k_F$, as in Sec.~\ref{secfinplasmon}. For $\omega_p > \mu$, which requires $\alpha > 3\,\pi/2$ for every direction, since $\omega_p$ does not depend on $\lambda_{\bs{q}}$, the zero is pinned just below the interband onset. Its distance from the onset is controlled by the weight $(1 - \lambda_{\bs{q}})/3$ of the logarithm of $\mu^2 - \omega^2$ in Eq.~\eqref{eq:h-einter}. For $q > q_{\rm max}$, the zero of $\Re \epsilon^{\rm RPA}$ lies inside the intraband continuum, where $\Im \epsilon^{\rm RPA} \neq 0$.

Beyond the expansion, the existence and the damping of the plasmon follow from the supports of Eq.~\eqref{eq:h-impi}. For $0 < q < k_F$, the window $v_0\,q < \omega < \mu$ lies above the intraband continuum and below both interband continua, since $2\,\mu - v_0\,q > \mu$, and $\Im \epsilon^{\rm RPA}$ vanishes in it. As in Sec.~\ref{secfinplasmon}, $\Re \epsilon^{\rm RPA}$ tends to $-\,\infty$ for $\omega \to v_0\,q^+$ and to $+\,\infty$ for $\omega \to \mu^-$, because of the finite jumps of $\Im \Pi$ at the edges (see Appendix~\ref{app:hfull-plasmon}). Hence the window contains at least one undamped plasmon for every $0 < q < k_F$. For $q \geq k_F$, every frequency $\omega > v_0\,q$ is above the onset $\mu$ of the transitions $0 \to +$, and $\Im \Pi_{0+} < 0$ there. The plasmon is therefore damped by the flat-band transitions as soon as $q \geq k_F$, and below $q = k_F$ it is damped by the intraband continuum only if it enters $\omega < v_0\,q$. The damping rate of a mode at $\omega_{\rm pl}$ in the damped region is $\gamma = \Im \epsilon^{\rm RPA}/\partial_\omega \Re \epsilon^{\rm RPA}$ at $\omega = \omega_{\rm pl}$, with $\Im \epsilon^{\rm RPA}$ from Eq.~\eqref{eq:h-impi}.

As shown in Sec.~\ref{sechdamp}, no region free of excitations exists for $k_F \leq q \leq 2\,k_F$, and there is hence no undamped mode. For $q > 2\,k_F$, the free region $\omega < \min \lbrace \mu, v_0\,(q - 2\,k_F) \rbrace$ contains no zero of $\Re \epsilon^{\rm RPA}$ for $q \ll \Lambda$, since $\Re \epsilon^{\rm RPA} > 1$ there, and this holds for every direction of $\bs{q}$.

The above also applies to the TSM. The overlap factors of Eq.~\eqref{eq:lin-ff} coincide with Eq.~\eqref{eq:h-fl} at $\lambda = 0$, and the rotational symmetry makes $\lambda = 0$ independent of the direction of $\bs{q}$. All expressions of this section at $\lambda = 0$ therefore reproduce the polarisability of the TSM of Sec.~\ref{secfinite} for arbitrary $q$ and $\omega$.

\section{WSM at arbitrary momentum-transfers}
\label{secwsmhigh}

Ref.~\cite{ips-nr-plasmon} derived the plasmon of the untilted WSM at long wavelengths, and these results are collected in Sec.~\ref{seccompare}. We extend them to finite wavevectors, which gives access to the particle-hole continuum, the plasmon dispersion, and the static response that controls the Friedel oscillations~\cite{rhim-kim-friedel, rahmipoor-2d-nlsm}. We impose no relation between $q$ and $k_F$ and require only that both be much smaller than the cut-off $\Lambda$. As in Secs.~\ref{secfinite} and~\ref{sechfull}, the cut-off is implemented as an upper limit $\mathcal{R} = v_0\,\Lambda$ on the transition energy of the interband transitions. Secs.~\ref{secwim} and~\ref{secwre} give the imaginary part with the particle-hole continuum and the real part in closed form. The static polarisability, the screening, and the Friedel oscillations follow in Secs.~\ref{secwstat} and~\ref{secwfried}, and the plasmon is discussed in Sec.~\ref{secwexp}. The derivations are collected in Appendix~\ref{app-WSM}.

The Hamiltonian, the eigenvalues, and the overlap factors of Eq.~\eqref{eq:w-ff} are given in Sec.~\ref{sec:wsm}. The noninteracting polarisability at $T = 0$ follows from Eq.~\eqref{eq:pol} with $s, s' = \pm 1$, and the RPA dielectric function from Eq.~\eqref{eq:epsrpa}. We take $\mu > 0$, and the results for $\mu < 0$ follow upon replacing $\mu$ by $|\mu|$. The Fermi momentum is $k_F = \mu/v_0$, and the DOS at the Fermi level is $\rho (\mu) = g\,\mu^2/(2\,\pi^2\,v_0^3)$. We define the dimensionless parameter and the frequency in units of momentum as
\begin{align}
\label{eq:w-defs}
\bar\alpha = \frac{\alpha}{6\,\pi} = \frac{g\,e^2}{6\,\pi\,v_0}\,, \quad \varpi = \frac{\omega}{v_0}\,.
\end{align}

The occupations are $f_0 (\varepsilon_{-1}) = 1$ and $f_0 (\varepsilon_1) = \Theta (k_F - k)$. We split $\Pi = \Pi_{\rm vac} + \Pi_\mu$, where $\Pi_{\rm vac}$ is the polarisability at $\mu = 0$ and depends on $\Lambda$. The doped part $\Pi_\mu$ is linear in $\Theta (k_F - k)$, independent of $\Lambda$, and given by
\begin{align}
\label{eq:w-pimu}
\Pi_\mu (q, \omega) = g \int_{k < k_F} \frac{d^3 \bs{k}}{(2\,\pi)^3} \sum_{s' = \pm 1} \mathcal{F}_{1 s'} (\bs{k}, \bs{k}')
\left[ \frac{1}{\omega + \varepsilon_1 (\bs{k}) - \varepsilon_{s'} (\bs{k}') + i\,0^+} - \frac{1}{\omega - \varepsilon_1 (\bs{k}) + \varepsilon_{s'} (\bs{k}') + i\,0^+} \right].
\end{align}
The terms $s' = 1$ and $s' = -1$ are the intraband part $\Pi_{\rm intra}$ and the interband part, respectively. The latter removes the Pauli-blocked transitions from the vacuum polarisability. We replace the angle between $\bs{k}$ and $\bs{q}$ by the variables $\mathfrak{k}_\pm = k \pm k'$, with $k' = |\bs{k} + \bs{q}|$, which gives
\begin{align}
\label{eq:w-master}
\Pi_\mu (q, \omega) = -\,\frac{g}{16\,\pi^2\,v_0\,q} \iint_{\mathfrak{D}} d\mathfrak{k}_-\,d\mathfrak{k}_+ \left[ \frac{\left( \mathfrak{k}_+^2 - q^2 \right) \mathfrak{k}_-}{\left( \varpi + i\,0^+ \right)^2 - \mathfrak{k}_-^2} + \frac{\left( q^2 - \mathfrak{k}_-^2 \right) \mathfrak{k}_+}{\left( \varpi + i\,0^+ \right)^2 - \mathfrak{k}_+^2} \right].
\end{align}
The first and second terms are the intraband and interband parts, and the domain is $\mathfrak{D} = \lbrace |\mathfrak{k}_-| \leq q \leq \mathfrak{k}_+,\; \mathfrak{k}_- + \mathfrak{k}_+ \leq 2\,k_F \rbrace$. The vacuum part is minus the interband term on $|\mathfrak{k}_-| \leq q \leq \mathfrak{k}_+ \leq \Lambda$, where $\mathfrak{k}_+ \leq \Lambda$ is the upper limit $\mathcal{R}$ on the transition energy $v_0\,\mathfrak{k}_+$. For $q, \varpi \ll \Lambda$, it gives
\begin{align}
\label{eq:w-vac}
\Pi_{\rm vac} (q, \omega) = -\,\frac{g\,q^2}{24\,\pi^2\,v_0}\,\ln \!\left( \frac{\Lambda^2}{\left| \varpi^2 - q^2 \right|} \right), \quad
\Im \Pi_{\rm vac} (q, \omega) = -\,\frac{g\,q^2}{24\,\pi\,v_0}\,\Theta \left( \varpi - q \right).
\end{align}
The derivations are given in Appendix~\ref{app-WSM}.

Fig.~\ref{fig:wsm-plasmon-static}(a) shows $-\,\Im\Pi$ from Eqs.~\eqref{eq:w-imintra} and~\eqref{eq:w-iminter}, normalised by $g\,q^2/(6\,\pi\,v_0)$, together with the undamped plasmon branch, which is drawn up to $q_{\rm max}$. Beyond $q_{\rm max}$, the zero of $\Re\epsilon^{\rm RPA}$ either lies within $2 \times 10^{-3}\,v_0\,q$ of the edge $\omega = v_0\,q$ or has left the window through the edge $\omega = 2\,\mu - v_0\,q$. The solid black lines are the edges of the continuum, namely $\omega = v_0\,q$, $\omega = 2\,\mu - v_0\,q$ for $q < k_F$, and $\omega = v_0 \left( q - 2\,k_F \right)$ for $q > 2\,k_F$. The dashed line $\omega = 2\,\mu - v_0\,q$ for $k_F < q < 2\,k_F$ marks a kink of the intraband weight, and the dotted line $\omega = v_0\,q + 2\,\mu$ marks the point beyond which $\Im\Pi$ has reached its intrinsic value. The right panel shows the branch from Eq.~\eqref{eq:w-repi} (solid) and from Eq.~\eqref{eq:w-plasdisp} (dashed), restricted to the window between the black lines $\omega = v_0\,q$ and $\omega = 2\,\mu - v_0\,q$. The two solutions coincide at small $q$ and separate at larger $q$, and the branch ends where it meets the line $\omega = 2\,\mu - v_0\,q$ at $q < k_F$. Fig.~\ref{fig:wsm-plasmon-static}(b) shows the second derivative of $\Pi (q, 0)$ from Eq.~\eqref{eq:w-stat}, scaled by $k_F^2\,v_0/g$, which diverges logarithmically towards negative values at $q = 2\,k_F$ as in Eq.~\eqref{eq:w-sing}. The right panel shows the static dielectric function from Eq.~\eqref{eq:w-stat} (solid) and the Thomas-Fermi form of Eq.~\eqref{eq:w-tf} (dashed), which lies on top of the solid curve over the range of $q$ in which it is drawn. The parameter is $\mu = 0.01\,\text{eV}$, which gives $k_F = \mu/v_0 = 25\,\text{eV}$ and $v_0\,\Lambda/\mu = 6.4$, and hence $L = \ln 6.4$.

\begin{figure}[t!]
\centering
\subfigure[]{\includegraphics[width=0.75\linewidth]{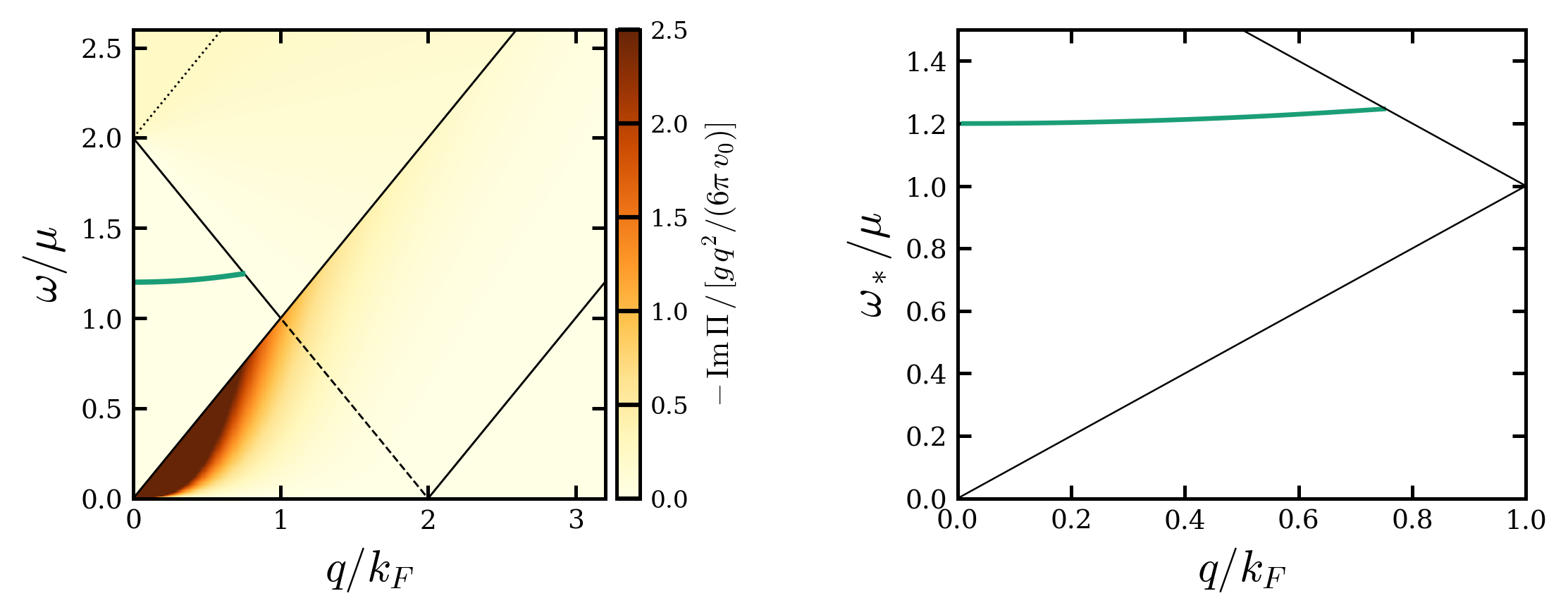}}
\subfigure[]{\includegraphics[width=0.75\linewidth]{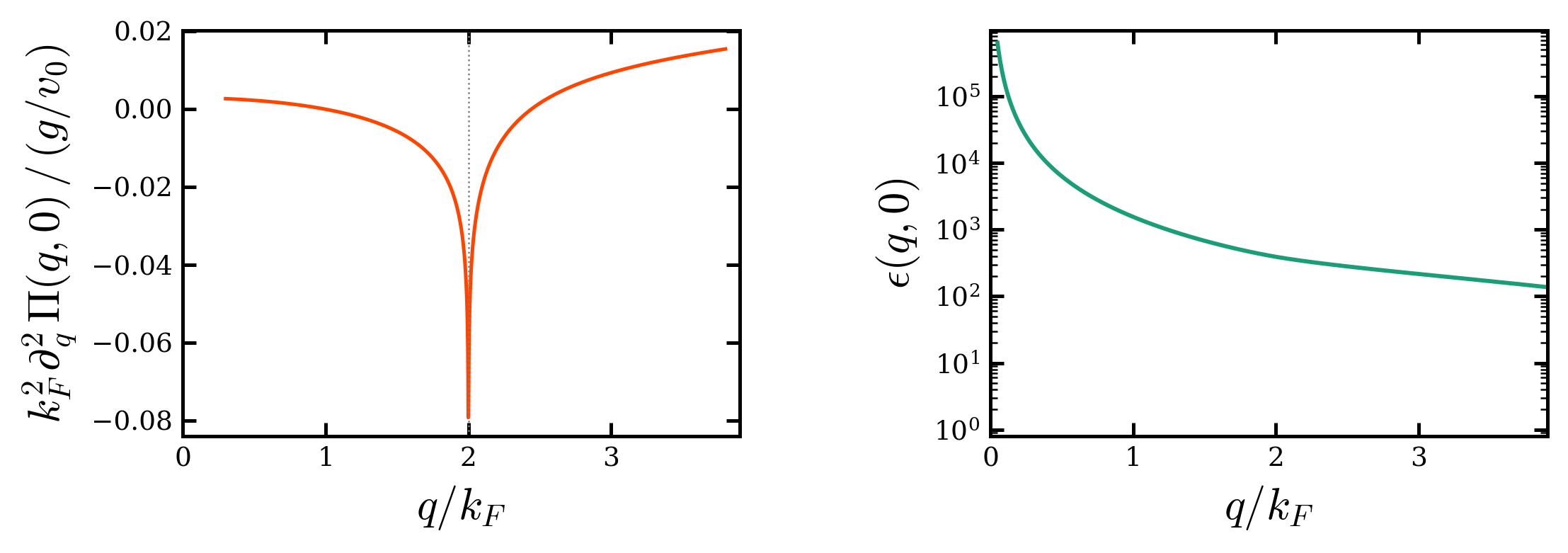}}
\caption{\label{fig:wsm-plasmon-static} WSM characteristics. (a) $-\,\Im\Pi$ normalised by $g\,q^2/(6\,\pi\,v_0)$, with the undamped plasmon branch (left), and the plasmon branch (right). (b) Second derivative of $\Pi (q, 0)$ (left) and static dielectric function (right). The parameter is $\mu = 0.01\,\text{eV}$.}
\end{figure}

\subsection{Imaginary part of polarisability: Particle-hole continua and their edges}
\label{secwim}

The imaginary part follows from the delta functions at the poles of Eq.~\eqref{eq:w-master}. For arbitrary $q$ and $\varpi > 0$, the intraband part is
\begin{align}
\label{eq:w-imintra}
\Im \Pi_{\rm intra} (q, \omega) &= -\,\frac{g\,\varpi}{16\,\pi\,v_0\,q} \left( 4\,k_F^2 - q^2 + \frac{\varpi^2}{3} \right) \Theta \!\left( \min \lbrace q, 2\,k_F - q \rbrace - \varpi \right) \nn
& \quad - \frac{g}{96\,\pi\,v_0\,q} \left( \varpi + 2\,k_F - q \right)^2 \left( \varpi + 2\,k_F + 2\,q \right) \Theta \!\left( \varpi - \max \lbrace 2\,k_F - q, q - 2\,k_F \rbrace \right) \Theta \left( q - \varpi \right).
\end{align}
The interband part, which contains the vacuum part, is
\begin{align}
\label{eq:w-iminter}
\Im \Pi_{\rm inter} (q, \omega) = -\,\frac{g}{96\,\pi\,v_0\,q}
\begin{cases}
0 & \text{ for } \varpi < \max \lbrace q, 2\,k_F - q \rbrace \\[4pt]
\left( \varpi + q - 2\,k_F \right)^2 \left( 2\,k_F + 2\,q - \varpi \right) & \text{ for } \max \lbrace q, 2\,k_F - q \rbrace < \varpi < 2\,k_F + q \\[4pt]
4\,q^3 & \text{ for } \varpi > 2\,k_F + q
\end{cases}.
\end{align}
Each piece of $\Im \Pi = \Im \Pi_{\rm intra} + \Im \Pi_{\rm inter}$ is non-positive. The imaginary part of the dielectric function, $\Im \epsilon^{\rm RPA} = -\,\mathcal{V} (q)\,\Im \Pi$, is
\begin{align}
\label{eq:w-imeps}
\Im \epsilon^{\rm RPA} (q, \omega) &= \frac{3\,\pi\,\bar\alpha\,\varpi}{2\,q^3} \left( 4\,k_F^2 - q^2 + \frac{\varpi^2}{3} \right) \Theta \!\left( \min \lbrace q, 2\,k_F - q \rbrace - \varpi \right) \nn
& \quad + \frac{\pi\,\bar\alpha}{4\,q^3} \left( \varpi + 2\,k_F - q \right)^2 \left( \varpi + 2\,k_F + 2\,q \right) \Theta \!\left( \varpi - \max \lbrace 2\,k_F - q, q - 2\,k_F \rbrace \right) \Theta \left( q - \varpi \right) + \pi\,\bar\alpha\,\mathcal{W}_q (\varpi)\,,
\end{align}
where $\mathcal{W}_q (\varpi) = 0$ for $\varpi < \max \lbrace q, 2\,k_F - q \rbrace$, $\mathcal{W}_q (\varpi) = \left( \varpi + q - 2\,k_F \right)^2 \left( 2\,k_F + 2\,q - \varpi \right)/(4\,q^3)$ for $\max \lbrace q, 2\,k_F - q \rbrace < \varpi < 2\,k_F + q$, and $\mathcal{W}_q (\varpi) = 1$ for $\varpi > 2\,k_F + q$.

In the intrinsic limit $k_F \to 0$, the intraband part vanishes and Eq.~\eqref{eq:w-iminter} reduces to the step
\begin{align}
\label{eq:w-Im0}
\Im \Pi^{(0)} (q, \omega) = -\,\frac{g\,q^2}{24\,\pi\,v_0}\,\Theta \left( \varpi - q \right),
\end{align}
which is $\Im \Pi_{\rm vac}$ of Eq.~\eqref{eq:w-vac}. In contrast to the HSM, there is no absorption below $\varpi = q$, since the Weyl cone has no flat-band and hence no transitions $0 \to +$. In the doped system, $\Im \Pi$ reaches this value for $\varpi > 2\,k_F + q$, where $\mathcal{W}_q = 1$ and the intraband part is absent. Pauli-blocking therefore modifies the imaginary part only below $\varpi = 2\,k_F + q$.

For $q \leq k_F$, we have $\max \lbrace q, 2\,k_F - q \rbrace = 2\,k_F - q$ and the second term of Eq.~\eqref{eq:w-imintra} is absent. The continuum consists of the intraband region $0 < \omega < v_0\,q$ and the interband region above the onset
\begin{align}
\label{eq:w-onset}
\omega_{\rm th} (q) = 2\,\mu - v_0\,q\,.
\end{align}
The function $\mathcal{W}_q$ is continuous at both ends of the middle window. Hence the interband absorption rises continuously for $2\,\mu - v_0\,q < \omega < 2\,\mu + v_0\,q$ and saturates at $\Im \epsilon^{\rm RPA} = \pi\,\bar\alpha = g\,e^2/(6\,v_0)$ above it. For $q \to 0$, this reproduces the onset $2\,\mu$ and the frequency-independent absorption of the long-wavelength result. The window $v_0\,q < \omega < 2\,\mu - v_0\,q$ of width $2 \left( \mu - v_0\,q \right)$ is free of damping, and an undamped plasmon can exist only there. The continuum and the plasmon branches are shown in Fig.~\ref{fig:wsm-plasmon-static}(a).

For $q > k_F$, the second term of Eq.~\eqref{eq:w-imintra} is present, and the interband onset is $\omega = v_0\,q$ instead of $\omega_{\rm th} (q)$. For $k_F < q < 2\,k_F$, the first term covers $0 < \varpi < 2\,k_F - q$, the second term covers $2\,k_F - q < \varpi < q$, and the interband part covers $\varpi > q$. Hence the continuum extends over all $\varpi > 0$ without a gap. For $q > 2\,k_F$, the intraband continuum is restricted to $q - 2\,k_F < \varpi < q$, and the region $0 < \varpi < q - 2\,k_F$ is free of single-particle excitations. The imaginary part is continuous at $\varpi = 2\,k_F - q$ and vanishes quadratically at $\varpi = q - 2\,k_F$. It jumps at $\varpi = q$ by
\begin{align}
\label{eq:w-jump}
\Im \Pi (q, \varpi \to q^+) - \Im \Pi (q, \varpi \to q^-) = \frac{g \left( 6\,k_F^2 - q^2 \right)}{24\,\pi\,v_0} \quad \text{ for } q \geq k_F\,,
\end{align}
which stems from the termination of the intraband continuum and the onset of the interband continuum.

\subsection{Real part of polarisability and RPA dielectric function}
\label{secwre}

We define $\mathcal{L} (\tilde x) = \ln \!\left( \left| (\varpi + \tilde x)/(\varpi - \tilde x) \right| \right)$. The real part follows from the dispersion relation with the cut-off $\mathcal{R} = v_0\,\Lambda$ on the transition energy. For the undoped polarisability, $\Im \Pi^{(0)}$ of Eq.~\eqref{eq:w-Im0} gives
\begin{align}
\label{eq:w-Re0}
\Re \Pi^{(0)} (q, \omega) = \Pi_{\rm vac} (q, \omega) = -\,\frac{g\,q^2}{24\,\pi^2\,v_0}\,\ln \!\left( \frac{v_0^2\,\Lambda^2}{\left| \omega^2 - v_0^2\,q^2 \right|} \right),
\end{align}
which is Eq.~\eqref{eq:w-vac} written in terms of $\omega$. Since $\Im \Pi - \Im \Pi^{(0)}$ vanishes above $\omega_{\rm w} = v_0\,q + 2\,\mu$, the doped part $\Pi_\mu = \Pi - \Pi_{\rm vac}$ is the finite window integral
\begin{align}
\label{eq:w-kk}
\Re \Pi_\mu (q, \omega) = \frac{2}{\pi}\,\mathrm{P} \!\! \int_0^{\omega_{\rm w}} d\omega'\,\frac{\omega' \left[ \Im \Pi (q, \omega') - \Im \Pi^{(0)} (q, \omega') \right]}{\omega'^2 - \omega^2}\,,
\end{align}
which is independent of $\Lambda$. Evaluating it in closed form for every $q$ and $\varpi > 0$, we obtain
\begin{align}
\label{eq:w-repi}
\Re \Pi (q, \omega) = \Pi_{\rm vac} (q, \omega) - \frac{g}{16\,\pi^2\,v_0\,q}\,\mathcal{T} (q, \varpi)\,,
\end{align}
with
\begin{align}
\label{eq:w-S}
\mathcal{T} &= \frac{16\,k_F^2\,q}{3} + \frac{2\,q^3}{3}\,\ln \!\left( \left| \varpi^2 - q^2 \right| \right) - \frac{\varpi \left( 12\,k_F^2 + \varpi^2 - 3\,q^2 \right)}{3}\,\mathcal{L} (q) \nn
& \quad + \frac{4\,k_F^3 + 3\,k_F\,\varpi^2 - 3\,k_F\,q^2 - q^3}{3}\,\ln \!\left( \left| \varpi^2 - \left( 2\,k_F + q \right)^2 \right| \right) - \frac{4\,k_F^3 + 3\,k_F\,\varpi^2 - 3\,k_F\,q^2 + q^3}{3}\,\ln \!\left( \left| \varpi^2 - \left( 2\,k_F - q \right)^2 \right| \right) \nn
& \quad + \frac{\varpi \left( 12\,k_F^2 + \varpi^2 - 3\,q^2 \right)}{6} \left[ \mathcal{L} \left( 2\,k_F + q \right) - \mathcal{L} \left( 2\,k_F - q \right) \right].
\end{align}
The logarithmic singularities at $\varpi = q$, $\varpi = \left| 2\,k_F - q \right|$, and $\varpi = 2\,k_F + q$ are the edges of the continuum in Eqs.~\eqref{eq:w-imintra} and~\eqref{eq:w-iminter}. For $q \leq k_F$, the function $\mathcal{T}$ is the sum of the intraband and interband integrals of Eq.~\eqref{eq:w-master}. For $q > k_F$, the two integrals change individually, and only their sum retains the form of Eq.~\eqref{eq:w-S}, as shown in Appendix~\ref{app:w-real}.

\subsection{Static polarisability and screening}
\label{secwstat}

At $\omega = 0$, Eq.~\eqref{eq:w-S} gives the doped part for all $q$,
\begin{align}
\label{eq:w-statmu}
\Pi_\mu (q, 0) = -\,\frac{g}{24\,\pi^2\,v_0\,q} \left[ 8\,k_F^2\,q + k_F \left( 4\,k_F^2 - 3\,q^2 \right) \ln \!\left( \left| \frac{2\,k_F + q}{2\,k_F - q} \right| \right) - q^3\,\ln \!\left( \left| \frac{4\,k_F^2 - q^2}{q^2} \right| \right) \right].
\end{align}
Adding $\Pi_{\rm vac} (q, 0)$ from Eq.~\eqref{eq:w-vac} cancels the logarithms in $q$ and yields
\begin{align}
\label{eq:w-stat}
\Pi (q, 0) = -\,\frac{g}{24\,\pi^2\,v_0} \left[ 8\,k_F^2 + \frac{k_F \left( 4\,k_F^2 - 3\,q^2 \right)}{q}\,\ln \!\left( \left| \frac{2\,k_F + q}{2\,k_F - q} \right| \right) + q^2\,\ln \!\left( \frac{\Lambda^2}{\left| 4\,k_F^2 - q^2 \right|} \right) \right],
\end{align}
which is analytic at $q = 0$. For $q \ll k_F$, it reduces to
\begin{align}
\label{eq:w-statlow}
\Pi (q, 0) = -\,\rho (\mu) - \frac{g\,q^2}{24\,\pi^2\,v_0} \left[ 2\,\ln \!\left( \frac{v_0\,\Lambda}{2\,\mu} \right) - \frac{8}{3} \right] - \frac{g\,q^4}{480\,\pi^2\,v_0\,k_F^2} + \mathcal{O} \left( q^6 \right),
\end{align}
where the leading term is the DOS at the Fermi level. For $q \gg k_F$, the doped part is a small correction to the vacuum part,
\begin{align}
\label{eq:w-stathigh}
\Pi (q, 0) = -\,\frac{g\,q^2}{24\,\pi^2\,v_0}\,\ln \!\left( \frac{\Lambda^2}{q^2} \right) - \frac{g}{24\,\pi^2\,v_0} \left[ \frac{8\,k_F^4}{q^2} + \frac{64\,k_F^6}{15\,q^4} + \mathcal{O} \left( \frac{k_F^8}{q^6} \right) \right],
\end{align}
since the terms of order $k_F^2$ cancel between the intraband and interband parts.

For $q \to 0$, Eq.~\eqref{eq:w-statlow} gives the Thomas-Fermi screening
\begin{align}
\label{eq:w-tf}
\epsilon^{\rm RPA} (q, 0) = 1 + \frac{q_{\rm TF}^2}{q^2} + \bar\alpha \left[ 2\,\ln \!\left( \frac{v_0\,\Lambda}{2\,\mu} \right) - \frac{8}{3} \right] + \mathcal{O} \left( q^2 \right), \quad q_{\rm TF}^2 = 4\,\pi\,e^2\,\rho (\mu) = 12\,\bar\alpha\,k_F^2\,.
\end{align}
The static background in the bracket differs from that of $\mathcal{E}_0$ in Eq.~\eqref{eq:w-epsexp} by the constant $-\,8\,\bar\alpha/3$, because the limits $q \to 0$ and $\omega \to 0$ do not commute. The steps are given in Appendix~\ref{app:w-stat}, and the static polarisability and dielectric function are shown in Fig.~\ref{fig:wsm-plasmon-static}(b).

\subsection{Friedel oscillations}
\label{secwfried}

The function $\Pi (q, 0)$ and its first derivative are continuous at $q = 2\,k_F$, since the coefficient of $\ln \!\left( \left| q - 2\,k_F \right| \right)$ in Eq.~\eqref{eq:w-stat} vanishes quadratically there. The non-analytic part is
\begin{align}
\label{eq:w-sing}
\Pi (q, 0) = \Pi_{\rm reg} (q) + \frac{g}{24\,\pi^2\,v_0}\,\frac{\left( q - 2\,k_F \right)^2 \left( q + k_F \right)}{q}\,\ln \!\left( \left| q - 2\,k_F \right| \right),
\end{align}
where $\Pi_{\rm reg}$ is analytic at $q = 2\,k_F$. Expanding the prefactor about $q = 2\,k_F$ gives
\begin{align}
\label{eq:w-singexp}
\Pi (q, 0) - \Pi_{\rm reg} (q) = \left[ \frac{g \left( q - 2\,k_F \right)^2}{16\,\pi^2\,v_0} - \frac{g \left( q - 2\,k_F \right)^3}{96\,\pi^2\,v_0\,k_F} + \cdots \right] \ln \!\left( \left| q - 2\,k_F \right| \right),
\end{align}
Hence the second derivative of $\Pi (q, 0)$ diverges logarithmically, as shown in Fig.~\ref{fig:wsm-plasmon-static}(b). At $q = 2\,k_F$, the function takes the value
\begin{align}
\label{eq:w-pi2kf}
\Pi (2\,k_F, 0) = -\,\frac{g\,k_F^2}{3\,\pi^2\,v_0} \left[ 1 + \ln \!\left( \frac{\Lambda}{4\,k_F} \right) \right],
\end{align}
and the static dielectric function is
\begin{align}
\label{eq:w-eps2kf}
\epsilon_{2 \, k_F} = 1 + 2\,\bar\alpha \left[ 1 + \ln \!\left( \frac{v_0\,\Lambda}{4\,\mu} \right) \right].
\end{align}

We consider a static point charge $Z\,e$ at the origin with the bare potential energy $\mathcal{U} (q) = -\,4\,\pi\,Z\,e^2/q^2$ of an electron. Within the RPA, the induced density is $\delta n (\bs{q}) = \Pi (q, 0)\,\mathcal{U} (q)/\epsilon^{\rm RPA} (q, 0)$, and its non-analytic part at $q = 2\,k_F$ is $\Pi_{\rm sing}\,\mathcal{U}/\epsilon^2$, with $\Pi_{\rm sing}$ the singular term of Eq.~\eqref{eq:w-sing}, $\mathcal{U} (2\,k_F) = -\,\pi\,Z\,e^2/k_F^2$, and $\epsilon = \epsilon_{2 \, k_F}$ in the smooth factors. The isotropy of the model allows us to use the radial Fourier transform, which gives for $k_F\,r \gg 1$
\begin{align}
\label{eq:w-friedel}
\delta n (r) \simeq -\,\frac{g\,Z\,e^2}{8\,\pi^2\,v_0\,k_F\,\epsilon_{2 \, k_F}^2}\,\frac{\sin \!\left( 2\,k_F\,r \right)}{r^4} = -\,\frac{3\,\bar\alpha\,Z}{4\,\pi\,k_F\,\epsilon_{2 \, k_F}^2}\,\frac{\sin \!\left( 2\,k_F\,r \right)}{r^4}\,.
\end{align}
The oscillations have the period $\pi/k_F$ and decay as $r^{-4}$, whereas the singularity $(q - 2\,k_F) \ln \!\left( \left| q - 2\,k_F \right| \right)$ of a parabolic band in 3d gives the decay $r^{-3}$. The weaker singularity of the WSM originates from the overlap factor $\mathcal{F}_{1,1} = (1 + \cos\theta)/2$, which vanishes quadratically in the deviation from backscattering. The amplitude depends on the doping through the prefactor $1/k_F$ and through the logarithm in $\epsilon_{2 \, k_F}$. The oscillations of all systems are compared in Fig.~\ref{fig:friedel-all}.

\subsection{Plasmon dispersion and Landau damping}
\label{secwexp}

For $v_0\,q \ll \omega$ and $|\omega - 2\,\mu| \gg v_0\,q$, we expand Eq.~\eqref{eq:w-repi} in $q$ at fixed $\varpi$ and include the expansion of $\Pi_{\rm vac}$,
\begin{align}
\label{eq:w-exptot}
\Re \Pi (q, \omega) = \frac{g\,q^2}{24\,\pi^2\,v_0} \left[ \frac{4\,k_F^2}{\varpi^2} - \ln \!\left( \frac{\Lambda^2}{\left| 4\,k_F^2 - \varpi^2 \right|} \right) \right] - \frac{g\,q^4}{24\,\pi^2\,v_0} \left[ \frac{1}{\varpi^2} + \frac{16\,k_F^4 \left( 5\,\varpi^2 - 12\,k_F^2 \right)}{5\,\varpi^4 \left( 4\,k_F^2 - \varpi^2 \right)^2} \right] + \mathcal{O} \left( q^6 \right).
\end{align}
The first bracket reproduces the long-wavelength results, namely the Drude term with $\omega_p^2 = 4\,\bar\alpha\,\mu^2 = 2\,g\,e^2\,\mu^2/(3\,\pi\,v_0)$ and the background with the onset $2\,\mu$. The dielectric function is
\begin{align}
\label{eq:w-epsexp}
& \Re \epsilon^{\rm RPA} (q, \omega) = \mathcal{E}_0 (\omega) + \left( v_0\,q \right)^2 \mathcal{E}_2 (\omega) + \mathcal{O} \left( q^4 \right), \nn
& \mathcal{E}_0 (\omega) = 1 + \bar\alpha\,\ln \!\left( \frac{v_0^2\,\Lambda^2}{\left| 4\,\mu^2 - \omega^2 \right|} \right) - \frac{4\,\bar\alpha\,\mu^2}{\omega^2}\,, \quad
\mathcal{E}_2 (\omega) = \bar\alpha \left[ \frac{1}{\omega^2} 
+ \frac{16\,\mu^4 \left( 5\,\omega^2 - 12\,\mu^2 \right)}{5\,\omega^4 \left( 4\,\mu^2 - \omega^2 \right)^2} \right].
\end{align}
The derivative $\partial_{\omega^2} \mathcal{E}_0 = \bar\alpha/(4\,\mu^2 - \omega^2) + 4\,\bar\alpha\,\mu^2/\omega^4$ is positive for $\omega < 2\,\mu$. Since $\mathcal{E}_0$ tends to $-\,\infty$ for $\omega \to 0$ and to $+\,\infty$ for $\omega \to 2\,\mu^-$, it has a unique zero $\omega_*$ below the onset. The undamped plasmon is the zero $\omega_* (q)$ of $\Re \epsilon^{\rm RPA}$ in the window $v_0\,q < \omega < 2\,\mu - v_0\,q$, and to order $q^2$ it is given by
\begin{align}
\label{eq:w-plasdisp}
\omega_*^2 (q) = \omega_*^2 - \left( v_0\,q \right)^2 \frac{\mathcal{E}_2 (\omega_*)}{\partial \mathcal{E}_0/\partial \omega^2 \at[\big]{\omega = \omega_*}}\,.
\end{align}
For $\omega \ll \mu$, the second term of $\mathcal{E}_2$ reduces to $-\,12\,\bar\alpha\,\mu^2/(5\,\omega^4)$, which is the correction $-\,(\omega_p^2/\omega^2)\,(3/5)/\omega^2$ of relative order $(v_0\,q/\omega)^2$ to the Drude term.
 The branch is bounded by the window $v_0\,q < \omega < \omega_{\rm th} (q)$ of Eq.~\eqref{eq:w-onset}. 
The branch leaves the window through $\omega_{\rm th} (q)$ from below, as shown in Fig.~\ref{fig:wsm-plasmon-static}(a).

For $q > k_F$, the RPA dielectric function has no undamped plasmon. For $k_F < q < 2\,k_F$, every term of Eq.~\eqref{eq:w-imeps} is non-negative and $\Im \epsilon^{\rm RPA} > 0$ for every $\varpi > 0$. For $q > 2\,k_F$, we have $\Im \epsilon^{\rm RPA} > 0$ for $\varpi > q - 2\,k_F$, and in the gap $0 < \varpi < q - 2\,k_F$ the dispersion relation gives $\Re \Pi (q, \omega) = (2/\pi) \int_{v_0 (q - 2 \, k_F)}^{\mathcal{R}} d\omega'\,\omega'\,\Im \Pi (q, \omega')/(\omega'^2 - \omega^2) \leq 0$. Hence $\Re \epsilon^{\rm RPA} \geq 1$ in the gap, and there is no zero. Together with Sec.~\ref{secwim}, the undamped plasmon exists only for $q < k_F$ and only in the window $v_0\,q < \omega < 2\,\mu - v_0\,q$.

A zero $\omega_r (q)$ of $\Re \epsilon^{\rm RPA}$ inside the continuum is a damped mode. Expanding $\epsilon^{\rm RPA} (q, \omega_r - i\,\gamma)$ to first order in $\gamma$ gives the damping rate
\begin{align}
\label{eq:w-gamma}
\gamma (q) = \frac{\Im \epsilon^{\rm RPA} (q, \omega_r)}{\partial \Re \epsilon^{\rm RPA}/\partial \omega \at[\big]{\omega = \omega_r}}\,,
\end{align}
which applies when $\gamma$ is small compared to the scale on which $\Im \epsilon^{\rm RPA}$ and $\Re \epsilon^{\rm RPA}$ vary. The numerator is given by Eq.~\eqref{eq:w-imeps} and is proportional to $\bar\alpha$. The damping stems from the intraband transitions for $\omega_r < v_0\,q$ and from the interband transitions for $\omega_r > v_0\,q$, and for $q > k_F$ every zero of $\Re \epsilon^{\rm RPA}$ is of this type.

In Table~\ref{tab:compare}, we use $\beta = \sqrt{1 + 4\,c\,\mu/v_0}$, $\mu_1 = 2\,\mu/(1 + \beta)$, $L = \ln \!\left( v_0\,\Lambda/\mu \right)$, and $L_1$ of Sec.~\ref{secquadtop}, and $\lambda = \lambda_{\bs{q}}$ takes the values $0$ along $\bs{\hat z}$ and $1/2$ in the plane. The QTSM entries refer to $c < 0$ unless the flat-band row or $c > 0$ is explicitly indicated. Constants of $\epsilon_b$ that depend on the cut-off scheme are omitted.

\begin{table}[t!]
\centering
\renewcommand{\arraystretch}{2.0}
\begin{tabular}{|c|c|c|c|c|}
\hline
Quantity & TSM ($c=0$) & QTSM & HSM & WSM \\ \hline
Chern number & $\pm 2$ (BC-monopole) & $\pm 2$ (BC-monopole) & $0$ (BC-dipole) & $\pm 1$ (BC-monopole) \\ \hline
Number of bands & three & three & three & two \\ \hline
\makecell{Interband pairs \\ at order $q^2$} & \makecell{flat-band to \\ upper band, $v_0\,k$} & \makecell{flat-band to \\ upper band, $v_0\,k$} & \makecell{flat-band to upper band, $v_0\,k$ \\ dispersive pair, $2\,v_0\,k$} & \makecell{dispersive pair, \\ $2\,v_0\,k$} \\ \hline
Fermi velocity $V_F$ & $v_0$ & $\beta\,v_0$ & $v_0$ & $v_0$ \\ \hline
$\omega_p^2$ & $\dfrac{2\,g\,e^2\,\mu^2}{3\,\pi\,v_0}$ & $\dfrac{2\,g\,e^2\,\beta\,\mu_1^2}{3\,\pi\,v_0}$ & $\dfrac{2\,g\,e^2\,\mu^2}{3\,\pi\,v_0}$ & $\dfrac{2\,g\,e^2\,\mu^2}{3\,\pi\,v_0}$ \\ \hline
\makecell{Flat-band term \\ of $\omega_p^2$ ($c>0$)} & --- & $8\,\pi\,e^2\,c\,v_0\,\tilde n_0$ & absent & absent \\ \hline
$q_{\rm TF}^2/\omega_p^2$ & $3/v_0^2$ & $3/V_F^2$ & $3/v_0^2$ & $3/v_0^2$ \\ \hline
$\omega_{\rm th}$ & $\mu$ & $\mu_1$ & \makecell{$\mu$, and $2\,\mu$ \\ for $\lambda\neq 0$} & $2\,\mu$ \\ \hline
$\epsilon_b-1$ & $\dfrac{4\,g\,e^2}{3\,\pi\,v_0}\,L$ & $\dfrac{4\,g\,e^2}{3\,\pi\,v_0}\,L_1$ & $\dfrac{(4-3\,\lambda)\,g\,e^2}{3\,\pi\,v_0}\,L$ & $\dfrac{g\,e^2}{3\,\pi\,v_0}\,L$ \\ \hline
\makecell{$\Im \epsilon^{\rm RPA}$ above \\ all onsets} & $\dfrac{2\,g\,e^2}{3\,v_0}$ & $\dfrac{2\,g\,e^2}{3\,v_0}$ & $\dfrac{(4-3\,\lambda)\,g\,e^2}{6\,v_0}$ & $\dfrac{g\,e^2}{6\,v_0}$ \\ \hline
$\omega_p^2/\omega_{\rm th}^2$ & $\dfrac{2\,g\,e^2}{3\,\pi\,v_0}$ & $\dfrac{2\,g\,e^2\,\beta}{3\,\pi\,v_0}$ & $\dfrac{2\,g\,e^2}{3\,\pi\,v_0}$ & $\dfrac{g\,e^2}{6\,\pi\,v_0}$ \\ \hline
Anisotropy of response & isotropic & isotropic & \makecell{through $\lambda_{\bs{q}}$ \\ in $\epsilon_b$} & isotropic \\ \hline
Undamped plasmon & \makecell{below $\mu$, ends \\ at $q_{\rm max}<k_F$} & \makecell{$q<q_{\rm max}$, with \\ $q_{\rm max}\leq q_1$ for $c>0$} & \makecell{below $\mu$, ends \\ at $q_{\rm max}<k_F$} & \makecell{below $2\,\mu$, ends \\ at $q_{\rm max}\leq k_F$} \\ \hline
Bohm-Gross velocity & $v_0$ & $V_{\rm BG}$ & $v_0$ & $v_0$ \\ \hline
Friedel decay & $r^{-5}$ & \makecell{$r^{-5}$ (top band); \\ $r^{-3}$ (flat-band, $c>0$)} & \makecell{$r^{-4}$ generic; \\ $r^{-5}$ along $\bs{\hat z}$} & $r^{-4}$ \\ \hline
\makecell{Doped part of $\Pi(q,0)$ \\ for $q\gg k_F$} & \makecell{constant, \\ $\dfrac{g\,k_F^2}{6\,\pi^2\,v_0}$} & \makecell{$\propto 1/(c\,q^2)$, with \\ $k_F^3+\Theta(c)\,(k_F^{(0)})^3$} & \makecell{constant, \\ $\dfrac{g\,(1-\lambda)\,k_F^2}{6\,\pi^2\,v_0}$} & $\propto k_F^4/q^2$ \\ \hline
\makecell{Closed form of \\ $\Pi(q,0)$ at all $q$} & available & \makecell{not available, \\ integral of $\Im\Pi$} & available & available \\ \hline
\end{tabular}
\caption{\label{tab:compare} Comparison of TSM, QTSM, HSM, and WSM.}
\end{table}

\section{Comparison across various systems}
\label{seccompare}

We compare the TSM of Sec.~\ref{seclinear} with the untilted WSM of Ref.~\cite{ips-nr-plasmon} and the untilted HSM of Ref.~\cite{ips-vnr-plasmon}, using the same velocity $v_0$ for all three. The main results are collected in Table~\ref{tab:compare}. The three systems have the same FS and the same band-velocity at the Fermi level. Their DOS, Drude weights, and bare plasmon frequencies therefore coincide, with $\omega_p \propto \mu \propto \tilde n^{1/3}$. The differences lie in the interband response, which is controlled by the quantum geometry of the eigenstates. The WSM has two bands, whereas the HSM and the TSM have the same three-band spectrum and differ only in their eigenstates. The QTSM of Sec.~\ref{secquad} is compared with the HSM and the WSM in a single subsection, since the quadratic correction changes the FS and the band-velocity at the Fermi level. The HSM and the WSM are also compared with each other. Each subsection refers to the continuum and the plasmon branches in panels~(a) and to the static response in panels~(b) of Figs.~\ref{fig:plasmon-static}, \ref{fig:qtsm-plasmon-static}, \ref{fig:qtsm-cpos-plasmon-static}, \ref{fig:hsm-plasmon-static}, and~\ref{fig:wsm-plasmon-static}. The Friedel oscillations of all systems are compared in Sec.~\ref{seccompfried}.

\subsection{TSM versus HSM}
\label{seccomphsm}

The TSM and the HSM share the spectrum, and hence their FS, Drude weights, bare plasmons, and the onset $\mu$ set by the flat-band. The eigenstates differ. The dispersive bands of the TSM carry the Chern number $2\,\chi$, and its pseudospin texture is invariant under all rotations. The HSM has a vanishing Chern number with a BC-dipole, and its texture is invariant only under rotations about $\bs{\hat z}$. This difference appears solely in the interband part.

In the TSM, the metric is isotropic and the transitions between the two dispersive bands carry no weight at order $q^2$. In the HSM, the metric of the flat-band transitions depends on the polar angle, and the dispersive pair contributes for in-plane wavevectors with the onset $2\,\mu$. Along $\bs{\hat z}$, the dispersive pair has no weight and the angular average of the flat-band metric equals that of the TSM. Hence the background and the absorption coincide with the isotropic TSM values. In the plane, they are smaller, with the coefficient of the logarithm in $\epsilon_b - 1$ equal to $5/8$ of the TSM value, and the absorption acquires a second step at $2\,\mu$.

The TSM plasmon is isotropic and equals the HSM plasmon along $\bs{\hat z}$, whereas the in-plane HSM plasmon has a higher static screened frequency. The anisotropy of the HSM was attributed to its BC-dipole~\cite{ips-vnr-plasmon}. The TSM has the same spectrum and a BC-monopole, and its response is isotropic. The anisotropy therefore originates from the eigenstates and not from the bands. At finite wavevectors, the two systems have the same regions free of single-particle excitations, since these follow from supports that do not depend on $\lambda_{\bs{q}}$. The undamped plasmon of both ends before $q = k_F$, as shown in Figs.~\ref{fig:plasmon-static}(a) and~\ref{fig:hsm-plasmon-static}(a). The Kohn singularity is of third order in the TSM and, off the axis, of second order in the HSM. The corresponding Friedel oscillations decay as $r^{-5}$ in the TSM and as $r^{-4}$ in the HSM, except along $\bs{\hat z}$, where the HSM decay is also $r^{-5}$.

\subsection{TSM versus WSM}
\label{seccomptsmwsm}

In the WSM, a single pair of bands contributes at order $q^2$, with the transition energy $2\,v_0\,k$ and the quantum metric $(\delta_{ij} - \hat k_i\,\hat k_j)/(4\,k^2)$. In the TSM, the flat-band is coupled to the upper band with the transition energy $v_0\,k$ and a metric of $(\delta_{ij} - \hat k_i\,\hat k_j)/(2\,k^2)$, which is twice as large. For a transition energy $\upsilon\,v_0\,k$, the radial integral contributes a factor $1/\upsilon$ to the polarisability. The interband polarisability of the TSM is therefore four times that of the WSM, with the same logarithmic dependence on the cut-off, and the background $\epsilon_b - 1$ and the absorption $\Im \epsilon^{\rm RPA}$ above the onset are enhanced by the same factor. The onset is halved, since the blocked region $k < k_F$ is the same while the transition energy is halved. The Drude weights coincide, and the ratio $\omega_p^2/\omega_{\rm th}^2$ is therefore four times larger for the TSM.

The monotonicity argument of Sec.~\ref{seclinscreen} applies to the WSM with $2\,\mu$ as the onset, and both systems then support a unique undamped plasmon below their onsets. At the same $\mu$ and $\Lambda$, the larger background of the TSM lowers its static screened frequency $\tilde\omega_p$ below that of the WSM, and its plasmon is confined to frequencies below $\mu$, whereas that of the WSM is bounded by $2\,\mu$. At finite wavevectors, the undamped plasmon of both systems exists only for $q < k_F$, in the windows $v_0\,q < \omega < \mu$ for the TSM and $v_0\,q < \omega < 2\,\mu - v_0\,q$ for the WSM, as shown in Figs.~\ref{fig:plasmon-static}(a) and~\ref{fig:wsm-plasmon-static}(a). Both systems have an absorption that is independent of $\omega$ above the onset, in contrast to the $1/\omega$ decay of the NRSs of Ref.~\cite{ips-nr-plasmon}. The Friedel oscillations decay as $r^{-5}$ in the TSM and as $r^{-4}$ in the WSM, as shown in Fig.~\ref{fig:friedel-all}.

\subsection{QTSM versus HSM and WSM}
\label{seccompqtsmhsm}
\label{seccompqtsmwsm}

The quadratic correction leaves the eigenstates unchanged, and the differences of the TSM from the HSM and the WSM therefore persist in the QTSM. The response of the QTSM remains isotropic for either sign of $c$, whereas that of the HSM depends on the direction of $\bs{q}$ through $\lambda_{\bs{q}}$. The correction removes the coincidence of the FS and of the band-velocity at the Fermi level, which underlies the comparison of the TSM with the HSM and the WSM. At the same $\mu$, the QTSM has the Fermi momentum $k_F = 2\,\mu/\left[ v_0 \left( 1 + \beta \right) \right]$ and the Fermi velocity $V_F = \beta\,v_0$ of Eq.~\eqref{eq:q-kf}, compared with $\mu/v_0$ and $v_0$ for the HSM and the WSM. For $c < 0$, only the top band has an FS, and the Drude weight of the QTSM is that of the HSM and the WSM multiplied by $4\,\beta/(1 + \beta)^2$, which does not exceed unity. The Thomas-Fermi wavevector obeys $q_{\rm TF}^2 = 3\,\omega_p^2/V_F^2$ for the QTSM with $c \leq 0$ and $q_{\rm TF}^2 = 3\,\omega_p^2/v_0^2$ for the HSM and the WSM.

The interband response depends on $c$ neither through the eigenstates and the overlap factors nor through the transition energy $v_0\,k$. For $c < 0$, it has the form of the TSM, with $\mu$ replaced by $\mu_1$ in the onset and in the logarithm. It has a single step at $\mu_1 = 2\,\mu/(1 + \beta)$, to be compared with the first onset $\mu$ of the HSM and the onset $2\,\mu$ of the WSM. There is no second step at $2\,\mu$, since the transitions between the two dispersive bands carry $\mathcal{F}_{1,-1} = \mathcal{O} \left( q^4 \right)$. The onset $\mu_1$ lies below the WSM onset $2\,\mu$ for every $c$ that admits an FS, since $\beta > 0$. The coefficient of $\epsilon_b - 1$ is $4\,g\,e^2/(3\,\pi\,v_0)$ and the absorption above the onset is $2\,g\,e^2/(3\,v_0)$. Both are independent of $c$ and four times the WSM values $g\,e^2/(3\,\pi\,v_0)$ and $g\,e^2/(6\,v_0)$. They are also $4/(4 - 3\,\lambda)$ times the HSM values, which is a factor of unity along $\bs{\hat z}$ and of $8/5$ in the plane. The ratio $\omega_p^2/\omega_{\rm th}^2$ of the QTSM is $\beta$ times that of the HSM and $4\,\beta$ times that of the WSM, and it reduces to the factor four of the TSM relative to the WSM for $\beta \to 1$.

For $c > 0$, the flat-band of the QTSM acquires an FS and contributes a Drude weight of the form of a parabolic band. The HSM has no such contribution, since its flat-band stays at zero energy and enters only the interband response, and the WSM has no flat-band. The contribution gives the term of Eq.~\eqref{eq:q-wp-dens} that is linear in $\tilde n_0$, in contrast to the scaling $\omega_p^2 \propto \tilde n^{2/3}$ of the HSM and the WSM, and it adds its DOS $\rho_0$ to $q_{\rm TF}^2$. The interband coefficients and the onset $\mu_1$ remain unchanged, because the transitions $-1 \to 0$ take over from the transitions $0 \to 1$ beyond $k_F^{(0)}$.

At finite wavevectors, the undamped plasmon of the HSM and the WSM can exist only for $q < k_F$, with $q_{\rm max} < k_F$ for the HSM and $q_{\rm max} \leq k_F$ for the WSM, as shown in Figs.~\ref{fig:hsm-plasmon-static}(a) and~\ref{fig:wsm-plasmon-static}(a). The WSM window is $v_0\,q < \omega < 2\,\mu - v_0\,q$, and the criteria that end its branch are given in Sec.~\ref{secwexp}. In the QTSM, the branch ends at the wavevector $q_{\rm max}$ at which the zero of $\Re\epsilon^{\rm RPA}$ reaches the upper edge of the intraband continuum or the lowest interband onset, as shown in Fig.~\ref{fig:qtsm-plasmon-static}(a) for $c < 0$. For $c > 0$, the transitions between the two FSs damp the mode at all frequencies for $q_1 \leq q \leq q_2$, and hence $q_{\rm max} \leq q_1 = k_F^{(0)} - k_F$. The correction to the HSM dispersion is isotropic, and the QTSM dispersion of Eq.~\eqref{eq:qf-wdisp} holds at order $q^2$ in the free window.

The overlap factor $\mathcal{F}_{11}$ of the top band of the QTSM vanishes quartically at backscattering, whereas that of the WSM vanishes quadratically. The Friedel oscillations of the top band of the QTSM therefore decay as $r^{-5}$ in every direction, compared with $r^{-4}$ for the WSM and for the HSM off the axis, and $r^{-5}$ for the HSM along $\bs{\hat z}$. For $c > 0$, the overlap factor of the flat-band is unity at backscattering, and its oscillations decay as $r^{-3}$, as in an ordinary electron gas, which is more slowly than those of the HSM and the WSM in any direction. They dominate at large distances, and the transitions between the two FSs add isotropic oscillations that decay as $r^{-4}$, as shown in Fig.~\ref{fig:friedel-all}.

The static response of all three systems takes the Thomas-Fermi form at small $q$, as shown in panels~(b) of Figs.~\ref{fig:qtsm-plasmon-static},~\ref{fig:qtsm-cpos-plasmon-static},~\ref{fig:hsm-plasmon-static}, and~\ref{fig:wsm-plasmon-static}. The static background of the HSM contains, besides the logarithm, a constant that depends on $\lambda_{\bs{q}}$. In the WSM, the logarithms in $q$ cancel between the intraband and interband parts of the closed form of Eq.~\eqref{eq:w-stat}. In the QTSM, the logarithm of the interband part is the same as in the TSM, and the intraband terms of order $q^2$ shift the background by $-\,(\alpha/\pi)\,\mathcal{S}$ of Eq.~\eqref{eq:qf-tf}, where $\mathcal{S}$ depends on $c\,k_F$ and, for $c > 0$, on the flat-band. For $c \to 0^-$, we have $\mathcal{S} \to 7/9$, which equals the shift of the HSM along $\bs{\hat z}$, whereas the in-plane shift of the HSM contains an additional term that depends on $\ln 2$. For $c > 0$, the term $\mathcal{S}$ acquires the contribution of the flat-band, which diverges for $c \to 0^+$. For $q \gg k_F$, the doped part of $\Pi (q, 0)$ of the HSM approaches a constant, which equals that of the linear model along $\bs{\hat z}$ and is reduced by the factor $1 - \lambda$ in general, whereas that of the WSM is smaller than its vacuum part and falls off as $k_F^4/q^2$. In the QTSM, this constant is replaced by the correction of Eq.~\eqref{eq:qf-large}, which decays as $q^{-2}$ and is controlled by the quadratic term. It is proportional to $k_F^3$ for $c < 0$ and to $k_F^3 + (k_F^{(0)})^3$ for $c > 0$, and inversely proportional to $c$. The limits $c \to 0$ and $q \to \infty$ therefore do not commute. The HSM and the WSM have the closed forms of Eqs.~\eqref{eq:h-pistat} and~\eqref{eq:w-stat} for all $q$, whereas for the QTSM only the small-$q$ and large-$q$ forms are available, and the value at an arbitrary wavevector follows from the integral of the imaginary part.

\subsection{HSM versus WSM}
\label{seccomphsmwsm}

The WSM has two bands and a BC-monopole with the Chern number $\pm 1$, whereas the HSM has three bands, a vanishing Chern number, and a BC-dipole. The two systems share the FS and the band-velocity at the Fermi level. Their DOS, Drude weights, and isotropic bare plasmon frequency therefore coincide, and $q_{\rm TF}^2 = 3\,\omega_p^2/v_0^2$ holds for both. The differences lie in the interband response. In Eq.~\eqref{eq:h-imeps0}, the absorption of the HSM is the sum of two terms. The term with the weight $\lambda = \lambda_{\bs{q}}$ and the onset $2\,\mu$ has exactly the form of the WSM absorption. The term with the weight $1 - \lambda$ and the onset $\mu$ comes from the flat-band and has the TSM form, which is four times the WSM value. The background of Sec.~\ref{sechstatic} has the same structure, and its coefficient is $\left[ 4\,(1 - \lambda) + \lambda \right]$ times the WSM coefficient, as is the absorption above $2\,\mu$. This factor equals four for $\lambda = 0$ and $5/2$ for $\lambda = 1/2$. Since $\lambda \leq 1/2$, the HSM has a larger background and a larger absorption than the WSM for every direction of $\bs{q}$. Its onset is $\mu$ instead of $2\,\mu$, and the ratio $\omega_p^2/\omega_{\rm th}^2$ is four times larger.

Both systems have a unique undamped plasmon below the onset at long wavelengths, below $\mu$ for the HSM and below $2\,\mu$ for the WSM. At the same $\mu$ and $\Lambda$, the larger background of the HSM lowers its static screened frequency below that of the WSM. The screened frequency $\tilde\omega_p = \omega_p/\sqrt{\epsilon_b}$ of the WSM is isotropic, whereas that of the HSM depends on the direction of $\bs{q}$ through $\epsilon_b (\lambda_{\bs{q}})$, which is the only source of its anisotropy at small $q$. The Bohm-Gross dispersion has the coefficient $3\,v_0^2/5$ and is isotropic in both systems.

At finite wavevectors, the undamped plasmon of both systems can exist only for $q < k_F$, as shown in Figs.~\ref{fig:hsm-plasmon-static}(a) and~\ref{fig:wsm-plasmon-static}(a). The branch of the HSM ends at a critical wavevector $q_{\rm max} < k_F$, whereas that of the WSM ends at $q_{\rm max} \leq k_F$. Both systems have a weak Kohn singularity, because backscattering is suppressed quadratically in the deviation from the backward direction, as shown in Figs.~\ref{fig:hsm-plasmon-static}(b) and~\ref{fig:wsm-plasmon-static}(b). Both have Friedel oscillations with the period $\pi/k_F$ that decay as $r^{-4}$. The amplitude of the WSM is isotropic, whereas that of the HSM is proportional to $\sin^2\vartheta_{\bs{r}}$ and vanishes along $\bs{\hat z}$, where the decay is $r^{-5}$ as in the TSM.

\subsection{Friedel oscillations}
\label{seccompfried}

\begin{figure}[t!]
\centering
\includegraphics[width=\linewidth]{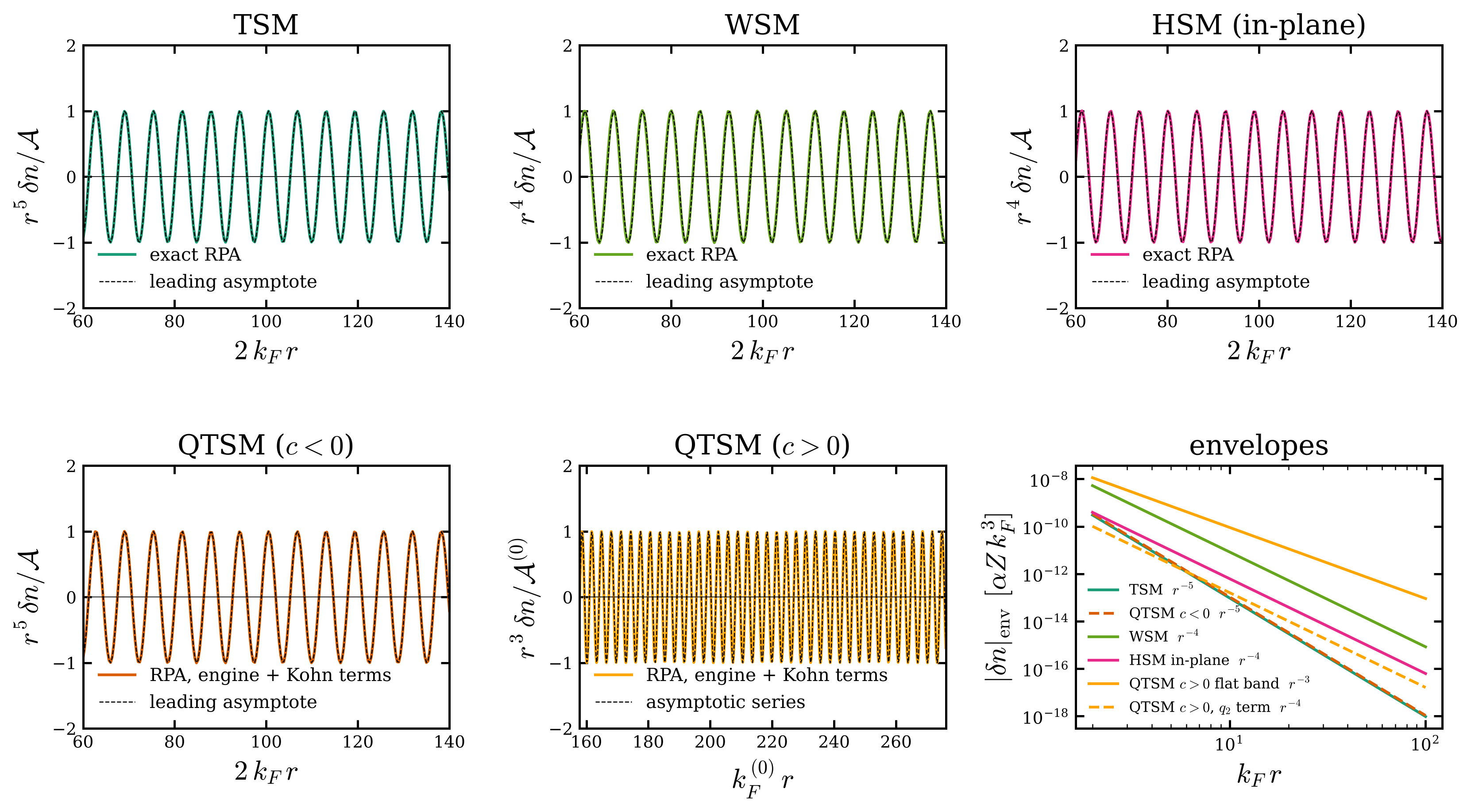}
\caption{\label{fig:friedel-all} Friedel oscillations of the induced charge density $\delta n (r)$ around a Coulomb impurity. Upper panels: TSM, WSM, and in-plane HSM. Lower panels: QTSM with $c < 0$ and $c > 0$, and the envelopes of all cases.}
\end{figure}

Fig.~\ref{fig:friedel-all} shows the induced charge density $\delta n(r)$ around a Coulomb impurity of charge $Z\,e$. The panels show $r^{\ell}\,\delta n(r)$, normalised by its asymptotic amplitude $\mathcal{A}$, for the TSM, the WSM, the in-plane HSM, and the QTSM with $c < 0$ and $c > 0$, and the last panel compares the envelopes. The exponent $\ell$ is fixed by the power with which the intraband overlap factor of the top band vanishes at backscattering, since this power sets the order of the Kohn singularity of $\Pi (q, 0)$ at $q = 2\,k_F$. A singularity $(q - 2\,k_F)^{\tilde \ell} \ln |q - 2\,k_F|$ gives $\delta n (r) \propto \sin \!\left( 2\,k_F\,r \right)/r^{\tilde \ell + 1}$ for even $\tilde \ell $ and $\cos \!\left( 2\,k_F\,r \right)/r^{ \tilde \ell +1}$ for odd $ \tilde \ell $, with $ \ell = \tilde \ell + 1$. For the TSM and the QTSM with $c < 0$, the factor $\mathcal{F}_{11} = (1 + \cos\theta)^2/ 4$ vanishes quartically at $\theta = \pi$, the singularity is cubic, and $ \ell = 5 $. For the HSM and the WSM, the factor vanishes quadratically, the singularity is quadratic, and $ \ell = 4$. For the QTSM with $c > 0$, the flat-band overlap factor $\mathcal{F}_{00} = \cos^2\theta$ equals unity at backscattering. The flat-band response then has the parabolic singularity $(q - 2\,k_F^{(0)}) \ln |q - 2\,k_F^{(0)}|$, which dominates at large $r$ and gives $ \ell = 3$. The Kohn singularities are visible as the logarithmic divergences of the derivatives of $\Pi (q, 0)$ in panels~(b) of Figs.~\ref{fig:plasmon-static}, \ref{fig:qtsm-plasmon-static}, \ref{fig:qtsm-cpos-plasmon-static}, \ref{fig:hsm-plasmon-static}, and~\ref{fig:wsm-plasmon-static}.

The upper panels show the exact RPA result for the TSM, the WSM, and the in-plane HSM. It is obtained from the radial Fourier transform of $\mathcal{U}(q)\,\Pi(q,0)/\epsilon^{\rm RPA}(q,0)$ with the closed forms of Eqs.~\eqref{eq:fin-static}, \eqref{eq:w-stat}, and~\eqref{eq:h-pistat}. A smooth window replaces the sharp cut-off at $\Lambda$ and lies below $\Lambda/k_F = v_0\,\Lambda/\mu$, up to which the linearised models hold. The integrand grows with $q$, and a sharp cut-off produces a boundary term that exceeds the Friedel tail by orders of magnitude. For the TSM, the Thomas-Fermi screening cloud decays as $\exp(-q_{\rm TF}\,r)$ with $q_{\rm TF} = \sqrt{2\,\alpha/\pi}\,k_F$, and it is negligible in the range shown. The Friedel tail is small in absolute terms, and the transforms are evaluated with high-precision arithmetic. The panels show the range $30<k_F\,r<70$. The dashed curves are the leading terms of Eqs.~\eqref{eq:fin-friedel}, \eqref{eq:w-friedel}, and~\eqref{eq:h-friedel}, with $\epsilon^{\rm RPA}(2\,k_F,0)$ taken from the same closed forms. The exact curves follow them in amplitude and phase. The HSM panel is drawn for the in-plane direction with $\lambda_{\bs q} = 1/2$, for which the amplitude $\sin^2\vartheta_{\bs r}$ of Eq.~\eqref{eq:h-friedel} is maximal. The radial transform with the in-plane $\Pi(q,0)$ reproduces the leading term of the in-plane direction, and the oscillations along $\bs{\hat z}$ decay one power faster.

The QTSM lacks a closed form for the full $\Pi(q,0)$. Only the intraband part of the flat-band for $c>0$ has one, up to a single radial integral, in Eq.~\eqref{eq:qf-pi00}. The remaining channels follow from Eq.~\eqref{eq:qf-kk}, and their numerical evaluation is limited by the accuracy of the quadrature over $\omega$. For $c<0$, the Friedel term in the window of $k_F\,r$ shown lies below this accuracy, and the radial transform of the numerical $\Pi(q,0)$ deviates strongly from the leading asymptote. We therefore write $\Pi(q,0)=\sum_i C_i\,(q-q_i)^{n_i}\ln|q-q_i|+P(q)$, where the analytic Kohn terms of Eqs.~\eqref{eq:qf-kohn1}, \eqref{eq:qf-kohn0}, and~\eqref{eq:qf-kohn01} are added to a smooth remainder $P(q)$ fitted by one Chebyshev polynomial to the numerical values. This construction is used for both signs of $c$, since it needs only a small number of evaluations of $\Pi(q,0)$. The label ``engine + Kohn terms'' in Fig.~\ref{fig:friedel-all} refers to this decomposition. The fit coefficients decay to the noise floor only if the Kohn coefficients are correct, which checks the Kohn terms. As a further check, the same procedure reproduces the TSM closed-form result. The lower left and middle panels show the radial transform of $\mathcal{U}(q)\,\Pi(q,0)/\epsilon^{\rm RPA}(q,0)$ for $c<0$ and $c>0$. The $c<0$ curve follows $+\cos(2\,k_F\,r)$ with the amplitude of Eq.~\eqref{eq:qf-friedel1}, as in the TSM. For $c>0$, the window is narrower because $k_F^{(0)}\gg k_F$, and we show $16<k_F\,r<28$, that is $158<k_F^{(0)}\,r<276$, where the curve follows the asymptotic series of Eqs.~\eqref{eq:qf-friedel1}, \eqref{eq:qf-friedel0}, and~\eqref{eq:qf-friedel01}, with the flat-band term dominating and the other terms a small modulation. The lower right panel compares the envelopes of all cases. The TSM and $c<0$ QTSM envelopes nearly coincide, while the HSM envelope lies below the WSM because of the larger numerical factor in the denominator of Eq.~\eqref{eq:h-friedel} and the stronger screening at $q=2\,k_F$.

In Fig.~\ref{fig:friedel-all}, the dashed curves in the QTSM panels are $+\cos \!\left( 2\,k_F\,r \right)$ for $c<0$ and, for $c>0$, the asymptotic series of the flat-band term, the term of the transitions between the two FSs, and the top-band term. The lower right panel shows the envelopes $|\mathcal{A}|/r^{\ell}$ in units of $g\,Z\,e^2\,k_F^3/v_0$, with $\epsilon^{\rm RPA} (q, 0)$ evaluated at $q = 2\,k_F$, $2\,k_F^{(0)}$, and $q_2$. The parameters of the TSM, WSM, and HSM are those of Figs.~\ref{fig:plasmon-static}, \ref{fig:wsm-plasmon-static}, and~\ref{fig:hsm-plasmon-static}, and those of the QTSM with $c<0$ and $c>0$ are those of Figs.~\ref{fig:qtsm-plasmon-static} and~\ref{fig:qtsm-cpos-plasmon-static}, respectively. These give $\xi = c\,\mu/v_0 = -\,0.025$ and $0.0105$, $\beta \simeq 0.949$ and $1.021$, and $k_F^{(0)}/k_F \simeq 9.86$ for $c > 0$. The condition $2\,c\,q < 1$ then holds up to $q = 2\,k_F^{(0)}$.

\section{Summary and outlook}
\label{secsum}

We have studied the plasmons of pseudospin-$1$ TSMs within the RPA, for the linear model and for the QTSM, in which a quadratic correction makes the flat-band dispersive. In the linear model, the flat-band affects only the interband response. The Drude weight and the bare plasmon frequency are fixed by the FS and the band-velocity of the top band and are isotropic. The intrinsic system has no zero of the dielectric function and hence no undamped plasmon. In the doped regime, the interband continuum starts at the transitions between the flat-band and the top band. Below this onset, the real part of the dielectric function is monotonic and has a unique zero, which is an undamped and isotropic plasmon. The interband screening depends logarithmically on the cut-off of the linearised description, and only its dependence on the chemical potential is therefore independent of the regularisation.

In the QTSM, the quadratic correction leaves the eigenstates and the overlap factors unchanged and acts only through the band energies. When the top band is the only band with an FS, the plasmon retains the structure of the linear model, with a renormalised Drude weight and a shifted interband onset. When the flat-band also has an FS, it adds a Drude weight of the form of a parabolic band. The interband response is insensitive to this addition, because the transitions into the flat-band take over from those out of it at its Fermi momentum. The screening and the onset therefore remain governed by the top band.

At generic wavevectors, the plasmon disperses upwards and remains undamped until it meets the edge of the intraband continuum, beyond which it is Landau damped. In the QTSM, the polarisability is assembled channel by channel, and the plasmon lies in the window free of excitations. The correction modifies the dispersion through a velocity that is averaged over the FSs with the Drude weights.
 When the flat-band has an FS, the critical wavevector cannot exceed the separation of the two FSs, because the transitions between them damp the mode at all frequencies in an intermediate range of wavevectors. The static response has the Thomas-Fermi form at small wavevectors, with a background that is shifted by the intraband terms. At large wavevectors, the doped system approaches the intrinsic one with a correction that is controlled by the quadratic term, and its limit does not commute with that of the linear model. The overlap factor of the top band suppresses backscattering, which weakens the Kohn anomaly, and the Friedel oscillations therefore decay faster than in an ordinary electron gas. When the flat-band has an FS, it restores the conventional decay at large distances, and the transitions between the two FSs add a further singularity.

The HSM and the untilted WSM have been treated at generic wavevectors with the same cut-off prescription. In both systems, the undamped plasmon exists only below the Fermi momentum, and the Kohn anomaly is weak for the same reason as in the TSM. The Friedel amplitude is isotropic for the WSM, whereas for the HSM it depends on the direction and the decay is faster along the $z$-axis, as in the TSM. At long wavelengths, the anisotropy of the HSM plasmon is contained entirely in the background dielectric function.

The comparison separates the roles of the spectrum and of the eigenstates. The TSM shares the FS and the band-velocity with the WSM, and hence its Drude part. The flat-band lowers the interband onset and enhances the interband coefficients relative to the WSM. The TSM shares the spectrum with the HSM and differs in the BC structure of the eigenstates. The BC-monopole of the TSM gives an isotropic response, whereas the BC-dipole of the HSM makes the interband screening anisotropic. The anisotropy of the HSM plasmon thus originates from the eigenstates and not from the bands. The quadratic correction removes the coincidence of the FS and of the band-velocity with those of the WSM and the HSM, but it does not change the eigenstates, and the QTSM therefore remains isotropic.

Several extensions are natural. A tilt of the node would make the FS anisotropic and introduce the distinction between type-I and type-II nodes, as for the WSM~\cite{ips-nr-plasmon, yadav23_magneto, ips-ruiz, ips-tilted}. Finite temperature, disorder, and interactions beyond the RPA remain to be explored. Another direction is the high-doping regime of the NRSs. The results of Refs.~\cite{ips-nr-plasmon} and~\cite{ips-vnr-plasmon} rely on the mapping onto decoupled two-dimensional Dirac cones, which requires $k_F = \mu/v_0$ to be much smaller than the ring radius $k_0$. At higher doping, where $k_F$ becomes comparable to $k_0$, this mapping fails, since the FS is no longer approximately cylindrical and the curvature of the ring enters the overlap factors. The full lattice Hamiltonian, or at least an expansion that retains the angular dependence of the band structure along the ring, must then be used. An analytic treatment may still be possible for the imaginary part of the polarisability, but the real part will in general require numerical evaluation of the Kramers-Kronig integral over the full Brillouin zone. Such a study would also give access to the crossover between the ring-dominated and the point-node-dominated regimes, and to the fate of the plasmon when the FS deforms into a spindle torus or a more general self-intersecting cyclide.
In our companion paper \cite{ips-plasmon-rsw-kw}, we have applied the same approach to other multifold nodes, viz. the pseudospin-$3/2$ Rarita-Schwinger-Weyl and Krame-Weyl nodes~\cite{bernevig, grushin-multifold, ips-exact-rsw, ips-exact-kwn}.

\section*{Data Availability Statement}
This manuscript has no associated data.

\appendix

\section{TSM at small momentum-transfers}
\label{app:tsm}

This appendix gives the intermediate steps behind Sec.~\ref{seclinear}, where $c = 0$ and $\chi = 1$. Throughout, $\varepsilon_s (\bs{k}) = s\,v_0\,k$ denotes the energy of the band $s$, and $k_F = \mu/v_0$ denotes the Fermi momentum for $\mu > 0$. The spin-$1$ matrices are $\bs{S} = (\bs{\sigma} \otimes \mathbb{I} + \mathbb{I} \otimes \bs{\sigma})/2$, and the spin-$1$ eigenstates along a unit vector $\bs{n}$ are the symmetric products of the spin-$1/2$ eigenstates $|{\uparrow_{\bs{n}}}\rangle$ and $|{\downarrow_{\bs{n}}}\rangle$ of $\bs{n} \cdot \bs{\sigma}$,
\begin{align}
|1, \bs{n} \rangle = |{\uparrow_{\bs{n}}} {\uparrow_{\bs{n}}} \rangle\,, \quad
|0, \bs{n} \rangle = \frac{|{\uparrow_{\bs{n}}} {\downarrow_{\bs{n}}} \rangle + |{\downarrow_{\bs{n}}} {\uparrow_{\bs{n}}} \rangle}{\sqrt{2}}\,, \quad
|-1, \bs{n} \rangle = |{\downarrow_{\bs{n}}} {\downarrow_{\bs{n}}} \rangle\,.
\end{align}
These states satisfy $\bs{n} \cdot \bs{S}\,|s, \bs{n} \rangle = s\,|s, \bs{n} \rangle$. For $\mathcal{H}_\chi$, we have $\bs{n} = \hat{\bs{\tilde k}}$ and $\bs{\tilde k} \cdot \bs{\tilde k}' = \bs{k} \cdot \bs{k}'$ because $\chi^2 = 1$, and the overlap factors therefore depend only on $\cos\theta = \hat{\bs{k}} \cdot \hat{\bs{k}}'$.

Let $\bs{n}$ and $\bs{n}'$ enclose the angle $\theta$. The two spin-$1/2$ bases are related by an SU(2) matrix, which fixes
\begin{align}
\langle {\uparrow'} | {\uparrow} \rangle = \mathfrak{u}\,, \quad
\langle {\downarrow'} | {\uparrow} \rangle = \mathfrak{w}\,, \quad
\langle {\uparrow'} | {\downarrow} \rangle = -\,\mathfrak{w}^*\,, \quad
\langle {\downarrow'} | {\downarrow} \rangle = \mathfrak{u}^*\,, \quad
|\mathfrak{u}|^2 = \cos^2 \!\left( \frac{\theta}{2} \right)\,, \quad |\mathfrak{w}|^2 = \sin^2 \!\left( \frac{\theta}{2} \right)\,.
\end{align}
The squared overlaps are independent of the phase convention, and the amplitudes between the spin-$1$ states are
\begin{align}
\langle 1' | 1 \rangle = \mathfrak{u}^2\,, \quad
\langle -1' | 1 \rangle = \mathfrak{w}^2\,, \quad
\langle 0' | 1 \rangle = \sqrt{2}\,\mathfrak{u}\,\mathfrak{w}\,, \quad
\langle 0' | 0 \rangle = |\mathfrak{u}|^2 - |\mathfrak{w}|^2 = \cos\theta\,,
\end{align}
and the remaining ones follow by symmetry. For example, $\langle 0' | 1 \rangle = \left( \langle {\uparrow'} | {\uparrow} \rangle \langle {\downarrow'} | {\uparrow} \rangle + \langle {\downarrow'} | {\uparrow} \rangle \langle {\uparrow'} | {\uparrow} \rangle \right)/\sqrt{2}$. Taking the squared moduli gives
\begin{align}
\mathcal{F}_{11} = \cos^4 \!\left( \frac{\theta}{2} \right)\,, \quad
\mathcal{F}_{1,-1} = \sin^4 \!\left( \frac{\theta}{2} \right)\,, \quad
\mathcal{F}_{10} = 2\,\cos^2 \!\left( \frac{\theta}{2} \right)\,\sin^2 \!\left( \frac{\theta}{2} \right) = \frac{\sin^2\theta}{2}\,, \quad
\mathcal{F}_{00} = \cos^2\theta\,,
\end{align}
which are the entries of Eq.~\eqref{eq:lin-ff} and satisfy the sum rule $\mathcal{F}_{11} + \mathcal{F}_{10} + \mathcal{F}_{1,-1} = \left[ (1 + \cos\theta)^2 + 2\,\sin^2\theta + (1 - \cos\theta)^2 \right]/4 = 1$.

For $\bs{k}' = \bs{k} + \bs{q}$, the sine of the angle is $\sin\theta = |\hat{\bs{k}} \times \hat{\bs{k}}'| = |\bs{k} \times \bs{q}|/(k\,k')$ with $k' = |\bs{k} + \bs{q}|$. For $q \ll k$,
\begin{align}
\sin^2\theta \simeq \frac{q^2 - \left( \bs{q} \cdot \hat{\bs{k}} \right)^2}{k^2} = \frac{q_i\,q_j \left( \delta_{ij} - \hat k_i\,\hat k_j \right)}{k^2}\,.
\end{align}
Hence $\mathcal{F}_{10} \simeq q_i\,q_j\,Q_{ij}$ with $Q_{ij}$ of Eq.~\eqref{eq:lin-qm}. The factor $\mathcal{F}_{1,-1} \simeq \theta^4/16$ is of order $q^4$, and the intraband factor is $\mathcal{F}_{11} = 1 - \sin^2\theta/2 + \mathcal{O} \left( q^4 \right)$. The same expansions hold for the pair $(0, -1)$ with the roles of $s = \pm 1$ exchanged.

The angular average follows from $\int d\tilde\Omega\,\hat k_i\,\hat k_j = 4\,\pi\,\delta_{ij}/3$ and $\int d\tilde\Omega = 4\,\pi$,
\begin{align}
\label{eq:lin-ang}
\int d\tilde\Omega\;k^2\,Q_{ij} = \frac{1}{2} \left( 4\,\pi - \frac{4\,\pi}{3} \right) \delta_{ij} = \frac{4\,\pi}{3}\,\delta_{ij}\,.
\end{align}

\subsection{Intrinsic limit}
\label{app:lin-intr}

We give the steps behind Eqs.~\eqref{eq:lin-pi0} to~\eqref{eq:lin-eps0}. For $\mu \to 0^+$ at $T = 0$, the occupations are $f_0 (\varepsilon_{-1}) = f_0 (\varepsilon_0) = 1$ and $f_0 (\varepsilon_1) = 0$. The occupation factor in Eq.~\eqref{eq:pol} then vanishes for $s = s'$ and for the pairs $(-1, 0)$ and $(0, -1)$. The pairs $(-1, 1)$ and $(1, -1)$ carry $\mathcal{F}_{1,-1} = \mathcal{O} \left( q^4 \right)$ and are dropped. The pair $(0, 1)$ has the occupation factor $+1$ and the denominator $\omega - v_0\,k'$, while the pair $(1, 0)$ has the factor $-1$ and the denominator $\omega + v_0\,k$. Since $\mathcal{F}_{10}$ is already of order $q^2$, we set $k' = k$ in the denominators, and the sum becomes
\begin{align}
\mathcal{F}_{10} \left( \frac{1}{\omega - v_0\,k + i\,0^+} - \frac{1}{\omega + v_0\,k + i\,0^+} \right) = \mathcal{F}_{10}\,\frac{2\,v_0\,k}{(\omega + i\,0^+)^2 - v_0^2\,k^2}\,.
\end{align}
This replacement requires $q \ll k$ at the momenta that dominate the integral, which are $k = \omega/v_0$ for the imaginary part and $k \gtrsim \omega/v_0$ for the real part. The condition therefore reads $v_0\,q \ll \omega$.

With the measure $d^3 \bs{k} = k^2\,dk\,d\tilde\Omega$, the angular integral of Eq.~\eqref{eq:lin-ang} yields the prefactor $g\,q_i\,q_j\,(4\,\pi/3)\,\delta_{ij}/(2\,\pi)^3 = g\,q^2/(6\,\pi^2)$ of Eq.~\eqref{eq:lin-pi0}. For the real part of the radial integral, we use $\partial_k \ln \!\left( |\omega^2 - v_0^2\,k^2| \right) = -\,2\,v_0^2\,k/(\omega^2 - v_0^2\,k^2)$. The principal value is
\begin{align}
\Re \int_0^{\Lambda} dk\, \frac{2\,v_0\,k}{\omega^2 - v_0^2\,k^2} = -\,\frac{1}{v_0} \left[ \ln \!\left( \left| \omega^2 - v_0^2\,k^2 \right| \right) \right]_0^{\Lambda} = -\,\frac{1}{v_0}\,\ln \!\left( \frac{v_0^2\,\Lambda^2 - \omega^2}{\omega^2} \right) \simeq -\,\frac{1}{v_0}\,\ln \!\left( \frac{v_0^2\,\Lambda^2}{\omega^2} \right),
\end{align}
where the last step uses $\omega \ll v_0\,\Lambda$. For the imaginary part, only the first term of the bracket has a pole for $\omega > 0$, and the relation $\Im \left[ 1/(\omega - v_0\,k + i\,0^+) \right] = -\,\pi\,\delta (\omega - v_0\,k)$ then gives
\begin{align}
\Im \int_0^{\Lambda} dk\, \frac{2\,v_0\,k}{(\omega + i\,0^+)^2 - v_0^2\,k^2} = -\,\pi \int_0^{\Lambda} dk\, \delta (\omega - v_0\,k) = -\,\frac{\pi}{v_0}\,,
\end{align}
since $k = \omega/v_0$ lies in $(0, \Lambda)$. Multiplying these results by $g\,q^2/(6\,\pi^2)$ gives Eq.~\eqref{eq:lin-pi0b}. Because $\Pi_{\rm intra} = 0$ at the intrinsic point, Eq.~\eqref{eq:lin-eps0} follows from Eq.~\eqref{eq:epsrpa} through $\Re \epsilon^{\rm RPA} = 1 - 4\,\pi\,e^2\,\Re \Pi/q^2$ and $\Im \epsilon^{\rm RPA} = -\,4\,\pi\,e^2\,\Im \Pi/q^2$, where the second relation gives $4\,\pi\,e^2\,g/(6\,\pi\,v_0) = 2\,g\,e^2/(3\,v_0)$.

\subsection{Doped regime: DOS and Drude weights}
\label{app:lin-drude}

We give the steps behind Eqs.~\eqref{eq:lin-dos} and~\eqref{eq:lin-wp}. For $\mu > 0$, the bands $s = -1$ and $s = 0$, at $\varepsilon_{-1} = -\,v_0\,k \leq 0$ and $\varepsilon_0 = 0$, are filled. Only the band $s = 1$ has an FS, with $\delta (\varepsilon_1 - \mu) = \delta (k - k_F)/v_0$. The DOS, $\rho = g \int d^3 \bs{k}/(2\,\pi)^3\,\delta (\varepsilon_1 - \mu)$, is
\begin{align}
\rho (\mu) = \frac{g}{(2\,\pi)^3}\, \frac{4\,\pi\,k_F^2}{v_0} = \frac{g\,\mu^2}{2\,\pi^2\,v_0^3}\,.
\end{align}
The density measured from $\mu = 0$ is $\tilde n = \int_0^\mu d\mu'\,\rho (\mu') = g\,\mu^3/(6\,\pi^2\,v_0^3)$, which equals $g\,(4\,\pi\,k_F^3/3)/(2\,\pi)^3$, the number of states inside the Fermi sphere. With $\bs{V}^{(1)} = v_0\,\hat{\bs{k}}$ and $\langle \hat k_i\,\hat k_j \rangle = \delta_{ij}/3$, Eq.~\eqref{eq:drude} gives
\begin{align}
D_{ij} = \pi\,e^2\,\rho\,v_0^2\,\frac{\delta_{ij}}{3} = \frac{g\,e^2\,\mu^2}{6\,\pi\,v_0}\,\delta_{ij}\,.
\end{align}

We now derive Eq.~\eqref{eq:intra} for the present case. Only the band $s = 1$ is partially occupied, and the occupation differences of the filled bands vanish. With $\mathcal{F}_{11} = 1 - \mathcal{O} \left( q^2 \right)$, we expand in $q$ at fixed $\omega$. At $T = 0$,
\begin{align}
f_0 (\varepsilon_1 (\bs{k})) - f_0 (\varepsilon_1 (\bs{k}')) \simeq q_i\,V_i^{(1)}\,\delta (\varepsilon_1 - \mu)\,, \quad
\omega + \varepsilon_1 (\bs{k}) - \varepsilon_1 (\bs{k}') \simeq \omega - q_j\,V_j^{(1)}\,.
\end{align}
Expanding $1/(\omega - q_j\,V_j^{(1)})$ in powers of $q/\omega$ gives
\begin{align}
\Pi_{\rm intra} (\bs{q}, \omega) \simeq g \int \frac{d^3 \bs{k}}{(2\,\pi)^3}\,\delta (\varepsilon_1 - \mu) \left[ \frac{q_i\,V_i^{(1)}}{\omega} + \frac{q_i\,q_j\,V_i^{(1)}\,V_j^{(1)}}{\omega^2} \right].
\end{align}
The first term vanishes, because it is odd under $\hat{\bs{k}} \to -\,\hat{\bs{k}}$. The second term equals $q_i\,q_j\,D_{ij}/(\pi\,e^2\,\omega^2)$, which is Eq.~\eqref{eq:intra}. Since $D_{ij}$ is proportional to $\delta_{ij}$, we have $q_i\,q_j\,D_{ij} = q^2\,D_{ii}$ with no sum. Hence
\begin{align}
\epsilon^{\rm RPA} (\bs{q}, \omega) = 1 - \frac{4\,\pi\,e^2}{q^2}\,\frac{q^2\,D_{ii}}{\pi\,e^2\,\omega^2} = 1 - \frac{4\,D_{ii}}{\omega^2}\,,
\end{align}
and the zero gives $\omega_p^2 = 4\,D_{ii} = 2\,g\,e^2\,\mu^2/(3\,\pi\,v_0)$, as in Eq.~\eqref{eq:lin-wp}.

\subsection{Interband screening and plasmons}
\label{app:lin-screen}

We provide the steps behind Eqs.~\eqref{eq:lin-eps} and~\eqref{eq:lin-eb} and behind the bound on $\omega_*$ in Sec.~\ref{seclinscreen}, assuming $\mu > 0$, $v_0\,q \ll \omega$, and $q \ll k_F$ for the expansion of the overlap factors. The occupations are $f_0 (\varepsilon_{-1}) = f_0 (\varepsilon_0) = 1$ and $f_0 (\varepsilon_1) = \Theta (k_F - k)$. The pairs $(-1, 0)$ and $(0, -1)$ remain blocked at every $k$, and the pairs involving $s = \pm 1$ are dropped as in Appendix~\ref{app:lin-intr}. For the pairs $(0, 1)$ and $(1, 0)$, the occupation factors are $1 - \Theta (k_F - k) = \Theta (k - k_F)$ and its negative, respectively. The result of Appendix~\ref{app:lin-intr} therefore holds with the lower limit of the radial integral replaced by $k_F$. The real part becomes
\begin{align}
\Re \int_{k_F}^{\Lambda} dk\, \frac{2\,v_0\,k}{\omega^2 - v_0^2\,k^2} = -\,\frac{1}{v_0}\,\ln \!\left( \frac{v_0^2\,\Lambda^2 - \omega^2}{|\mu^2 - \omega^2|} \right) \simeq -\,\frac{1}{v_0}\,\ln \!\left( \frac{v_0^2\,\Lambda^2}{|\mu^2 - \omega^2|} \right),
\end{align}
where $v_0\,k_F = \mu$ has been used. The imaginary part is $-\,\pi\,\Theta (\omega - \mu)/v_0$, because the delta function fixes $k = \omega/v_0$, which lies above $k_F$ if and only if $\omega > \mu$. After adding $\Pi_{\rm intra} = q^2\,D_{ii}/(\pi\,e^2\,\omega^2)$ from Appendix~\ref{app:lin-drude}, Eq.~\eqref{eq:epsrpa} gives Eq.~\eqref{eq:lin-eps}. In the static limit $\omega \to 0$, the logarithm tends to $2\,\ln \!\left( v_0\,\Lambda/\mu \right)$, which gives Eq.~\eqref{eq:lin-eb}.

For the zero of the real part, we write $\Re \epsilon^{\rm RPA} = \epsilon_b (\omega) - \omega_p^2/\omega^2$ for $0 < \omega < \mu$, with
\begin{align}
\epsilon_b (\omega) = 1 + \frac{2\,g\,e^2}{3\,\pi\,v_0}\,\ln \!\left( \frac{v_0^2\,\Lambda^2}{\mu^2 - \omega^2} \right).
\end{align}
The derivative $\partial_\omega \epsilon_b (\omega) = 4\,g\,e^2\,\omega/[3\,\pi\,v_0\,(\mu^2 - \omega^2)]$ and the derivative of the Drude term, $2\,\omega_p^2/\omega^3$, are both positive on $(0, \mu)$, and the real part is therefore continuous and strictly increasing. It tends to $-\,\infty$ for $\omega \to 0$ and to $+\,\infty$ for $\omega \to \mu^-$, which gives exactly one zero $\omega_*$ in $(0, \mu)$, since $\alpha > 0$. Since $\epsilon_b (\omega) \geq \epsilon_b$, we have
\begin{align}
0 = \Re \epsilon^{\rm RPA} (\omega_*) \geq \epsilon_b - \frac{\omega_p^2}{\omega_*^2} \quad \Rightarrow \quad \omega_*^2 \leq \frac{\omega_p^2}{\epsilon_b} = \tilde\omega_p^2\,,
\end{align}
which is the bound stated in Sec.~\ref{seclinscreen}. The imaginary part of $\epsilon^{\rm RPA}$ vanishes below $\omega_{\rm th} = \mu$, and the mode at $\omega_*$ is therefore undamped.

\section{TSM at arbitrary momentum-transfers}
\label{app:fin}

We give the steps behind Eqs.~\eqref{eq:fin-Ima}--\eqref{eq:fin-wdisp}. 

\subsection{Imaginary part of polarisability: Particle-hole continua and their edges}
\label{app:fin-im}

We write $\tilde x = k'$ with $\bs{k}' = \bs{k} + \bs{q}$. At fixed $q$, the angular integral is converted into an integral over $\tilde x$, with $d^3\bs{k} = (2\,\pi/q)\,k\,dk\,\tilde x\,d\tilde x$ and $|k - q| \leq \tilde x \leq k + q$. The law of cosines gives $1 + \cos\theta = \left( (k + \tilde x)^2 - q^2 \right)/(2\,k\,\tilde x)$ and $1 - \cos\theta = \left( q^2 - (k - \tilde x)^2 \right)/(2\,k\,\tilde x)$. Equation~\eqref{eq:lin-ff} becomes
\begin{align}
\label{eq:app-fin-ff}
\mathcal{F}_{11} = \frac{\left( (k + \tilde x)^2 - q^2 \right)^2}{16\,k^2\,\tilde x^2}\,, \quad
\mathcal{F}_{10} = \frac{\left( q^2 - (k - \tilde x)^2 \right) \left( (k + \tilde x)^2 - q^2 \right)}{8\,k^2\,\tilde x^2}\,, \quad
\mathcal{F}_{1,-1} = \frac{\left( q^2 - (k - \tilde x)^2 \right)^2}{16\,k^2\,\tilde x^2}\,.
\end{align}
For $\varpi > 0$, each group of transitions has a single energy-conserving delta function.

The intraband transitions require $\tilde x - k = \varpi$ and $\Theta (k_F - k) - \Theta (k_F - \tilde x) = 1$, which restricts $k$ to $k_F - \varpi < k < k_F$. The triangle inequality $\tilde x - k \leq q \leq \tilde x + k$ gives $\varpi \leq q$ and $k \geq (q - \varpi)/2$. This fixes $\check{k}_-$ and the range in Eq.~\eqref{eq:fin-Ima}. The remaining integral is
\begin{align}
\label{eq:app-fin-Ima}
\Im \Pi^{\rm L}_{11} = -\,\frac{g}{4\,\pi\,v_0\,q} \int_{\check{k}_-}^{k_F} dk\, \frac{\left( (2\,k + \varpi)^2 - q^2 \right)^2}{16\,k \left( k + \varpi \right)}\,,
\end{align}
and $\mathcal{G}^{\rm L}_{11}$ in Eq.~\eqref{eq:fin-G} is its antiderivative. The transitions $0 \to 1$ require $\tilde x = \varpi$, with $\tilde x > k_F$ imposed by Pauli-blocking. The integral over $k \in [|\varpi - q|, \varpi + q]$ is
\begin{align}
\label{eq:app-fin-Imb}
\Im \Pi^{\rm L}_{01} = -\,\frac{g\,\varpi}{4\,\pi\,v_0\,q} \int_{|\varpi - q|}^{\varpi + q} dk \left[ \frac{k}{2} - \frac{\left( k^2 + \varpi^2 - q^2 \right)^2}{8\,\varpi^2\,k} \right]
= -\,\frac{g}{32\,\pi\,v_0} \left[ 2 \left( \varpi^2 + q^2 \right) - \frac{\left( \varpi^2 - q^2 \right)^2}{\varpi\,q}\,\ln \!\left( \frac{\varpi + q}{|\varpi - q|} \right) \right].
\end{align}
The transitions $-1 \to 1$ require $k + \tilde x = \varpi$ with $\tilde x > k_F$. Labelling the integration variable by the momentum in the band $s = 1$, we obtain $\varpi \geq q$, $\tilde x \leq (\varpi + q)/2$, and $\tilde x \geq \max\lbrace k_F, (\varpi - q)/2 \rbrace = \check{k}_+$. Hence
\begin{align}
\label{eq:app-fin-Imc}
\Im \Pi^{\rm L}_{-1,1} = -\,\frac{g}{4\,\pi\,v_0\,q} \int_{\check{k}_+}^{(\varpi + q)/2} d\tilde x\, \frac{\left( q^2 - (2\,\tilde x - \varpi)^2 \right)^2}{16\,\tilde x \left( \varpi - \tilde x \right)}\,,
\end{align}
and $\mathcal{G}^{\rm L}_{-1,1}$ is its antiderivative. This range is nonempty only for $\varpi > 2\,k_F - q$. For $k_F \to 0$, we have $\check{k}_+ = (\varpi - q)/2$ and
\begin{align}
\mathcal{G}^{\rm L}_{-1,1} \!\left( \frac{\varpi + q}{2} \right) 
- \mathcal{G}^{\rm L}_{-1,1} \!\left( \frac{\varpi - q}{2} \right) = \frac{q \left( 5\,q^2 - 3\,\varpi^2 \right)}{12} + \frac{\left( \varpi^2 - q^2 \right)^2}{8\,\varpi}
\ln \!\left( \frac{\varpi + q}{\varpi - q} \right).
\end{align}
Adding $\Im \Pi^{\rm L}_{01}$ cancels the logarithm and the powers of $\varpi$. At $\varpi \to q^+$, the logarithmic term vanishes and $\Im \Pi^{\rm L}_{-1,1} = -\,g\,q^2/(24\,\pi\,v_0)$. For $\varpi > q + 2\,k_F$, the lower limit is $\check{k}_+ = (\varpi - q)/2$, and the doped result therefore coincides with the intrinsic one.

\subsection{Real part of polarisability and RPA dielectric function}
\label{app:fin-re}

The retarded function is analytic in the upper half-plane. With the cut-off $\mathcal{R}$ on the transition energy, it obeys $\Re \Pi (q, \omega) = (2/\pi)\,\mathrm{P} \!\! \int_0^{\mathcal{R}} d\omega'\, \omega'\,\Im \Pi (q, \omega')/(\omega'^2 - \omega^2)$. We subtract the intrinsic reference $\Im \Pi_{\rm ref} = -\,g\,q^2\,\Theta (\omega' - v_0\,q)/(6\,\pi\,v_0)$, whose contribution is
$-\,\frac{g\,q^2}{6\,\pi^2\,v_0}\,\ln \!\left( \frac{\mathcal{R}^2}{|\omega^2 - v_0^2\,q^2|} \right)$
for $\mathcal{R} \gg \omega$. The difference $\Im \Pi - \Im \Pi_{\rm ref}$ vanishes for $\omega' > v_0\,q + 2\,\mu$, which gives Eq.~\eqref{eq:fin-kk}.

In the intrinsic case, the difference is nonzero only for $\omega' < v_0\,q$, where it equals $\Im \Pi^{\rm L}_{01}$. With $\mathfrak{v} = \omega'/(v_0\,q)$ and $\ell (\mathfrak{v}) = \ln \!\left( (1 + \mathfrak{v})/(1 - \mathfrak{v}) \right)$, the window integral is
\begin{align}
-\,\frac{g\,q^2}{16\,\pi^2\,v_0} \int_0^1 d\mathfrak{v}\, \frac{2\,\mathfrak{v} \left( \mathfrak{v}^2 + 1 \right) - \left( \mathfrak{v}^2 - 1 \right)^2 \ell (\mathfrak{v})}{\mathfrak{v}^2 - W^2}\,.
\end{align}
We use $\left( \mathfrak{v}^2 - 1 \right)^2/(\mathfrak{v}^2 - W^2) = \mathfrak{v}^2 + W^2 - 2 + \left( W^2 - 1 \right)^2/(\mathfrak{v}^2 - W^2)$ and the integrals $\int_0^1 \ell\,d\mathfrak{v} = 2 \ln 2$, $\int_0^1 \mathfrak{v}^2\,\ell\,d\mathfrak{v} = (1 + 2 \ln 2)/3$, and $\int_0^1 2\,\mathfrak{v} \left( \mathfrak{v}^2 + 1 \right)/(\mathfrak{v}^2 - W^2)\,d\mathfrak{v} = 1 + (1 + W^2) \ln \!\left( (W^2 - 1)/W^2 \right)$. The last integral is
\begin{align}
\int_0^1 d\mathfrak{v}\, \frac{\ell (\mathfrak{v})}{\mathfrak{v}^2 - W^2} = \frac{1}{2\,W} \left[ \mathrm{Li}_2 \!\left( -\,\frac{W + 1}{W - 1} \right) - \mathrm{Li}_2 \!\left( -\,\frac{W - 1}{W + 1} \right) \right],
\end{align}
which follows from the substitution $\mathfrak{v} = (1 - \tilde y)/(1 + \tilde y)$, under which $\ell = -\ln {\tilde y}$ and the denominator becomes quadratic in $\tilde  y$ with the roots $-(W + 1)/(W - 1)$ and its inverse. Collecting the terms gives Eq.~\eqref{eq:fin-Re0}.

\subsection{Static polarisability and screening}
\label{app:fin-static}

We write $\Pi = \Pi^{(0)} + \delta \Pi$, where $\delta \Pi$ is the change caused by filling the band $s = 1$ up to $k_F$. Only the terms of Eq.~\eqref{eq:pol} containing $f_0 (\varepsilon_1)$ contribute, and
\begin{align}
\delta \Pi (q, \omega) = \frac{g}{4\,\pi^2\,v_0\,q} \int_0^{k_F} dk 
\int_{|k - q|}^{k + q} d{\tilde x} \sum_{\check\sigma = \pm} \sum_s 
\frac{k\,\tilde  x\,\mathcal{F}_{1 s}\!\left( k, \tilde  x \right)}
{\check\sigma\,\varpi + k - s\,\tilde  x}\,, \quad s \in \lbrace 1, 0, -1 \rbrace.
\end{align}
At $\varpi = 0$, only the line $\tilde x = k$ is singular, and the order of the principal-value integrations is immaterial. The sum over $s$ and $\check\sigma$ simplifies to $ 2\left [
-\,\frac{{\tilde x}^3}{8\,k^2} - \frac{\tilde x \left( 3\,k^2 - 2\,q^2 \right)}{8\,k^2} 
- \frac{\left( 4\,k^2 - q^2 \right)^2}{16\,k^2} \left( \frac{1}{\tilde x + k} 
+ \frac{1}{\tilde x - k} \right) \right ]$.
The integral over $\tilde x$ gives
\begin{align}
-\,k\,q + \frac{q^3}{4\,k} - \frac{\left( 4\,k^2 - q^2 \right)^2}{16\,k^2}
\ln \!\left( \frac{q + 2\,k}{|q - 2\,k|} \right),
\end{align}
which is finite for $k \to 0$. With $u = 2\,k$ and $\ell_u = \ln \!\left( (q + u)/|q - u| \right)$, the remaining integral is
\begin{align}
\delta \Pi (q, 0) = \frac{g}{4\,\pi^2\,v_0\,q} \int_0^{2\,k_F} du \left[ -\,\frac{q\,u}{2} + \frac{q^3}{2\,u} - \frac{\left( u^2 - q^2 \right)^2}{4\,u^2}\,\ell_u \right].
\end{align}
The antiderivatives are $\int \ell_u\,du = (q + u) \ln \!\left( q + u \right) - (u - q) \ln \!\left( |u - q| \right)$, $\int u^2\,\ell_u\,du = u^3\,\ell_u/3 + q\,u^2/3 + (q^3/3) \ln \!\left( |q^2 - u^2| \right)$, and $\int \ell_u\,du/u^2 = -\,\ell_u/u + (2/q) \ln u - (1/q) \ln \!\left( |q^2 - u^2| \right)$. The terms with $\ln u$ cancel at the lower limit, and we find
\begin{align}
\delta \Pi (q, 0) = \frac{g}{4\,\pi^2\,v_0} \left[ -\,\frac{4\,k_F^2}{3} - \frac{q^2}{2} + \frac{2\,q^2}{3}\,\ln \!\left( \frac{|4\,k_F^2 - q^2|}{q^2} \right) - \frac{1}{q} \left( \frac{2\,k_F^3}{3} - k_F\,q^2 - \frac{q^4}{8\,k_F} \right) \ln \!\left( \frac{2\,k_F + q}{|2\,k_F - q|} \right) \right].
\end{align}
The intrinsic part follows from Eq.~\eqref{eq:fin-kk} at $\omega = 0$. The window $\omega' < v_0\,q$ contributes
\begin{align}
-\,\frac{g\,q^2}{16\,\pi^2\,v_0} \int_0^1 d\mathfrak{v} \left[ \frac{2 \left( \mathfrak{v}^2 + 1 \right)}{\mathfrak{v}} - \frac{\left( \mathfrak{v}^2 - 1 \right)^2 \ell (\mathfrak{v})}{\mathfrak{v}^2} \right] = -\,\frac{g\,q^2}{16\,\pi^2\,v_0}\,\frac{16 \ln 2 - 4}{3}\,.
\end{align}
Together with the logarithm of Eq.~\eqref{eq:fin-kk}, this gives
\begin{align}
\Pi^{(0)} (q, 0) = -\,\frac{g\,q^2}{3\,\pi^2\,v_0} \left[ \ln \!\left( \frac{2\,\Lambda}{q} \right) - \frac{1}{4} \right].
\end{align}
Adding $\delta \Pi$ gives Eq.~\eqref{eq:fin-static}. For $q \to 0$, we expand $\ln \!\left( (2\,k_F + q)/(2\,k_F - q) \right) = q/k_F + q^3/(12\,k_F^3) + \dots$, which gives $\Pi \to -\,(g/(4\,\pi^2\,v_0)) \left[ 2\,k_F^2 + q^2 \left( (4/3) \ln \!\left( \Lambda/k_F \right) - 7/9 \right) \right]$ and hence Eq.~\eqref{eq:fin-tf}. For $q \gg k_F$, the bracket of $\delta \Pi$ tends to $2\,k_F^2/3$, which gives Eq.~\eqref{eq:fin-large}.

\subsection{Kohn singularity and Friedel oscillations}
\label{app:fin-friedel}

Let $\mathfrak{c} (q) = 2\,q^2/3 + \left( 2\,k_F^3/3 - k_F\,q^2 - q^4/(8\,k_F) \right)/q$. The coefficient of $\ln \!\left( |2\,k_F - q| \right)$ in Eq.~\eqref{eq:fin-static} is $+\,(g/(4\,\pi^2\,v_0))\,\mathfrak{c} (q)$ after combining the two logarithms. We find $\mathfrak{c} (2\,k_F) = \mathfrak{c}' (2\,k_F) = \mathfrak{c}'' (2\,k_F) = 0$ and $\mathfrak{c}''' (2\,k_F) = -\,1/k_F$. Hence $\mathfrak{c} (q) \simeq -\,\delta^3/(6\,k_F)$ with $\delta = q - 2\,k_F$. This gives Eq.~\eqref{eq:fin-kohn}.
The radial Fourier transform of an isotropic function is $\delta n (r) = (1/(2\,\pi^2\,r)) \int_0^{\infty} dq\, q\,\sin \!\left( q\,r \right) \delta n (q)$. Only the singularity at $q = 2\,k_F$ is nonanalytic, because $\Pi$, $\mathcal{U}$, and $\epsilon^{\rm RPA}$ are analytic functions of $q^2$ elsewhere. Writing $\Pi = \Pi_{\rm reg} + \Pi_{\rm sing}$ and $\epsilon^{\rm RPA} = \epsilon_{\rm reg} - \mathcal{V}\,\Pi_{\rm sing}$ with $\epsilon_{\rm reg} = 1 - \mathcal{V}\,\Pi_{\rm reg}$, the part of $\delta n$ linear in $\Pi_{\rm sing}$ is $\mathcal{U}\,\Pi_{\rm sing} \left( \epsilon_{\rm reg} + \mathcal{V}\,\Pi_{\rm reg} \right)/\epsilon_{\rm reg}^2 = \mathcal{U}\,\Pi_{\rm sing}/\epsilon_{\rm reg}^2$. Let $\delta n_{\rm sing} (q) = C\,\delta^3 \ln |\delta|$ with $C = -\,g\,\mathcal{U} (2\,k_F)/(24\,\pi^2\,v_0\,k_F\,\left[ \epsilon^{\rm RPA} (2\,k_F, 0) \right]^2)$. From $\int d\delta\, \ln |\delta| e^{i\,\delta\,r} = -\,\pi/|r|$, we obtain $\int d\delta\, \delta^3 \ln |\delta| e^{i\,\delta\,r} = 6\,\pi\,i/r^4$ for $r > 0$. Setting $q \simeq 2\,k_F$ in the prefactor gives $\delta n (r) \simeq 6\,k_F\,C\,\cos \!\left( 2\,k_F\,r \right)/(\pi\,r^5)$, which is Eq.~\eqref{eq:fin-friedel}. The corrections are of relative order $1/r$.

\subsection{Plasmon dispersion and Landau damping}
\label{app:fin-wdisp}

Let $q < \varpi < k_F$. We insert the ranges of Eqs.~\eqref{eq:fin-Ima}--\eqref{eq:fin-Imc} into Eq.~\eqref{eq:fin-kk}. The logarithm of Eq.~\eqref{eq:fin-kk} combines with the integral over $q < \omega' < \mu/v_0$, where the reference is the only nonzero term, into $-\,(g\,q^2/(6\,\pi^2\,v_0)) \ln \!\left( v_0^2\,\Lambda^2/(\mu^2 - \omega^2) \right)$. The remaining windows are $0 < \varpi' < q$ for the intraband transitions, $k_F < \varpi' < 2\,k_F + q$ for the transitions $0 \to 1$, and $2\,k_F - q < \varpi' < 2\,k_F + q$ for the transitions $-1 \to 1$. We denote the contributions of these three channels to the integral of Eq.~\eqref{eq:fin-kk} over their windows by $\mathscr{S}^{\rm L}_{11}$, $\mathscr{S}^{\rm L}_{01}$, and $\mathscr{S}^{\rm L}_{-1,1}$, respectively. In units of $g/v_0$, we expand the closed forms of Eqs.~\eqref{eq:fin-Ima}--\eqref{eq:fin-Imc} in $q$ and integrate term by term. The result is
\begin{align}
\mathscr{S}^{\rm L}_{11} = \frac{k_F^2\,q^2}{12\,\pi\,\varpi^2} + q^4 \left( \frac{k_F^2}{20\,\pi\,\varpi^4} - \frac{1}{40\,\pi\,\varpi^2} \right), \quad
\mathscr{S}^{\rm L}_{01} = \frac{q^4}{60\,\pi\,\varpi^2}\,\ln \!\left( \frac{4\,k_F^2 - \varpi^2}{4 \left( k_F^2 - \varpi^2 \right)} \right) + \frac{q^5}{60\,\pi\,k_F \left( 4\,k_F^2 - \varpi^2 \right)}\,,
\end{align}
and $\mathscr{S}^{\rm L}_{-1,1} = -\,q^5/(60\,\pi\,k_F \left( 4\,k_F^2 - \varpi^2 \right))$. The terms of order $q^5$ cancel. For $\mathscr{S}^{\rm L}_{01}$, we used $\Im \Pi^{\rm L}_{01} + g\,q^2/(6\,\pi\,v_0) \simeq g\,q^4/(30\,\pi\,v_0\,\varpi'^2)$ for $q \ll \varpi'$. The sum $(2/\pi) \left( \mathscr{S}^{\rm L}_{11} + \mathscr{S}^{\rm L}_{01} + \mathscr{S}^{\rm L}_{-1,1} \right)$ gives Eq.~\eqref{eq:fin-pi24}. The zero of $\Re \epsilon^{\rm RPA} = 1 - 4\,\pi\,e^2 \left( \Pi_2 + q^2\,\Pi_4 \right)$ obeys $\omega^2\,\epsilon_b (\omega) - \omega_p^2 = (4\,\alpha\,v_0^2\,q^2/\pi) \left[ \dots \right]$, where the bracket is that of Eq.~\eqref{eq:fin-wdisp}. Linearising around $\omega_*^2 (0)$ with $\partial_{\omega^2} \left( \omega^2\,\epsilon_b \right) = \Phi$ gives Eq.~\eqref{eq:fin-wdisp}. The critical wavevector $q_{\rm max}$ is defined as in Sec.~\ref{secfinplasmon}, and the zero of $\Re \epsilon^{\rm RPA}$ is found with $\Re \Pi$ evaluated from Eq.~\eqref{eq:fin-kk}.

\section{QTSM at small momentum-transfers}
\label{app:quad-top}

This appendix contains the derivations behind Sec.~\ref{secquad}. The term $c\,v_0\,k^2\,\mathbb{I}_{3\times3}$ commutes with $v_0\,\bs{\tilde k} \cdot \bs{S}$. The eigenvectors, the overlap factors, the quantum metric, and the angular average of Appendix~\ref{app:tsm} therefore apply unchanged.

For a pair of bands $s < s'$, Eq.~\eqref{eq:pol} contains the two terms
\begin{align}
\Pi_{ss'} = g \int \frac{d^3\bs{k}}{(2\,\pi)^3}\,\mathcal{F}_{ss'} (\bs{k}, \bs{k}') \left\lbrace \frac{n_s (\bs{k}) - n_{s'} (\bs{k}')}{\omega + \varepsilon_s (\bs{k}) - \varepsilon_{s'} (\bs{k}')} + \frac{n_{s'} (\bs{k}) - n_s (\bs{k}')}{\omega + \varepsilon_{s'} (\bs{k}) - \varepsilon_s (\bs{k}')} \right\rbrace.
\end{align}
The overlap factor depends only on the angle between $\bs{k}$ and $\bs{k}'$, and it is therefore unchanged by the substitution $\bs{k} \to -\,\bs{k} - \bs{q}$, which exchanges $k$ and $k'$. In the second term, this substitution produces the numerator $n_{s'} (\bs{k}') - n_s (\bs{k})$ and the denominator $\omega + X_{ss'}$, whereas the first term has the denominator $\omega - X_{ss'}$. Their sum is
\begin{align}
\left[ n_s (\bs{k}) - n_{s'} (\bs{k}') \right] \left( \frac{1}{\omega - X_{ss'}} - \frac{1}{\omega + X_{ss'}} \right) = \left[ n_s (\bs{k}) - n_{s'} (\bs{k}') \right] \frac{2\,X_{ss'}}{\omega^2 - X_{ss'}^2}\,,
\end{align}
which is Eq.~\eqref{eq:q-pair} and holds exactly.

To establish Eq.~\eqref{eq:q-master}, we first take $\omega$ with a fixed positive imaginary part $\omega_{\rm I}$. The function $G (X) = 2\,X/(\omega^2 - X^2)$ and its derivative are then bounded on the real axis. Writing $X^0 = (s' - s)\,v_0\,k$, we have $X_{ss'} - X^0 = \varepsilon_{s'} (k') - \varepsilon_{s'} (k)$. The mean-value theorem gives $|X_{ss'} - X^0| \leq v_0 \left( 1 + 2\,|c|\,\Lambda \right) q$, since $|\partial_k \varepsilon_{s'}| \leq v_0 \left( 1 + 2\,|c|\,\Lambda \right)$ and $|k' - k| \leq q$. Hence $G (X_{ss'}) - G (X^0) = \mathcal{O} (q)$ uniformly.

Since $\mathcal{F}_{10} = \sin^2\theta/2$ and $\sin\theta = |\bs{k} \times \bs{q}|/(k\,k')$ with $|\bs{k} \times \bs{q}| = |\bs{k}' \times \bs{q}| \leq k'\,q$, we have $\mathcal{F}_{10} \leq \min \lbrace 1, q^2/(2\,k^2) \rbrace$. The integral of $\mathcal{F}_{10}$ over $k < \Lambda$ is therefore $\mathcal{O} (q^2)$, and replacing $G (X_{ss'})$ by $G (X^0)$ changes Eq.~\eqref{eq:q-pair} by $\mathcal{O} (q^3)$. For $k \geq 2\,q$, we have $\bs{k} \cdot \bs{k}' \geq k^2 - k\,q > 0$ and $k' \geq k/2$. The difference between $\sin^2\theta = |\bs{k} \times \bs{q}|^2/(k^2\,k'^2)$ and its leading term $|\bs{k} \times \bs{q}|^2/k^4$ is bounded by $|\bs{k} \times \bs{q}|^2 \left| k^2 - k'^2 \right|/(k^4\,k'^2) \leq 12\,q^3/k^3$. Thus $\mathcal{F}_{10} = q_i\,q_j\,Q_{ij} + \mathcal{O} \left( q^3/k^3 \right)$ for $k \geq 2\,q$, and the integral of the remainder over $2\,q \leq k \leq \Lambda$ is $\mathcal{O} \left( q^3 \ln \!\left( \Lambda/q \right) \right)$. For $k < 2\,q$, we use $\mathcal{F}_{ss'} \leq 1$ and $q_i\,q_j\,Q_{ij} \leq q^2/(2\,k^2)$, which gives a contribution of $\mathcal{O} (q^3)$. The factor $\mathcal{F}_{1,-1} = \sin^4 \left( \theta/2 \right) = \sin^4\theta/\left[ 4 \left( 1 + \cos\theta \right)^2 \right]$ is bounded by $q^4/(4\,k^4)$ for $k \geq 2\,q$, and its integral is again $\mathcal{O} (q^3)$.

The occupation difference splits as $n_s (\bs{k}) - n_{s'} (\bs{k}') = \left[ n_s (\bs{k}) - n_{s'} (\bs{k}) \right] + \left[ n_{s'} (\bs{k}) - n_{s'} (\bs{k}') \right]$. The second bracket is nonzero only if $k$ and $k'$ lie on opposite sides of a Fermi momentum $k_{s'} > 0$ of the band $s'$, which requires $|k - k_{s'}| \leq q$. This shell has a volume of order $q$, and $\mathcal{F}_{ss'} = \mathcal{O} (q^2)$ on it. Its contribution is therefore $\mathcal{O} (q^3)$, and it is absent in the intrinsic limit.

Combining these estimates, Eq.~\eqref{eq:q-pair} equals the right-hand side of Eq.~\eqref{eq:q-master} up to $\mathcal{O} \left( q^3 \ln \!\left( \Lambda/q \right) \right)$, which is $o (q^2)$. The coefficient of $q^2$ is the integral at $\bs{k}' = \bs{k}$ with $\mathcal{F}_{ss'} \to q_i\,q_j\,Q_{ij}$. At real frequencies, it is the limit $\omega_{\rm I} \to 0^+$ of this coefficient, which is the radial integral of Appendix~\ref{app:lin-intr} with the lower limit set by the occupations.

The imaginary part provides an independent check at real frequencies, since it does not rely on the expansion above. We consider $c < 0$, $\mu \to 0^+$, and the pair $(0, 1)$ for $0 < \varpi < 1/(4\,|c|)$ with $\varpi = \omega/v_0$, using Eq.~\eqref{eq:qf-im01}. The substitution $\Sigma = 2\,\varpi + \tilde x$ turns the two brackets of the numerator into $\mathfrak{a} + \mathfrak{f}_+\,\tilde x$ and $\mathfrak{a} - \mathfrak{f}_-\,\tilde x$, with $\mathfrak{a} = q \left( 1 + 4\,c\,\varpi \right)$, $\mathfrak{f}_+ = 1 + 2\,c\,q$, and $\mathfrak{f}_- = 1 - 2\,c\,q$. Their product is $\mathfrak{a}^2 - \tilde x^2 + \mathcal{O} (q^3)$ for $\tilde x = \mathcal{O} (q)$. The limits $\Sigma_- = (2\,\varpi - q)/(1 + 2\,c\,q)$ and $\Sigma_+ = (2\,\varpi + q)/(1 - 2\,c\,q)$ correspond to $\tilde x_- = -\,\mathfrak{a}/\mathfrak{f}_+$ and $\tilde x_+ = \mathfrak{a}/\mathfrak{f}_-$. The conditions $\Sigma \geq q$ and $\Sigma \leq \sqrt{\varpi/|c|}$ hold at $\Sigma = 2\,\varpi$ because $4\,|c|\,\varpi < 1$. At this point, the three denominator factors equal $1 + 4\,c\,\varpi$, $\varpi \left( 1 + 4\,c\,\varpi \right)$, and $\varpi \left( 1 + 4\,c\,\varpi \right)$, and $\Sigma^2 - q^2 = 4\,\varpi^2 + \mathcal{O} (q)$. The integral therefore equals
\begin{align}
\frac{4\,\varpi^2}{\varpi^2 \left( 1 + 4\,c\,\varpi \right)^3}\,\frac{4\,\mathfrak{a}^3}{3} \left( 1 + \mathcal{O} (q) \right) = \frac{16\,q^3}{3} \left( 1 + \mathcal{O} (q) \right),
\end{align}
and Eq.~\eqref{eq:qf-im01} gives $\Im \Pi_{01} = -\,g\,q^2/(6\,\pi\,v_0)$ at order $q^2$, independently of $c$. For $c > 0$ and the pair $(-1, 0)$, the integrand of Eq.~\eqref{eq:qf-imm10} is that of Eq.~\eqref{eq:qf-im01} with $c \to -\,c$, and the same steps apply for $\varpi < 1/(4\,c)$. The additional lower bound $\Sigma \geq k_F^{(0)} - \varkappa \left( \mu/v_0 - \varpi \right)$ reduces to $\Sigma \geq \left( 1 - \sqrt{1 - 4\,c\,\varpi} \right)/(2\,c)$ for $\mu \to 0^+$. This bound lies below $2\,\varpi$ and does not restrict the range. The result is again $-\,g\,q^2/(6\,\pi\,v_0)$, in agreement with Eq.~\eqref{eq:lin-pi0b}.

\subsection{Intrinsic limit}
\label{app:quad-intr}

As $\mu \to 0^+$, we have $\beta \to 1$, and Eqs.~\eqref{eq:q-dos} and~\eqref{eq:q-dos0} show that $D^{(1)}_{ij} \propto \mu^2$ and $D^{(0)}_{ij} \propto \mu^{3/2}$ vanish. The bare plasmon frequency therefore vanishes as well, and the response is purely interband.

For $0 < k < \Lambda$ and $|c|\,\Lambda < 1$, the sign of each energy is fixed. We have $\varepsilon_{-1} = -\,v_0\,k \left( 1 - c\,k \right) < 0$ and $\varepsilon_1 = v_0\,k \left( 1 + c\,k \right) > 0$ for both signs of $c$, whereas $\varepsilon_0 = c\,v_0\,k^2$ is negative for $c < 0$ and positive for $c > 0$. At $T = 0$ and $\mu \to 0^+$, every band is thus completely filled or completely empty, and the pairs of bands with equal occupation are blocked. These are $(-1, 0)$ for $c < 0$ and $(0, 1)$ for $c > 0$, in addition to all intraband pairs. The pair $(-1, 1)$ carries $\mathcal{F}_{1,-1} = \mathcal{O} \left( q^4 \right)$ and is dropped. The only remaining pair is $(0, 1)$ with $n_0 - n_1 = 1$ for $c < 0$, and $(-1, 0)$ with $n_{-1} - n_0 = 1$ for $c > 0$. Both have $s' - s = 1$ and the overlap factor $\mathcal{F}_{10}$. Eq.~\eqref{eq:q-master} and the angular average of Eq.~\eqref{eq:lin-ang} then give, for both signs of $c$,
\begin{align}
\Pi_{\rm inter} (\bs{q}, \omega) = \frac{g\,q^2}{6\,\pi^2} \int_0^{\Lambda} dk\, \frac{2\,v_0\,k}{\left( \omega + i\,0^+ \right)^2 - v_0^2\,k^2}\,,
\end{align}
which is Eq.~\eqref{eq:lin-pi0}. The radial integral is evaluated in Appendix~\ref{app:lin-intr} and yields Eq.~\eqref{eq:lin-pi0b}. Since $\Pi_{\rm intra} = 0$ at this point, Eq.~\eqref{eq:epsrpa} gives Eq.~\eqref{eq:lin-eps0}, with $\Re \epsilon^{\rm RPA} > 1$ for $\omega < v_0\,\Lambda$ and $\Im \epsilon^{\rm RPA} = 2\,g\,e^2/(3\,v_0)$. The dielectric function therefore has no zero, and no undamped plasmon exists. This result agrees with the limit $\mu \to 0^+$ of Eqs.~\eqref{eq:q-eps} and~\eqref{eq:q-pi-flat}, in which $\mu_1, \mu_0 \to 0$.

\subsection{Fermi momentum, DOS, and Drude weight}
\label{app:quad-fs}

The condition $\varepsilon_1 (k_F) = \mu$ is the quadratic equation $c\,k_F^2 + k_F - \mu/v_0 = 0$, whose roots $(-1 \pm \beta)/(2\,c)$ are real for $1 + 4\,\xi \geq 0$ and $c \neq 0$. With $\beta^2 - 1 = 4\,c\,\mu/v_0$, the root $(\beta - 1)/(2\,c) = (\beta^2 - 1)/[2\,c\,(\beta + 1)] = 2\,\mu/[v_0\,(\beta + 1)]$ vanishes for $\mu \to 0$. It is positive for both signs of $c$, because $\beta - 1$ has the sign of $c$. At the FS, $2\,c\,k_F = \beta - 1$, and the band-velocity $v_0 \left( 1 + 2\,c\,k_F \right)$ equals $V_F = v_0\,\beta$. This proves Eq.~\eqref{eq:q-kf}. For $c < 0$, the velocity vanishes at $\mu = v_0/(4\,|c|)$, which is the maximum of the band.

At $T = 0$, we have $\delta (\varepsilon_1 - \mu) = \delta (k - k_F)/V_F$ for $V_F > 0$. With $\mu_1 = v_0\,k_F$, the DOS and the density are
\begin{align}
\rho_1 (\mu) = \frac{g}{(2\,\pi)^3}\,\frac{4\,\pi\,k_F^2}{V_F} = \frac{g\,\mu_1^2}{2\,\pi^2\,v_0^3\,\beta}\,, \quad
\tilde n_1 = \frac{g}{(2\,\pi)^3}\,\frac{4\,\pi\,k_F^3}{3} = \frac{g\,\mu_1^3}{6\,\pi^2\,v_0^3}\,.
\end{align}
With $\bs{V}^{(1)} = V_F\,\hat{\bs{k}}$ and $\langle \hat k_i\,\hat k_j \rangle = \delta_{ij}/3$, Eq.~\eqref{eq:drude} gives
\begin{align}
D^{(1)}_{ij} = \pi\,e^2\,\rho_1\,V_F^2\,\frac{\delta_{ij}}{3} = \frac{g\,e^2\,k_F^2\,V_F}{6\,\pi}\,\delta_{ij} = \frac{g\,e^2\,\beta\,\mu_1^2}{6\,\pi\,v_0}\,\delta_{ij}\,.
\end{align}
The second form of Eq.~\eqref{eq:q-dos} follows from $\mu_1 = 2\,\mu/(1 + \beta)$, and the ratio to the linear-model value $g\,e^2\,\mu^2/(6\,\pi\,v_0)$ is
\begin{align}
\frac{4\,\beta}{\left( 1 + \beta \right)^2} = 1 - \left( \frac{\beta - 1}{\beta + 1} \right)^2.
\end{align}
This ratio does not exceed unity and equals unity only for $\beta = 1$. Since $\beta - 1 = 2\,\xi + \mathcal{O} (\xi^2)$ and $\beta + 1 = 2 + \mathcal{O} (\xi)$, its expansion is $1 - \xi^2 + \mathcal{O} (\xi^3)$.

The intraband expansion of Appendix~\ref{app:lin-drude} uses only $\delta (\varepsilon_1 - \mu)$, the band-velocity, and $\mathcal{F}_{11} = 1 - \mathcal{O} (q^2)$. It therefore holds for $c \neq 0$, and $\Pi_{\rm intra} = q_i\,q_j\,D^{(1)}_{ij}/(\pi\,e^2\,\omega^2)$.

\subsection{Interband response and screening for $c < 0$}
\label{app:quad-inter}

For $c < 0$ and $\mu > 0$, the bands $s = 0$ and $s = -1$ have negative energies and are filled at every $\bs{k}$. The band $s = 1$ increases up to $k = 1/(2\,|c|)$ and decreases beyond it. It is occupied for $k < k_F$ and empty for $k_F < k < k_2 = (1 + \beta)/(2\,|c|)$, where $k_2$ is the second root of $\varepsilon_1 = \mu$. For $\Lambda < k_2$, it is therefore empty on the whole range $k_F < k < \Lambda$.

As in Appendix~\ref{app:quad-intr}, the pairs of bands with equal occupation are blocked, and the pair $(-1, 1)$ is dropped. The only remaining pair is $(0, 1)$ with $n_0 - n_1 = \Theta (k - k_F)$ and $s' - s = 1$. Eq.~\eqref{eq:q-master} then reproduces the linear-model integrand of Eq.~\eqref{eq:lin-pi0} with the lower limit $k_F$. The replacement $\bs{k}' = \bs{k}$ in the transition energy and in the occupations is justified in Appendix~\ref{app:quad-top} for frequencies away from the onset. The steps of Appendix~\ref{app:lin-screen} therefore apply with $v_0\,k_F = \mu_1$ in place of $\mu$. They give the logarithm of $v_0^2\,\Lambda^2/|\mu_1^2 - \omega^2|$, the onset $\omega_{\rm th} = \mu_1$, and the imaginary part proportional to $\Theta (\omega - \mu_1)$. Adding $\Pi_{\rm intra} = q_i\,q_j\,D^{(1)}_{ij}/(\pi\,e^2\,\omega^2)$ from Appendix~\ref{app:quad-fs} proves Eqs.~\eqref{eq:q-eps} and~\eqref{eq:q-eb}, with $\omega_p^2 = 4\,D^{(1)}_{ii}$ as in Eq.~\eqref{eq:q-wp}.

The argument of Appendix~\ref{app:lin-screen} for the zero carries over with $\mu$ replaced by $\mu_1$. For $0 < \omega < \mu_1$, we write $\Re \epsilon^{\rm RPA} = \epsilon_b (\omega) - \omega_p^2/\omega^2$ with $\epsilon_b (\omega) = 1 + \left( 2\,g\,e^2/(3\,\pi\,v_0) \right) \ln \!\left( v_0^2\,\Lambda^2/(\mu_1^2 - \omega^2) \right)$. Both terms increase with $\omega$, and $\Re \epsilon^{\rm RPA}$ runs from $-\,\infty$ at $\omega \to 0$ to $+\,\infty$ at $\omega \to \mu_1^-$, which gives exactly one zero $\omega_*$ in $(0, \mu_1)$. Since $\epsilon_b (\omega) \geq \epsilon_b$, we have $\omega_*^2 \leq \omega_p^2/\epsilon_b$. The imaginary part vanishes below $\mu_1$, and the plasmon is undamped.

\subsection{Flat-band at the Fermi level for $c > 0$}
\label{app:quad-flat}

The condition $\varepsilon_0 (k_F^{(0)}) = \mu$ reads $c\,v_0\,k^2 = \mu$, with the positive root $k_F^{(0)} = \sqrt{\mu/(c\,v_0)} = \sqrt{\xi}/c$. Since $\varepsilon_1 (k) - \varepsilon_0 (k) = v_0\,k > 0$, we have $\varepsilon_1 (k_F^{(0)}) > \mu$, and the monotonic increase of the band $s = 1$ for $c > 0$ gives $k_F < k_F^{(0)}$. The band $s = -1$ has $\varepsilon_{-1} = v_0\,k \left( c\,k - 1 \right)$ and lies below $\mu$ for $0 \leq k < k_2$, where $k_2 = (1 + \beta)/(2\,c)$ is the positive root of $c\,v_0\,k^2 - v_0\,k - \mu = 0$. The window $k_F^{(0)} < k_2$ is nonempty because $2\,\sqrt{\xi} < 1 + \beta$ follows from $\left( 1 + \beta \right)^2 - 4\,\xi = 2 \left( 1 + \beta \right) > 0$, and for $\Lambda < k_2$ the band $s = -1$ is filled at every $k < \Lambda$.

At $T = 0$, we have $\delta (\varepsilon_0 - \mu) = \delta (k - k_F^{(0)})/V_0$ with $V_0 = 2\,c\,v_0\,k_F^{(0)} = 2\,v_0\,\sqrt{\xi}$. The DOS and the density are
\begin{align}
\rho_0 (\mu) = \frac{g}{(2\,\pi)^3}\,\frac{4\,\pi \left( k_F^{(0)} \right)^2}{V_0} = \frac{g\,\mu_0^2}{4\,\pi^2\,v_0^3\,\sqrt{\xi}}\,, \quad
\tilde n_0 = \frac{g}{(2\,\pi)^3}\,\frac{4\,\pi \left( k_F^{(0)} \right)^3}{3} = \frac{g\,\mu_0^3}{6\,\pi^2\,v_0^3}\,,
\end{align}
where $\mu_0 = v_0\,k_F^{(0)}$ has been used. With $\bs{V}^{(0)} = V_0\,\hat{\bs{k}}$ and $\langle \hat k_i\,\hat k_j \rangle = \delta_{ij}/3$, Eq.~\eqref{eq:drude} gives
\begin{align}
D^{(0)}_{ij} = \pi\,e^2\,\rho_0\,V_0^2\,\frac{\delta_{ij}}{3} = \frac{g\,e^2 \left( k_F^{(0)} \right)^2 V_0}{6\,\pi}\,\delta_{ij} = \frac{g\,e^2\,\sqrt{\xi}\,\mu_0^2}{3\,\pi\,v_0}\,\delta_{ij}\,.
\end{align}
Using $\mu_0^2 = \mu^2/\xi$, the last expression equals $2/\sqrt{\xi}$ times the linear-model value $g\,e^2\,\mu^2/(6\,\pi\,v_0)$. Dividing by the factor $4\,\beta/(1 + \beta)^2$ of Appendix~\ref{app:quad-fs} gives the ratio $D^{(0)}_{ii}/D^{(1)}_{ii} = (1 + \beta)^2/(2\,\beta\,\sqrt{\xi})$. With $4\,\xi = \beta^2 - 1$, its square is
\begin{align}
\frac{\left( 1 + \beta \right)^4}{\beta^2 \left( \beta^2 - 1 \right)} = \frac{\left( 1 + \beta \right)^3}{\beta^2 \left( \beta - 1 \right)}\,.
\end{align}
The difference $\left( 1 + \beta \right)^3 - \beta^2 \left( \beta - 1 \right) = 1 + 3\,\beta + 4\,\beta^2$ is positive, and the ratio therefore exceeds unity and tends to unity for $\beta \to \infty$. As in Appendix~\ref{app:quad-fs}, the intraband expansion applies to each partially filled band separately, and the two intraband terms add up to $\Pi_{\rm intra} = q_i\,q_j \left( D^{(1)}_{ij} + D^{(0)}_{ij} \right)/(\pi\,e^2\,\omega^2)$, and Eq.~\eqref{eq:plasmon} then gives Eq.~\eqref{eq:q-wp0}, whose two terms are $4\,D^{(1)}_{ii}$ of Eq.~\eqref{eq:q-wp} and $4\,D^{(0)}_{ii}$.

For the interband part, the occupations at $T = 0$ are $f_0 (\varepsilon_{-1}) = 1$, $f_0 (\varepsilon_0) = \Theta (k_F^{(0)} - k)$, and $f_0 (\varepsilon_1) = \Theta (k_F - k)$. The pairs $(-1, 1)$ and $(1, -1)$ carry $\mathcal{F}_{1,-1} = \mathcal{O} (q^4)$ and are dropped. As in Appendix~\ref{app:quad-inter}, we set $k' = k$ in the denominators and in the blocking functions, which is justified for $\omega < \mu_1$ because the integrand has no pole there. The remaining four pairs all carry $\mathcal{F}_{10} = \mathcal{F}_{0,-1}$ of Eq.~\eqref{eq:lin-ff}. The pair $(0, 1)$ has the occupation factor $\Theta (k_F^{(0)} - k) - \Theta (k_F - k) = \Theta (k - k_F)\,\Theta (k_F^{(0)} - k)$ and the denominator $\omega - v_0\,k$, while the pair $(1, 0)$ has the opposite factor and the denominator $\omega + v_0\,k$. Likewise, the pair $(-1, 0)$ has the factor $1 - \Theta (k_F^{(0)} - k) = \Theta (k - k_F^{(0)})$ and the denominator $\omega - v_0\,k$, while the pair $(0, -1)$ has the opposite factor and the denominator $\omega + v_0\,k$. The sum is
\begin{align}
\mathcal{F}_{10} \left[ \Theta (k - k_F)\,\Theta (k_F^{(0)} - k) + \Theta (k - k_F^{(0)}) \right] \frac{2\,v_0\,k}{(\omega + i\,0^+)^2 - v_0^2\,k^2} = \mathcal{F}_{10}\,\Theta (k - k_F)\,\frac{2\,v_0\,k}{(\omega + i\,0^+)^2 - v_0^2\,k^2}\,,
\end{align}
where the equality uses $k_F < k_F^{(0)}$. The integrand thus coincides with that of Appendix~\ref{app:quad-inter}, because the blocking of the transitions $0 \to 1$ above $k_F^{(0)}$ is exactly compensated by the opening of the transitions $-1 \to 0$. The interband part therefore gives the logarithm $\ln \!\left( v_0^2\,\Lambda^2/|\mu_1^2 - \omega^2| \right)$, the onset $\omega_{\rm th} = \mu_1$, and the imaginary part $\Theta (\omega - \mu_1)$, which proves Eqs.~\eqref{eq:q-eps} and~\eqref{eq:q-eb} for $c > 0$.

For the separate contributions of the two groups of transitions, we apply the antiderivative of Appendix~\ref{app:lin-intr} to each range. The transitions $0 \to 1$ have $k_F < k < k_F^{(0)}$, and the transitions $-1 \to 0$ have $k_F^{(0)} < k < \Lambda$. The real parts of the radial integrals are
\begin{align}
\Re \int_{k_F}^{k_F^{(0)}} dk\, \frac{2\,v_0\,k}{\omega^2 - v_0^2\,k^2} = -\,\frac{1}{v_0}\,\ln \!\left( \frac{|\mu_0^2 - \omega^2|}{|\mu_1^2 - \omega^2|} \right), \quad
\Re \int_{k_F^{(0)}}^{\Lambda} dk\, \frac{2\,v_0\,k}{\omega^2 - v_0^2\,k^2} \simeq -\,\frac{1}{v_0}\,\ln \!\left( \frac{v_0^2\,\Lambda^2}{|\mu_0^2 - \omega^2|} \right),
\end{align}
where $v_0\,k_F = \mu_1$ and $v_0\,k_F^{(0)} = \mu_0$ have been used. The delta function $\delta (\omega - v_0\,k)$ fixes $k = \omega/v_0$. This lies in the first range for $\mu_1 < \omega < \mu_0$ and in the second range for $\omega > \mu_0$. The imaginary parts are therefore $-\,(\pi/v_0)\,\Theta (\omega - \mu_1)\,\Theta (\mu_0 - \omega)$ and $-\,(\pi/v_0)\,\Theta (\omega - \mu_0)$, respectively. Multiplication by $g\,q^2/(6\,\pi^2)$ gives Eq.~\eqref{eq:q-pi-flat}. The static limits, with $\epsilon_b - 1 = -\,4\,\pi\,e^2\,\Pi/q^2$, are $(4\,g\,e^2/3\,\pi\,v_0)\,\ln \!\left( \mu_0/\mu_1 \right)$ and $(4\,g\,e^2/3\,\pi\,v_0)\,\ln \!\left( v_0\,\Lambda/\mu_0 \right)$. Their sum is $(4\,g\,e^2/3\,\pi\,v_0)\,\ln \!\left( v_0\,\Lambda/\mu_1 \right)$, as in Eq.~\eqref{eq:q-eb}.

The corrections to the replacement $k' = k$ arise from the shell of width of order $q$ around $k_F^{(0)}$, in which the blocking functions of the two pairs differ from their leading forms. The integrand is smooth there for $\omega < \mu_1$, and the corrections are suppressed by $q/k_F$. The zero of $\Re\epsilon^{\rm RPA} = \epsilon_b (\omega) - \omega_p^2/\omega^2$ follows from the monotonicity argument of Appendix~\ref{app:quad-inter}, with $\omega_p^2$ of Eq.~\eqref{eq:q-wp0}, and it is unique in $(0, \mu_1)$ with $\omega_*^2 \leq \omega_p^2/\epsilon_b$. The imaginary part vanishes below $\mu_1$, and the plasmon is therefore undamped. For the ratio $\omega_p^2/\omega_{\rm th}^2$, we use $\mu_0 = \mu/\sqrt{\xi}$ and $\mu_1 = 2\,\mu/(1 + \beta)$, which give $2\,\sqrt{\xi}\,\mu_0^2 = \mu_1^2 \left( 1 + \beta \right)^2/(2\,\sqrt{\xi})$. Dividing Eq.~\eqref{eq:q-wp0} by $\mu_1^2$ yields the expression quoted in Sec.~\ref{secquadflat}.

\subsection{Density dependence and static screening}
\label{app:quad-dens}

We give the steps behind Eqs.~\eqref{eq:q-wp-dens} to~\eqref{eq:q-wp-rho}. For the top band, Eq.~\eqref{eq:q-wp} reads $4\,D^{(1)}_{ii} = 2\,g\,e^2\,\beta\,\mu_1^2/(3\,\pi\,v_0)$. With $\mu_1 = v_0\,k_F$, this equals $2\,g\,e^2\,v_0\,\beta\,k_F^2/(3\,\pi)$. The relation $\tilde n_1 = g\,k_F^3/(6\,\pi^2)$ gives $k_F^2 = \left( 6\,\pi^2\,\tilde n_1/g \right)^{2/3}$, and Eq.~\eqref{eq:q-kf} gives $\beta = 1 + 2\,c\,k_F$. This is the first term of Eq.~\eqref{eq:q-wp-dens}. For the flat-band, Eq.~\eqref{eq:q-wp0} gives $4\,D^{(0)}_{ii} = 4\,g\,e^2\,\sqrt{\xi}\,\mu_0^2/(3\,\pi\,v_0)$. With $\mu_0 = v_0\,k_F^{(0)}$ and $\sqrt{\xi} = c\,k_F^{(0)}$ from Eq.~\eqref{eq:q-kf0}, this equals $4\,g\,e^2\,c\,v_0 \left( k_F^{(0)} \right)^3/(3\,\pi)$. Using $\tilde n_0 = g \left( k_F^{(0)} \right)^3/(6\,\pi^2)$, we obtain $8\,\pi\,e^2\,c\,v_0\,\tilde n_0$, which is the second term. It equals $4\,\pi\,e^2\,\tilde n_0/m_*$ for $m_* = 1/(2\,c\,v_0)$.

For the static limit, we take $q \to 0$ at fixed temperature $T > 0$ and $\omega = 0$, and then $T \to 0$. The intraband part of a band $s$ is
\begin{align}
\Pi_{ss} (\bs{q}, 0) = g \int \frac{d^3\bs{k}}{(2\,\pi)^3}\,\mathcal{F}_{ss} (\bs{k}, \bs{k}')\,\frac{f_0 (\varepsilon_s (\bs{k})) - f_0 (\varepsilon_s (\bs{k}'))}{\varepsilon_s (\bs{k}) - \varepsilon_s (\bs{k}')}\,.
\end{align}
By the mean-value theorem, the difference quotient equals $f_0' (\varepsilon_{\rm m})$ for some $\varepsilon_{\rm m}$ between $\varepsilon_s (\bs{k})$ and $\varepsilon_s (\bs{k}')$. Since $|\varepsilon_s (\bs{k}') - \varepsilon_s (\bs{k})| \leq v_0 \left( 1 + 2\,|c|\,\Lambda \right) q$ and $f_0'$ is uniformly continuous at $T > 0$, the quotient tends to $f_0' (\varepsilon_s (\bs{k}))$ uniformly for $q \to 0$. For $k \geq 2\,q$, the factors $\mathcal{F}_{11}$ and $\mathcal{F}_{00}$ differ from unity by at most $q^2/k^2$, as follows from Eq.~\eqref{eq:lin-ff} and $\sin\theta \leq q/k$. The region $k < 2\,q$ has the volume $\mathcal{O} (q^3)$. Hence $\Pi_{ss} (\bs{q} \to 0, 0) = g \int d^3\bs{k}/(2\,\pi)^3\,f_0' (\varepsilon_s)$, which tends to $-\,g \int d^3\bs{k}/(2\,\pi)^3\,\delta (\varepsilon_s - \mu) = -\,\rho_s$ for $T \to 0$. Only the bands with an FS contribute, namely $s = 1$ for $c \leq 0$, and $s = 1$ and $s = 0$ for $c > 0$.

For the interband pairs, we use Eq.~\eqref{eq:q-pair} at $\omega = 0$, where the last factor equals $-\,2/X_{ss'}$. At $T = 0$, the occupation difference is nonzero only for the unblocked transitions. For $q$ below $q_1 = k_F^{(0)} - k_F$ if $c > 0$, and for every small $q$ if $c < 0$, all nonzero terms have $n_s (\bs{k}) - n_{s'} (\bs{k}') = +\,1$. The opposite sign requires a transition from the filled upper band of the pair into the empty lower band. Such transitions are blocked for $c < 0$, and they exist only for $q \geq q_1$ for $c > 0$ (Eq.~\eqref{eq:qf-th-10}). By Eqs.~\eqref{eq:qf-th-neg}, \eqref{eq:qf-th-pos}, and~\eqref{eq:qf-th-m0}, the energy $X_{ss'}$ is bounded from below by $v_0\,\varpi_{\rm th} (q)$, which tends to $v_0\,k_F$ for the pair $(0, 1)$ and to $v_0\,k_F^{(0)}$ for the pair $(-1, 0)$ for $q \to 0$. Both limits are positive. Therefore $|2/X_{ss'}|$ is bounded, and $|\Pi_{ss'}| \leq C \int d^3\bs{k}\,\mathcal{F}_{ss'} = \mathcal{O} (q^2)$, where the last step uses $\mathcal{F}_{ss'} \leq \min \lbrace 1, q^2/(2\,k^2) \rbrace$ as in Appendix~\ref{app:quad-top}. The pair $(-1, 1)$ has $\mathcal{F}_{1,-1} \leq q^4/(4\,k^4)$ for $k \geq 2\,q$ and the onset $\varpi_{\rm th}^{(-1,1)} \to 2\,k_F$ of Eq.~\eqref{eq:qf-th-m1}, and its contribution is therefore of order $q^3$. This proves Eq.~\eqref{eq:q-static}.

Since $\epsilon^{\rm RPA} = 1 - 4\,\pi\,e^2\,\Pi/q^2$, we have $q^2\,\epsilon^{\rm RPA} (q, 0) = q^2 - 4\,\pi\,e^2\,\Pi (q, 0) \to 4\,\pi\,e^2\,\rho_{\rm tot}$. The DOS of Eqs.~\eqref{eq:q-dos} and~\eqref{eq:q-dos0} give $4\,\pi\,e^2\,\rho_1 = 2\,g\,e^2\,k_F^2/(\pi\,v_0\,\beta)$ and $4\,\pi\,e^2\,\rho_0 = g\,e^2 \left( k_F^{(0)} \right)^2/(\pi\,v_0\,\sqrt{\xi})$, and their sum is Eq.~\eqref{eq:q-tf}. For the isotropic FSs, Eq.~\eqref{eq:drude} gives $D^{(s)}_{ii} = \pi\,e^2\,\rho_s\,V_s^2/3$ without a sum over $i$, and $\omega_p^2 = 4\,D_{ii}$ gives Eq.~\eqref{eq:q-wp-rho}. For $c \leq 0$, we have $\rho_0$ absent, and $\omega_p^2 = (4\,\pi\,e^2\,\rho_1/3)\,V_F^2 = q_{\rm TF}^2\,V_F^2/3$. For $c \to 0$, $\beta \to 1$, and $V_F \to v_0$ give $q_{\rm TF}^2 = 3\,\omega_p^2/v_0^2$.

\section{QTSM at arbitrary momentum-transfers}
\label{app:qf}

This appendix gives the intermediate steps behind the results of Sec.~\ref{sec-qtsm-full}.

\subsection{Imaginary part of polarisability: Particle-hole continua and their edges}
\label{app:qf-cont}

\subsubsection{Continuum edges}
\label{app:qf-edges}

At fixed $\bs{k}$ with $k < k_F$, the final momentum $k' = |\bs{k} + \bs{q}|$ obeys $|k - q| \leq k' \leq k + q$ and must satisfy $k' > k_F$. Since $\varepsilon_1$ is monotonic, the energy $\varepsilon_1 (k') - \varepsilon_1 (k)$ is largest for $k' = k + q$, which is the collinear configuration. It is smallest for $k' = \max \lbrace k_F, |k - q| \rbrace$.

For the upper edge, the collinear energy is $g_+ (k) = \varepsilon_1 (k + q) - \varepsilon_1 (k) = v_0\,q \left[ 1 + c \left( 2\,k + q \right) \right]$. It is linear in $k$ with the slope $2\,c\,v_0\,q$. The allowed range is $\max \lbrace 0, k_F - q \rbrace \leq k \leq k_F$, where the lower limit follows from $k + q \geq k_F$. The position of the maximum depends on the sign of $c$.
\begin{itemize}
\item For $c < 0$, the maximum is at the lower limit.
\item For $c > 0$, the maximum is at $k = k_F$.
\end{itemize}
This gives $\varpi_+$ of Eq.~\eqref{eq:qf-edges}.

For the lower edge with $q \leq 2\,k_F$, the configuration $k = k' = k_F$ is allowed, and the minimum is zero. For $q > 2\,k_F$, we have $k' \geq q - k > k_F$, and the minimum is attained for $k' = q - k$. The energy $\varepsilon_1 (q - k) - \varepsilon_1 (k)$ decreases with $k$, and its minimum at $k = k_F$ equals $\varepsilon_1 (q - k_F) - \varepsilon_1 (k_F) = v_0 \left( q - 2\,k_F \right) \left( 1 + c\,q \right)$, which is $\varpi_-$ of Eq.~\eqref{eq:qf-edges}. Every energy between the two edges is reached by continuity in $k$.

For fixed $k$, the energy $v_0 \left[ k' + c \left( k'^2 - k^2 \right) \right]$ increases with $k'$, since $1 + 2\,c\,k' > 0$. It is therefore smallest for the smallest admissible $k'$, namely $k' = \max \lbrace k_F, |k - q| \rbrace$. The value $k' = k_F$ is reachable only for $|q - k_F| \leq k \leq q + k_F$, because the triangle inequality $|k - q| \leq k' \leq k + q$ must hold. For $k < |q - k_F|$ and $q < k_F$, the state $k'$ lies inside the Fermi sphere, and no transition is allowed. When $k_F$ is reachable, the energy is $\varepsilon_1 (k_F) - \varepsilon_0 (k) = \mu - c\,v_0\,k^2$. The minimum over $k$ depends on the sign of $c$.
\begin{itemize}
\item For $c < 0$, the energy increases with $k$, and the minimum is at $k = |q - k_F|$, which gives Eq.~\eqref{eq:qf-th-neg}.
\item For $c > 0$, the energy decreases with $k$, and the minimum is at the upper limit $k = \min \lbrace q + k_F, k_F^{(0)} \rbrace$, where the upper limit of the occupied range of the band $s = 0$ enters.
\end{itemize}
Outside this range, we have $k' = |k - q|$. For $k > q + k_F$, we have $k' = k - q$, and the derivative of the energy with respect to $k$ is $v_0 \left( 1 - 2\,c\,q \right) > 0$. For $k < q - k_F$, we have $k' = q - k$, and the derivative is $-\,v_0 \left( 1 + 2\,c\,q \right) < 0$. Both signs follow from $\left| 2\,c\,q \right| < 1$ within the range of the linearised description. In both regions, the minimum is therefore at the boundary of the range with $k' = k_F$, and no lower value arises.

For $c > 0$ and $q + k_F \leq k_F^{(0)}$, that is $q \leq q_1$, the minimum is at $k = q + k_F$ and equals $\mu - c\,v_0 \left( q + k_F \right)^2$, which is the first line of Eq.~\eqref{eq:qf-th-pos}. For $q_1 \leq q \leq q_2$, the minimum is at $k = k_F^{(0)}$, which is allowed since $q - k_F \leq k_F^{(0)} \leq q + k_F$. Because $\varepsilon_0 \left( k_F^{(0)} \right) = \mu = \varepsilon_1 (k_F)$, the energy vanishes. For $q \geq q_2$, the sphere $k = k_F^{(0)}$ no longer reaches $k_F$, and the minimum is at $k = k_F^{(0)}$ with $k' = q - k_F^{(0)}$. It equals $\varepsilon_1 \left( q - k_F^{(0)} \right) - \mu$, and with $\mu = c\,v_0 \left( k_F^{(0)} \right)^2$ this is the last line of Eq.~\eqref{eq:qf-th-pos}. The three lines match continuously at $q_1$ and $q_2$.

For the transitions $-1 \to 1$, we use the variables $\Sigma = k + k'$ and $\check{D} = k' - k$. The energy is $v_0\,\Sigma \left( 1 + c\,\check{D} \right)$, with $1 + c\,\check{D} > 0$. The triangle inequality $|\check{D}| \leq q \leq \Sigma$ and the condition $k' = (\Sigma + \check{D})/2 \geq k_F$ define the allowed region. For fixed $\check{D}$, the energy increases with $\Sigma$, and the smallest admissible value is $\Sigma = \max \lbrace q, 2\,k_F - \check{D} \rbrace$. For $q \leq k_F$, we have $2\,k_F - \check{D} > q$ for all $\check{D} \leq q$, and therefore $\Sigma = 2\,k_F - \check{D}$ throughout. The energy $(2\,k_F - \check{D})(1 + c\,\check{D})$ has the second derivative $-\,2\,c$.
\begin{itemize}
\item For $c < 0$, the first derivative is $-\,1 + 2\,c\,k_F - 2\,c\,\check{D} < 0$, since $2\,|c|\,q < 1$. The minimum is then at $\check{D} = q$, which is the same value as for $c > 0$ below.
\item For $c > 0$, the energy is concave in $\check{D} \in [-q, q]$, and the minimum is at an endpoint. The endpoint $\check{D} = q$ gives $(2\,k_F - q)(1 + c\,q)$. The endpoint $\check{D} = -\,q$ gives $(2\,k_F + q)(1 - c\,q)$, and the difference of the two is $2\,q \left( 1 - 2\,c\,k_F \right) > 0$.
\end{itemize}
For $q \geq k_F$, the region has two parts. For $\check{D} \leq 2\,k_F - q$, we have $\Sigma = 2\,k_F - \check{D}$, and for $\check{D} \geq 2\,k_F - q$, we have $\Sigma = q$. In the second part, the energy $q \left( 1 + c\,\check{D} \right)$ is linear in $\check{D}$. In the first part, the endpoint $\check{D} = 2\,k_F - q$ gives the same value as the second part.
\begin{itemize}
\item For $c < 0$, the energy of the second part decreases with $\check{D}$, and the minimum is at $\check{D} = q$. The first part decreases with $\check{D}$, as for $q \leq k_F$. Its endpoint $\check{D} = 2\,k_F - q$ gives $q \left[ 1 + c \left( 2\,k_F - q \right) \right]$, which exceeds $q \left( 1 + c\,q \right)$ since $c \left( 2\,k_F - 2\,q \right) \geq 0$ for $c < 0$ and $q \geq k_F$.
\item For $c > 0$, the energy of the second part increases with $\check{D}$, and the minimum is at $\check{D} = 2\,k_F - q$. In the first part, the other endpoint $\check{D} = -\,q$ gives $(2\,k_F + q)(1 - c\,q)$. The difference from the value at $\check{D} = 2\,k_F - q$ is $2\,k_F \left( 1 - 2\,c\,q \right) > 0$.
\end{itemize}
This gives Eq.~\eqref{eq:qf-th-m1}.

For the transitions $-1 \to 0$ with $c > 0$, the energy at fixed $k$ increases with $k'$, and the smallest admissible value is $k' = \max \lbrace k_F^{(0)}, |k - q| \rbrace$. The value $k' = k_F^{(0)}$ is reachable for $|q - k_F^{(0)}| \leq k \leq q + k_F^{(0)}$. The energy is then $G (k) = k + c \left( \left( k_F^{(0)} \right)^2 - k^2 \right)$, which is concave in $k$. The minimum is at an endpoint. The endpoints differ by $G (q + k_F^{(0)}) - G (|q - k_F^{(0)}|) = 2\,\min \lbrace q, k_F^{(0)} \rbrace \left( 1 - 2\,c\,\max \lbrace q, k_F^{(0)} \rbrace \right) > 0$, and the minimum is therefore at $k = |q - k_F^{(0)}|$. For $q \leq k_F^{(0)}$, no other region contributes, since the states $k' > k_F^{(0)}$ are unreachable for $k < k_F^{(0)} - q$. This gives the first line of Eq.~\eqref{eq:qf-th-m0}. For $q \geq k_F^{(0)}$ and $k < q - k_F^{(0)}$, we have $k' = q - k$ and the energy $k \left( 1 - 2\,c\,q \right) + c\,q^2$, which increases with $k$. The minimum is at $k = 0$ and equals $c\,q^2$. It lies below the value at $k = q - k_F^{(0)}$. For $k > q + k_F^{(0)}$, we have $k' = k - q$, and the energy again increases with $k$. This gives the second line of Eq.~\eqref{eq:qf-th-m0}.

For the transitions $1 \to 0$ with $c > 0$, the energy at fixed $k < k_F$ increases with $k'$, and $k' > k_F^{(0)}$. Since $k' \leq k + q < k_F + q$, no transition exists for $q < k_F^{(0)} - k_F = q_1$. For $q_1 \leq q \leq q_2$, the pair $k = k_F$, $k' = k_F^{(0)}$ is allowed. The energy $\varepsilon_0 (k') - \varepsilon_1 (k)$ is non-negative, since $\varepsilon_0 (k') \geq \mu \geq \varepsilon_1 (k)$, and it vanishes for this pair. For $q > q_2$, the smallest admissible final momentum is $k' = q - k$, and the energy $c\,q^2 - \left( 1 + 2\,c\,q \right) k$ decreases with $k$. Its minimum at $k = k_F$ gives the second line of Eq.~\eqref{eq:qf-th-10}.

At fixed $\bs{k}$ with $k < k_F^{(0)}$, the energy $\varepsilon_0 (k') - \varepsilon_0 (k)$ increases with $k'$, and $|k - q| \leq k' \leq k + q$. For the upper edge, the collinear energy $c\,v_0\,q \left( 2\,k + q \right)$ increases with $k$, and its maximum at $k = k_F^{(0)}$ gives $\varpi_+^{(0)}$. For the lower edge with $q \leq 2\,k_F^{(0)}$, the configuration $k = k' = k_F^{(0)}$ is allowed, and the minimum is zero. For $q > 2\,k_F^{(0)}$, the smallest admissible final momentum is $k' = q - k$, and the energy $c\,v_0 \left[ (q - k)^2 - k^2 \right] = c\,v_0\,q \left( q - 2\,k \right)$ decreases with $k$. Its minimum at $k = k_F^{(0)}$ gives $\varpi_-^{(0)}$. Every energy between the edges is reached by continuity in $k$. For the comparison with the top band, $\varpi_+ = q \left( \beta + c\,q \right)$ for $c > 0$, and $2\,c\,k_F^{(0)} = 2\,\sqrt{\xi}$ follows from $c \left( k_F^{(0)} \right)^2 = \mu/v_0 = \xi/c$.

\subsubsection{Spectral weight of individual channels}
\label{app:qf-chan}

For $\omega > 0$, we have $f_0 (\varepsilon_1 (\bs{k})) - f_0 (\varepsilon_1 (\bs{k}')) = \Theta (k_F - k)\,\Theta (k' - k_F)$, and Eq.~\eqref{eq:pol} gives
\begin{align}
\Im \Pi_{11} = -\,\pi\,g \int \frac{d^3\bs{k}}{(2\,\pi)^3}\,\mathcal{F}_{11}\,\Theta (k_F - k)\,\Theta (k' - k_F)\,\delta \left( \omega + \varepsilon_1 (k) - \varepsilon_1 (k') \right).
\end{align}
The measure is $d^3\bs{k} = 2\,\pi\,k\,k'\,dk\,dk'/q$ on the region $|k - q| \leq k' \leq k + q$. Using $\cos\theta = \left( k^2 + k'^2 - q^2 \right)/(2\,k\,k')$, we write $\mathcal{F}_{11} = \left( (k + k')^2 - q^2 \right)^2/(16\,k^2\,k'^2)$. The delta function removes the integral over $k'$ and leaves the factor $1/\left[ v_0 \left( 1 + 2\,c\,k' \right) \right]$. Thus
\begin{align}
\Im \Pi_{11} = -\,\frac{g}{4\,\pi\,v_0\,q} \int dk\; \frac{\left( \Sigma^2 - q^2 \right)^2}{16\,k\,k' \left( 1 + 2\,c\,k' \right)}\,,
\end{align}
where $k'$ is the solution of $\left( k' - k \right) \left[ 1 + c \left( k' + k \right) \right] = \varpi$. With $\check{D} = k' - k$, the condition reads $\check{D} \left( 1 + c\,\Sigma \right) = \varpi$, and $k = (\Sigma - \check{D})/2$, $k' = (\Sigma + \check{D})/2$. The product is $k\,k' = (\Sigma^2 - \check{D}^2)/4$. The Jacobian is $dk = \tfrac{1}{2} \left( 1 - d\check{D}/d\Sigma \right) d\Sigma$ with $d\check{D}/d\Sigma = -\,c\,\check{D}/(1 + c\,\Sigma)$. It follows that $dk = \left( 1 + 2\,c\,k' \right) d\Sigma/\left[ 2 \left( 1 + c\,\Sigma \right) \right]$, and the factor $1 + 2\,c\,k'$ cancels. We obtain
\begin{align}
\Im \Pi_{11} = -\,\frac{g}{32\,\pi\,v_0\,q} \int d\Sigma\; \frac{\left( \Sigma^2 - q^2 \right)^2}{\left( 1 + c\,\Sigma \right) \left( \Sigma^2 - \check{D}^2 \right)}\,.
\end{align}
Inserting $\check{D} = \varpi/(1 + c\,\Sigma)$ gives $\left( 1 + c\,\Sigma \right) \left( \Sigma^2 - \check{D}^2 \right) = \left[ \left( \Sigma + c\,\Sigma^2 \right)^2 - \varpi^2 \right]/(1 + c\,\Sigma)$, which is Eq.~\eqref{eq:qf-im11}. The limits follow from the constraints on the pair $(k, k')$. The condition $k \leq k_F$ gives $\Sigma \leq \Sigma_+ = k_F + \varkappa \left( \mu/v_0 + \varpi \right)$, where $k' = \varkappa \left( \mu/v_0 + \varpi \right)$ at $k = k_F$. The condition $k' \geq k_F$ gives $\Sigma \geq k_F + \varkappa \left( \mu/v_0 - \varpi \right)$, where $k = \varkappa \left( \mu/v_0 - \varpi \right)$ at $k' = k_F$. The triangle inequality $k + k' \geq q$ gives $\Sigma \geq q$, and $k' - k \leq q$ gives $\check{D} = \varpi/(1 + c\,\Sigma) \leq q$, that is $c\,q\,\Sigma \geq \varpi - q$.
\begin{itemize}
\item For $c < 0$, this is an upper bound on $\Sigma$.
\item For $c > 0$, this is a lower bound on $\Sigma$.
\end{itemize}
The condition $k' - k \geq -\,q$ holds since $\check{D} > 0$. The range is nonempty exactly for $\varpi_- < \varpi < \varpi_+$. The integrand is regular, because $\Sigma^2 - \check{D}^2 = 4\,k\,k' > 0$ on the interior.

For the evaluation of the integral in Eq.~\eqref{eq:qf-im11}, we denote the numerator by $N (\Sigma) = \left( 1 + c\,\Sigma \right) \left( \Sigma^2 - q^2 \right)^2$ and the denominator by $P (\Sigma) = \left( c\,\Sigma^2 + \Sigma \right)^2 - \varpi^2$. The numerator has degree five and the denominator has degree four. Polynomial division gives $N/P = \Sigma/c - 1/c^2 + \left( N - \left( \Sigma/c - 1/c^2 \right) P \right)/P$, where the remainder is a cubic polynomial. The denominator factorises into $\left( c\,\Sigma^2 + \Sigma - \varpi \right) \left( c\,\Sigma^2 + \Sigma + \varpi \right)$, with the four simple roots $\Sigma_{\tau\check\sigma}$ of $c\,\Sigma^2 + \Sigma = \tau\,\varpi$. The partial fractions then read $N/P = \Sigma/c - 1/c^2 + \sum_{\tau, \check\sigma} \bar\varrho_{\tau\check\sigma}/(\Sigma - \Sigma_{\tau\check\sigma})$, with $\bar\varrho_{\tau\check\sigma} = N (\Sigma_{\tau\check\sigma})/P' (\Sigma_{\tau\check\sigma})$. At a root, we have $1 + 2\,c\,\Sigma_{\tau\check\sigma} = \check\sigma\,\check{r}_\tau$ and $1 + c\,\Sigma_{\tau\check\sigma} = \left( 1 + \check\sigma\,\check{r}_\tau \right)/2$. Also $P' = 2 \left( c\,\Sigma^2 + \Sigma \right) \left( 1 + 2\,c\,\Sigma \right) = 2\,\tau\,\varpi\,\check\sigma\,\check{r}_\tau$. This gives $\bar\varrho_{\tau\check\sigma}$ of the main text. Integration yields Eq.~\eqref{eq:qf-im11-cf}. The root $\Sigma_{+ +} = \varkappa (\varpi)$ lies below the range of integration, since $\Sigma^2 > \check{D}^2$ is equivalent to $\Sigma \left( 1 + c\,\Sigma \right) > \varpi$. None of the four roots lies in the range, and the logarithms are regular.

For $c \to 0$, the roots $\Sigma_{\tau, -} = -\,(1 + \check{r}_\tau)/(2\,c)$ tend to infinity, while $\Sigma_{\tau, +} \to \tau\,\varpi$. The terms in $1/c$ and $1/c^2$ cancel between the polynomial part and the logarithms of the far roots, and the remainder is the linear-model result.

For $\omega > 0$, the occupation factor of the transitions $0 \to 1$ is $\Theta (k' - k_F)$ times the filling of the band $s = 0$ at $k$. For every pair of bands $(s, s')$ of the remaining channels, Eq.~\eqref{eq:pol} gives $\Im \Pi_{ss'} = -\,\pi\,g \int d^3\bs{k}/(2\,\pi)^3\;\mathcal{F}_{ss'}\,\delta \left( \omega + \varepsilon_s (k) - \varepsilon_{s'} (k') \right)$ on the domain fixed by the occupations. With $d^3\bs{k} = 2\,\pi\,k\,k'\,dk\,dk'/q$, we need $k\,k'\,\mathcal{F}_{ss'}$. For the transitions $0 \to 1$, we use $\sin^2\theta = \left[ q^2 - (k - k')^2 \right] \left[ (k + k')^2 - q^2 \right]/(4\,k^2\,k'^2)$ and obtain $k\,k'\,\mathcal{F}_{01} = \left( q^2 - \check{D}^2 \right) \left( \Sigma^2 - q^2 \right)/\left[ 2 \left( \Sigma^2 - \check{D}^2 \right) \right]$, where $k\,k' = (\Sigma^2 - \check{D}^2)/4$. The delta function removes the integral over $k'$ and leaves $1/\left[ v_0 \left( 1 + 2\,c\,k' \right) \right]$.

The energy condition in terms of $\Sigma$ and $\check{D}$ reads $(\Sigma + \check{D})/2 + c\,\check{D}\,\Sigma = \varpi$, which gives $\check{D} (\Sigma)$ above. We write $\mathfrak{b} = 1 + 2\,c\,\Sigma$. The derivative is $d\check{D}/d\Sigma = -\left( 1 + 4\,c\,\varpi \right)/\mathfrak{b}^2$, and the Jacobian is $dk/d\Sigma = \left[ 1 + 2\,c\,\Sigma + 2\,c^2 \Sigma^2 + 2\,c\,\varpi \right]/\mathfrak{b}^2$. The final-state factor is $1 + 2\,c\,k' = \left[ 1 + 2\,c\,\Sigma + 2\,c^2 \Sigma^2 + 2\,c\,\varpi \right]/\mathfrak{b}$. The brackets cancel, and $dk/(1 + 2\,c\,k') = d\Sigma/\mathfrak{b}$. Thus
\begin{align}
\Im \Pi_{01} = -\,\frac{g}{8\,\pi\,v_0\,q} \int d\Sigma\; \frac{\left( q^2 - \check{D}^2 \right) \left( \Sigma^2 - q^2 \right)}{\left( \Sigma^2 - \check{D}^2 \right) \mathfrak{b}}\,.
\end{align}
Inserting $\check{D}$, the ratio of the first and the third factors becomes $\left( q^2 \mathfrak{b}^2 - (2\,\varpi - \Sigma)^2 \right)/\left( \Sigma^2 \mathfrak{b}^2 - (2\,\varpi - \Sigma)^2 \right)$. The denominator factorises as $4 \left( c\,\Sigma^2 + \Sigma - \varpi \right) \left( c\,\Sigma^2 + \varpi \right)$, and the numerator as $\left[ q - 2\,\varpi + (1 + 2\,c\,q) \Sigma \right] \left[ q + 2\,\varpi - (1 - 2\,c\,q) \Sigma \right]$. This gives Eq.~\eqref{eq:qf-im01}.

For the limits, $\check{D}$ decreases with $\Sigma$, since $d\check{D}/d\Sigma < 0$, which turns $|\check{D}| \leq q$ into the two bounds of the main text. The momentum $k = (\Sigma - \check{D})/2$ increases with $\Sigma$. The behaviour of $k'$ depends on the sign of $c$.
\begin{itemize}
\item For $c > 0$, the momentum $k' = (\Sigma + \check{D})/2$ also increases with $\Sigma$, because $dk'/d\Sigma \propto c \left( \Sigma + c\,\Sigma^2 - \varpi \right) > 0$ in the interior. The conditions $k' \geq k_F$ and $k \leq k_F^{(0)}$ are then lower and upper bounds on $\Sigma$. At $k' = k_F$, we have $c\,k^2 = \mu/v_0 - \varpi$. At $k = k_F^{(0)}$, we have $k' = \varkappa (\mu/v_0 + \varpi)$.
\item For $c < 0$, the momentum $k'$ decreases with $\Sigma$ by the same expression, and $k' \geq k_F$ is an upper bound. At $k' = k_F$, we have $|c|\,k^2 = \varpi - \mu/v_0$, which requires $\varpi \geq \mu/v_0$.
\end{itemize}

For the evaluation of the integral in Eq.~\eqref{eq:qf-im01}, we denote the integrand by $\iota (\Sigma)$ and the numerator by $N (\Sigma)$. The numerator has degree four and the denominator has degree five, and therefore $\iota (\Sigma)$ has no polynomial part. The factorisation of the denominator in the derivation above gives $4 \left( c\,\Sigma^2 + \Sigma - \varpi \right) \left( c\,\Sigma^2 + \varpi \right) = \mathfrak{b}^2 \left( \Sigma^2 - \check{D}^2 \right)$, where $\Sigma^2 - \check{D}^2 = 4\,k\,k' > 0$ in the interior of the range. The product of the last two denominator factors is thus positive there. For $c > 0$, we have $\mathfrak{b} > 0$. For $c < 0$, the upper bound $\Sigma \leq (2\,\varpi + q)/(1 + 2\,|c|\,q)$ lies below $1/(2\,|c|)$, because $4\,|c|\,\varpi < 1$, and $\mathfrak{b} > 0$ again. None of the poles lies in the range, and the logarithms are regular. The partial fractions read
\begin{align}
\iota (\Sigma) = \frac{\varrho_{\rm W}}{\Sigma - \Sigma_\star} + \sum_{\check\sigma = \pm 1} \frac{\varrho_{\check\sigma}}{\Sigma - \Sigma_{+\check\sigma}} + \varrho_{\rm q} \left[ \frac{1}{\Sigma - \Sigma_{\rm q}} + \frac{1}{\Sigma + \Sigma_{\rm q}} \right],
\end{align}
where $\Sigma_\star = -\,1/(2\,c)$ is the root of $\mathfrak{b}$, and $\Sigma_{\rm q}^2 = -\,\varpi/c$. The residues are evaluated one pole at a time.

At $\Sigma_\star$, we have $c\,\Sigma_\star^2 + \Sigma_\star = -\,1/(4\,c)$, which gives $c\,\Sigma_\star^2 + \Sigma_\star - \varpi = -\,\check{r}_+^2/(4\,c)$ and $c\,\Sigma_\star^2 + \varpi = \check{r}_+^2/(4\,c)$. The two square brackets of the numerator become $-\,\check{r}_+^2/(2\,c)$ and $\check{r}_+^2/(2\,c)$. The factors $\check{r}_+^4$ cancel, and $\varrho_{\rm W} = 2 \left( \Sigma_\star^2 - q^2 \right)/c$, which is the value in the main text. At $\Sigma_{+\check\sigma}$, the derivative of $c\,\Sigma^2 + \Sigma - \varpi$ is $\mathfrak{b}$, with $\mathfrak{b} = \check\sigma\,\check{r}_+$. The residue is therefore $N/\left[ \mathfrak{b}^2 \left( c\,\Sigma^2 + \varpi \right) \right]$. At the root, $c\,\Sigma^2 = \varpi - \Sigma$ gives $2\,\varpi - \Sigma = \Sigma\,\mathfrak{b}$ and $c\,\Sigma^2 + \varpi = \Sigma\,\mathfrak{b}$. The two square brackets of the numerator are $q\,\mathfrak{b} \pm \left( 2\,\varpi - \Sigma \right)$, and their product is $\mathfrak{b}^2 \left( q^2 - \Sigma^2 \right)$. Hence $N = -\,\mathfrak{b}^2 \left( \Sigma^2 - q^2 \right)^2$, and $\varrho_{\check\sigma} = -\,\left( \Sigma_{+\check\sigma}^2 - q^2 \right)^2/\left( \check\sigma\,\check{r}_+\,\Sigma_{+\check\sigma} \right)$.

At $\Sigma = \Sigma_{\rm q}$, we have $c\,\Sigma^2 = -\,\varpi$, and the residue is $N/\left[ 2\,c\,\Sigma\,\mathfrak{b} \left( \Sigma - 2\,\varpi \right) \right]$. We define $u = 1 - 4\,c\,\varpi + 4\,c\,\Sigma$. The product of the two square brackets of the numerator is $q^2 \mathfrak{b}^2 - \left( 2\,\varpi - \Sigma \right)^2 = \left( c\,q^2 + \varpi \right) u/c$. In addition, $\Sigma^2 - q^2 = -\left( c\,q^2 + \varpi \right)/c$ and $\Sigma\,\mathfrak{b} \left( \Sigma - 2\,\varpi \right) = -\,\varpi\,u/c$. The factors $u$ cancel, and the residue is $\left( c\,q^2 + \varpi \right)^2/(2\,c^2\,\varpi) = \varrho_{\rm q}$. It is the same at $\Sigma = \Sigma_{\rm q}$ and at $\Sigma = -\,\Sigma_{\rm q}$. The two terms then combine into $2\,\varrho_{\rm q}\,\Sigma/\left( \Sigma^2 - \Sigma_{\rm q}^2 \right) = 2\,c\,\varrho_{\rm q}\,\Sigma/\left( c\,\Sigma^2 + \varpi \right)$. This is the logarithmic derivative of $c\,\Sigma^2 + \varpi$ times $\varrho_{\rm q}$, and the term proportional to $1/\left( c\,\Sigma^2 + \varpi \right)$, which would give the arctangent for $c > 0$, is absent. This holds for both signs of $c$. Integration of the partial fractions yields Eq.~\eqref{eq:qf-im01-cf}. The constant $-\,\ln \left| 2\,c \right|$ from $\ln \left| \Sigma - \Sigma_\star \right|$ drops out of the difference of $\mathcal{G}_{01}$ at the two limits.

For $\omega > 0$, the occupation factor of the transitions $-1 \to 1$ is $\Theta (k' - k_F)$, because the band $s = -1$ is filled at every $k$. Using $1 - \cos\theta = \left( q^2 - \check{D}^2 \right)/(2\,k\,k')$ and $k\,k' = (\Sigma^2 - \check{D}^2)/4$, we obtain $k\,k'\,\mathcal{F}_{-1,1} = \left( q^2 - \check{D}^2 \right)^2/\left[ 4 \left( \Sigma^2 - \check{D}^2 \right) \right]$. The delta function removes the integral over $k'$ and leaves $1/\left[ v_0 \left( 1 + 2\,c\,k' \right) \right]$.

The energy condition in terms of $\Sigma$ and $\check{D}$ reads $\Sigma \left( 1 + c\,\check{D} \right) = \varpi$, which gives $\Sigma (\check{D}) = \varpi/(1 + c\,\check{D})$. Then $k = (\Sigma - \check{D})/2$ has the derivative $dk/d\check{D} = -\left[ \left( 1 + c\,\check{D} \right)^2 + c\,\varpi \right]/\left[ 2 \left( 1 + c\,\check{D} \right)^2 \right]$. The final-state factor is $1 + 2\,c\,k' = 1 + c\,\Sigma + c\,\check{D} = \left[ \left( 1 + c\,\check{D} \right)^2 + c\,\varpi \right]/(1 + c\,\check{D})$. The brackets cancel, and $|dk|/(1 + 2\,c\,k') = d\check{D}/\left[ 2 \left( 1 + c\,\check{D} \right) \right]$. Since $|\check{D}| \leq q$ and $2\,|c|\,q < 1$, we have $1 + c\,\check{D} > 1/2$. We also have $\left( 1 + c\,\check{D} \right)^2 + c\,\varpi > 0$ for both signs of $c$, since $\left( 1 + c\,\check{D} \right)^2 > 1/4 > |c|\,\varpi$ for $\varpi < 1/(4\,|c|)$. The momentum $k$ therefore decreases with $\check{D}$, and the range of $\check{D}$ is traversed with a positive measure. Using $\Sigma^2 - \check{D}^2 = \left[ \varpi^2 - \check{D}^2 \left( 1 + c\,\check{D} \right)^2 \right]/\left( 1 + c\,\check{D} \right)^2$, we obtain Eq.~\eqref{eq:qf-imm11}.

For the limits, the derivative of $k' = (\Sigma + \check{D})/2$ is $dk'/d\check{D} = \left[ \left( 1 + c\,\check{D} \right)^2 - c\,\varpi \right]/\left[ 2 \left( 1 + c\,\check{D} \right)^2 \right]$, which is positive by the same inequality. The condition $k' \geq k_F$ is then a lower bound on $\check{D}$. At $k' = k_F$, the energy condition reads $k - c\,k^2 = \varpi - \mu/v_0$, which gives $k = -\,\varkappa \left( \mu/v_0 - \varpi \right)$ and $\check{D} = k_F + \varkappa \left( \mu/v_0 - \varpi \right)$. This is the bound of the main text. The triangle inequality $k + k' \geq q$ gives $\Sigma \geq q$, that is $q \left( 1 + c\,\check{D} \right) \leq \varpi$, or $c\,q\,\check{D} \leq \varpi - q$. For $c > 0$, this is an upper bound on $\check{D}$. For $c < 0$, it is a lower bound. The inequality $|\check{D}| \leq q$ gives the remaining bounds. The momentum $k \geq 0$ follows from $\Sigma \geq q \geq |\check{D}|$. For $c \to 0$ and $\varpi > q$, the bound from $\Sigma \geq q$ disappears, and the range reduces to $\max \lbrace -q, 2\,k_F - \varpi \rbrace \leq \check{D} \leq q$. For $\varpi < q$, the bound from $\Sigma \geq q$ makes the range empty. The integrand then reduces to $\left( \check{D}^2 - q^2 \right)^2/\left( \varpi^2 - \check{D}^2 \right)$, which is the integrand of Eq.~\eqref{eq:fin-Imc} in the variable $\check{D}$. For the evaluation of the integral, we note that $\varpi^2 - \left( \check{D} + c\,\check{D}^2 \right)^2 = -\,P (\check{D})$ and $\left( 1 + c\,\check{D} \right) \left( \check{D}^2 - q^2 \right)^2 = N (\check{D})$, with $N$ and $P$ defined in Appendix~\ref{app:qf-chan}. The integral of $N/P$ is $\mathcal{G}_1$ of Eq.~\eqref{eq:qf-im11-cf}, and the sign reversal of the denominator gives the overall sign of Eq.~\eqref{eq:qf-imm11-cf}. The roots $\Sigma_{\tau\check\sigma}$ are the zeros of $P$. They do not lie in the range of $\check{D}$, because $P (\check{D}) = -\left( 1 + c\,\check{D} \right)^2 \left( \Sigma^2 - \check{D}^2 \right) = -\left( 1 + c\,\check{D} \right)^2\,4\,k\,k' < 0$ there. For $c \to 0$, the roots $\Sigma_{\tau, -}$ tend to infinity, and the roots $\Sigma_{\tau, +} \to \tau\,\varpi$ lie outside the range $|\check{D}| \leq q$ for $\varpi > q$. The terms in $1/c$ and $1/c^2$ cancel as in Appendix~\ref{app:qf-chan}.

For $c > 0$ and $\varpi > 1/(4\,c)$, the expressions $dk'/d\check{D}$ and $dk/d\check{D}$ of the derivation above keep their signs if $\left( 1 + c\,\check{D} \right)^2 > c\,\varpi$ on the range of $\check{D}$. The smallest value of the left side is $\left( 1 - c\,q \right)^2$ at $\check{D} = -\,q$, and the condition therefore holds for $\varpi < \left( 1 - c\,q \right)^2/c$. It is equivalent to $2\,c\,k < 1$. The equation $k - c\,k^2 = \varpi - \mu/v_0$ for the bound from $k' \geq k_F$ has a real root only for $\varpi - \mu/v_0 \leq 1/(4\,c)$, because $k - c\,k^2 \leq 1/(4\,c)$. For larger $\varpi$, the energy $\varepsilon_1 (k_F) - \varepsilon_{-1} (k) = v_0 \left( \mu/v_0 + k - c\,k^2 \right)$ lies below $v_0\,\varpi$ for every $k$, and the energy condition requires $k' > k_F$. The bound is then absent. Since $\check{r}_- = i\sqrt{4\,c\,\varpi - 1}$, the roots are $\Sigma_{-,\pm} = \Sigma_{\rm R} \pm i\,\Sigma_{\rm I}$ with $\Sigma_{\rm R} = -\,1/(2\,c)$ and $\Sigma_{\rm I} = \sqrt{4\,c\,\varpi - 1}/(2\,c)$. We have $\bar\varrho_{--} = \bar\varrho_{-+}^*$, since $P$ and $N$ are real. The two corresponding terms of the partial fractions add up to $2\,\Re \left[ \bar\varrho_{-+}/(\check{D} - \Sigma_{\rm R} - i\,\Sigma_{\rm I}) \right]$. The antiderivative of $1/(\check{D} - \Sigma_{\rm R} - i\,\Sigma_{\rm I})$ is $\tfrac{1}{2}\,\ln \left[ \left( \check{D} - \Sigma_{\rm R} \right)^2 + \Sigma_{\rm I}^2 \right] + i\,\arctan \!\left( (\check{D} - \Sigma_{\rm R})/\Sigma_{\rm I} \right)$. Using $\left( \check{D} - \Sigma_{\rm R} \right)^2 + \Sigma_{\rm I}^2 = \left( c\,\check{D}^2 + \check{D} + \varpi \right)/c$ and $(\check{D} - \Sigma_{\rm R})/\Sigma_{\rm I} = (1 + 2\,c\,\check{D})/\sqrt{4\,c\,\varpi - 1}$, we obtain the last two terms of Eq.~\eqref{eq:qf-imm11-cf2}. The argument of the logarithm is positive for every $\check{D}$, and the arctangent is regular.

For $\omega > 0$, the occupation factor of the transitions $-1 \to 0$ is $\Theta (k' - k_F^{(0)})$, because the band $s = -1$ is filled at every $k$. The overlap factor is the same as for the transitions $0 \to 1$, and $k\,k'\,\mathcal{F}_{-1,0} = \left( q^2 - \check{D}^2 \right) \left( \Sigma^2 - q^2 \right)/\left[ 2 \left( \Sigma^2 - \check{D}^2 \right) \right]$. The energy $\varepsilon_0 (k') - \varepsilon_{-1} (k) = v_0 \left[ k + c \left( k'^2 - k^2 \right) \right]$ has the derivative $2\,c\,v_0\,k'$ with respect to $k'$. The delta function removes the integral over $k'$ and leaves $1/\left( 2\,c\,v_0\,k' \right)$.

The energy condition in terms of $\Sigma$ and $\check{D}$ reads $(\Sigma - \check{D})/2 + c\,\Sigma\,\check{D} = \varpi$, which gives $\check{D} (\Sigma)$ above. We write $U = 1 - 2\,c\,\Sigma$. The derivative is $d\check{D}/d\Sigma = \left( 1 - 4\,c\,\varpi \right)/U^2$, and $dk/d\Sigma = \left( 1 - d\check{D}/d\Sigma \right)/2 = -\,2\,c \left( \Sigma - c\,\Sigma^2 - \varpi \right)/U^2$. The final-state factor is $2\,c\,k' = c \left( \Sigma + \check{D} \right) = 2\,c \left( \Sigma - c\,\Sigma^2 - \varpi \right)/U$. The brackets cancel, and $|dk|/\left( 2\,c\,k' \right) = d\Sigma/U$. Thus
\begin{align}
\Im \Pi_{-1,0} = -\,\frac{g}{8\,\pi\,v_0\,q} \int d\Sigma\; \frac{\left( q^2 - \check{D}^2 \right) \left( \Sigma^2 - q^2 \right)}{\left( \Sigma^2 - \check{D}^2 \right) U}\,.
\end{align}
Inserting $\check{D}$, the ratio of the first and the third factors becomes $\left( q^2\,U^2 - (\Sigma - 2\,\varpi)^2 \right)/\left( \Sigma^2\,U^2 - (\Sigma - 2\,\varpi)^2 \right)$. The denominator factorises as $4 \left( \varpi - c\,\Sigma^2 \right) \left( \Sigma - c\,\Sigma^2 - \varpi \right)$, and the numerator as $\left[ q + 2\,\varpi - (1 + 2\,c\,q) \Sigma \right] \left[ q - 2\,\varpi + (1 - 2\,c\,q) \Sigma \right]$. This gives Eq.~\eqref{eq:qf-imm10}.

For the limits, the condition $|\check{D}| \leq q$ is $|\Sigma - 2\,\varpi| \leq q\,|1 - 2\,c\,\Sigma|$. For $\Sigma < 1/(2\,c)$, it gives $(2\,\varpi - q)/(1 - 2\,c\,q) \leq \Sigma \leq (2\,\varpi + q)/(1 + 2\,c\,q)$, and this range is nonempty for $\varpi \leq 1/(4\,c)$. For $\Sigma > 1/(2\,c)$, the same condition gives $(2\,\varpi + q)/(1 + 2\,c\,q) \leq \Sigma \leq (2\,\varpi - q)/(1 - 2\,c\,q)$, and this range is nonempty for $\varpi \geq 1/(4\,c)$ only. For $\varpi < 1/(4\,c)$, we therefore have $U > 0$. From $2\,c\,k' > 0$, it follows that $\Sigma - c\,\Sigma^2 - \varpi > 0$. The relation $\Sigma^2 - \check{D}^2 = 4 \left( \varpi - c\,\Sigma^2 \right) \left( \Sigma - c\,\Sigma^2 - \varpi \right)/U^2 > 0$ then gives $\varpi - c\,\Sigma^2 > 0$. The three denominator factors of Eq.~\eqref{eq:qf-imm10} are thus positive. At fixed energy, $k' = \sqrt{\left( \varpi - k + c\,k^2 \right)/c}$ decreases with $k$, and $d\Sigma/dk = -\,U/(2\,c\,k') < 0$. The momentum $k'$ therefore increases with $\Sigma$, since $dk'/d\Sigma = \left( 1 - 2\,c\,k \right)/U > 0$. The condition $k' \geq k_F^{(0)}$ is then a lower bound on $\Sigma$. At $k' = k_F^{(0)}$, we have $c \left( k_F^{(0)} \right)^2 = \mu/v_0$, and the energy condition reads $k - c\,k^2 = \varpi - \mu/v_0$. This gives $k = -\,\varkappa \left( \mu/v_0 - \varpi \right)$ and $\Sigma = k_F^{(0)} - \varkappa \left( \mu/v_0 - \varpi \right)$. The value of $k$ is nonnegative only for $\varpi \geq \mu/v_0$. For $\varpi < \mu/v_0$, the value $k' = \sqrt{\varpi/c}$ at $k = 0$ lies below $k_F^{(0)}$, and no transition exists. The triangle inequality $k + k' \geq q$ is $\Sigma \geq q$. The momentum $k \geq 0$ follows from $\Sigma \geq q \geq |\check{D}|$.

For the evaluation of the integral, we note that the integrand of Eq.~\eqref{eq:qf-imm10} is that of Eq.~\eqref{eq:qf-im01} with $c \to -\,c$, since $\left( 1 + 2\,c\,\Sigma \right) \left( c\,\Sigma^2 + \Sigma - \varpi \right) \left( c\,\Sigma^2 + \varpi \right) \to \left( 1 - 2\,c\,\Sigma \right) \left( \Sigma - c\,\Sigma^2 - \varpi \right) \left( \varpi - c\,\Sigma^2 \right)$, and the brackets of the numerator are exchanged. The partial fractions of Appendix~\ref{app:qf-chan} hold for either sign of $c$, and the antiderivative follows from $\mathcal{G}_{01}$ with $c \to -\,c$. Then $\varrho_{\rm W} \to -\,\varrho_{\rm W}$, and $\ln \left| 1 + 2\,c\,\Sigma \right|$ becomes $\ln \left| 1 - 2\,c\,\Sigma \right|$. The root $\Sigma_{+\check\sigma}$ becomes $-\,\Sigma_{-\check\sigma}$, and $\check{r}_+$ becomes $\check{r}_-$. Hence $\varrho_{\check\sigma}$ becomes $\tilde\varrho_{\check\sigma}$ and $\ln \left| \Sigma - \Sigma_{+\check\sigma} \right|$ becomes $\ln \left| \Sigma + \Sigma_{-\check\sigma} \right|$. Finally, $\varrho_{\rm q} \to \tilde\varrho_{\rm q}$, and the logarithm becomes $\ln \left| \varpi - c\,\Sigma^2 \right|$. This gives Eq.~\eqref{eq:qf-imm10-cf}.

For $\omega > 0$, the occupation factor of the transitions $0 \to 0$ is $\Theta (k_F^{(0)} - k) - \Theta (k_F^{(0)} - k')$, which equals unity for $k < k_F^{(0)} < k'$. We have $k^2 + k'^2 = (\Sigma^2 + \check{D}^2)/2$ and $2\,k\,k' = (\Sigma^2 - \check{D}^2)/2$, and therefore $\cos\theta = \left( \Sigma^2 + \check{D}^2 - 2\,q^2 \right)/\left( \Sigma^2 - \check{D}^2 \right)$. It follows that $k\,k'\,\mathcal{F}_{00} = \left( \Sigma^2 + \check{D}^2 - 2\,q^2 \right)^2/\left[ 4 \left( \Sigma^2 - \check{D}^2 \right) \right]$. The delta function removes the integral over $k'$ and leaves $1/\left( 2\,c\,v_0\,k' \right)$.

The energy condition $c\,\Sigma\,\check{D} = \varpi$ gives $\check{D} = \varpi/(c\,\Sigma)$. The derivative is $dk/d\Sigma = \left( 1 - d\check{D}/d\Sigma \right)/2 = \left( c\,\Sigma^2 + \varpi \right)/\left( 2\,c\,\Sigma^2 \right)$. The final-state factor is $2\,c\,k' = c \left( \Sigma + \check{D} \right) = \left( c\,\Sigma^2 + \varpi \right)/\Sigma$. The brackets cancel, and $dk/\left( 2\,c\,k' \right) = d\Sigma/(2\,c\,\Sigma)$. Thus
\begin{align}
\Im \Pi_{00} = -\,\frac{g}{32\,\pi\,c\,v_0\,q} \int d\Sigma\; \frac{\left( \Sigma^2 + \check{D}^2 - 2\,q^2 \right)^2}{\Sigma \left( \Sigma^2 - \check{D}^2 \right)}\,.
\end{align}
Inserting $\check{D}$, we have $\Sigma^2 + \check{D}^2 - 2\,q^2 = \left( c^2\,\Sigma^4 - 2\,c^2\,q^2\,\Sigma^2 + \varpi^2 \right)/\left( c^2\,\Sigma^2 \right)$ and $\Sigma^2 - \check{D}^2 = \left( c^2\,\Sigma^4 - \varpi^2 \right)/\left( c^2\,\Sigma^2 \right)$. This gives Eq.~\eqref{eq:qf-im00}.

For the limits, the condition $\check{D} \leq q$ gives $\Sigma \geq \varpi/(c\,q)$, and the triangle inequality gives $\Sigma \geq q$. The condition $\check{D} \geq -\,q$ holds since $\check{D} > 0$. The momentum $k = (\Sigma - \check{D})/2$ increases with $\Sigma$, since $dk/d\Sigma > 0$. The momentum $k' = (\Sigma + \check{D})/2$ has the derivative $\left( c\,\Sigma^2 - \varpi \right)/\left( 2\,c\,\Sigma^2 \right)$, which is positive in the interior, where $\Sigma^2 > \check{D}^2$ is equivalent to $c\,\Sigma^2 > \varpi$. The conditions $k < k_F^{(0)}$ and $k' > k_F^{(0)}$ are then an upper and a lower bound on $\Sigma$. At $k = k_F^{(0)}$, we have $c\,\Sigma^2 - 2\,c\,k_F^{(0)}\,\Sigma - \varpi = 0$, which gives $\Sigma_+$. At $k' = k_F^{(0)}$, we have $c\,\Sigma^2 - 2\,c\,k_F^{(0)}\,\Sigma + \varpi = 0$. Its roots are real only for $\varpi \leq c \left( k_F^{(0)} \right)^2$, and their product is $\varpi/c$. The smaller root lies below $\sqrt{\varpi/c}$, outside the range, and the larger root gives the third entry of $\Sigma_-$. For $\varpi > c \left( k_F^{(0)} \right)^2$, we have $k' > \sqrt{\varpi/c} > k_F^{(0)}$ at every $\Sigma$. The range closes at $\varpi = \varpi_+^{(0)}$, where $\Sigma_+ = \varpi/(c\,q)$. For $q > 2\,k_F^{(0)}$, it also closes at $\varpi = \varpi_-^{(0)}$, where $\Sigma_+ = q$. This reproduces the edges of Eq.~\eqref{eq:qf-edges0}.

For the evaluation of the integral, we change the variable to $u = \Sigma^2$, with $d\Sigma/\Sigma^3 = du/(2\,u^2)$. The integrand becomes $f (u) = \left( c^2 u^2 - 2\,c^2 q^2 u + \varpi^2 \right)^2/\left[ u^2 \left( c^2 u^2 - \varpi^2 \right) \right]$. The numerator and the denominator have degree four, and the polynomial part is the constant $c^2$. The poles are at $u = 0$ with order two, and at $u = \pm\,\varpi/c$. The coefficient of $1/u^2$ is $\varpi^4/\left( -\,\varpi^2 \right) = -\,\varpi^2$, and the coefficient of $1/u$ is $4\,c^2 q^2$. The residues at $u = \pm\,\varpi/c$ give the partial fractions
\begin{align}
f (u) = c^2 + \frac{4\,c^2 q^2}{u} - \frac{\varpi^2}{u^2} + \frac{2\,c^2 \left( c\,q^2 - \varpi \right)^2}{\varpi \left( c\,u - \varpi \right)} - \frac{2\,c^2 \left( c\,q^2 + \varpi \right)^2}{\varpi \left( c\,u + \varpi \right)}\,.
\end{align}
Integration gives $c^2\,\mathcal{G}_{00}$ with $u = \Sigma^2$, and the prefactor $1/2$ from $du$ turns $-\,g/(32\,\pi\,c^3\,v_0\,q)$ into $-\,g/(64\,\pi\,c\,v_0\,q)$. This is Eq.~\eqref{eq:qf-im00-cf}. The argument $c\,\Sigma^2 + \varpi$ is positive, and $c\,\Sigma^2 - \varpi$ does not vanish on the range of integration.

For $\omega > 0$, the occupation factor of the transitions $1 \to 0$ is $\Theta (k_F - k)\,\Theta (k' - k_F^{(0)})$. The overlap factor is that of the transitions $0 \to 1$, and $k\,k'\,\mathcal{F}_{10} = \left( q^2 - \check{D}^2 \right) \left( \Sigma^2 - q^2 \right)/\left[ 2 \left( \Sigma^2 - \check{D}^2 \right) \right]$. The energy $\varepsilon_0 (k') - \varepsilon_1 (k) = v_0 \left[ c \left( k'^2 - k^2 \right) - k \right]$ has the derivative $2\,c\,v_0\,k'$ with respect to $k'$. The delta function removes the integral over $k'$ and leaves $1/\left( 2\,c\,v_0\,k' \right)$.

The energy condition in terms of $\Sigma$ and $\check{D}$ reads $c\,\Sigma\,\check{D} - (\Sigma - \check{D})/2 = \varpi$, which gives $\check{D} (\Sigma)$ above. We write $\mathfrak{b} = 1 + 2\,c\,\Sigma$. The derivative is $d\check{D}/d\Sigma = \left( 1 - 4\,c\,\varpi \right)/\mathfrak{b}^2$, and $dk/d\Sigma = \left( 1 - d\check{D}/d\Sigma \right)/2 = 2\,c \left( c\,\Sigma^2 + \Sigma + \varpi \right)/\mathfrak{b}^2$. The final-state factor is $2\,c\,k' = c \left( \Sigma + \check{D} \right) = 2\,c \left( c\,\Sigma^2 + \Sigma + \varpi \right)/\mathfrak{b}$. The brackets cancel, and $dk/\left( 2\,c\,k' \right) = d\Sigma/\mathfrak{b}$. The result has the form of the transitions $0 \to 1$,
\begin{align}
\Im \Pi_{10} = -\,\frac{g}{8\,\pi\,v_0\,q} \int d\Sigma\; \frac{\left( q^2 - \check{D}^2 \right) \left( \Sigma^2 - q^2 \right)}{\left( \Sigma^2 - \check{D}^2 \right) \mathfrak{b}}\,.
\end{align}
Inserting $\check{D}$, we have $\Sigma^2\,\mathfrak{b}^2 - (\Sigma + 2\,\varpi)^2 = 4 \left( c\,\Sigma^2 - \varpi \right) \left( c\,\Sigma^2 + \Sigma + \varpi \right)$ and $q^2\,\mathfrak{b}^2 - (\Sigma + 2\,\varpi)^2 = \left[ q - 2\,\varpi - (1 - 2\,c\,q) \Sigma \right] \left[ q + 2\,\varpi + (1 + 2\,c\,q) \Sigma \right]$. This gives Eq.~\eqref{eq:qf-im10}.

For the limits, we use that $\varpi = c\,k'^2 - k - c\,k^2$ and $k' \leq k + q$ give $\varpi \leq c\,q^2 + k \left( 2\,c\,q - 1 \right) \leq c\,q^2$. Hence $\varpi < 1/(4\,c)$ for $2\,c\,q < 1$, and $d\check{D}/d\Sigma > 0$. At fixed $\varpi$, the momentum $k' = \sqrt{\left( \varpi + k + c\,k^2 \right)/c}$ and the sum $\Sigma = k + k'$ increase with $k$. The condition $k < k_F$ is then an upper bound on $\Sigma$. At $k = k_F$, we have $k_F + c\,k_F^2 = \mu/v_0$, which gives $\Sigma = k_F + \sqrt{\left( \mu/v_0 + \varpi \right)/c}$. The condition $k' > k_F^{(0)}$ is $c\,k^2 + k > \mu/v_0 - \varpi$, which gives a lower bound $k > \varkappa \left( \mu/v_0 - \varpi \right)$ for $\varpi < \mu/v_0$, and $\Sigma = k_F^{(0)} + \varkappa \left( \mu/v_0 - \varpi \right)$ at the bound. For $\varpi > \mu/v_0$, the condition is automatic. The condition $\check{D} \leq q$ gives $\Sigma \left( 1 - 2\,c\,q \right) \leq q - 2\,\varpi$, which is an upper bound since $\check{D}$ increases with $\Sigma$. The triangle inequality gives $\Sigma \geq q$. The momentum $k \geq 0$ follows from $\Sigma \geq q \geq \check{D}$, and this is equivalent to $c\,\Sigma^2 \geq \varpi$. The three denominator factors of Eq.~\eqref{eq:qf-im10} are therefore positive on the range of integration.

For the evaluation of the integral, we note that the integrand of Eq.~\eqref{eq:qf-im10} is that of Eq.~\eqref{eq:qf-im01} with $\varpi \to -\,\varpi$, since the denominator $\left( 1 + 2\,c\,\Sigma \right) \left( c\,\Sigma^2 + \Sigma - \varpi \right) \left( c\,\Sigma^2 + \varpi \right)$ becomes $\left( 1 + 2\,c\,\Sigma \right) \left( c\,\Sigma^2 + \Sigma + \varpi \right) \left( c\,\Sigma^2 - \varpi \right)$, and the numerator brackets map onto those above. The partial fractions of Appendix~\ref{app:qf-chan} are algebraic identities and hold for this replacement. Then $\varrho_{\rm W}$ is unchanged, $\check{r}_+$ becomes $\check{r}_-$, and the root $\Sigma_{+\check\sigma}$ becomes $\Sigma_{-\check\sigma}$. Hence $\varrho_{\check\sigma}$ becomes $\hat\varrho_{\check\sigma}$, $\varrho_{\rm q}$ becomes $\hat\varrho_{\rm q}$, and $\ln \left| c\,\Sigma^2 + \varpi \right|$ becomes $\ln \left| c\,\Sigma^2 - \varpi \right|$. This gives Eq.~\eqref{eq:qf-im10-cf}.

\subsection{Real part of polarisability and RPA dielectric function}
\label{app:qf-re}

We give the steps behind Eqs.~\eqref{eq:qf-kk}--\eqref{eq:qf-revac}. Within the regularisation of Sec.~\ref{secfinite}, Eq.~\eqref{eq:pol} contains only the terms with $\left| \varepsilon_{s'} (\bs{k}') - \varepsilon_s (\bs{k}) \right| < \mathcal{R}$. For a pair of bands $s < s'$, the two orderings combine into Eq.~\eqref{eq:q-pair}, which is exact. For $s = s'$, the single term has the denominator $\omega - X$ with $X = \varepsilon_s (\bs{k}') - \varepsilon_s (\bs{k})$. The substitution $\bs{k} \to -\,\bs{k} - \bs{q}$ exchanges $k$ and $k'$ and leaves $\mathcal{F}_{ss}$ invariant, and it turns the term into one with the denominator $\omega + X$ and the opposite occupation difference. Taking the mean of the two forms, we obtain
\begin{align}
\frac{1}{2}\,\mathcal{F}_{ss} \left[ n_s (\bs{k}) - n_s (\bs{k}') \right] \frac{2\,X}{\left( \omega + i\,0^+ \right)^2 - X^2}\,.
\end{align}
Every contribution to $\Pi$ therefore has the form $\mathfrak{M}\,[\,1/(\omega - X + i\,0^+) - 1/(\omega + X + i\,0^+)\,]$, with a real weight $\mathfrak{M}$ and a finite $X \in (-\,\mathcal{R}, \mathcal{R})$. The weights are integrable, since $\mathcal{F}_{ss'} \leq 1$ and, within the range of the linearised description where $|c|\,k < 1$, the region of $\bs{k}$ with a nonzero occupation difference and a transition energy below $\mathcal{R}$ is bounded. We can therefore write $\Pi (q, \omega) = \int_{-\mathcal{R}}^{\mathcal{R}} d\tilde x\,\mathcal{Y} (\tilde x)/(\omega - \tilde x + i\,0^+)$, with a real spectral function $\mathcal{Y}$ that is odd in $\tilde x$ and has a finite integral of its modulus. Then $\Im \Pi (q, \omega) = -\,\pi\,\mathcal{Y} (\omega)$, and
\begin{align}
\Re \Pi (q, \omega) = \mathrm{P} \!\! \int_{-\mathcal{R}}^{\mathcal{R}} d\tilde x\,\frac{\mathcal{Y} (\tilde x)}{\omega - \tilde x} = \int_0^{\mathcal{R}} d\tilde x\,\mathcal{Y} (\tilde x) \left( \frac{1}{\omega - \tilde x} - \frac{1}{\omega + \tilde x} \right) = \frac{2}{\pi}\,\mathrm{P} \!\! \int_0^{\mathcal{R}} d\tilde x\,\frac{\tilde x\,\Im \Pi (q, \tilde x)}{\tilde x^2 - \omega^2}\,,
\end{align}
which is Eq.~\eqref{eq:qf-kk}. No additional term arises, because $\Pi$ decays as $1/\omega^2$ for $|\omega| \to \infty$.

For the support of $\Delta \Im \Pi$, we compare the occupations of the doped system with those of the undoped one, which are given in Sec.~\ref{secquadintr}. They differ only for the states that lie inside the Fermi spheres.
\begin{itemize}
\item For $c < 0$, these are the states of the band $s = 1$ with $k < k_F$.
\item For $c > 0$, these are the states of the band $s = 1$ with $k < k_F$ and the states of the band $s = 0$ with $k < k_F^{(0)}$, where $k_F < k_F^{(0)}$.
\end{itemize}
In both cases, the differing states have a momentum below $k_m$ of Eq.~\eqref{eq:qf-omc}. A pair of states connected by $\bs{q}$ contributes to $\Delta \Im \Pi$ only if at least one of them is such a state. Since $|k - k'| \leq q$, both momenta then obey $k, k' \leq k_m + q$. The transition energy obeys
\begin{align}
\left| \varepsilon_{s'} (k') - \varepsilon_s (k) \right| = v_0 \left| s'\,k' - s\,k + c \left( k' + k \right) \left( k' - k \right) \right| \leq v_0 \left( k + k' \right) \left( 1 + |c|\,q \right) \leq \omega_{\rm w}\,.
\end{align}
The imaginary part is a sum of delta functions at the transition energies, and $\Delta \Im \Pi$ therefore vanishes for $\omega > \omega_{\rm w}$. Splitting $\Im \Pi = \Im \Pi_{\rm vac}^{(c)} + \Delta \Im \Pi$ in Eq.~\eqref{eq:qf-kk} and using $\omega_{\rm w} < \mathcal{R}$ gives Eq.~\eqref{eq:qf-re}. The channels that enter $\Im \Pi_{\rm vac}^{(c)}$ are those listed in Sec.~\ref{secqfre}, which gives Eq.~\eqref{eq:qf-imvac}. The limits $\mu \to 0^+$ of the integration ranges follow from the conditions stated in Appendix~\ref{app:qf-chan}. The lower bounds that involve $\varkappa \left( \mu/v_0 - \varpi \right)$ are absent for every $\varpi > 0$, since $\varpi > \mu/v_0$ is then automatic.

Finally, Eq.~\eqref{eq:qf-revac} follows by writing $\Im \Pi_{\rm vac}^{(c)} = \Im \Pi^{(0)} + \left( \Im \Pi_{\rm vac}^{(c)} - \Im \Pi^{(0)} \right)$ in Eq.~\eqref{eq:qf-kk}. The first part gives $\Re \Pi^{(0)}$, which is Eq.~\eqref{eq:fin-Re0} for $\omega > v_0\,q$ with the same cut-off, as derived in Appendix~\ref{app:fin-re}. The second part is the remaining integral. At $\omega = 0$, we have $\omega'/(\omega'^2 - \omega^2) = 1/\omega'$, which gives the static form quoted in Sec.~\ref{secqfstatic}.

\subsection{Static polarisability and screening}
\label{app:qf-static}

At $\omega = 0$, the pairs of states connected by $\bs{q}$ split into intraband pairs, for the bands $s$ with an FS, and interband pairs. We treat them separately for $q \ll k_F$.

For an intraband pair, the identity $\left[ \Theta (\mu - \varepsilon) - \Theta (\mu - \varepsilon') \right]/(\varepsilon - \varepsilon') = -\int_0^1 d\nu\,\delta \!\left( \mu - (1 - \nu)\,\varepsilon - \nu\,\varepsilon' \right)$ holds, with $\varepsilon = \varepsilon_s (k)$ and $\varepsilon' = \varepsilon_s (k')$. It removes the principal value. With $\Delta = \varepsilon' - \varepsilon$, we expand the delta function in $\Delta$ and integrate over $\nu$, which gives
\begin{align}
\Pi_{ss} (q, 0) = -\,\frac{g}{4\,\pi^2} \sum_{m = 0}^{3} \frac{1}{\left( m + 1 \right)!} \int dk\,k^2 \int_{-1}^{1} dx\,\mathcal{F}_{ss}\,\Delta^m\,\delta^{(m)} \!\left( \varepsilon_s (k) - \mu \right) + \dots\,,
\end{align}
where $\tilde x = \cos\theta_{\bs{k},\bs{q}}$ and $k' = \sqrt{k^2 + 2\,k\,q\,\tilde x + q^2}$. We expand $\mathcal{F}_{ss}$ and $\Delta$ in $q$, integrate over $\tilde x$, and use $\int dk\,G (k)\,\delta^{(m)} \!\left( \varepsilon_s - \mu \right) = \left( -\,1 \right)^m \left[ \left( \partial_k \varepsilon_s \right)^{-1} \partial_k \right]^m \left( G/\partial_k \varepsilon_s \right) \big|_{k = k_s}$. The terms of order $q^0$ and $q^2$ of the two bands are
\begin{align}
\Pi_{11} (q, 0) = -\,\frac{g}{4\,\pi^2\,v_0} \left[ \frac{2\,k_F^2}{\beta} 
- \frac{30\,c^2\,k_F^2 + 30\,c\,k_F + 7}{9\,\beta^3}\,q^2 \right], \quad
\Pi_{00} (q, 0) = -\,\frac{g}{4\,\pi^2\,v_0} \left[ \frac{k_F^{(0)}}{c} 
- \frac{3\,q^2}{4\,c\,k_F^{(0)}} \right],
\end{align}
and the odd orders vanish.
\begin{itemize}
\item The top band $s = 1$ contributes for both signs of $c$. The first term of $\Pi_{11}$ is $-\,\rho_1$ of Eq.~\eqref{eq:q-dos}, and the second term gives the first part of $\mathcal{S}$. For $c = 0$, it is the expansion of Eq.~\eqref{eq:fin-static}.
\item The flat-band $s = 0$ contributes for $c > 0$ only. The first term of $\Pi_{00}$ is $-\,\rho_0$ of Eq.~\eqref{eq:q-dos0}, and the second term gives the part of $\mathcal{S}$ proportional to $1/\sqrt{\xi}$, with $c\,k_F^{(0)} = \sqrt{\xi}$.
\end{itemize}

For the interband pairs, the overlap factors $\mathcal{F}_{ss'}$ with $s \neq s'$ are of order $q^2$ for $k \gg q$, with $\mathcal{F}_{01} = \mathcal{F}_{-1, 0} = q^2 \left( 1 - \tilde x^2 \right)/(2\,k^2)$, and $\mathcal{F}_{-1, 1}$ is of order $q^4$. The transition energy is $v_0\,k$ up to corrections of order $q$. For $k < k_F$, all pairs are blocked for both signs of $c$, since the bands $s = -1$, $0$, and $1$ are filled. For $k > k_F$, the pair with a filled initial state and an empty final state at order $q^2$ depends on the sign of $c$.
\begin{itemize}
\item For $c < 0$, the pair $0 \to 1$ is the only one for all $k > k_F$.
\item For $c > 0$, the pairs between the bands $s = -1$ and $s = 0$ are blocked for $k_F < k < k_F^{(0)}$, where the pair $0 \to 1$ is the only one. For $k > k_F^{(0)}$, the pair $-1 \to 0$ is the only one.
\end{itemize}
With $\int_{-1}^{1} d\tilde x\,\mathcal{F} = 2\,q^2/(3\,k^2)$, the pair at a given $k > k_F$ contributes $-\,(g/(4\,\pi^2\,v_0))\,(4\,q^2/3)\,dk/k$ in the static limit. The weight is the same for both signs of $c$. The integral from $k_F$ to $\Lambda$ gives $-\,(g\,q^2/(4\,\pi^2\,v_0))\,(4/3) \ln \!\left( \Lambda/k_F \right)$. The boundary layers of width $q$ around $k = k_F$ and $k = k_F^{(0)}$ give contributions of order $q^3$. The cut-off on the transition energy changes the upper limit by terms that vanish for $\Lambda \to \infty$, as in Appendix~\ref{app:qf-wdisp}. Adding the intraband and interband terms gives Eq.~\eqref{eq:qf-static}, and $\epsilon^{\rm RPA} (q, 0) = 1 - 4\,\pi\,e^2\,\Pi (q, 0)/q^2$ gives Eq.~\eqref{eq:qf-tf}, with $4\,\pi\,e^2\,g/(4\,\pi^2\,v_0) = \alpha/\pi$.

For $q \gg k_F, k_F^{(0)}$, we write $\Pi = \Pi_{\rm vac}^{(c)} + \Delta\Pi$ and use Eq.~\eqref{eq:qf-ball} at $\omega = 0$, which gives $\Delta\Pi (q, 0) = 2\,g \sum_s \int_{k < k_s} \frac{d^3\bs{k}}{(2\,\pi)^3} \sum_{s'} \mathcal{F}_{s s'}/Y$. For $k \ll q$, we have $k' \simeq q$ and $\cos\theta \simeq \tilde x$, and $\mathcal{F}_{s s'}$ therefore depends only on $\tilde x$, with an average of $1/3$ over $\tilde x$ for each $s'$. The energies are $Y_{s, 1} \simeq -\,v_0\,q \left( 1 + c\,q \right)$, $Y_{s, 0} \simeq -\,c\,v_0\,q^2$, and $Y_{s, -1} \simeq v_0\,q \left( 1 - c\,q \right)$, for $s = 1$ and $s = 0$, where the corrections are of relative order $k_s/q$ and, for $Y_{1, 0}$, $k_F/\left( c\,q^2 \right)$. With $\int_{k < k_s} d^3\bs{k}/(2\,\pi)^3 = k_s^3/(6\,\pi^2)$, we obtain
\begin{align}
\Delta\Pi (q, 0) = \frac{g \left[ k_F^3 + \Theta (c) \left( k_F^{(0)} \right)^3 \right]}{9\,\pi^2\,v_0} \left[ \frac{2\,c}{1 - c^2\,q^2} \ - \frac{1}{c\,q^2} \right],
\end{align}
which is Eq.~\eqref{eq:qf-large}, since $2\,c/(1 - c^2\,q^2) - 1/(c\,q^2) = \left( 3\,c^2\,q^2 - 1 \right)/\left( c\,q^2 \left( 1 - c^2\,q^2 \right) \right)$. For $c = 0$, the channel $1 \to 0$ has $Y_{1, 0} = v_0\,k$, and $2\,g \int_{k < k_F} \frac{d^3\bs{k}}{(2\,\pi)^3}\,(1/3)/(v_0\,k) = g\,k_F^2/(6\,\pi^2\,v_0)$.

For $c > 0$ and arbitrary $q$, we evaluate the flat-band intraband part $\Pi_{00}$ as follows. The substitution $\bs{k} \to -\,\bs{k} - \bs{q}$ shows that the two terms of $f_0 (k) - f_0 (k')$ are equal, as in Appendix~\ref{app:qf-kohn0}. With $\varepsilon_0 (k) - \varepsilon_0 (k') = c\,v_0 \left( k^2 - k'^2 \right)$, $k'^2 = k^2 + 2\,k\,q\,\tilde x + q^2$, and $\mathcal{F}_{00} = \cos^2\theta = (k + q\,\tilde x)^2/k'^2$, we obtain
\begin{align}
\Pi_{00} (q, 0) = -\,\frac{g}{2\,\pi^2\,c\,v_0} \int_0^{k_F^{(0)}} dk\,k^2 \int_{-1}^{1} d\tilde x\;\frac{\left( k + q\,\tilde x \right)^2}{k'^2 \left( q^2 + 2\,k\,q\,\tilde x \right)}\,,
\end{align}
with the principal value at $\tilde x = -\,q/(2\,k)$ for $k > q/2$. We change the variable to $u = k'^2$, which gives $d\tilde x = du/(2\,k\,q)$, $k + q\,\tilde x = (u + a)/(2\,k)$, and $q^2 + 2\,k\,q\,\tilde x = u - k^2$, with $a = k^2 - q^2$ and $\left( k - q \right)^2 \leq u \leq \left( k + q \right)^2$. The decomposition
\begin{align}
\frac{\left( u + a \right)^2}{u \left( u - k^2 \right)} = 1 - \frac{a^2}{k^2\,u} + \frac{\left( 2\,k^2 - q^2 \right)^2}{k^2 \left( u - k^2 \right)}\,,
\end{align}
and the integrals $\int du = 4\,k\,q$, $\int du/u = 2 \ln \!\left| (k + q)/(k - q) \right|$, and $\int du/(u - k^2) = \ln \!\left| (2\,k + q)/(2\,k - q) \right|$ in the principal-value sense, give $\int d\tilde x\,(\dots) = \mathcal{B} (k, q)/(8\,k^3\,q)$ with $\mathcal{B}$ of Eq.~\eqref{eq:qf-pi00}. The remaining integral over $k$ gives Eq.~\eqref{eq:qf-pi00}.

We next extract the behaviour of Eq.~\eqref{eq:qf-pi00} for $q \ll k_F^{(0)}$. For $k \gg q$, the expansion of the logarithms in $q/k$ gives $\mathcal{B} = 4\,k\,q + 3\,q^3/k + \dots$. This expansion fails for $k \sim q$, where the integral over $k$ could acquire a term of order $q^2$, and hence $\Pi_{00}$ a term linear in $q$. We show that no such term arises. With $k = q\,\mathfrak{s}$, we write $\mathcal{B} (k, q) = q^2\,\hat{\mathcal{B}} (\mathfrak{s})$, which gives
\begin{align}
\int_0^{k_F^{(0)}} \frac{dk}{k}\,\mathcal{B} (k, q) = q^2 \int_0^{S} \frac{d\mathfrak{s}}{\mathfrak{s}}\,\hat{\mathcal{B}} (\mathfrak{s})\,, \quad
S = \frac{k_F^{(0)}}{q}\,, \quad
\hat{\mathcal{B}} (\mathfrak{s}) = 4\,\mathfrak{s} - \frac{2 \left( \mathfrak{s}^2 - 1 \right)^2}{\mathfrak{s}^2} \ln \!\left| \frac{\mathfrak{s} + 1}{\mathfrak{s} - 1} \right| + \frac{\left( 2\,\mathfrak{s}^2 - 1 \right)^2}{\mathfrak{s}^2} \ln \!\left| \frac{2\,\mathfrak{s} + 1}{2\,\mathfrak{s} - 1} \right|.
\end{align}
For $\mathfrak{s} \gg 1$, we have $\hat{\mathcal{B}} (\mathfrak{s}) = 4\,\mathfrak{s} + 3/\mathfrak{s} + \mathcal{O} (\mathfrak{s}^{-3})$, which contains odd powers of $\mathfrak{s}$ only. The function $\hat{\mathcal{B}} (\mathfrak{s})/\mathfrak{s} - 4$ therefore decays as $3/\mathfrak{s}^2$, and it is integrable at $\mathfrak{s} \to 0$ since $\hat{\mathcal{B}} (\mathfrak{s}) \propto \mathfrak{s}^2$. We obtain
\begin{align}
\int_0^{S} \frac{d\mathfrak{s}}{\mathfrak{s}}\,\hat{\mathcal{B}} (\mathfrak{s}) = 4\,S + C - \frac{3}{S} + \mathcal{O} \!\left( S^{-3} \right), \quad
C = \int_0^\infty d\mathfrak{s} \left[ \frac{\hat{\mathcal{B}} (\mathfrak{s})}{\mathfrak{s}} - 4 \right].
\end{align}
The absence of even powers of $\mathfrak{s}$ in $\hat{\mathcal{B}} (\mathfrak{s})$ excludes a term proportional to $\ln S$. The constant $C$ is the value at $z = 0$ of the analytic continuation of the Mellin transform $M (z) = \int_0^\infty d\mathfrak{s}\,\mathfrak{s}^{z - 1}\,\hat{\mathcal{B}} (\mathfrak{s})$, which converges for $-2 < z < -1$. The continuation across the pole at $z = -1$, which arises from the term $4\,\mathfrak{s}$, removes this term. We use
\begin{align}
\int_0^\infty d\mathfrak{s}\,\mathfrak{s}^{w - 1} \ln \!\left| \frac{1 + \mathfrak{s}}{1 - \mathfrak{s}} \right| = \frac{\pi}{w} \tan \!\left( \frac{\pi\,w}{2} \right), \quad |w| < 1\,,
\end{align}
and the replacement $\mathfrak{s} \to 2\,\mathfrak{s}$ for the second logarithm, which gives the factor $2^{-w}$. We write
\begin{align}
\hat{\mathcal{B}} (\mathfrak{s}) = 4\,\mathfrak{s} - 2 \left( \mathfrak{s}^2 - 2 + \mathfrak{s}^{-2} \right) \ln \!\left| \frac{\mathfrak{s} + 1}{\mathfrak{s} - 1} \right| + \left( 4\,\mathfrak{s}^2 - 4 + \mathfrak{s}^{-2} \right) \ln \!\left| \frac{2\,\mathfrak{s} + 1}{2\,\mathfrak{s} - 1} \right|
\end{align}
and use $\tan \!\left( \pi \left( z \pm 2 \right)/2 \right) = \tan \!\left( \pi\,z/2 \right)$, which gives
\begin{align}
M (z) = \pi \tan \!\left( \frac{\pi\,z}{2} \right) \left\{ -\,2 \left[ \frac{1}{z + 2} - \frac{2}{z} + \frac{1}{z - 2} \right] + 2^{-z} \left[ \frac{1}{z + 2} - \frac{4}{z} + \frac{4}{z - 2} \right] \right\}.
\end{align}
The poles at $z = 0$ of the two brackets cancel, since they are $4/z$ and $-\,4/z$, and the braces therefore have a finite limit for $z \to 0$. The prefactor vanishes linearly in this limit, and hence $C = M (0) = 0$. Inserting the result into Eq.~\eqref{eq:qf-pi00}, we obtain
\begin{align}
\Pi_{00} (q, 0) = -\,\frac{g}{4\,\pi^2\,v_0} \left[ \frac{k_F^{(0)}}{c} - \frac{3\,q^2}{4\,c\,k_F^{(0)}} + \mathcal{O} \!\left( q^4 \right) \right],
\end{align}
which is the expression of the flat-band quoted above, as obtained from the expansion of the delta function. The first term is $-\,\rho_0$ of Eq.~\eqref{eq:q-dos0}. The second term gives the part of $\mathcal{S}$ proportional to $1/\sqrt{\xi}$ in Eq.~\eqref{eq:qf-static}, since $c\,k_F^{(0)} = \sqrt{\xi}$.


\subsection{Friedel oscillations}
\label{app:qf-fried}

\subsubsection{Kohn singularity and Friedel oscillations of the top band}
\label{app:qf-kohn1}

At $\omega = 0$, the substitution $\bs{k} \to -\,\bs{k} - \bs{q}$ shows that the two terms of $f_0 (k) - f_0 (k')$ in $\Pi_{11}$ are equal, as in Appendix~\ref{app:qf-kohn0}. The energy difference is $\varepsilon_1 (k') - \varepsilon_1 (k) = v_0\,\check{D} \left( 1 + c\,\Sigma \right)$ with $\Sigma = k + k'$ and $\check{D} = k' - k$. Using $d^3\bs{k} = 2\,\pi\,k\,k'\,dk\,dk'/q$, $dk\,dk' = d\Sigma\,d\check{D}/2$, $k\,k' = (\Sigma^2 - \check{D}^2)/4$, and $\mathcal{F}_{11} = \left( \Sigma^2 - q^2 \right)^2/(16\,k^2\,k'^2)$, we obtain
\begin{align}
\Pi_{11} (q, 0) = -\,\frac{g}{16\,\pi^2\,v_0\,q} \int d\Sigma\,d\check{D}\;\frac{\left( \Sigma^2 - q^2 \right)^2}{\left( \Sigma^2 - \check{D}^2 \right) \check{D} \left( 1 + c\,\Sigma \right)}\,,
\end{align}
with the principal value at $\check{D} = 0$. The conditions $k \leq k_F$, $q \leq \Sigma$, and $|\check{D}| \leq q$ give $q \leq \Sigma \leq q + 2\,k_F$ and $\check{D}_- \leq \check{D} \leq q$, with $\check{D}_- = \max \lbrace -\,q, \Sigma - 2\,k_F \rbrace$. For $q > k_F$, we have $\check{D}_- = \Sigma - 2\,k_F$ on the whole range. The integral over $\check{D}$ is elementary, since $1/\left[ (\Sigma^2 - \check{D}^2)\,\check{D} \right] = (1/\Sigma^2) \left[ 1/\check{D} + \check{D}/(\Sigma^2 - \check{D}^2) \right]$, and
\begin{align}
\Pi_{11} (q, 0) = -\,\frac{g}{16\,\pi^2\,v_0\,q} \int_q^{q + 2 \, k_F} d\Sigma\;w (\Sigma, q) \left[ \ln \!\left( \frac{q}{\sqrt{\Sigma^2 - q^2}} \right) - \ln \!\left( \frac{|\Sigma - 2\,k_F|}{\sqrt{\Sigma^2 - (\Sigma - 2\,k_F)^2}} \right) \right],
\end{align}
with $w (\Sigma, q) = \left( \Sigma^2 - q^2 \right)^2/\left[ \Sigma^2 \left( 1 + c\,\Sigma \right) \right]$. The parameter $c$ enters only through $w$.

For $q = 2\,k_F + \delta$, the only logarithm that becomes singular at the integration limits is $\ln \left| \Sigma - 2\,k_F \right|$, since it is singular at $\Sigma = 2\,k_F$, which is the lower limit $\Sigma = q$ for $\delta = 0$. The other logarithms are singular at $\Sigma = q$ and at $\check{D} = \pm \Sigma$, which are fixed points of the integration range. With $\zeta = \Sigma - q$, we have $w = w_2\,\zeta^2 + w_3\,\zeta^3 + \dots$ and $\ln \left| \Sigma - 2\,k_F \right| = \ln \left| \zeta + \delta \right|$. The non-analytic part of $\int_0^L d\zeta\,\zeta^n \ln \left| \zeta + \delta \right|$ comes from the lower limit $\mathfrak{g} = \zeta + \delta = \delta$ of the substitution $\mathfrak{g} = \zeta + \delta$ and equals $-\,(-1)^n\,\delta^{n+1}\ln \left| \delta \right|/(n + 1)$. The leading term is $n = 2$, and it equals $-\,\delta^3 \ln \left| \delta \right|/3$. At $q = 2\,k_F$, we have $w_2 = 4/(1 + 2\,c\,k_F) = 4/\beta$. The bracket contains $-\,\ln \left| \Sigma - 2\,k_F \right|$, and the prefactor is $-\,g/(16\,\pi^2\,v_0\,q)$. The singular part is therefore the product of the prefactor, the factor $-1$, $w_2$, and $-\,\delta^3 \ln \left| \delta \right|/3$. This gives
\begin{align}
\Pi_{11}^{\rm sing} = -\,\frac{g}{16\,\pi^2\,v_0\,q}\,\frac{4}{3\,\beta}\,\delta^3 \ln \left| \delta \right| = -\,\frac{g}{24\,\pi^2\,\beta\,v_0\,k_F}\,\delta^3 \ln \left| \delta \right|
\end{align}
at $q = 2\,k_F$, which is Eq.~\eqref{eq:qf-kohn1}. The corrections are of the order $\delta^4 \ln \left| \delta \right|$. For $c = 0$, we recover Eq.~\eqref{eq:fin-kohn}. The interband channels have no pair of states at the Fermi energy connected by $q = 2\,k_F$, and they do not contribute to the singularity.

The Fourier transform follows that of Appendix~\ref{app:fin-friedel}. With $\delta n_{\rm sing} (q) = C\,\delta^3 \ln \left| \delta \right|$ and $C = -\,g\,\mathcal{U} (2\,k_F)/\left[ 24\,\pi^2\,\beta\,v_0\,k_F \left( \epsilon^{\rm RPA} (2\,k_F, 0) \right)^2 \right]$, we obtain $\delta n (r) \simeq 6\,k_F\,C\,\cos \!\left( 2\,k_F\,r \right)/(\pi\,r^5)$, which is Eq.~\eqref{eq:qf-friedel1}.

\subsubsection{Kohn singularity and Friedel oscillations of the flat-band for $c > 0$}
\label{app:qf-kohn0}

At $\omega = 0$, the intraband term of the flat-band is $\Pi_{00} = g \int d^3\bs{k}/(2\,\pi)^3\,\cos^2\theta\,\left[ f_0 (k) - f_0 (k') \right]/\left[ c\,v_0 \left( k^2 - k'^2 \right) \right]$. The substitution $\bs{k} \to -\,\bs{k} - \bs{q}$ exchanges $k$ and $k'$ and leaves $\cos^2\theta$ unchanged. Hence the two terms are equal, and
\begin{align}
\Pi_{00} (q, 0) = -\,\frac{2\,g}{c\,v_0} \int \frac{d^3\bs{k}}{(2\,\pi)^3}\,\Theta \left( k_F^{(0)} - k \right) \frac{\cos^2\theta}{q^2 + 2\,k\,q\,\tilde x}\,,
\end{align}
where $\tilde x$ is the cosine of the angle between $\bs{k}$ and $\bs{q}$, and the integral over $\tilde x$ is a principal value. We have $k'^2 = k^2 + q^2 + 2\,k\,q\,\tilde x$ and $\cos^2\theta = (k + q\,\tilde x)^2/k'^2$. With $u = q^2 + 2\,k\,q\,\tilde x$ and $A = 2\,k^2 - q^2$, the integrand in $u$ is $(u + A)^2/\left[ 4\,k^2\,u \left( u + k^2 \right) \right]$, which is decomposed as
\begin{align}
\frac{(u + A)^2}{u \left( u + k^2 \right)} = 1 + \frac{A^2}{k^2\,u} - \frac{\left( k^2 - q^2 \right)^2}{k^2 \left( u + k^2 \right)}\,.
\end{align}
The range is $q \left( q - 2\,k \right) \leq u \leq q \left( q + 2\,k \right)$, and $u + k^2 = (k \pm q)^2$ at the end points. Using $dx = du/(2\,k\,q)$, we obtain
\begin{align}
\Pi_{00} (q, 0) = -\,\frac{g}{16\,\pi^2\,c\,v_0\,q} \int_0^{k_F^{(0)}} \frac{dk}{k} \left[ 4\,k\,q + \frac{\left( 2\,k^2 - q^2 \right)^2}{k^2}\,\ln \!\left| \frac{q + 2\,k}{q - 2\,k} \right| - \frac{\left( k^2 - q^2 \right)^2}{k^2}\,\ln \!\left( \frac{\left( k + q \right)^2}{\left( k - q \right)^2} \right) \right].
\end{align}
The integrand is finite for $k \to 0$. For $q \to 0$, this gives $-\,g\,k_F^{(0)}/(4\,\pi^2\,c\,v_0)$, the DOS of the flat-band.

A non-analytic dependence on $q$ arises when the singularity of a logarithm reaches the end point $k = k_F^{(0)}$ of the integration. For $q = 2\,k_F^{(0)} + \delta$, this occurs only for $\ln \left| q - 2\,k \right|$. The logarithm $\ln \left| k - q \right|$ is regular, since $|k - q| = k_F^{(0)}$ at the end point. The coefficient of $\ln \left| q - 2\,k \right|$ in the integrand is $h (k) = g \left( 2\,k^2 - q^2 \right)^2/(16\,\pi^2\,c\,v_0\,q\,k^3)$, and $h \left( k_F^{(0)} \right) = g/(8\,\pi^2\,c\,v_0)$ at $q = 2\,k_F^{(0)}$. With $\zeta = k_F^{(0)} - k$ and $\mathfrak{g} = \delta + 2\,\zeta$, we have $\int_0^L d\zeta\,\ln \left| \delta + 2\,\zeta \right| = (1/2) \left[ \mathfrak{g} \ln \left| \mathfrak{g} \right| - \mathfrak{g} \right]_{\mathfrak{g} = \delta}^{\mathfrak{g} = \delta + 2 \bar\Lambda}$. The upper limit is analytic in $\delta$, and the lower limit gives the non-analytic part $-\,(1/2)\,\delta \ln \left| \delta \right|$. Multiplying by the value of $h$ at the end point gives $-\,(h/2)\,\delta \ln \left| \delta \right|$, and the variation of $h$ with $k$ contributes only terms of the order $\delta^2 \ln \left| \delta \right|$. This gives Eq.~\eqref{eq:qf-kohn0}. It coincides with the singularity of the Lindhard function of a parabolic band, $+\,g\,m\,k_F \left( 1 - \tilde x \right) \ln \left| 1 - \tilde x \right|/(4\,\pi^2)$ with $\tilde x = q/(2\,k_F)$ and $m = 1/(2\,c\,v_0)$.

The radial Fourier transform of an isotropic function is $\delta n (r) = (1/(2\,\pi^2\,r)) \int_0^{\infty} dq\,q\,\sin \!\left( q\,r \right) \delta n (q)$. Let $\delta n_{\rm sing} (q) = C\,\delta \ln \left| \delta \right|$ with $C = -\,g\,\mathcal{U} \left( 2\,k_F^{(0)} \right)/\left[ 16\,\pi^2\,c\,v_0 \left( \epsilon^{\rm RPA} \left( 2\,k_F^{(0)}, 0 \right) \right)^2 \right]$. From $\int d\delta\,\ln \left| \delta \right|\,e^{i\,\delta\,r} = -\,\pi/|r|$, we obtain $\int d\delta\,\delta \ln \left| \delta \right|\,e^{i\,\delta\,r} = -\,i\,\pi/r^2$ for $r > 0$. Setting $q \simeq 2\,k_F^{(0)}$ in the prefactor, we find $\delta n (r) \simeq -\,k_F^{(0)}\,C\,\cos \!\left( 2\,k_F^{(0)}\,r \right)/(\pi\,r^3)$, which is Eq.~\eqref{eq:qf-friedel0}. The corrections are of relative order $1/\left( k_F^{(0)}\,r \right)$.

\subsubsection{Kohn singularities from transitions between two Fermi surfaces for $c > 0$}
\label{app:qf-kohn01}

At $\omega = 0$, the two terms of the transitions $0 \to 1$ and $1 \to 0$ are exchanged by $\bs{k} \to -\,\bs{k} - \bs{q}$, as in Appendix~\ref{app:qf-kohn0}. Only two combinations of the occupation factors survive, and
\begin{align}
\Pi_{01} (q, 0) + \Pi_{10} (q, 0) = 2\,g \int \frac{d^3\bs{k}}{(2\,\pi)^3}\,\mathcal{F}_{01}\,\frac{\Theta (k_F^{(0)} - k)\,\Theta (k' - k_F) - \Theta (k - k_F^{(0)})\,\Theta (k_F - k')}{\varepsilon_0 (k) - \varepsilon_1 (k')}\,.
\end{align}
The denominator does not vanish on either domain, except at the point $k = k_F^{(0)}$, $k' = k_F$. With $\Sigma = k + k'$ and $\check{D} = k' - k$, we have $\varepsilon_0 (k) - \varepsilon_1 (k') = -\left( v_0/2 \right) \left( \Sigma + \check{D} + 2\,c\,\Sigma\,\check{D} \right)$. Using $d^3\bs{k} = 2\,\pi\,k\,k'\,dk\,dk'/q$, $dk\,dk' = d\Sigma\,d\check{D}/2$, and $\mathcal{F}_{01} = \left( q^2 - \check{D}^2 \right) \left( \Sigma^2 - q^2 \right)/\left[ 8\,k^2\,k'^2 \right]$, we obtain
\begin{align}
\Pi_{01} (q, 0) + \Pi_{10} (q, 0) = -\,\frac{g}{4\,\pi^2\,v_0\,q} \int d\Sigma\,d\check{D}\;\frac{\left( q^2 - \check{D}^2 \right) \left( \Sigma^2 - q^2 \right)}{\left( \Sigma^2 - \check{D}^2 \right) \left( \Sigma + \check{D} + 2\,c\,\Sigma\,\check{D} \right)} \left[ \Theta_{+} - \Theta_{-} \right],
\end{align}
where $\Theta_+$ and $\Theta_-$ are the two products of step functions. The region of integration is a polygon bounded by the straight lines $k = k_F^{(0)}$, $k' = k_F$, $|\check{D}| = q$, and $\Sigma = q$. A non-analytic dependence on $q$ arises when a vertex of this polygon reaches the singular point. We use the deviations $\mathfrak{d}_0 = k - k_F^{(0)}$ and $\mathfrak{d}_1 = k' - k_F$, and the velocities $u_0 = 2\,c\,k_F^{(0)}$ and $u_1 = 1 + 2\,c\,k_F$ of the two bands at their FSs, in units of $v_0$. Close to the singular point, $\varepsilon_0 (k) - \varepsilon_1 (k') \simeq v_0 \left( u_0\,\mathfrak{d}_0 - u_1\,\mathfrak{d}_1 \right)$. The step functions are $\Theta_+ = \Theta (-\mathfrak{d}_0)\,\Theta (\mathfrak{d}_1)$ and $\Theta_- = \Theta (\mathfrak{d}_0)\,\Theta (-\mathfrak{d}_1)$. We write $\delta = q - q_i$.

At $q_2$, the pair $(k, k')$ is antiparallel, with $\check{D} \to -\,q_1$ and $\Sigma \to q_2$. The vertex is the line $\Sigma = q$, which is the triangle inequality $k + k' \geq q$, that is $\mathfrak{d}_0 + \mathfrak{d}_1 \geq \delta$. Then $\Sigma - q = \mathfrak{d}_0 + \mathfrak{d}_1 - \delta$, $q^2 - \check{D}^2 \to q_2^2 - q_1^2 = 4\,k_F\,k_F^{(0)}$, and $\Sigma^2 - q^2 \to 2\,q_2 \left( \mathfrak{d}_0 + \mathfrak{d}_1 - \delta \right)$. This gives $\mathcal{F}_{01} \simeq q_2 \left( \mathfrak{d}_0 + \mathfrak{d}_1 - \delta \right)/\left( k_F\,k_F^{(0)} \right)$. Setting $k\,k' \to k_F\,k_F^{(0)}$ in the measure, we have
\begin{align}
\Pi_{01} + \Pi_{10} \simeq -\,\frac{g}{2\,\pi^2\,v_0}\,J (\delta), \quad
J (\delta) = \int_0^\infty d\mathfrak{m}_0 \int_0^\infty d\mathfrak{m}_1\;\frac{\left( \mathfrak{m}_1 - \mathfrak{m}_0 - \delta \right)_+ + \left( \mathfrak{m}_0 - \mathfrak{m}_1 - \delta \right)_+}{u_0\,\mathfrak{m}_0 + u_1\,\mathfrak{m}_1}\,,
\end{align}
where $(\tilde x)_+ = \tilde x\,\Theta (\tilde x)$. The first term comes from $\Theta_+$, with $\mathfrak{d}_0 = -\,\mathfrak{m}_0$ and $\mathfrak{d}_1 = \mathfrak{m}_1$, and the second from $\Theta_-$, with $\mathfrak{d}_0 = \mathfrak{m}_0$ and $\mathfrak{d}_1 = -\,\mathfrak{m}_1$. The integral is cut off at a large value $\Lambda$ of $\mathfrak{m}_0$ and $\mathfrak{m}_1$, and the dependence on $\Lambda$ is analytic in $\delta$. With $\tilde x = \mathfrak{m}_1 - \mathfrak{m}_0$, we have $\left( \tilde x - \delta \right)_+ + \left( -\,\tilde x - \delta \right)_+ = -\,\delta + \max \lbrace |\delta|, |\tilde x| \rbrace$. The second derivative with respect to $\delta$ of this expression is $\delta_{\rm D} \left( |\delta| - |\tilde x| \right)$ for both signs of $\delta$, where $\delta_{\rm D}$ is the Dirac delta function. It is supported on the lines $\mathfrak{m}_1 = \mathfrak{m}_0 \pm |\delta|$, and we obtain
\begin{align}
J'' (\delta) = \int_0^\Lambda \frac{d\mathfrak{m}_0}{\left( u_0 + u_1 \right) \mathfrak{m}_0 + u_1 |\delta|} + \int_0^\Lambda \frac{d\mathfrak{m}_1}{\left( u_0 + u_1 \right) \mathfrak{m}_1 + u_0 |\delta|} = -\,\frac{2}{u_0 + u_1} \ln \left| \delta \right| + \text{const}\,.
\end{align}
The corrections are of higher order in $\delta$. Two integrations give the non-analytic part $J^{\rm sing} (\delta) = -\,\delta^2 \ln \left| \delta \right|/\left( u_0 + u_1 \right)$. Since $u_0 + u_1 = 1 + 2\,c \left( k_F^{(0)} + k_F \right) = 1 + 2\,c\,q_2$, this is Eq.~\eqref{eq:qf-kohn01}. We also have $c\,k_F^{(0)} = \sqrt{\xi}$ with $\xi = c\,k_F \left( 1 + c\,k_F \right)$, and therefore $1 + 2\,c\,q_2 = 1 + 2\,c\,k_F + 2\,\sqrt{c\,k_F \left( 1 + c\,k_F \right)} = \left( \sqrt{c\,k_F} + \sqrt{1 + c\,k_F} \right)^2$.

At $q_1$, the pair is parallel, with $\check{D} \to -\,q_1$ and $\Sigma \to q_2$. The vertex is the triangle inequality $k - k' \leq q$, that is $\mathfrak{d}_0 - \mathfrak{d}_1 \leq \delta$. Then $q + \check{D} = \delta - \mathfrak{d}_0 + \mathfrak{d}_1$, $q - \check{D} \to 2\,q_1$, and $\Sigma^2 - q^2 \to 4\,k_F\,k_F^{(0)}$. This gives $\mathcal{F}_{01} \simeq q_1 \left( \delta - \mathfrak{d}_0 + \mathfrak{d}_1 \right)/\left( k_F\,k_F^{(0)} \right)$, and the same steps lead to
\begin{align}
\Pi_{01} + \Pi_{10} \simeq -\,\frac{g}{2\,\pi^2\,v_0}\,J_1 (\delta), \quad
J_1 (\delta) = \int_0^\infty d\mathfrak{m}_0 \int_0^\infty d\mathfrak{m}_1\;\frac{\left( \delta + \mathfrak{m}_0 + \mathfrak{m}_1 \right)_+ + \left( \delta - \mathfrak{m}_0 - \mathfrak{m}_1 \right)_+}{u_0\,\mathfrak{m}_0 + u_1\,\mathfrak{m}_1}\,.
\end{align}
With $\mathfrak{m} = \mathfrak{m}_0 + \mathfrak{m}_1$, we have $\left( \delta + \mathfrak{m} \right)_+ + \left( \delta - \mathfrak{m} \right)_+ = \delta + \max \lbrace |\delta|, \mathfrak{m} \rbrace$. Its second derivative is $\delta_{\rm D} \left( |\delta| - \mathfrak{m} \right)$, and
\begin{align}
J_1'' (\delta) = \int_0^{|\delta|} \frac{d\mathfrak{m}_0}{u_0\,\mathfrak{m}_0 + u_1 \left( |\delta| - \mathfrak{m}_0 \right)} = \frac{\ln \!\left( u_0/u_1 \right)}{u_0 - u_1}\,.
\end{align}
This is independent of $\delta$, and $J_1$ is analytic at this order. The two terms of $J_1$ have second derivatives with a jump at $\delta = 0$, and the jumps cancel in the sum.

The Fourier transform follows that of Appendix~\ref{app:fin-friedel}. With $\delta n_{\rm sing} (q) = C\,\delta^2 \ln \left| \delta \right|$ and $C = g\,\mathcal{U} (q_2)/\left[ 2\,\pi^2\,v_0 \left( 1 + 2\,c\,q_2 \right) \left( \epsilon^{\rm RPA} (q_2, 0) \right)^2 \right]$, we use $\int d\delta\,\delta^2 \ln \left| \delta \right|\,e^{i\,\delta\,r} = 2\,\pi/r^3$ for $r > 0$. This follows from $\int d\delta\,\ln \left| \delta \right|\,e^{i\,\delta\,r} = -\,\pi/|r|$ by two derivatives with respect to $r$. Setting $q \simeq q_2$ in the prefactor, we find $\delta n (r) \simeq q_2\,C\,\sin \!\left( q_2\,r \right)/(\pi\,r^4)$, which is Eq.~\eqref{eq:qf-friedel01}.

\subsection{Plasmon dispersion and Landau damping}
\label{app:qf-wdisp}

Let $\varpi$ lie in the window of Sec.~\ref{secqfplasmon}. We write $\Pi = \Pi_{\rm vac} + \Delta\Pi$, with $\Pi_{\rm vac}$ the polarisability for $\mu \to 0^+$ at fixed $c$. The filled bands of this system and the extra occupation of the doped system depend on the sign of $c$.
\begin{itemize}

\item For $c < 0$, the filled bands are $s = -1$ and $s = 0$. The extra occupation is $\delta_s (k) = \Theta (k_s - k)$ for the band $s = 1$ with the Fermi momentum $k_1 = k_F$.

\item For $c > 0$, the filled band is $s = -1$. The extra occupation is $\delta_s (k) = \Theta (k_s - k)$ for the bands $s = 1$ and $s = 0$, with the Fermi momentum $k_s$ taking the values $k_F$ and $k_F^{(0)}$, respectively.
\end{itemize}
Writing $n_s = n_s^{\rm vac} + \delta_s$ in Eq.~\eqref{eq:pol} and relabelling $\bs{k} \to -\bs{k} - \bs{q}$ in the terms with $\delta_{s'} (k')$, we obtain
\begin{align}
\label{eq:qf-ball}
\Delta\Pi = -\,2\,g \sum_{s} \int_{k < k_s} \frac{d^3\bs{k}}{(2\,\pi)^3} \sum_{s'} \mathcal{F}_{s s'}\,\frac{Y}{\omega^2 - Y^2}\,, \quad
Y = \varepsilon_s (k) - \varepsilon_{s'} (k')\,,
\end{align}
and, by the pairing of Appendix~\ref{app:qf-re}, $\Pi_{\rm vac} = g \int \frac{d^3\bs{k}}{(2\,\pi)^3} \sum_{(s, s')} \mathcal{F}_{ss'}\,2\,X/(\omega^2 - X^2)$, with $X = \varepsilon_{s'} (k') - \varepsilon_s (k)$. The sum runs over the following pairs.
\begin{itemize}
\item For $c < 0$, the pairs $(-1, 1)$ and $(0, 1)$.
\item For $c > 0$, the pairs $(-1, 1)$ and $(-1, 0)$.
\end{itemize}
The upper limit of the integral is set by $X < \mathcal{R}$.

We write $k' = \sqrt{k^2 + 2\,k\,q\,\tilde x + q^2}$ and expand the integrands in $q$ at fixed $k$ and $\tilde x = \cos\theta$. The odd orders vanish after the integration over $\tilde x$. With $d^3\bs{k}/(2\,\pi)^3 = k^2\,dk\,d\tilde x/(4\,\pi^2)$, we obtain $\Pi_n = (g/(4\,\pi^2\,v_0))\,\mathcal{I}_n$ with $\mathcal{I}_n = \int dk\,\mathcal{J}_n (k)$. The integrand $\mathcal{J}_n$ is the sum of the coefficients of $q^n$ of all terms, multiplied by $k^2$. Its form depends on the region. The occupied bands change at $k = k_F$ and, for $c > 0$, at $k = k_F^{(0)}$. For $n = 2$, we find
\begin{align}
\mathcal{J}_2 = \begin{cases}
4\,k \left( 1 + 3\,c\,k \right)/(3\,\varpi^2) & \text{ for } k < k_F \text{ and } c < 0 \\
4\,k/\left( 3 \left( \varpi^2 - k^2 \right) \right) & \text{ for } k > k_F \text{ and } c < 0 \\
4\,k \left( 1 + 6\,c\,k \right)/(3\,\varpi^2) & \text{ for } k < k_F \text{ and } c > 0 \\
4\,c\,k^2/\varpi^2 + 4\,k/\left( 3 \left( \varpi^2 - k^2 \right) \right) & \text{ for } k_F < k < k_F^{(0)} \text{ and } c > 0 \\
4\,k/\left( 3 \left( \varpi^2 - k^2 \right) \right) & \text{ for } k > k_F^{(0)} \text{ and } c > 0
\end{cases}.
\end{align}
For $n = 4$, the partial-fraction decomposition in $k$ gives, with $\bar\varpi_\pm = \varpi \pm k$ and for $c < 0$,
\begin{align}
\mathcal{J}_4 = \begin{cases} 
\frac{16\,c^3\,k^4}{\varpi^4} + \frac{96\,c^2\,k^3}{5\,\varpi^4} 
+ \frac{36\,c\,k^2}{5\,\varpi^4} + \frac{4\,k}{5\,\varpi^4} - \frac{4\,c}{15\,\varpi^2} - \frac{4\,c}{15} \left( \frac{1}{\bar\varpi_+^2} 
+ \frac{1}{\bar\varpi_-^2} \right) + \frac{2}{15} \left( \frac{4\,c\,\varpi - 1}{\bar\varpi_+^3} + \frac{4\,c\,\varpi + 1}{\bar\varpi_-^3} \right) & \text{ for } k < k_F \\
\frac{2\,c \left( 8\,c\,\varpi - 3 \right)}{15\,\bar\varpi_+^2} - \frac{2\,c \left( 8\,c\,\varpi + 3 \right)}{15\,\bar\varpi_-^2} - \frac{2 \left( 2\,c\,\varpi - 1 \right)^2}{15\,\bar\varpi_+^3} + \frac{2 \left( 2\,c\,\varpi + 1 \right)^2}{15\,\bar\varpi_-^3} & \\
- \frac{2 \left( 4\,c^2\,\varpi^2 - 1 \right)}{15\,\varpi^2} 
\left( \frac{1}{\bar\varpi_+} - \frac{1}{\bar\varpi_-} \right) - \frac{4}{15\,\varpi^2} 
\left( \frac{1}{\varpi + 2\,k} - \frac{1}{\varpi - 2\,k} \right)
& \text{ for } k > k_F
\end{cases}.
\end{align}
For $c > 0$, we find
\begin{align}
\mathcal{J}_4 = \begin{cases}
\frac{32\,c^3\,k^4}{\varpi^4} + \frac{96\,c^2\,k^3}{5\,\varpi^4} + \frac{36\,c\,k^2}{5\,\varpi^4} + \frac{4\,k}{5\,\varpi^4} - \frac{4\,c}{5\,\varpi^2} - \frac{8\,c}{15} \left( \frac{1}{\bar\varpi_+^2} + \frac{1}{\bar\varpi_-^2} \right) + \frac{16\,c\,\varpi}{15} \left( \frac{1}{\bar\varpi_+^3} + \frac{1}{\bar\varpi_-^3} \right) & \text{ for } k < k_F \\
\frac{16\,c^3\,k^4}{\varpi^4} - \frac{8\,c}{15\,\varpi^2} - \frac{8\,c\,\varpi \left( c\,\varpi - 2 \right)}{15\,\bar\varpi_+^3} + \frac{8\,c\,\varpi \left( c\,\varpi + 2 \right)}
{15\,\bar\varpi_-^3} + \frac{2\,c \left( 8\,c\,\varpi - 5 \right)}
{15\,\bar\varpi_+^2} - \frac{2\,c \left( 8\,c\,\varpi + 5 \right)}{15\,\bar\varpi_-^2} & \\
- \frac{2 \left( 4\,c^2\,\varpi^2 - 1 \right)}{15\,\varpi^2} \left( \frac{1}{\bar\varpi_+} - \frac{1}{\bar\varpi_-} \right) - \frac{4}{15\,\varpi^2} \left( \frac{1}{\varpi + 2\,k} - \frac{1}{\varpi - 2\,k} \right) & \text{ for } k_F < k < k_F^{(0)} \\
 -\,\frac{8\,c^2\,\varpi^2}{15} \left( \frac{1}{\bar\varpi_+^3} - \frac{1}{\bar\varpi_-^3} \right) + \frac{2\,c \left( 8\,c\,\varpi + 1 \right)}{15\,\bar\varpi_+^2} - \frac{2\,c \left( 8\,c\,\varpi - 1 \right)}{15\,\bar\varpi_-^2} & \\
- \frac{2 \left( 4\,c^2\,\varpi^2 - 1 \right)}{15\,\varpi^2} \left( \frac{1}{\bar\varpi_+} - \frac{1}{\bar\varpi_-} \right) - \frac{4}{15\,\varpi^2} \left( \frac{1}{\varpi + 2\,k} - \frac{1}{\varpi - 2\,k} \right) & \text{ for } k > k_F^{(0)}
\end{cases}.
\end{align}
The poles at $k = \varpi$ and $k = \varpi/2$ arise from the expansion of principal-value integrals, and the integrals over $k$ are finite-part integrals. For $\varpi < k_F$, the pole at $k = \varpi$ lies in the first region, and the antiderivative is continuous across all boundaries. The poles at $k = \varpi/2$ lie below $k_F$, where the terms in $1/(\varpi \pm 2\,k)$ are absent. For $c \to 0$, the integrands reduce to those of Appendix~\ref{app:fin-wdisp}.

The integration over $k$ from $0$ to $\Lambda$ gives the finite-part values of the antiderivatives at the boundaries. The terms with powers of $\Lambda$ cancel between the regions, and only $\ln \Lambda$ remains. Up to terms that vanish for $\Lambda \to \infty$, this gives $\mathcal{I}_2 = (2/3)\,[\,\dots\,]$ with the bracket of Eq.~\eqref{eq:qf-pi24}, which proves Eq.~\eqref{eq:qf-pi24}, and
\begin{align}
\label{eq:qf-i4}
\mathcal{I}_4 = \frac{2}{15\,\varpi^2} \ln \!\left( \frac{4\,k_F^2 - \varpi^2}{4 \left( k_F^2 - \varpi^2 \right)} \right) - \frac{8\,c^2}{15} \ln \!\left( \frac{\Lambda^2}{k_F^2 - \varpi^2} \right) + \mathfrak{r}\,.
\end{align}
The rational part is $\mathfrak{r} = \mathfrak{r}_1 + \Theta (c)\,\mathfrak{r}_0$, with
\begin{align}
\mathfrak{r}_1 &= \frac{2\,\beta^3\,k_F^2}{5\,\varpi^4} - \frac{3 + 4\,c\,k_F}{15\,\varpi^2} - \frac{4\,c^2\,k_F^2}{15} \left[ \frac{1}{\left( k_F + \varpi \right)^2} + \frac{1}{\left( k_F - \varpi \right)^2} \right] - \frac{2\,c \left( 1 + 4\,c\,k_F \right)}{15} \left( \frac{1}{k_F + \varpi} + \frac{1}{k_F - \varpi} \right), \nn
\mathfrak{r}_0 &= -\,\frac{8\,\sqrt{\xi}}{15\,\varpi^2} + \frac{16\,\xi^{3/2} \left( k_F^{(0)} \right)^2}{5\,\varpi^4} + \frac{16\,\sqrt{\xi} \left[ \left( k_F^{(0)} \right)^2 + \varpi^2 \right]}{15 \left[ \left( k_F^{(0)} \right)^2 - \varpi^2 \right]^2} + \frac{8\,\sqrt{\xi}}{15 \left[ \left( k_F^{(0)} \right)^2 - \varpi^2 \right]}\,.
\end{align}
The finite-part integrals alone give $\mathfrak{r}^{(k)} = \mathfrak{r} + 8\,c^2/5 + \left( 1 + 2\,\Theta (c) \right)/(15\,\varpi^2)$. The two additional terms have different origins, and we derive them in turn. Both enter Eq.~\eqref{eq:qf-i4}, from which Eq.~\eqref{eq:qf-pi24} follows through $\mathcal{B} = \mathcal{B}_1 + \Theta (c)\,\mathcal{B}_0 = \varpi^2\,\mathcal{I}_4/4$.

The first term comes from the cut-off. The condition $\varepsilon_{\rm tr} < \mathcal{R}$ on the transition energy limits the integration over $k$ in the pairs $(0, 1)$ for $c < 0$ and $(-1, 0)$ for $c > 0$ to $k < k_{\rm c} (\tilde x, q)$, where $\varepsilon_{\rm tr} (k_{\rm c}, \tilde x, q) = \mathcal{R}$. To first order in $q$, we have the following.
\begin{itemize}
\item For $c < 0$, $k_{\rm c} = \Lambda - q\,\tilde x \left( 1 + 2\,c\,\Lambda \right)$ for the pair $(0, 1)$.
\item For $c > 0$, $k_{\rm c} = \Lambda - 2\,c\,\Lambda\,q\,\tilde x$ for the pair $(-1, 0)$.
\end{itemize}
The pair $(-1, 1)$ converges for $\Lambda \to \infty$ and is not affected. For large $k$, the integrand multiplied by $k^2$ is $-\,2\,k^2\,\mathcal{F}/\varepsilon_{\rm tr}$ with $\mathcal{F} = \sin^2\theta/2$. We integrate it from $\Lambda$ to $k_{\rm c} (\tilde x, q)$, expand in $q$ at fixed $\tilde x$, integrate over $\tilde x$, and take $\Lambda \to \infty$. The order $q^2$ gives zero, and the order $q^4$ gives $-\,8\,c^2/5$ for both pairs. This is the first additional term.

The second term comes from the region $k \lesssim q$, where the expansion at fixed $k$ fails. The coefficients of $q^4$ of the individual pairs diverge as $1/k$ for $k \to 0$, and the divergences cancel in the sum, since the sum of the coefficients of $1/k$ is $\left( 4/(15\,\varpi^2) \right) \left( 1 - 1 \right) = 0$ for both signs of $c$. The exact integral over the region $k < \mathfrak{n}\,q$ then differs from its expansion by a constant times $q^4$, which does not depend on $\mathfrak{n}$ for $\mathfrak{n} \to \infty$. In this region, $c$ enters the energies only in the combination $c\,q$, and the constant is therefore independent of $c$ for $q \to 0$ and of the sign of $c$ within each pair. It depends only on the pair. \begin{itemize}
\item For $c < 0$, we fix it by comparison with Eq.~\eqref{eq:fin-pi24}. For $c = 0$, Eq.~\eqref{eq:qf-i4} with $\mathfrak{r}^{(k)}$ gives $2 \left( 3\,k_F^2 - \varpi^2 \right)/(15\,\varpi^4)$ for the rational part, whereas Eq.~\eqref{eq:fin-pi24} requires $2\,k_F^2/(5\,\varpi^4) - 1/(5\,\varpi^2)$. The difference is $-\,1/(15\,\varpi^2)$.
\item For $c > 0$, the pair $(-1, 0)$ replaces the pair $(0, 1)$. The two pairs are related by $k \leftrightarrow k'$, which leaves $\mathcal{F}_{01} = \mathcal{F}_{-1,0} = \sin^2\theta/2$ and the measure unchanged. Their exact integrals therefore coincide for $c = 0$ up to a contribution from the cut-off boundary that vanishes for $\Lambda \to \infty$. The constants then differ by minus the difference of their finite-part integrals. At $c = 0$, the difference of the integrands of $q^4$ is $-\,4\,k \left( k^2 + 3\,\varpi^2 \right)/\left( 15 \left( \varpi^2 - k^2 \right)^3 \right)$, with the finite-part integral $2/(15\,\varpi^2)$ from $0$ to $\infty$. The constant for the pair $(-1, 0)$ is therefore $-\,1/(15\,\varpi^2) - 2/(15\,\varpi^2) = -\,3/(15\,\varpi^2)$. This is the second additional term.
\end{itemize}

For the zero of $\Re\epsilon^{\rm RPA} = 1 - 4\,\pi\,e^2 \left( \Pi_2 + q^2\,\Pi_4 \right)$, we use $1 - 4\,\pi\,e^2\,\Pi_2 = \epsilon_b (\omega) - \omega_p^2/\omega^2$ and $4\,\pi\,e^2\,q^2\,\Pi_4 = 4\,\alpha\,v_0^2\,q^2\,\mathcal{B}/(\pi\,\omega^2)$. Setting the sum to zero and multiplying by $\omega^2$ gives the equation quoted in Sec.~\ref{secqfplasmon}. Define $F (\omega^2, q^2) = \omega^2\,\epsilon_b (\omega) - \omega_p^2 - \left( 4\,\alpha\,v_0^2\,q^2/\pi \right) \mathcal{B}$. It is analytic in $(\omega^2, q^2)$ for $\omega < \mu_1$, since the poles of $\mathcal{B}$ lie at $\varpi = k_F$, $2\,k_F$, $k_F^{(0)}$ and the logarithms are regular for $\varpi < k_F$. At $q = 0$, we have $\partial_{\omega^2} F = \Phi (\omega_* (0)) > 0$, because $\epsilon_b > 1$ for $\Lambda > k_F$. By the implicit-function theorem, the zero continues to an analytic function $\omega_*^2 (q^2)$ for small $q$, which is unique near $\omega_* (0)$. Its first correction is Eq.~\eqref{eq:qf-wdisp}. The limits of the free region at small $q$ are $\varpi_+ \to \beta\,q$ and the lowest onset $\to k_F$. Hence $\omega_* (q)$ lies inside for small $q$, and the plasmon is undamped.

For $\omega_p \ll \mu_1$, we have $\omega_*^2 (0) = \omega_p^2/\epsilon_b$ with $\epsilon_b - 1 = \mathcal{O} (\alpha)$ and $\omega_p^2 = \mathcal{O} (\alpha)$. The terms of $\mathcal{B}$ proportional to $\varpi^{-2}$ are $\left( \beta^3\,k_F^2 + 8\,\xi^{3/2}\,\Theta (c) \left( k_F^{(0)} \right)^2 \right)/(10\,\varpi^2)$. Dividing by $\varpi^2 = \omega_p^2/v_0^2 = (2\,\alpha/(3\,\pi)) \left( \beta\,k_F^2 + 2\,\sqrt{\xi}\,\Theta (c) \left( k_F^{(0)} \right)^2 \right)$ gives Eq.~\eqref{eq:qf-bg}. The remaining terms of $\mathcal{B}$ are of order unity and give the correction of relative order $\alpha$. The weights $\beta\,k_F^2$ and $2\,\sqrt{\xi} \left( k_F^{(0)} \right)^2$ are proportional to the Drude weights of Eqs.~\eqref{eq:q-dos} and~\eqref{eq:q-dos0}, and the ratio is the average of $V^2$ weighted with them, with $V_F^2 = \beta^2\,v_0^2$ and $V_0^2 = 4\,\xi\,v_0^2$. It therefore lies between $V_0^2$ and $V_F^2$.

For the damping, we note that $\Im\Pi < 0$ at every point of the continuum, since the weights $\mathcal{F}_{ss'}$ are positive almost everywhere on the surface of constant transition energy. Beyond $q_{\rm max}$, the zero of $\Re\epsilon^{\rm RPA}$ lies in the continuum and has $\Im\epsilon^{\rm RPA} \neq 0$. The rate follows by expanding $\epsilon^{\rm RPA}$ around $\Re\,\omega_*$ to first order in $\Im\,\omega_*$. For $c > 0$ and $q_1 \leq q \leq q_2$, the continuum of Sec.~\ref{secqfedges} extends to $\varpi = 0$ and no free region exists, in agreement with $q_{\rm max} \leq q_1$.


\section{HSM at arbitrary momentum-transfers}
\label{app:hfull}

This appendix gives the intermediate steps behind the results of Sec.~\ref{sechfull}. We write $\varpi = \omega/v_0$ and $k' = |\bs{k} + \bs{q}|$, and we use $\lambda$ for $\lambda_{\bs{q}}$ whenever the direction of $\bs{q}$ is fixed. With $\mathfrak{e}_1 = (k_+, i\,k_z)^T$ and $\mathfrak{e}_2 = (i\,k_z, k_-)^T$, the states of Eq.~\eqref{eq:h-states} read $|\pm, \bs{k} \rangle = (1, \pm\,\mathfrak{e}_1/k)^T/\sqrt{2}$ and $|0, \bs{k} \rangle = (0, \mathfrak{e}_2/k)^T$. The Hamiltonian of Eq.~\eqref{eq:hhopf} has the block form $\mathcal{H}_{\rm H} = v_0 \begin{pmatrix} 0 & \mathfrak{e}_1^\dagger \\ \mathfrak{e}_1 & 0 \end{pmatrix}$. Acting on $(1, \pm\,\mathfrak{e}_1/k)^T$ gives $\pm\,v_0\,(\mathfrak{e}_1^\dagger \mathfrak{e}_1/k, \mathfrak{e}_1)^T$, and $\mathfrak{e}_1^\dagger \mathfrak{e}_1 = k_x^2 + k_y^2 + k_z^2 = k^2$ shows that these are eigenstates with the energies $\pm\,v_0\,k$. The vector $(0, \mathfrak{e}_2)^T$ is annihilated because $\mathfrak{e}_1^\dagger \mathfrak{e}_2 = k_-\,i\,k_z - i\,k_z\,k_- = 0$. The inner products are
\begin{align}
\langle s', \bs{k}' | s, \bs{k} \rangle &= \frac{1}{2} \left( 1 + s\,s'\,\frac{\mathfrak{e}_1'^\dagger \mathfrak{e}_1}{k\,k'} \right) = \frac{1 + s\,s'\,\mathcal{Z}}{2}\,, \quad s, s' = \pm 1, \nn
\langle 0, \bs{k}' | s, \bs{k} \rangle &= \frac{s}{\sqrt{2}\,k\,k'}\,\mathfrak{e}_2'^\dagger \mathfrak{e}_1\,, \quad
\langle 0, \bs{k}' | 0, \bs{k} \rangle = \frac{\mathfrak{e}_2'^\dagger \mathfrak{e}_2}{k\,k'}\,,
\end{align}
with $\mathfrak{e}_1'^\dagger \mathfrak{e}_1 = k'_-\,k_+ + k_z\,k'_z = k\,k'\,\mathcal{Z}$ and $\mathfrak{e}_2'^\dagger \mathfrak{e}_2 = k_z\,k'_z + k'_+\,k_- = k\,k'\,\mathcal{Z}^*$. Since $\mathfrak{e}_1'^\dagger \mathfrak{e}_2' = 0$, the vectors $\mathfrak{e}_1'/k'$ and $\mathfrak{e}_2'/k'$ form an orthonormal basis of $\mathbb{C}^2$, and $|\mathfrak{e}_1'^\dagger \mathfrak{e}_1|^2 + |\mathfrak{e}_2'^\dagger \mathfrak{e}_1|^2 = k^2\,k'^2$. Hence $|\mathfrak{e}_2'^\dagger \mathfrak{e}_1|^2/(k\,k')^2 = 1 - |\mathcal{Z}|^2$, and the squared moduli give Eq.~\eqref{eq:h-ff}. The real and imaginary parts of $\mathcal{Z}$ follow from $k'_-\,k_+ = \bs{k}_\perp \cdot \bs{k}'_\perp + i\,(k'_x\,k_y - k'_y\,k_x)$ and $\bs{k} \times \bs{k}' = \bs{k} \times \bs{q}$, which give Eq.~\eqref{eq:h-z}. The sum rule is $\mathcal{F}_{++} + \mathcal{F}_{+-} + \mathcal{F}_{+0} = (1 + |\mathcal{Z}|^2)/2 + (1 - |\mathcal{Z}|^2)/2 = 1$.

For the average of $\Upsilon^2$, we decompose $\bs{k} = k_\parallel\,\bs{\hat q} + \bs{k}_\perp$. At fixed $k$ and $k'$, both $k_\parallel$ and $|\bs{k}_\perp|$ are fixed, and the azimuthal angle $\phi$ of $\bs{k}_\perp$ about $\bs{\hat q}$ is uniformly distributed. The vector $\bs{k} \times \bs{q} = q\,\bs{k}_\perp \times \bs{\hat q}$ has the fixed modulus $q\,|\bs{k}_\perp|$ and lies in the plane perpendicular to $\bs{\hat q}$, where it rotates uniformly with $\phi$. Its $z$-component squared averages to $\tfrac{1}{2}\,|\bs{k} \times \bs{q}|^2\,[1 - (\bs{\hat q} \cdot \bs{\hat z})^2]$. With $|\bs{k} \times \bs{q}|^2/(k\,k')^2 = \sin^2\theta$ and $\bs{\hat q} \cdot \bs{\hat z} = \cos\vartheta_{\bs{q}}$, this is Eq.~\eqref{eq:h-avg}. Inserting it in Eq.~\eqref{eq:h-ff} and using $\sin^2\theta = 1 - \cos^2\theta$ gives Eq.~\eqref{eq:h-fl}. The sum rule is preserved, since the $\lambda_{\bs{q}}$-dependent terms of $\mathcal{F}_{++} + \mathcal{F}_{+-} + \mathcal{F}_{+0}$ are $\lambda_{\bs{q}}\,(1 - \cos^2\theta)\,(1/4 + 1/4 - 1/2) = 0$.

At $T = 0$ and $\mu > 0$, the occupations are $f_0 (\varepsilon_-) = f_0 (\varepsilon_0) = 1$ and $f_0 (\varepsilon_+) = \Theta (k_F - k)$. We write $\Pi = \Pi_{\rm vac} + \Pi_{\rm dop}$, where $\Pi_{\rm vac}$ is the response with $f_0 (\varepsilon_+) = 0$. In Eq.~\eqref{eq:pol}, the pairs $(0, +)$ and $(-, +)$ and their reverses contribute to $\Pi_{\rm vac}$, and the pair $(-, 0)$ is blocked. We find
\begin{align}
\Pi_{\rm vac} = g \sum_{s \in \lbrace 0, - \rbrace} \int \frac{d^3 k}{(2\,\pi)^3}\; \mathcal{F}_{s+} \left[ \frac{1}{\omega + \varepsilon_s (\bs{k}) - v_0\,k'} - \frac{1}{\omega + v_0\,k - \varepsilon_s (\bs{k}')} \right].
\end{align}
In the second term, we substitute $\bs{k} \to -\,\bs{k} - \bs{q}$, which exchanges $k$ and $k'$ and leaves $\mathcal{F}_{s+}$ invariant, since it depends on $\cos\theta$ and on $\Upsilon^2$ only, and $\Upsilon^2$ is invariant under the substitution. The bracket becomes $1/(\omega - X) - 1/(\omega + X) = 2\,X/(\omega^2 - X^2)$ with $X = v_0\,k' - \varepsilon_s (\bs{k})$. For $s = 0$, we exchange $k$ and $k'$ once more. This gives
\begin{align}
\Pi_{\rm vac} = -\,2\,g \int \frac{d^3 k}{(2\,\pi)^3} \sum_{s' \in \lbrace 0, - \rbrace} \frac{\mathcal{F}_{+ s'}\,X_{s'}\,\Theta (v_0\,\Lambda - X_{s'})}{X_{s'}^2 - \omega^2}\,.
\end{align}
The step function implements the cut-off on the transition energy. It is inactive for $k < k_F$, since $q$ and $k_F$ are much smaller than $\Lambda$.
For the doping part, the terms that contain $\Theta (k_F - k)$ or $\Theta (k_F - k')$ are
\begin{align}
\Pi_{\rm dop} = g \int \frac{d^3 k}{(2\,\pi)^3} \sum_{s'} \left[ \frac{\mathcal{F}_{+ s'}\,\Theta (k_F - k)}{\omega + v_0\,k - \varepsilon_{s'} (\bs{k}')} - \frac{\mathcal{F}_{s' +}\,\Theta (k_F - k')}{\omega + \varepsilon_{s'} (\bs{k}) - v_0\,k'} \right],
\end{align}
where the sum runs over $s' \in \lbrace +, 0, - \rbrace$. The same substitution in the second term turns it into $\mathcal{F}_{+ s'}\,\Theta (k_F - k)/(X_{s'} - \omega)$. The two terms add up to $2\,X_{s'}/(X_{s'}^2 - \omega^2)$, with $X_{s'} = v_0\,k - \varepsilon_{s'} (\bs{k}')$. This gives $X_+ = v_0\,(k - k')$, $X_0 = v_0\,k$, and $X_- = v_0\,(k + k')$. In the sum $\Pi_{\rm vac} + \Pi_{\rm dop}$, the contributions of $s' \in \lbrace 0, - \rbrace$ inside the ball $k < k_F$ cancel, and Eq.~\eqref{eq:h-split} remains. The prescription $\omega \to \omega + i\,0^+$ follows from Eq.~\eqref{eq:pol}.

For the explicit evaluation at fixed $\bs{q}$, the integration over $\bs{k}$ is replaced by one over $k$ and $k'$. Since $k'^2 = k^2 + q^2 + 2\,k\,q\,\cos\vartheta_{\bs{k} \bs{q}}$, where $\vartheta_{\bs{k} \bs{q}}$ is the angle between $\bs{k}$ and $\bs{q}$, we have $k'\,dk' = k\,q\,d\cos\vartheta_{\bs{k} \bs{q}}$. Hence
\begin{align}
\int \frac{d^3 k}{(2\,\pi)^3}\,h (k, k') = \frac{1}{4\,\pi^2\,q} \int dk\,k \int_{|k - q|}^{k + q} dk'\,k'\,h (k, k')\,.
\end{align}

\subsection{Imaginary part of polarisability: Particle-hole continua and their edges}
\label{app:hfull-im}

For $\omega > 0$, the imaginary part of Eq.~\eqref{eq:pol} is $\Im \Pi = -\,\pi\,g \sum_{ss'} \int d^3 k/(2\,\pi)^3\,\mathcal{F}_{ss'}\,[f_0 (\varepsilon_s (\bs{k})) - f_0 (\varepsilon_{s'} (\bs{k}'))]\,\delta (\omega + \varepsilon_s (\bs{k}) - \varepsilon_{s'} (\bs{k}'))$. Only three pairs have a pole at $\omega > 0$ with a non-vanishing occupation factor. These are $(+, +)$, $(0, +)$, and $(-, +)$. The pair $(+, -)$ has the occupation factor $-\,\Theta (k - k_F)$ but its denominator $\omega + v_0\,(k + k')$ does not vanish, and the pair $(-, 0)$ is blocked. With the radial measure of Appendix~\ref{app:hfull}, we obtain $\Im \Pi_{ss'} = -\,[g/(4\,\pi\,v_0\,q)] \int dk\,k\,dk'\,k'\,\mathcal{F}_{ss'}\,[\dots]\,\delta (\varpi + \dots)$, where $|k - k'| \leq q \leq k + k'$.

For the intraband pair, the delta function sets $k' = k + \varpi$, and the occupation factor is $\Theta (k_F - k)\,\Theta (k + \varpi - k_F)$. The condition $|k - k'| \leq q \leq k + k'$ becomes $\varpi \leq q$ and $k \geq (q - \varpi)/2$. The allowed range is $\check{k}_- < k < k_F$ with $\check{k}_- = \max \lbrace k_F - \varpi, (q - \varpi)/2 \rbrace$, and it is non-empty for $\varpi > q - 2\,k_F$. With $u = k\,(k + \varpi)$ and $\Delta = q^2 - \varpi^2$, we have $(2\,k + \varpi)^2 = 4\,u + \varpi^2$ and $1 + \cos\theta = [(2\,k + \varpi)^2 - q^2]/(2\,u)$. Using Eq.~\eqref{eq:h-fl}, we find
\begin{align}
k\,k'\,\mathcal{F}_{++} = \frac{(4\,u - \Delta)\,\left[ 4\,u - (1 - \lambda)\,\Delta \right]}{16\,u} = u - \frac{(2 - \lambda)\,\Delta}{4} + \frac{(1 - \lambda)\,\Delta^2}{16\,u}\,.
\end{align}
The integral over $k$ uses $\int dk/[k\,(k + \varpi)] = \ln \!\left[ k/(k + \varpi) \right]/\varpi$ and gives $\mathcal{P}_{++}$ of Eq.~\eqref{eq:h-pab}.

For the pair $(0, +)$, the delta function sets $k' = \varpi$ and the occupation factor is $\Theta (\varpi - k_F)$. The range of $k$ is $|\varpi - q| \leq k \leq \varpi + q$. Using $1 - \cos^2\theta = [(k + \varpi)^2 - q^2]\,[q^2 - (k - \varpi)^2]/(4\,k^2\,\varpi^2)$, we obtain
\begin{align}
k\,\varpi\,\mathcal{F}_{+0} = \frac{(1 - \lambda)\,\left[ 4\,k^2\,\varpi^2 - (k^2 + \varpi^2 - q^2)^2 \right]}{8\,k\,\varpi}\,.
\end{align}
The antiderivative is $(1 - \lambda)\,[(\varpi^2 + q^2)\,k^2 - k^4/4 - (\varpi^2 - q^2)^2 \ln k]/(8\,\varpi)$. With $(\varpi + q)^2 - (\varpi - q)^2 = 4\,\varpi\,q$ and $(\varpi + q)^4 - (\varpi - q)^4 = 8\,\varpi\,q\,(\varpi^2 + q^2)$, the evaluation at the limits gives the second line of Eq.~\eqref{eq:h-impi}. For $q \to 0$, we use $\ln \!\left| (\varpi + q)/(\varpi - q) \right| = 2\,q/\varpi + 2\,q^3/(3\,\varpi^3) + \dots$, and the bracket reduces to $16\,q^3\,\varpi/3$, which gives $\Im \Pi_{0+} \to -\,g\,(1 - \lambda)\,q^2/(6\,\pi\,v_0)$.

For the pair $(-, +)$, the delta function sets $k' = \varpi - k$, and the occupation factor is $\Theta (k' - k_F)$. The conditions $|k - k'| \leq q \leq k + k'$ become $\varpi \geq q$ and $(\varpi - q)/2 \leq k \leq (\varpi + q)/2$. The condition $k' \geq k_F$ adds $k \leq \varpi - k_F$. The range is non-empty for $\varpi > 2\,k_F - q$. With $u = k\,(\varpi - k)$ and $\tilde{D} = \varpi^2 - q^2$, we have $1 + \cos\theta = \tilde{D}/(2\,u)$ and $1 - \cos\theta = (4\,u - \tilde{D})/(2\,u)$. Hence
\begin{align}
k\,k'\,\mathcal{F}_{+-} = \frac{(4\,u - \tilde{D})\,\left[ 4\,u - (1 - \lambda)\,\tilde{D} \right]}{16\,u} = u - \frac{(2 - \lambda)\,\tilde{D}}{4} + \frac{(1 - \lambda)\,\tilde{D}^2}{16\,u}\,.
\end{align}
The integral over $k$ gives $\mathcal{P}_{-+}$ of Eq.~\eqref{eq:h-pab}. For $q \to 0$, the range of $k$ shrinks to the width $q$ about $\varpi/2$, where $4\,u - \tilde{D} = q^2 - (\varpi - 2\,k)^2$ and $4\,u - (1 - \lambda)\,\tilde{D} \simeq \lambda\,\varpi^2$. The integrand is then $\lambda\,[q^2 - (\varpi - 2\,k)^2]/4$, its integral is $\lambda\,q^3/6$, and $\Im \Pi_{-+} \to -\,g\,\lambda\,q^2/(24\,\pi\,v_0)$. Multiplying by $-\,4\,\pi\,e^2/q^2$ gives Eq.~\eqref{eq:h-imeps0}.

For the intrinsic limit $k_F \to 0$, the step function in $\Im \Pi_{0+}$ becomes unity, and for $\varpi < q$ this gives Eq.~\eqref{eq:h-Im0b}. For $\varpi > q$, the limits of $\mathcal{P}_{-+}$ in $\Im \Pi_{-+}$ are $k_1 = (\varpi - q)/2$ and $k_2 = (\varpi + q)/2$. We have $k_2 - k_1 = q$, $k_2^2 - k_1^2 = \varpi\,q$, and $k_2^3 - k_1^3 = q\,(3\,\varpi^2 + q^2)/4$. Moreover, $\ln \!\left[ k_2/(\varpi - k_2) \right] = -\,\ln \!\left[ k_1/(\varpi - k_1) \right] = \ln \!\left[ (\varpi + q)/(\varpi - q) \right]$. Eq.~\eqref{eq:h-pab} then gives
\begin{align}
\mathcal{P}_{-+} (k_2) - \mathcal{P}_{-+} (k_1) = \frac{q \left[ (5 - 3\,\lambda)\,q^2 - 3\,(1 - \lambda)\,\varpi^2 \right]}{12} + \frac{(1 - \lambda)\,(\varpi^2 - q^2)^2}{8\,\varpi}\,\ln \!\left( \frac{\varpi + q}{\varpi - q} \right).
\end{align}
The corresponding $\Im \Pi_{-+}^{(0)}$ contains the logarithm with the coefficient $-\,g\,(1 - \lambda)\,(\varpi^2 - q^2)^2/(32\,\pi\,v_0\,q\,\varpi)$. The second line of Eq.~\eqref{eq:h-impi} contains the same logarithm with the opposite sign. Hence the logarithms cancel in the sum. The remaining polynomial terms are
\begin{align}
-\,\frac{g\,(1 - \lambda)\,(\varpi^2 + q^2)}{16\,\pi\,v_0} - \frac{g \left[ (5 - 3\,\lambda)\,q^2 - 3\,(1 - \lambda)\,\varpi^2 \right]}{48\,\pi\,v_0} = -\,\frac{g\,(4 - 3\,\lambda)\,q^2}{24\,\pi\,v_0}\,.
\end{align}
This is Eq.~\eqref{eq:h-Im0}. At $\varpi \to q^+$, the logarithmic terms vanish. The bracket of the second line of Eq.~\eqref{eq:h-impi} equals $4\,q^4$, which gives $\Im \Pi_{0+}^{(0)} = -\,g\,(1 - \lambda)\,q^2/(8\,\pi\,v_0)$. The polynomial part of $\mathcal{P}_{-+} (k_2) - \mathcal{P}_{-+} (k_1)$ equals $q^3/6$, which gives $\Im \Pi_{-+}^{(0)} = -\,g\,q^2/(24\,\pi\,v_0)$. The two values add up to Eq.~\eqref{eq:h-Im0}, and the second one is the step of the transitions $- \to +$. It is independent of $\lambda$, since $\cos\theta = -1$ at $\varpi = q$ and Eq.~\eqref{eq:h-fl} gives $\mathcal{F}_{+-} = 1$ there.

For $\varpi > q + 2\,k_F$, we have $\varpi - k_F > (\varpi + q)/2$, and hence $k_2 = (\varpi + q)/2$ in Eq.~\eqref{eq:h-impi}. The intraband term vanishes for $\varpi > q$, and $\varpi > k_F$. Therefore, the doped imaginary part coincides with Eq.~\eqref{eq:h-Im0}.

\subsection{Real part of polarisability and RPA dielectric function}
\label{app:hfull-re}

As in Appendix~\ref{app:fin-re}, the Kramers-Kronig relation with the cut-off $\mathcal{R}$ on the transition energy reads $\Re \Pi (\bs{q}, \omega) = (2/\pi)\,\mathrm{P} \!\! \int_0^{\mathcal{R}} d\omega'\, \omega'\,\Im \Pi (\bs{q}, \omega')/(\omega'^2 - \omega^2)$. We subtract the reference $\Im \Pi_{\rm ref} = -\,g\,(4 - 3\,\lambda)\,q^2\,\Theta (\omega' - v_0\,q)/(24\,\pi\,v_0)$, whose contribution is
$-\,\frac{g\,(4 - 3\,\lambda)\,q^2}{24\,\pi^2\,v_0}\,\ln \!\left( \frac{\mathcal{R}^2}{|\omega^2 - v_0^2\,q^2|} \right) $
for $\mathcal{R} \gg \omega$. The difference $\Im \Pi - \Im \Pi_{\rm ref}$ vanishes for $\omega' > \omega_{\rm w}$, which gives Eq.~\eqref{eq:h-kk}.

In the intrinsic case, the difference is nonzero only for $\omega' < v_0\,q$, where it equals $\Im \Pi_{0+}^{(0)}$. By Eq.~\eqref{eq:h-impi}, this is $(1 - \lambda)$ times the expression of Eq.~\eqref{eq:app-fin-Imb}. The window integral is therefore $(1 - \lambda)$ times that of Appendix~\ref{app:fin-re}, which gives Eq.~\eqref{eq:h-Re0}. The doping-induced remainder of Eq.~\eqref{eq:h-dop} follows by writing $\Im \Pi = \Im \Pi_{++} + \Im \Pi_{0+} + \Im \Pi_{-+}$ and subtracting the intrinsic parts. We have $\Im \Pi_{0+} - \Im \Pi_{0+}^{(0)} = -\,\Theta (k_F - \varpi)\,\Im \Pi_{0+}^{(0)}$. For $\varpi < q$, both $\Im \Pi_{-+}$ and $\Im \Pi_{-+}^{(0)}$ vanish. For $q < \varpi < q + 2\,k_F$, the difference $\Im \Pi_{-+} - \Im \Pi_{-+}^{(0)}$ is nonzero, and it vanishes for larger $\varpi$.

For the cut-off constant, we consider $\omega = 0$ and the intrinsic system. The relation $\Pi_{\rm vac} (\bs{q}, 0) = (2/\pi) \int_0^{\mathcal{R}} d\omega'\,\Im \Pi^{(0)} (\bs{q}, \omega')/\omega'$ holds, since the cut-off acts on $\omega'$. The region $\omega' > v_0\,q$ gives $-\,[g\,(4 - 3\,\lambda)\,q^2/(24\,\pi\,v_0)]\,\ln \!\left( \Lambda/q \right)$ through Eq.~\eqref{eq:h-Im0}. For $\omega' < v_0\,q$, we write $\mathfrak{v} = \varpi'/q$ and $\ell (\mathfrak{v}) = \ln \!\left( (1 + \mathfrak{v})/(1 - \mathfrak{v}) \right)$. This region gives $-\,[g\,(1 - \lambda)\,q^2/(16\,\pi^2\,v_0)]\,J$ with
\begin{align}
J = \int_0^1 d\mathfrak{v} \left[ \frac{2\,(1 + \mathfrak{v}^2)}{\mathfrak{v}} - \frac{(1 - \mathfrak{v}^2)^2\,\ell (\mathfrak{v})}{\mathfrak{v}^2} \right] = \int_0^1 d\mathfrak{v} \left[ 2\,\mathfrak{v} + \frac{2}{\mathfrak{v}} - \frac{\ell (\mathfrak{v})}{\mathfrak{v}^2} + 2\,\ell (\mathfrak{v}) - \mathfrak{v}^2\,\ell (\mathfrak{v}) \right].
\end{align}
An antiderivative of $2/\mathfrak{v} - \ell/\mathfrak{v}^2$ is $\ell/\mathfrak{v} + \ln \!\left( 1 - \mathfrak{v}^2 \right)$, which takes the values $2\,\ln 2$ at $\mathfrak{v} = 1$ and $2$ at $\mathfrak{v} \to 0$. With $\int_0^1 \ell\,d\mathfrak{v} = 2\,\ln 2$ and $\int_0^1 \mathfrak{v}^2\,\ell\,d\mathfrak{v} = (1 + 2\,\ln 2)/3$, we obtain $J = 1 + 2\,\ln 2 - 2 + 4\,\ln 2 - (1 + 2\,\ln 2)/3 = (16\,\ln 2 - 4)/3$. Hence
\begin{align}
\Pi_{\rm vac} (\bs{q}, 0) = -\,\frac{g\,q^2}{2\,\pi^2\,v_0} \left[ \frac{4 - 3\,\lambda}{6}\,\ln \!\left( \frac{\Lambda}{q} \right) + \frac{(1 - \lambda)\,(4\,\ln 2 - 1)}{6} \right],
\end{align}
which is Eq.~\eqref{eq:h-pivac} with Eq.~\eqref{eq:h-b}. For $\lambda = 0$, this reproduces the large-$q$ limit of Eq.~\eqref{eq:fin-large}, since $3\,b_0/2 = \ln 2 - 1/4$.

With the radial measure, at $\omega = 0$ we have $\mathcal{R}/(\mathcal{R}^2 - \omega^2) = 1/\mathcal{R}$. Adding and subtracting the interband terms inside the ball $k < k_F$ in Eq.~\eqref{eq:h-split} gives
\begin{align}
\Pi (\bs{q}, 0) = \frac{g}{2\,\pi^2\,v_0\,q} \left[ \int_0^{k_F} dk\,k\,\upsilon (k) - \mathcal{I}_\Lambda (q) \right],
\end{align}
where $\mathcal{I}_\Lambda (q) = -\,2\,\pi^2\,v_0\,q\,\Pi_{\rm vac} (\bs{q}, 0)/g$ is the integral of $k\,\upsilon_{\rm inter} (k)$ over all $k$ with the cut-off on the transition energy,
and $\upsilon = \upsilon_{\rm intra} + \upsilon_{\rm inter}$ with
\begin{align}
\upsilon_{\rm intra} (k) = \int_{|k - q|}^{k + q} dk'\,\frac{k'\,\mathcal{F}_{++}}{k - k'}\,, \quad
\upsilon_{\rm inter} (k) = \int_{|k - q|}^{k + q} dk'\,k' \left[ \frac{\mathcal{F}_{+0}}{k} + \frac{\mathcal{F}_{+-}}{k + k'} \right].
\end{align}
The singularity of $\upsilon_{\rm intra}$ at $k' = k$ is a simple pole inside the integration range, and it is understood as a principal value. This follows from the combination of the two time orderings in Eq.~\eqref{eq:h-split}.

For the sum $k'\,[\mathcal{F}_{++}/(k - k') + \mathcal{F}_{+0}/k + \mathcal{F}_{+-}/(k + k')]$, with $\cos\theta$ expressed through $k$, $k'$, and $q$ by Eq.~\eqref{eq:h-fl}, the terms proportional to $1/k'$ cancel in the partial-fraction decomposition. An antiderivative with respect to $k'$ is
\begin{align}
G (k, k') = \frac{(\lambda - 1)\,k'^4 - 2\,k'^2 \left[ (\lambda + 3)\,k^2 + 2\,(\lambda - 1)\,q^2 \right] - 2\,(4\,k^2 - q^2) \left[ 4\,k^2 - (1 - \lambda)\,q^2 \right] \ln \!\left| k'^2 - k^2 \right|}{32\,k^3}\,.
\end{align}
We evaluate $G$ at the limits. The values $k' = k + q$ and $k' = |k - q|$ give $k'^2 - k^2 = q\,(2\,k + q)$ and $|k'^2 - k^2| = q\,|2\,k - q|$, and the differences of the powers are $(k + q)^4 - (k - q)^4 = 8\,k\,q\,(k^2 + q^2)$ and $(k + q)^2 - (k - q)^2 = 4\,k\,q$. The result is
\begin{align}
\upsilon (k) = -\,q + \frac{(1 - \lambda)\,q^3}{4\,k^2} - \frac{(4\,k^2 - q^2)\,\left[ 4\,k^2 - (1 - \lambda)\,q^2 \right]}{16\,k^3}\,\ln \!\left| \frac{2\,k + q}{2\,k - q} \right|.
\end{align}
At small $k$, the terms of order $1/k$ cancel, and $k\,\upsilon (k)$ is therefore integrable at $k = 0$. With $\tilde y = 2\,k/q$, we have $k\,\upsilon\,dk = q^3\,(\tilde y/4)\,\hat\upsilon (\tilde y)\,d\tilde y$, and an antiderivative of $(\tilde y/4)\,\hat\upsilon (\tilde y)$ is
\begin{align}
\mathcal{A} (\tilde y) = -\,\frac{\tilde y^2}{6} - \frac{3\,\lambda - 4}{12}\,\ln \!\left| \tilde y^2 - 1 \right| - \frac{\tilde y^4 + (3\,\lambda - 6)\,\tilde y^2 + 3\,\lambda - 3}{24\,\tilde y}\,\ln \!\left| \frac{\tilde y + 1}{\tilde y - 1} \right|.
\end{align}
The terms in $\ln \!\left| \tilde y - 1 \right|$ cancel at $\tilde y = 1$, and $\mathcal{A}$ is therefore continuous there. We have $\mathcal{A} (0) = (1 - \lambda)/4$, and hence $\int_0^{k_F} dk\,k\,\upsilon (k) = q^3\,[\mathcal{A} (2\,k_F/q) - (1 - \lambda)/4]$. Combining the logarithms, $\mathcal{A} (2\,k_F/q) - (1 - \lambda)/4 - (4 - 3\,\lambda)\,\ln \!\left( 2\,k_F/q \right)/6 = \Psi_\lambda (q, k_F)$ of Eq.~\eqref{eq:h-phi}.

For the interband term, we expand $k\,\upsilon_{\rm inter} (k)$ at $k \gg q$ using the corresponding antiderivative. The terms of order $k^2$ and $k^0$ cancel, and
\begin{align}
k\,\upsilon_{\rm inter} (k) = \frac{(4 - 3\,\lambda)\,q^3}{6\,k} + \mathcal{O} \left( \frac{q^5}{k^3} \right).
\end{align}
This fixes the coefficient of the logarithm, whereas the finite constant depends on the way in which the cut-off is imposed. For the cut-off on the transition energy, we obtain $\mathcal{I}_\Lambda (q) = q^3\,[(4 - 3\,\lambda)\,\ln \!\left( \Lambda/q \right)/6 + b_\lambda] + \mathcal{O} (q^5/\Lambda^2)$ with $b_\lambda$ of Eq.~\eqref{eq:h-b}, which was derived above. Writing $\ln \!\left( \Lambda/q \right) = \ln \!\left( \Lambda/2\,k_F \right) + \ln \!\left( 2\,k_F/q \right)$, we obtain Eq.~\eqref{eq:h-pistat}. The intrinsic response is the same expression without $\int_0^{k_F}$, which is Eq.~\eqref{eq:h-pivac}. The doping part $\Pi - \Pi_{\rm vac}$ is finite and does not depend on the cut-off. It equals $g\,q^2\,[\Psi_\lambda (q, k_F) + (4 - 3\,\lambda)\,\ln \!\left( 2\,k_F/q \right)/6]/(2\,\pi^2\,v_0)$.

The limits follow from the expansions of $\Psi_\lambda$. For $q \ll k_F$, the logarithms are expanded in $q/k_F$ and give $\Psi_\lambda (q, k_F) = -\,k_F^2/q^2 + 2/9 + (4 - 5\,\lambda)\,q^2/(240\,k_F^2) + \mathcal{O} (q^4/k_F^4)$. For $q \gg k_F$, we obtain $\mathcal{A} (2\,k_F/q) - (1 - \lambda)/4 = (1 - \lambda)\,k_F^2/(3\,q^2) + \mathcal{O} (k_F^4/q^4)$, which gives the constant difference quoted after Eq.~\eqref{eq:h-tf}. Using $\ln \!\left( \Lambda/2\,k_F \right) = \ln \!\left( v_0\,\Lambda/\mu \right) - \ln 2$ in Eq.~\eqref{eq:h-tf} gives the second form of $\epsilon_b (\lambda)$. The coefficient of $\ln \!\left( v_0\,\Lambda/\mu \right)$ in $\epsilon_b - 1$ is $g\,e^2\,(4 - 3\,\lambda)/(3\,\pi\,v_0)$, which equals $4\,g\,e^2/(3\,\pi\,v_0)$ for $\lambda = 0$ and $5\,g\,e^2/(6\,\pi\,v_0)$ for $\lambda = 1/2$.

\subsection{Kohn singularity and Friedel oscillations}
\label{app:hfull-friedel}

All terms of $\Psi_\lambda (q, k_F)$ are analytic at $q = 2\,k_F$ except the two logarithms. Writing $\ln \!\left| 1 - q^2/(4\,k_F^2) \right| = \ln \!\left| q - 2\,k_F \right| + \ln \!\left( q + 2\,k_F \right) - \ln \!\left( 4\,k_F^2 \right)$ and $\ln \!\left| (2\,k_F + q)/(2\,k_F - q) \right| = \ln \!\left( 2\,k_F + q \right) - \ln \!\left| q - 2\,k_F \right|$, the coefficient of $\ln \!\left| q - 2\,k_F \right|$ is
\begin{align}
\frac{4 - 3\,\lambda}{12} + \frac{q}{48\,k_F} \left[ \frac{16\,k_F^4}{q^4} + \frac{4\,(3\,\lambda - 6)\,k_F^2}{q^2} + 3\,\lambda - 3 \right] = \frac{(q - 2\,k_F)^2 \left[ 4\,k_F^2 + 4\,k_F\,q + (3\,\lambda - 3)\,q^2 \right]}{48\,k_F\,q^3}\,.
\end{align}
The factorisation is checked by multiplying out $(2\,k_F - q)^2 \left[ 4\,k_F^2 + 4\,k_F\,q + (3\,\lambda - 3)\,q^2 \right] = 16\,k_F^4 + 4\,(3\,\lambda - 6)\,k_F^2\,q^2 + 4\,(4 - 3\,\lambda)\,k_F\,q^3 + (3\,\lambda - 3)\,q^4$. Multiplying this coefficient by the prefactor $g\,q^2/(2\,\pi^2\,v_0)$ of Eq.~\eqref{eq:h-pistat} and expanding $\left[ 4\,k_F^2 + 4\,k_F\,q + (3\,\lambda - 3)\,q^2 \right]/(48\,k_F\,q)$ about $q = 2\,k_F$ gives $\lambda/8 - (4 - 3\,\lambda)\,(q - 2\,k_F)/(24\,q) + \mathcal{O} \left( (q - 2\,k_F)^2 \right)$, which is Eq.~\eqref{eq:h-sing}. With ${\tilde x} = q - 2\,k_F$, the singular part is $\Pi_{\rm sing} = g\,\lambda\, {\tilde x}^2\,\ln \!\left| {\tilde x} \right|/(16\,\pi^2\,v_0) + \dots$, where the dots denote terms with ${\tilde x}^3\,\ln \!\left| {\tilde x} \right|$ and analytic terms. The replacement of $\ln \!\left| {\tilde x}/q \right|$ by $\ln \!\left| {\tilde x} \right|$ changes only the analytic part. The next term follows from the factor $-\,(4 - 3\,\lambda)\,{\tilde x}/(24\,q)$ with $q \to 2\,k_F$ and equals $-\,g\,(4 - 3\,\lambda)\,{\tilde x}^3\,\ln \!\left| {\tilde x} \right|/(96\,\pi^2\,v_0\,k_F)$. For $\lambda = 0$, it is $-\,g\,{\tilde x}^3\,\ln \!\left| {\tilde x} \right|/(24\,\pi^2\,v_0\,k_F)$, which is Eq.~\eqref{eq:h-kohnz}. The second derivative of $\Pi_{\rm sing}$ contains $2\,\ln \!\left| {\tilde x} \right|$ times the coefficient of ${\tilde x}^2\,\ln \!\left| {\tilde x} \right|$, which gives the divergent part quoted after Eq.~\eqref{eq:h-sing}.

The induced density is $\delta n (\bs{q}) = \Pi\,\mathcal{U}/(1 - 4\,\pi\,e^2\,\Pi/q^2)$. Its non-analytic part at $q = 2\,k_F$ is obtained by differentiating with respect to $\Pi$, which gives $\delta n_{\rm sing} = \Pi_{\rm sing}\,\mathcal{U}/\epsilon^2$, with $\mathcal{U} (2\,k_F) = -\,\pi\,Z\,e^2/k_F^2$. The value of $\epsilon$ at $q = 2\,k_F$ follows from $\Psi_\lambda (1) = (4 - 3\,\lambda)\,\ln 2/6 - 5/12 + \lambda/4$ and $\epsilon_b$ of Eq.~\eqref{eq:h-tf}, which gives Eq.~\eqref{eq:h-eps2kf}. We write $\delta n (\bs{r}) = \int d^3 q\,e^{i \bs{q} \cdot \bs{r}}\,\delta n (\bs{q})/(2\,\pi)^3$ and consider a term $\mathfrak{j} (\vartheta_{\bs{q}})\,{\tilde x}^2\,\ln \!\left| {\tilde x} \right|$. The angular integral is done by stationary phase at $\bs{q} \parallel \pm\,\bs{\hat r}$, which gives
\begin{align}
\int d\tilde\Omega_{\bs{q}}\,e^{i \bs{q} \cdot \bs{r}}\,\mathfrak{j} (\vartheta_{\bs{q}}) \simeq \frac{2\,\pi}{i\,q\,r} \left[ \mathfrak{j} (\vartheta_{\bs{r}})\,e^{i \, q \, r} - \mathfrak{j} (\pi - \vartheta_{\bs{r}})\,e^{-\,i \, q \, r} \right] = \frac{4\,\pi\,\mathfrak{j} (\vartheta_{\bs{r}})}{q\,r}\,\sin \!\left( q\,r \right)\,,
\end{align}
since $\mathfrak{j}$ depends on $\sin^2\vartheta_{\bs{q}}$ only. The radial integral then involves $\int dq\,q\,\sin \!\left( q\,r \right)\,{\tilde x}^2\,\ln \!\left| {\tilde x} \right|$. We use the generalised Fourier transform $\int d\tilde x\,e^{i\, {\tilde x} \, r}\,\ln \!\left| {\tilde x} \right| = -\,\pi/r$ for $r > 0$, from which $\int d\tilde x\,e^{i \, {\tilde x} \, r}\,{\tilde x}^n\,\ln \!\left| {\tilde x} \right| = (-\,i\,\partial_r)^n (-\,\pi/r)$. For $n = 2$, this gives $2\,\pi/r^3$. With $q = 2\,k_F + {\tilde x}$, the leading term is $2\,k_F\,\Im \left[ e^{2 \,i\, k_F\, r}\,2\,\pi/r^3 \right] = 4\,\pi\,k_F\,\sin \!\left( 2\,k_F\,r \right)/r^3$. The remaining factors of $q$ and the smooth parts of the amplitude contribute at higher orders in $1/(k_F\,r)$. Altogether,
\begin{align}
\delta n (\bs{r}) \simeq \frac{1}{(2\,\pi)^3}\,\frac{4\,\pi}{r}\,
\mathfrak{j} (\vartheta_{\bs{r}})\,\frac{4\,\pi\,k_F\,\sin \!\left( 2\,k_F\,r \right)}{r^3} 
= \frac{2\,k_F\,\mathfrak{j} (\vartheta_{\bs{r}})}{\pi\,r^4}\,\sin \!\left( 2\,k_F\,r \right)\,,
\end{align}
with $\mathfrak{j} = -\,[g\,\sin^2\vartheta/(32\,\pi^2\,v_0)]\,(\pi\,Z\,e^2/k_F^2)/\epsilon_{2\, k_F}^2$. This is Eq.~\eqref{eq:h-friedel}. 

Near the $z$-axis, the stationary-phase result is not sufficient, and we use the exact angular integral. We write $\sin^2\vartheta_{\bs{q}} = (2/3) \left[ 1 - P_2 (\cos\vartheta_{\bs{q}}) \right]$ and use $\int d\tilde\Omega_{\bs{q}}\,e^{i \bs{q} \cdot \bs{r}}\,P_l (\cos\vartheta_{\bs{q}}) = 4\,\pi\,i^l\,j_l (q\,r)\,P_l (\cos\vartheta_{\bs{r}})$, where $j_l$ is the spherical Bessel function. This gives
\begin{align}
\int d\tilde\Omega_{\bs{q}}\,e^{i \bs{q} \cdot \bs{r}}\,\sin^2\vartheta_{\bs{q}} = \frac{8\,\pi}{3} \left[ j_0 (q\,r) + j_2 (q\,r)\,P_2 (\cos\vartheta_{\bs{r}}) \right].
\end{align}
For $\tilde y = q\,r \gg 1$, we have $j_0 = \sin \tilde y/\tilde y$ and $j_2 = -\,\sin \tilde y/\tilde y - 3\,\cos \tilde y/\tilde y^2 + \mathcal{O} (\tilde y^{-3})$. With $1 - P_2 (\cos\vartheta_{\bs{r}}) = 3\,\sin^2\vartheta_{\bs{r}}/2$, the right-hand side becomes $4\,\pi\,\sin^2\vartheta_{\bs{r}}\,\sin \tilde y/\tilde y - 8\,\pi\,P_2 (\cos\vartheta_{\bs{r}})\,\cos \tilde y/\tilde y^2$. The first term is the stationary-phase result used above. The second term is smaller by $1/\tilde y$, and it matters only for $\sin^2\vartheta_{\bs{r}} \lesssim 1/(k_F\,r)$. For the cubic term, the angular integral is $4\,\pi\,\sin \tilde y/\tilde y$ at $\lambda = 0$.

Let $\delta n_{\rm sing} (\bs{q}) = \left ( \mathcal{C}_2\,\sin^2\vartheta_{\bs{q}}\,{\tilde x}^2 + \mathcal{C}_3\,{\tilde x}^3 \right )
 \ln \!\left| {\tilde x} \right|$, with $\mathcal{C}_2 = 
 g\,\mathcal{U} (2\,k_F)/(32\,\pi^2\,v_0\,\epsilon_{2 \, k_F}^2 (0))$ and $\mathcal{C}_3 = -\,g\,\mathcal{U} (2\,k_F)/(24\,\pi^2\,v_0\,k_F\,\epsilon_{2 \, k_F}^2 (0))$. The dependence of $\epsilon$ and of the coefficient of the cubic term on $\lambda_{\bs{q}}$ gives relative corrections of order $\vartheta_{\bs{r}}^2$ near the axis. For the contributions with $\sin \tilde y$ and $\cos \tilde y$, we use $\int d\tilde x\,e^{i \, {\tilde x} \, r}\,{\tilde x}^2\,\ln \!\left| {\tilde x} \right| = 2\,\pi/r^3$ and $\int d{\tilde x}\,e^{i \, {\tilde x} \, r}\,{\tilde x}^3\,\ln \!\left| {\tilde x} \right| = 6\,\pi\,i/r^4$, with $q = 2\,k_F + {\tilde x}$. The radial integrals are
\begin{align}
& \int dq\,q\,{\tilde x}^2\,\ln \!\left| {\tilde x} \right|\,\sin \!\left( q\,r \right) 
\simeq \frac{4\,\pi\,k_F\,\sin \!\left( 2\,k_F\,r \right)}{r^3}\,, \quad
\int dq\,\tilde x^2\,\ln \!\left| {\tilde x} \right|\,\cos \!\left( q\,r \right) 
\simeq \frac{2\,\pi\,\cos \!\left( 2\,k_F\,r \right)}{r^3}\,, \nn &
\int dq\,q\,{\tilde x}^3\,\ln \!\left| \tilde x \right|\,\sin \!\left( q\,r \right) 
\simeq \frac{12\,\pi\,k_F\,\cos \!\left( 2\,k_F\,r \right)}{r^4}\,.
\end{align}
Collecting the three terms with the prefactor $1/(2\,\pi)^3$, we obtain
\begin{align}
\delta n (\bs{r}) \simeq \frac{1}{\pi} \left[ \frac{2\,k_F\,\mathcal{C}_2
\sin^2\vartheta_{\bs{r}}\,\sin \!\left( 2\,k_F\,r \right)}{r^4} 
+ \left( 6\,k_F\,\mathcal{C}_3 - 2\,\mathcal{C}_2\,P_2 (\cos\vartheta_{\bs{r}}) \right) \frac{\cos \!\left( 2\,k_F\,r \right)}{r^5} \right].
\end{align}
Inserting $\mathcal{U} (2\,k_F) = -\,\pi\,Z\,e^2/k_F^2$ gives Eq.~\eqref{eq:h-friedel-axis}. Along the $z$-axis, $P_2 = 1$ and $6\,k_F\,\mathcal{C}_3 - 2\,\mathcal{C}_2 = -\,5\,g\,\mathcal{U}/(16\,\pi^2\,v_0\,\epsilon_{2 \, k_F}^2 (0))$.

\subsection{Plasmon dispersion and Landau damping}
\label{app:hfull-plasmon}

The real part of $\Pi$ follows from the Kramers-Kronig relation $\Re \Pi (\bs{q}, \omega) = (2/\pi)\,\mathcal{P} \int_0^\infty d\varpi'\,\varpi'\,\Im \Pi (\bs{q}, \varpi')/(\varpi'^2 - \varpi^2)$, since $\Im \Pi$ is odd in $\omega$. We consider $q < k_F$ and $\varpi > q$, where the intraband support is $\varpi' < q$, and we expand $1/(\varpi'^2 - \varpi^2) = -\,\varpi^{-2}\,[1 + \varpi'^2/\varpi^2 + \dots]$. For $q < k_F$ and $\varpi' < q$, we have $\check{k}_- = k_F - \varpi'$, since $\varpi' < q < 2\,k_F - q$. Expanding $\Im \Pi_{++}$ in $q/k_F$ gives $\Im \Pi_{++} = -\,g\,k_F^2\,\varpi'/(4\,\pi\,v_0\,q)\,[1 + \mathcal{O} (q^2/k_F^2)]$. The moments are
\begin{align}
{\tilde \mu}_n = \int_0^q d\varpi'\,\varpi'^n\,\Im \Pi_{++} = -\,\frac{g\,k_F^2\,q^{n + 1}}{4\,\pi\,v_0\,(n + 2)}\,[1 + \mathcal{O} (q^2/k_F^2)]\,,
\end{align}
which gives ${\tilde \mu}_1 = -\,g\,k_F^2\,q^2/(12\,\pi\,v_0)$ and ${\tilde \mu}_3 = -\,g\,k_F^2\,q^4/(20\,\pi\,v_0)$. The intraband part is $\Re \Pi_{\rm intra} = -\,(2/(\pi\,\varpi^2))\,[{\tilde \mu}_1 + {\tilde \mu}_3/\varpi^2 + \dots]$, and the ratio ${\tilde \mu}_3/{\tilde \mu}_1 = 3\,q^2/5$ gives Eq.~\eqref{eq:h-drude}, after restoring $\varpi = \omega/v_0$.

The interband contribution to $\Re \Pi$ at $q \to 0$ follows from the Kramers-Kronig relation applied to Eq.~\eqref{eq:h-imeps0}. It gives $\epsilon_{\rm inter} (\omega)$ of Eq.~\eqref{eq:h-einter} without a further constant, since the cut-off acts on the transition energy, as in Sec.~\ref{sechre}. At $q \to 0$, the zero $\omega_*$ of $\epsilon_{\rm inter} (\omega) - \omega_p^2/\omega^2$ obeys $\epsilon_{\rm inter} (\omega_*)\,\omega_*^2 = \omega_p^2$. Including the term of order $q^2$ of Eq.~\eqref{eq:h-drude}, the zero of $\epsilon_{\rm inter}\,\omega^2 - \omega_p^2 - (3/5)\,\omega_p^2\,v_0^2\,q^2/\omega^2$ is shifted by $\delta\omega^2 = (3/5)\,v_0^2\,q^2/(1 + \eta)$, where
\begin{align}
\eta = \frac{\omega_*^2}{\epsilon_{\rm inter} (\omega_*)}\,\frac{d\epsilon_{\rm inter}}{d\omega^2} \Big|_{\omega_*} = \frac{\omega_*^2}{\epsilon_{\rm inter} (\omega_*)}\,\frac{2\,g\,e^2}{\pi\,v_0} \left[ \frac{1 - \lambda}{3\,(\mu^2 - \omega_*^2)} + \frac{\lambda}{12\,(4\,\mu^2 - \omega_*^2)} \right].
\end{align}
Since $\omega_*^2 \leq \omega_p^2/\epsilon_{\rm inter} (\omega_*)$ and $\omega_p^2/\mu^2 = 2\,\alpha/(3\,\pi)$, we have $\eta = \mathcal{O} (\alpha^2)$ for $\omega_p \ll \mu$.
 The terms of order $q^2/k_F^2$ in Eq.~\eqref{eq:h-drude} and the terms of order $q^2$ of the interband part shift $\omega^2$ by terms of order $\alpha\,v_0^2\,q^2$. This gives Eq.~\eqref{eq:h-disp}.

For Eq.~\eqref{eq:h-pi24}, we expand the Kramers-Kronig integral in $q$ at fixed $\omega < \mu$, with the cut-off $\mathcal{R}$ on the transition energy. Three parts contribute. For the intraband part, we keep the first correction in Eq.~\eqref{eq:h-impi} for $\varpi' < q$ and $\mathfrak{v} = \varpi'/q$,
\begin{align}
\Im \Pi_{++} = -\,\frac{g\,k_F^2\,\varpi'}{4\,\pi\,v_0\,q} \left[ 1 - \frac{q^2}{2\,k_F^2} \left( 1 - \frac{2\,\mathfrak{v}^2}{3} - \frac{\lambda\,(1 - \mathfrak{v}^2)}{2} \right) + \mathcal{O} \left( \frac{q^4}{k_F^4} \right) \right].
\end{align}
The moments are then ${\tilde \mu}_1 = -\,(g\,k_F^2\,q^2/(12\,\pi\,v_0))\,[1 - (3 - \lambda)\,q^2/(10\,k_F^2)]$ and ${\tilde \mu}_3 = -\,g\,k_F^2\,q^4/(20\,\pi\,v_0)\,[1 + \mathcal{O} (q^2/k_F^2)]$. The relation $\Re \Pi_{\rm intra} = -\,(2/(\pi\,\varpi^2))\,[{\tilde \mu}_1 + {\tilde \mu}_3/\varpi^2]$ gives the Drude term of $\Pi_2$, and the terms $\mu^2/(10\,\omega^2) - (3 - \lambda)/60$ of $\Pi_4$.

For the transitions $0 \to +$, which exist for $\varpi' > k_F \gg q$, the expansion of Eq.~\eqref{eq:h-impi} gives
\begin{align}
\Im \Pi_{0+} = -\,\frac{g\,(1 - \lambda)}{\pi\,v_0} \left[ \frac{q^2}{6} - \frac{q^4}{30\,\varpi'^2} + \mathcal{O} (q^6) \right].
\end{align}
The first term gives the logarithm of $\mu^2 - \omega^2$ in $\Pi_2$. For the second term, we use $\int_{k_F}^\infty d\varpi'/[\varpi'\,(\varpi'^2 - \varpi^2)] = -\,\ln \!\left( 1 - \varpi^2/k_F^2 \right)/(2\,\varpi^2)$, which gives $-\,(1 - \lambda)\,\ln \!\left( 1 - \omega^2/\mu^2 \right)/30$ in the bracket of $\Pi_4$.

For the transitions $- \to +$, the doped imaginary part differs from the intrinsic one in the window $2\,k_F - q < \varpi' < 2\,k_F + q$. With $\varpi' = 2\,k_F + q\,\mathfrak{o}$ and $-1 < \mathfrak{o} < 1$, we obtain $\Im \Pi_{-+} = g\,\lambda\,q^2\,(\mathfrak{o}^3 - 3\,\mathfrak{o} - 2)/(96\,\pi\,v_0) + \mathcal{O} (q^4)$, which equals the intrinsic value $-\,g\,\lambda\,q^2/(24\,\pi\,v_0)$ at $\mathfrak{o} = 1$. Above the window, we use the intrinsic expression, $\Im \Pi_{-+}^{(0)} = -\,g\,\lambda\,q^2/(24\,\pi\,v_0) - g\,(1 - \lambda)\,q^4/(30\,\pi\,v_0\,\varpi'^2) + \mathcal{O} (q^6)$, which follows from Eq.~\eqref{eq:h-Im0} and the expansion of $\Im \Pi_{0+}$. The first term of the intrinsic part gives the logarithm of $4\,\mu^2 - \omega^2$ in $\Pi_2$. Its lower limit $2\,k_F + q$ and the window contribute at order $q^3$, and these two contributions cancel. At order $q^4$, the lower limit and the expansion of $\varpi'/(\varpi'^2 - \varpi^2)$ around $\varpi' = 2\,k_F$ in the window leave $-\,\lambda\,\omega^2 \left( 4\,\mu^2 + \omega^2 \right)/[120 \left( 4\,\mu^2 - \omega^2 \right)^2]$ in the bracket of $\Pi_4$. The second term of the intrinsic part gives $(1 - \lambda)\,\ln \!\left( (4\,\mu^2 - \omega^2)/(4\,\mu^2) \right)/30$. Together with the contribution of the transitions $0 \to +$, this gives the logarithm in the bracket of $\Pi_4$. No odd powers of $q$ remain.

For Eq.~\eqref{eq:h-wdisp}, we write $\epsilon^{\rm RPA} = E (\omega) - 4\,\pi\,e^2\,q^2\,\Pi_4$, with $E (\omega) = \epsilon_{\rm inter} (\omega) - \omega_p^2/\omega^2$. The zero obeys $E (\omega_*) = 4\,\pi\,e^2\,q^2\,\Pi_4$, and hence $\delta \omega^2 = 4\,\pi\,e^2\,q^2\,\Pi_4/(dE/d\omega^2)$. Using $E (\omega_*) = 0$, we have $dE/d\omega^2 = \Xi (\omega_*)/\omega_*^2$, which gives Eq.~\eqref{eq:h-wdisp}. The bracket is positive, since its first two terms exceed $1/10 - 3/60$ for $\omega < \mu$, the logarithm is nonnegative, and the last term is bounded in magnitude by $5\,\lambda/1080$. The function $\Xi$ is positive, since $\epsilon_{\rm inter} > 1$ for $v_0\,\Lambda > \mu$. For $\omega_p > \mu$, the zero obeys $\epsilon_{\rm inter} (\omega_*)\,\omega_*^2 = \omega_p^2$ with $\epsilon_{\rm inter} \to \infty$ at $\omega \to \mu^-$. The zero is thus pinned below $\mu$, at a distance set by the coefficient $(1 - \lambda)/3$ of the logarithm.

For the existence of the zero, we consider $0 < q < k_F$. In the window $v_0\,q < \omega < \mu$, all three supports of Eq.~\eqref{eq:h-impi} are excluded. The intraband support ends at $\varpi = q$, the onset of $\Im \Pi_{0+}$ is $\varpi = k_F$, and the onset of $\Im \Pi_{-+}$ is $\max \lbrace q, 2\,k_F - q \rbrace = 2\,k_F - q > k_F$. Therefore $\Im \epsilon^{\rm RPA} = 0$ in the window, and $\Re \epsilon^{\rm RPA}$ is continuous. At the upper end of the intraband support, $\Im \Pi_{++} (\varpi \to q^-) = -\,[g/(4\,\pi\,v_0\,q)] \int_{k_F - q}^{k_F} dk\,k\,(k + q) \equiv -\,A < 0$, since $\Delta = 0$ at $\varpi = q$. The Kramers-Kronig integral near this edge is $(1/\pi) \int^q d\varpi'\,\Im \Pi_{++}/(\varpi' - \varpi) \simeq -\,(A/\pi)\,\ln \!\left| q - \varpi \right|$, which tends to $+\,\infty$ for $\varpi \to q^+$. Hence $\Re \Pi \to +\,\infty$ and $\Re \epsilon^{\rm RPA} = 1 - 4\,\pi\,e^2\,\Re \Pi/q^2 \to -\,\infty$. At the lower end of the support of $\Im \Pi_{0+}$, the integrand of the second line of Eq.~\eqref{eq:h-impi} is $(1 - \lambda)\,[(k + \varpi)^2 - q^2]\,[q^2 - (k - \varpi)^2]/(8\,k\,\varpi)$, which is positive for $|\varpi - q| < k < \varpi + q$, and $1 - \lambda \geq 1/2$. Hence $\Im \Pi_{0+} (\varpi \to k_F^+) = -\,B < 0$. The Kramers-Kronig integral near this edge is $(1/\pi) \int_{k_F} d\varpi'\,(-\,B)/(\varpi' - \varpi) \simeq (B/\pi)\,\ln \!\left| k_F - \varpi \right|$, which tends to $-\,\infty$ for $\varpi \to k_F^-$. Hence $\Re \Pi \to -\,\infty$ and $\Re \epsilon^{\rm RPA} \to +\,\infty$. All other parts of $\Im \Pi$ are continuous at these two frequencies. Hence they do not change the sign of the divergences. By the intermediate-value theorem, $\Re \epsilon^{\rm RPA}$ has at least one zero in the window. Near the lower edge, $\Re \epsilon^{\rm RPA}$ reaches the logarithmic divergence only at distances from $\varpi = q$ that are exponentially small. For $q > q_{\rm max}$, the function $\Re \epsilon^{\rm RPA}$ is positive at a finite distance from the edge, and hence the zero lies exponentially close to the edge.

For $q \geq k_F$, every $\varpi > q$ obeys $\varpi > k_F$, and the integrand of $\Im \Pi_{0+}$ is positive on $|\varpi - q| < k < \varpi + q$. Therefore $\Im \Pi_{0+} < 0$ and $\Im \epsilon^{\rm RPA} > 0$ at every $\omega > v_0\,q$, which excludes an undamped mode in this range.

For the free regions, we use the supports of Eq.~\eqref{eq:h-impi}. The intraband part is nonzero for $\max \lbrace 0, q - 2\,k_F \rbrace < \varpi < q$, the part $\Pi_{0+}$ for $\varpi > k_F$, and the part $\Pi_{-+}$ for $\varpi > \max \lbrace q, 2\,k_F - q \rbrace$. For $q < k_F$, the supports are those of the existence argument above, and the free region is therefore $q < \varpi < k_F$. For $k_F \leq q \leq 2\,k_F$, the intraband part covers $0 < \varpi < q$, and every $\varpi > q$ exceeds $k_F$. Hence $\Im \Pi \neq 0$ for every $\varpi > 0$. For $q > 2\,k_F$, the intraband part covers $q - 2\,k_F < \varpi < q$, the onset of $\Pi_{0+}$ is $k_F$, and the onset of $\Pi_{-+}$ is $q$. The free region is therefore $\varpi < \varpi_f = \min \lbrace k_F, q - 2\,k_F \rbrace$.

In the region $\varpi < \varpi_f$ of the last case, the Kramers-Kronig relation gives $\Re \Pi (\bs{q}, \omega) = (2/\pi)\,\mathrm{P} \int_{v_0\,\varpi_f}^{\mathcal{R}} d\omega'\,\omega'\,\Im \Pi (\bs{q}, \omega')/(\omega'^2 - \omega^2)$, since $\Im \Pi$ vanishes below $v_0\,\varpi_f$. The denominator is positive. Each part of $\Im \Pi$ is a negative prefactor times the integral of a non-negative function. For $\Pi_{++}$ and $\Pi_{-+}$, this follows from $k\,k'\,\mathcal{F}_{++} \geq 0$ and $k\,k'\,\mathcal{F}_{+-} \geq 0$ in Eq.~\eqref{eq:h-fl}, because $0 \leq \lambda \leq 1/2$ and $1 - \cos^2\theta \geq 0$. For $\Pi_{0+}$, it follows from $(1 - \lambda)\,(1 - \cos^2\theta) \geq 0$. Hence $\Im \Pi \leq 0$ and $\Re \Pi \leq 0$. The relation $\Re \epsilon^{\rm RPA} = 1 - 4\,\pi\,e^2\,\Re \Pi/q^2$ then gives $\Re \epsilon^{\rm RPA} \geq 1$. The inequality is strict, because $\Im \Pi < 0$ on a set of nonzero measure above $v_0\,\varpi_f$. The condition $q \ll \Lambda$ ensures that $v_0\,\varpi_f$ lies below the cut-off $\mathcal{R}$. There is therefore no zero of $\Re \epsilon^{\rm RPA}$ in the free region.

\section{WSM at arbitrary momentum-transfers}
\label{app-WSM}

We insert $f_0 (\varepsilon_{-1}) = 1$ and $f_0 (\varepsilon_1) = \Theta (k_F - k)$ into Eq.~\eqref{eq:pol}. The pair $(1, 1)$ has the occupation factor $\Theta (k_F - k) - \Theta (k_F - k')$. Relabelling $\bs{k} \leftrightarrow \bs{k}'$ in the second part replaces $\bs{q}$ by $-\,\bs{q}$, which is allowed because the polarisability depends only on $|\bs{q}|$ and $\mathcal{F}_{11}$ is symmetric. This gives the term $s' = 1$ of Eq.~\eqref{eq:w-pimu}. The pairs $(-1, 1)$ and $(1, -1)$ have the occupation factors $1 - \Theta (k_F - k')$ and $\Theta (k_F - k) - 1$. The parts proportional to $1$ form $\Pi_{\rm vac}$, which is minus the interband bracket of Eq.~\eqref{eq:w-pimu} summed over all $\bs{k}$. After the same relabelling, the parts proportional to $\Theta$ give the term $s' = -1$ of Eq.~\eqref{eq:w-pimu}, because $\varepsilon_1 (k) - \varepsilon_{-1} (k') = v_0 \left( k + k' \right)$ and $\mathcal{F}_{-1,1} = \mathcal{F}_{1,-1}$.

At fixed $q$, the momentum $k' = |\bs{k} + \bs{q}|$ obeys $|k - q| \leq k' \leq k + q$. With $\mathfrak{k}_\pm = k \pm k'$, the Jacobian is $|\partial (k, k')/\partial (\mathfrak{k}_-, \mathfrak{k}_+)| = 1/2$. Using $d^3 \bs{k} = 2\,\pi\,k\,k'\,dk\,dk'/q$ and $k\,k' = (\mathfrak{k}_+^2 - \mathfrak{k}_-^2)/4$, the measure becomes
\begin{align}
d^3 \bs{k} = \frac{\pi \left( \mathfrak{k}_+^2 - \mathfrak{k}_-^2 \right)}{4\,q}\,d\mathfrak{k}_-\,d\mathfrak{k}_+ \,.
\end{align}
The domain is $\mathfrak{D} = \lbrace |\mathfrak{k}_-| \leq q \leq \mathfrak{k}_+ \,,\; \mathfrak{k}_- + \mathfrak{k}_+ \leq 2\,k_F \rbrace$, where the first condition is the triangle inequality and the second is $k < k_F$. The relations $\cos\theta = (k^2 + k'^2 - q^2)/(2\,k\,k')$ and $k^2 + k'^2 = (\mathfrak{k}_+^2 + \mathfrak{k}_-^2)/2$ turn the overlap factors of Eq.~\eqref{eq:w-ff} into
\begin{align}
\mathcal{F}_{1,1} = \frac{\mathfrak{k}_+^2 - q^2}{\mathfrak{k}_+^2 - \mathfrak{k}_-^2} \,, \quad \mathcal{F}_{1,-1} = \frac{q^2 - \mathfrak{k}_-^2}{\mathfrak{k}_+^2 - \mathfrak{k}_-^2} \,.
\end{align}
The energy differences are $\varepsilon_1 (k) - \varepsilon_1 (k') = v_0\,\mathfrak{k}_-$ and $\varepsilon_1 (k) - \varepsilon_{-1} (k') = v_0\,\mathfrak{k}_+$, and the two time orderings of each pair combine into
\begin{align}
\frac{1}{\omega + v_0\,\mathfrak{k}_- + i\,0^+} - \frac{1}{\omega - v_0\,\mathfrak{k}_- + i\,0^+} = -\,\frac{2\,v_0\,\mathfrak{k}_-}{(\omega + i\,0^+)^2 - v_0^2\,\mathfrak{k}_-^2} \,,
\end{align}
with $\mathfrak{k}_- \to \mathfrak{k}_+$ for the interband pair. The factor $\mathfrak{k}_+^2 - \mathfrak{k}_-^2$ of the measure cancels against the denominator of the overlap factors, and the overall prefactor is $[g/(2\,\pi)^3]\,[\pi/(4\,q)]\,(-\,2/v_0) = -\,g/(16\,\pi^2\,v_0\,q)$, which gives Eq.~\eqref{eq:w-master}.

For the vacuum part, the $\mathfrak{k}_-$ integral of the interband term runs over $|\mathfrak{k}_-| \leq q$ and gives $\int_{-q}^{q} d\mathfrak{k}_- \left( q^2 - \mathfrak{k}_-^2 \right) = 4\,q^3/3$. For $q, \varpi \ll \Lambda$, the $\mathfrak{k}_+$ integral from $q$ to $\Lambda$ gives
\begin{align}
\int_q^{\Lambda} d\mathfrak{k}_+ \frac{\mathfrak{k}_+}{\left( \varpi + i\,0^+ \right)^2 - \mathfrak{k}_+^2} = -\,\frac{1}{2} \left[ \ln \!\left( \frac{\Lambda^2}{\left| \varpi^2 - q^2 \right|} \right) + i\,\pi\,\Theta \left( \varpi - q \right) \right],
\end{align}
and minus the product with the prefactor of Eq.~\eqref{eq:w-master} is Eq.~\eqref{eq:w-vac}.

\subsection{Imaginary part of polarisability: Particle-hole continua and their edges}
\label{app:w-imag}

For $\varpi > 0$, the imaginary part of Eq.~\eqref{eq:pol} is $\Im \Pi = -\,\pi\,g \sum_{ss'} \int d^3 k/(2\,\pi)^3\,\mathcal{F}_{ss'}\,[f_0 (\varepsilon_s (\bs{k})) - f_0 (\varepsilon_{s'} (\bs{k}'))]\,\delta (\omega + \varepsilon_s (\bs{k}) - \varepsilon_{s'} (\bs{k}'))$. Only the pairs $(1, 1)$ and $(-1, 1)$ have a pole at $\omega > 0$ and a non-vanishing occupation factor. The pair $(1, -1)$ has the occupation factor $-\,\Theta (k - k_F)$, but its denominator $\omega + v_0\,(k + k')$ does not vanish. With $k'\,dk' = k\,q\,d\cos\vartheta_{\bs{k} \bs{q}}$, we have $\int d^3 k/(2\,\pi)^3 = (4\,\pi^2\,q)^{-1} \int dk\,k \int dk'\,k'$ with $|k - k'| \leq q \leq k + k'$, and
\begin{align}
\Im \Pi_{ss'} = -\,\frac{g}{4\,\pi\,v_0\,q} \int dk\,dk'\,k\,k'\,\mathcal{F}_{ss'}\,[f_0 (\varepsilon_s (\bs{k})) - f_0 (\varepsilon_{s'} (\bs{k}'))]\,\delta (\varpi + s\,k - s'\,k')\,.
\end{align}
The overlap factors of Eq.~\eqref{eq:w-ff} give $k\,k'\,\mathcal{F}_{1,1} = [(k + k')^2 - q^2]/4$ and $k\,k'\,\mathcal{F}_{1,-1} = [q^2 - (k - k')^2]/4$.

For the intraband pair, the delta function sets $k' = k + \varpi$, and the occupation factor is $\Theta (k_F - k)\,\Theta (k + \varpi - k_F)$. The condition $|k - k'| \leq q \leq k + k'$ becomes $\varpi \leq q$ and $k \geq (q - \varpi)/2$. The range is $\check{k}_- < k < k_F$ with $\check{k}_- = \max \lbrace k_F - \varpi, (q - \varpi)/2 \rbrace$, which is non-empty for $\varpi > q - 2\,k_F$. The antiderivative of $[(2\,k + \varpi)^2 - q^2]/4$ is $\mathcal{G} (k) = (2\,k + \varpi)^3/24 - q^2\,k/4$. For $\varpi < 2\,k_F - q$, we have $\check{k}_- = k_F - \varpi$ and
\begin{align}
\mathcal{G} (k_F) - \mathcal{G} (k_F - \varpi) = \frac{(2\,k_F + \varpi)^3 - (2\,k_F - \varpi)^3}{24} - \frac{q^2\,\varpi}{4} = \frac{\varpi}{4} \left( 4\,k_F^2 - q^2 + \frac{\varpi^2}{3} \right).
\end{align}
For $\varpi > 2\,k_F - q$, we have $\check{k}_- = (q - \varpi)/2$ and $2\,\check{k}_- + \varpi = q$. Hence
\begin{align}
\mathcal{G} (k_F) - \mathcal{G} \!\left( \frac{q - \varpi}{2} \right) = \frac{(2\,k_F + \varpi)^3 - q^3}{24} - \frac{q^2 \left( 2\,k_F + \varpi - q \right)}{8} = \frac{\left( 2\,k_F + \varpi - q \right)^2 \left( 2\,k_F + \varpi + 2\,q \right)}{24}\,,
\end{align}
where $\tilde x^2 + \tilde x\,q - 2\,q^2 = (\tilde x - q)(\tilde x + 2\,q)$ with $\tilde x = 2\,k_F + \varpi$ has been used. Multiplication by $-\,g/(4\,\pi\,v_0\,q)$ gives the two lines of Eq.~\eqref{eq:w-imintra}. The two expressions coincide at $\varpi = 2\,k_F - q$, and the second vanishes quadratically at $\varpi = q - 2\,k_F$. For $q \leq k_F$, only the first applies, since $\varpi \leq q \leq 2\,k_F - q$.

For the interband pair, the delta function sets $k' = \varpi - k$, and the occupation factor is $\Theta (k' - k_F)$. The condition $|k - k'| \leq q \leq k + k'$ becomes $\varpi \geq q$ and $(\varpi - q)/2 \leq k \leq (\varpi + q)/2$, and $k' \geq k_F$ adds $k \leq \varpi - k_F$. With $\tilde y = 2\,k - \varpi$, we have $k\,k'\,\mathcal{F}_{1,-1} = (q^2 - \tilde y^2)/4$ and $dk = d\tilde y/2$. The range is $-\,q \leq \tilde y \leq \tilde y_u$ with $\tilde y_u = \min \lbrace q, \varpi - 2\,k_F \rbrace$, which is non-empty for $\varpi > 2\,k_F - q$, and together with $\varpi \geq q$ the onset is $\varpi = \max \lbrace q, 2\,k_F - q \rbrace$. The integral is
\begin{align}
\frac{1}{8} \int_{-q}^{\tilde y_u} d\tilde y \left( q^2 - \tilde y^2 \right) = \frac{\left( q + \tilde y_u \right)^2 \left( 2\,q - \tilde y_u \right)}{24}\,.
\end{align}
It equals $q^3/6$ for $\tilde y_u = q$, that is for $\varpi > 2\,k_F + q$, and gives the middle line of Eq.~\eqref{eq:w-iminter} for $\tilde y_u = \varpi - 2\,k_F$. The case $k_F \to 0$ is the step $-\,g\,q^2/(24\,\pi\,v_0)$ at $\varpi > q$ of Eq.~\eqref{eq:w-vac}.

Each piece of $\Im \Pi$ is non-positive, since the integrands are non-negative. For $q \geq k_F$, the jump at $\varpi = q$ is the interband value at $\varpi = q^+$ minus the intraband value at $\varpi = q^-$. These are $-\,g\,(q - k_F)^2 (2\,k_F + q)/(24\,\pi\,v_0\,q)$ and $-\,g\,k_F^2 (2\,k_F + 3\,q)/(24\,\pi\,v_0\,q)$, and their difference is $g\,(6\,k_F^2 - q^2)/(24\,\pi\,v_0)$, which is Eq.~\eqref{eq:w-jump}.

\subsection{Real part of polarisability and RPA dielectric function}
\label{app:w-real}

The doped part $\Pi_\mu$ of Eq.~\eqref{eq:w-master} is independent of $\Lambda$, analytic in the upper half-plane, and vanishes for $\omega \to \infty$. It therefore obeys the dispersion relation of Eq.~\eqref{eq:w-kk}, with $\Im \Pi_\mu = \Im \Pi - \Im \Pi^{(0)}$. We evaluate the master integral directly as $\Re \Pi_\mu = -\,g\,(\mathfrak{I}_1 + \mathfrak{I}_2)/(16\,\pi^2\,v_0\,q)$, with the principal-value integrals
\begin{align}
\mathfrak{I}_1 = \mathrm{P} \iint_{\mathfrak{D}} d\mathfrak{k}_-\,d\mathfrak{k}_+ \frac{\left( \mathfrak{k}_+^2 - q^2 \right) \mathfrak{k}_-}{\varpi^2 - \mathfrak{k}_-^2} \,, \quad
\mathfrak{I}_2 = \mathrm{P} \iint_{\mathfrak{D}} d\mathfrak{k}_-\,d\mathfrak{k}_+ \frac{\left( q^2 - \mathfrak{k}_-^2 \right) \mathfrak{k}_+}{\varpi^2 - \mathfrak{k}_+^2} \,.
\end{align}
We take $q \leq k_F$ first. For $\mathfrak{I}_1$, the $\mathfrak{k}_+$ integral runs from $q$ to $2\,k_F - \mathfrak{k}_-$ for all $|\mathfrak{k}_-| \leq q$, which gives
\begin{align}
\mathfrak{I}_1 = \mathrm{P} \int_{-q}^{q} d\mathfrak{k}_- \frac{\mathfrak{k}_-}{\varpi^2 - \mathfrak{k}_-^2} \left[ \frac{\left( 2\,k_F - \mathfrak{k}_- \right)^3}{3} - q^2 \left( 2\,k_F - \mathfrak{k}_- \right) + \frac{2\,q^3}{3} \right].
\end{align}
Only the part of the bracket that is odd in $\mathfrak{k}_-$ survives, and it equals $-\,\mathfrak{k}_- \left( 4\,k_F^2 - q^2 + \mathfrak{k}_-^2/3 \right)$. Using $\mathfrak{k}^2/(\varpi^2 - \mathfrak{k}^2) = -\,1 + \varpi^2/(\varpi^2 - \mathfrak{k}^2)$, $\mathfrak{k}^4/(\varpi^2 - \mathfrak{k}^2) = -\,\mathfrak{k}^2 - \varpi^2 + \varpi^4/(\varpi^2 - \mathfrak{k}^2)$, and $\mathrm{P} \int_{-q}^{q} d\mathfrak{k}/(\varpi^2 - \mathfrak{k}^2) = \mathcal{L} (q)/\varpi$, we obtain
\begin{align}
\label{eq:w-A}
\mathfrak{I}_1 = 2\,q \left( 4\,k_F^2 - q^2 \right) + \frac{2\,q^3}{9} + \frac{2\,q\,\varpi^2}{3} - \varpi \left( 4\,k_F^2 - q^2 + \frac{\varpi^2}{3} \right) \mathcal{L} (q)\,.
\end{align}

For $\mathfrak{I}_2$, we integrate over $\mathfrak{k}_-$ first. At fixed $\mathfrak{k}_+$, the range is $-q \leq \mathfrak{k}_- \leq \min \lbrace q, 2\,k_F - \mathfrak{k}_+ \rbrace$, and
\begin{align}
\mathfrak{I}_2 = \mathrm{P} \int_q^{2 \, k_F + q} d\mathfrak{k}_+ \frac{\mathfrak{k}_+}{\varpi^2 - \mathfrak{k}_+^2}\,\mathfrak{h} (\mathfrak{k}_+) \,, \quad
\mathfrak{h} (\mathfrak{k}_+) = \begin{cases}
4\,q^3/3 & \text{ for } \mathfrak{k}_+ < 2\,k_F - q \\[4pt]
\left( \mathfrak{k}_+ - 2\,k_F - q \right)^2 \left( \mathfrak{k}_+ + 2\,q - 2\,k_F \right)/3 & \text{ for } 2\,k_F - q < \mathfrak{k}_+ < 2\,k_F + q
\end{cases}.
\end{align}
The first range gives $-\,(2\,q^3/3)\,\ln \!\left( \left| \left( \varpi^2 - (2\,k_F - q)^2 \right)/\left( \varpi^2 - q^2 \right) \right| \right)$. In the second range, polynomial division gives $\mathfrak{k}\,\mathfrak{h} (\mathfrak{k}) = \left( \varpi^2 - \mathfrak{k}^2 \right) \mathcal{X} (\mathfrak{k}) + \mathfrak{p}_1\,\mathfrak{k} + \mathfrak{p}_0$ with
\begin{align}
\mathcal{X} (\mathfrak{k}) = -\,\frac{\mathfrak{k}^2 - 6\,k_F\,\mathfrak{k} + 12\,k_F^2 + \varpi^2 - 3\,q^2}{3}\,, \quad
\mathfrak{p}_1 = -\,\frac{2}{3} \left( 4\,k_F^3 + 3\,k_F\,\varpi^2 - 3\,k_F\,q^2 - q^3 \right), \quad
\mathfrak{p}_0 = \frac{\varpi^2 \left( 12\,k_F^2 + \varpi^2 - 3\,q^2 \right)}{3}\,.
\end{align}
The polynomial part integrates to $\int_{2 \, k_F - q}^{2 \, k_F + q} d\mathfrak{k}\,\mathcal{X} (\mathfrak{k}) = 2\,q \left( 8\,q^2 - 12\,k_F^2 - 3\,\varpi^2 \right)/9$, and the remainder to
\begin{align}
\int_{2 \, k_F - q}^{2 \, k_F + q} d\mathfrak{k}\,\frac{\mathfrak{p}_1\,\mathfrak{k} + \mathfrak{p}_0}{\varpi^2 - \mathfrak{k}^2} = \left[ -\,\frac{\mathfrak{p}_1}{2}\,\ln \!\left( \left| \varpi^2 - \mathfrak{k}^2 \right| \right) + \frac{\mathfrak{p}_0}{2\,\varpi}\,\mathcal{L} (\mathfrak{k}) \right]_{2 \, k_F - q}^{2 \, k_F + q}.
\end{align}
Collecting the terms, we obtain
\begin{align}
\label{eq:w-B}
\mathfrak{I}_2 &= -\,\frac{8\,k_F^2\,q}{3} + \frac{16\,q^3}{9} - \frac{2\,q\,\varpi^2}{3} + \frac{2\,q^3}{3}\,\ln \!\left( \left| \varpi^2 - q^2 \right| \right) \nn
& \quad + \frac{4\,k_F^3 + 3\,k_F\,\varpi^2 - 3\,k_F\,q^2 - q^3}{3}\,\ln \!\left( \left| \varpi^2 - \left( 2\,k_F + q \right)^2 \right| \right) - \frac{4\,k_F^3 + 3\,k_F\,\varpi^2 - 3\,k_F\,q^2 + q^3}{3}\,\ln \!\left( \left| \varpi^2 - \left( 2\,k_F - q \right)^2 \right| \right) \nn
& \quad + \frac{\varpi \left( 12\,k_F^2 + \varpi^2 - 3\,q^2 \right)}{6} \left[ \mathcal{L} \left( 2\,k_F + q \right) - \mathcal{L} \left( 2\,k_F - q \right) \right].
\end{align}
The sum $\mathfrak{I}_1 + \mathfrak{I}_2$ is the function $\mathcal{T}$ of Eq.~\eqref{eq:w-S}.

For $q > k_F$, we have $2\,k_F - q < q$ and the domain $\mathfrak{D}$ changes. In $\mathfrak{I}_1$, the $\mathfrak{k}_-$ integral runs from $-\,q$ to $2\,k_F - q$ with the same integrand, and the odd-part argument no longer applies. In $\mathfrak{I}_2$, the first range of $\mathfrak{h}$ is absent, and $\mathfrak{k}_+$ runs from $q$ to $2\,k_F + q$ with the second expression of $\mathfrak{h}$. Let $\mathfrak{I}_1^{<}$ and $\mathfrak{I}_2^{<}$ denote the analytic continuations in $q$ of Eqs.~\eqref{eq:w-A} and~\eqref{eq:w-B}. The actual integrals differ from them by integrals over $2\,k_F - q \leq \mathfrak{k} \leq q$,
\begin{align}
\mathfrak{I}_1 = \mathfrak{I}_1^{<} - \mathrm{P} \int_{2 \, k_F - q}^{q} d\mathfrak{k} \frac{\mathfrak{k}\,\mathfrak{A} (\mathfrak{k})}{\varpi^2 - \mathfrak{k}^2}\,, \quad
\mathfrak{I}_2 = \mathfrak{I}_2^{<} - \mathrm{P} \int_{2 \, k_F - q}^{q} d\mathfrak{k} \frac{\mathfrak{k} \left[ \mathfrak{h}_2 (\mathfrak{k}) - 4\,q^3/3 \right]}{\varpi^2 - \mathfrak{k}^2}\,,
\end{align}
where $\mathfrak{A} (\mathfrak{k}) = (2\,k_F - \mathfrak{k})^3/3 - q^2 \left( 2\,k_F - \mathfrak{k} \right) + 2\,q^3/3$ is the bracket in $\mathfrak{I}_1$ and $\mathfrak{h}_2$ is the second expression of $\mathfrak{h}$. With $u = 2\,k_F - \mathfrak{k}$, we have $\mathfrak{A} = (u - q)^2 (u + 2\,q)/3$ and $\mathfrak{h}_2 = (u + q)^2 (2\,q - u)/3$. Hence $\mathfrak{A} + \mathfrak{h}_2 = 4\,q^3/3$. The two integrands cancel in the sum, and $\mathfrak{I}_1 + \mathfrak{I}_2 = \mathfrak{I}_1^{<} + \mathfrak{I}_2^{<}$ is again the function $\mathcal{T}$ of Eq.~\eqref{eq:w-S}. This also holds for $q > 2\,k_F$, with the poles at $\mathfrak{k} = \pm\,\varpi$ treated as principal values.

\subsection{Static polarisability and screening}
\label{app:w-stat}

For $\varpi \to 0$, we have $\mathcal{L} ({\tilde x}) \to 2\,\varpi/ {\tilde x}$, and the terms of Eq.~\eqref{eq:w-S} proportional to $\varpi$ vanish. The remaining terms combine into
\begin{align}
\frac{3}{2}\,\mathcal{T} (q, 0) = 8\,k_F^2\,q + k_F \left( 4\,k_F^2 - 3\,q^2 \right) 
\ln \!\left( \left| \frac{2\,k_F + q}{2\,k_F - q} \right| \right) 
- q^3\,\ln \!\left( \left| \frac{4\,k_F^2 - q^2}{q^2} \right| \right),
\end{align}
which gives Eq.~\eqref{eq:w-statmu}. The vacuum part is Eq.~\eqref{eq:w-vac} at $\omega = 0$,
\begin{align}
\Pi_{\rm vac} (q, 0) = -\,\frac{g\,q^2}{24\,\pi^2\,v_0}
\ln \!\left( \frac{\Lambda^2}{q^2} \right),
\end{align}
and the terms $\pm\,q^2 \ln \!\left( q^2 \right)$ cancel in the sum, which gives Eq.~\eqref{eq:w-stat}. We write it as $\Pi (q, 0) = -\,g\,\mathcal{N}/(24\,\pi^2\,v_0)$ with
\begin{align}
\mathcal{N} = 8\,k_F^2 + \check{\mathcal{C}}\,\ln \!\left( \left| \frac{2\,k_F + q}{2\,k_F - q} \right| \right) + q^2\,\ln \!\left( \frac{\Lambda^2}{\left| 4\,k_F^2 - q^2 \right|} \right), \quad \check{\mathcal{C}} = \frac{k_F \left( 4\,k_F^2 - 3\,q^2 \right)}{q}\,.
\end{align}

For $q \ll k_F$, we use $\ln \!\left( (2\,k_F + q)/(2\,k_F - q) \right) = q/k_F + q^3/(12\,k_F^3) + q^5/(80\,k_F^5) + \cdots$ and $\ln \!\left( \Lambda^2/(4\,k_F^2 - q^2) \right) = 2\,\ln \!\left( \Lambda/(2\,k_F) \right) + q^2/(4\,k_F^2) + \cdots$. This gives $\mathcal{N} = 12\,k_F^2 + q^2 \left[ 2\,\ln \!\left( \Lambda/(2\,k_F) \right) - 8/3 \right] + q^4/(20\,k_F^2) + \cdots$, which is Eq.~\eqref{eq:w-statlow}. For $q \gg k_F$, we use $\ln \!\left( (q + 2\,k_F)/(q - 2\,k_F) \right) = 2 \left[ \tilde y + \tilde y^3/3 + \tilde y^5/5 + \cdots \right]$ with $\tilde y = 2\,k_F/q$ and $\ln \!\left( \Lambda^2/(q^2 - 4\,k_F^2) \right) = \ln \!\left( \Lambda^2/q^2 \right) + 4\,k_F^2/q^2 + 8\,k_F^4/q^4 + 64\,k_F^6/(3\,q^6) + \cdots$. The terms of order $k_F^2$ cancel, those of order $k_F^4/q^2$ add up to $8\,k_F^4/q^2$, and those of order $k_F^6/q^4$ add up to $64\,k_F^6/(15\,q^4)$, which is Eq.~\eqref{eq:w-stathigh}.

With $\epsilon^{\rm RPA} = 1 - 4\,\pi\,e^2\,\Pi/q^2$, Eq.~\eqref{eq:w-statlow} gives Eq.~\eqref{eq:w-tf}, where $4\,\pi\,e^2\,\rho (\mu) = 12\,\bar\alpha\,k_F^2$ follows from $\rho (\mu) = g\,k_F^2/(2\,\pi^2\,v_0)$.

\subsection{Kohn singularity and Friedel oscillations}
\label{app:w-friedel}

Near $q = 2\,k_F$, the only non-analytic function in $\mathcal{N}$ is $\ln \!\left( \left| q - 2\,k_F \right| \right)$, whose coefficient is $-\,(\check{\mathcal{C}} + q^2) = -\,(q^3 - 3\,k_F\,q^2 + 4\,k_F^3)/q = -\,(q - 2\,k_F)^2 (q + k_F)/q$. All remaining terms are analytic, which gives Eq.~\eqref{eq:w-sing}. Expanding $(q + k_F)/q = 3/2 - (q - 2\,k_F)/(4\,k_F) + \mathcal{O} \left( (q - 2\,k_F)^2 \right)$ gives Eq.~\eqref{eq:w-singexp}. The singularity is thus of the form $(q - 2\,k_F)^2 \ln \!\left( \left| q - 2\,k_F \right| \right)$, which is weaker than the $(q - 2\,k_F) \ln \!\left( \left| q - 2\,k_F \right| \right)$ of an ordinary electron gas.

To obtain the value at $q = 2\,k_F$, we write the logarithms of $\mathcal{N}$ as
\begin{align}
\check{\mathcal{C}} \left[ \ln \!\left( 2\,k_F + q \right) - \ln \!\left( \left| 2\,k_F - q \right| \right) \right] + q^2 \left[ \ln \!\left( \Lambda^2 \right) - \ln \!\left( 2\,k_F + q \right) - \ln \!\left( \left| 2\,k_F - q \right| \right) \right].
\end{align}
The total coefficient $-\,(\check{\mathcal{C}} + q^2)$ of $\ln \!\left( \left| 2\,k_F - q \right| \right)$ vanishes at $q = 2\,k_F$, where $\check{\mathcal{C}} = -\,4\,k_F^2$ and $q^2 = 4\,k_F^2$. The remaining terms give
\begin{align}
\mathcal{N} (2\,k_F) = 8\,k_F^2 + \left( \check{\mathcal{C}} - q^2 \right) 
\ln \!\left( 4\,k_F \right) + q^2 \ln \!\left( \Lambda^2 \right) 
= 8\,k_F^2 \left[ 1 + \ln \!\left( \frac{\Lambda}{4\,k_F} \right) \right],
\end{align}
which is Eq.~\eqref{eq:w-pi2kf}. With $\epsilon^{\rm RPA} = 1 - 4\,\pi\,e^2\,\Pi/q^2$ at $q = 2\,k_F$ and $g\,e^2/(3\,\pi\,v_0) = 2\,\bar\alpha$, this gives Eq.~\eqref{eq:w-eps2kf}.

The induced density is $\delta n (\bs{q}) = \Pi\,\mathcal{U}/(1 - 4\,\pi\,e^2\,\Pi/q^2)$, and differentiation with respect to $\Pi$ gives the singular part $\delta n_{\rm sing} = \Pi_{\rm sing}\,\mathcal{U}/\epsilon^2$ with the smooth factors evaluated at $q = 2\,k_F$. The coefficient of $\tilde x^2 \ln |\tilde x|$, with $\tilde x = q - 2\,k_F$, is $\mathfrak{j} = -\,[g/(16\,\pi^2\,v_0)]\,(\pi\,Z\,e^2/k_F^2)/\epsilon_{2 \, k_F}^2$. For an isotropic function, $\delta n (r) = (2\,\pi^2\,r)^{-1} \int_0^\infty dq\,q\,\sin \!\left( q\,r \right)\,\delta n (q)$. The radial integral of the term $\tilde x^2 \ln |\tilde x|$ is evaluated as in Appendix~\ref{app:hfull-friedel}, which gives $\int dq\,q\,\sin \!\left( q\,r \right)\,\tilde x^2 \ln |\tilde x| \simeq 4\,\pi\,k_F\,\sin \!\left( 2\,k_F\,r \right)/r^3$ with $\tilde x = q - 2\,k_F$. The remaining factors of $q$ and the term in $\tilde x^3 \ln |\tilde x|$ contribute at higher orders in $1/(k_F\,r)$. Hence
\begin{align}
\delta n (r) \simeq \frac{2\,k_F\,\mathfrak{j}}{\pi}\,\frac{\sin \!\left( 2\,k_F\,r \right)}{r^4}\,,
\end{align}
which is Eq.~\eqref{eq:w-friedel}.

\subsection{Plasmon dispersion and Landau damping}
\label{app:w-exp}

Expanding the logarithms of Eq.~\eqref{eq:w-S} in $q$ at fixed $\varpi$, we find for the doped part
\begin{align}
\label{eq:w-expmu}
\Pi_\mu (q, \omega) = \frac{g\,q^2}{24\,\pi^2\,v_0} \left[ \frac{4\,k_F^2}{\varpi^2} + \ln \!\left( \left| \frac{4\,k_F^2 - \varpi^2}{\varpi^2} \right| \right) \right] - \frac{2\,g\,q^4\,k_F^4 \left( 5\,\varpi^2 - 12\,k_F^2 \right)}{15\,\pi^2\,v_0\,\varpi^4 \left( 4\,k_F^2 - \varpi^2 \right)^2} + \mathcal{O} \left( q^6 \right).
\end{align}
For $q < \varpi$, the vacuum part of Eq.~\eqref{eq:w-vac} expands as
\begin{align}
\Pi_{\rm vac} (q, \omega) = -\,\frac{g\,q^2}{24\,\pi^2\,v_0} \left[ \ln \!\left( \frac{\Lambda^2}{\varpi^2} \right) + \frac{q^2}{\varpi^2} + \mathcal{O} \left( q^4 \right) \right].
\end{align}
The sum gives Eq.~\eqref{eq:w-exptot}. With $\epsilon^{\rm RPA} = 1 - \mathcal{V} (q)\,\Pi$ and $4\,\pi\,e^2\,g/(24\,\pi^2\,v_0) = \bar\alpha$, the coefficients of $q^0$ and $q^2$ are $\mathcal{E}_0$ and $\mathcal{E}_2$ of Eq.~\eqref{eq:w-epsexp}, after the replacements $\varpi \to \omega/v_0$ and $k_F \to \mu/v_0$.

The shift of the zero follows from $\mathcal{E}_0 (\omega_*) = 0$. For $\omega^2 = \omega_*^2 + \delta$, the condition $\Re \epsilon^{\rm RPA} = 0$ gives $\delta\,\partial \mathcal{E}_0/\partial \omega^2 + \left( v_0\,q \right)^2 \mathcal{E}_2 = 0$ to order $q^2$, which is Eq.~\eqref{eq:w-plasdisp}.


\bibliography{ref_pl}

\end{document}